\documentclass[11pt]{article}
\usepackage{verbatim,amsmath,amssymb}
\usepackage{epsfig,float,color}
\usepackage{geometry}
\usepackage{setspace}
\usepackage{natbib}
\usepackage{amsthm,mathrsfs,amsfonts,dsfont} 
\usepackage{hyperref}

\definecolor{darkblue}{rgb}{0,0,1}
\hypersetup{pdftex=true, colorlinks=true, breaklinks=true, linkcolor=darkblue, menucolor=darkblue, pagecolor=darkblue, citecolor=darkblue, urlcolor=darkblue}

\newcommand{\bitm}{\begin{itemize}}
\newcommand{\eitm}{\end{itemize}}
\newcommand{\bnumr}{\begin{enumerate}}
\newcommand{\enumr}{\end{enumerate}}

\newcommand {\sigab}{\sigma^{\alpha\beta}}

\newcommand {\aab}{a^{\alpha\beta}}

\newcommand {\agd}{a^{\gamma\delta}}

\newcommand {\Aab}{A^{\alpha\beta}}

\newcommand {\tauab}{\tau^{\alpha\beta}}

\newcommand{\mrT}{\mathrm{T}}

\newcommand {\eqb}[1]{\begin{equation}\begin{array}{#1}}
\newcommand {\eqe}{\end{array}\end{equation}}

\newcommand {\esb}[1]{\begin{equation*}\begin{array}{#1}}
\newcommand {\ese}{\end{array}\end{equation*}}
\newcommand {\ds}{\displaystyle}

\newcommand {\pa}[2]{\frac{\partial{#1}}{\partial{#2}}}

\newcommand {\back}{\! \! \!}
\newcommand {\is}{\back &=& \back}
\newcommand {\dis}{\back &:=& \back}

\newcommand {\plus}{\back &+& \back}
\newcommand {\mi}{\back &-& \back}

\newcommand {\norm}[1]{\|#1\|}

\newcommand {\dif}{\mathrm{d}}

\newcommand {\II}{{I\kern-.3em I}}
\newcommand {\III}{{I\kern-.3em I\kern-.3em I}}

\newcommand {\mrp}{\mathrm{p}}

\newcommand {\ma}{\mathbf{a}}

\newcommand {\mf}{\mathbf{f}}
\newcommand {\mg}{\mathbf{g}}

\newcommand {\mk}{\mathbf{k}}

\newcommand {\mq}{\mathbf{q}}

\newcommand {\muu}{\mathbf{u}}

\newcommand {\mx}{\mathbf{x}}

\newcommand {\ba}{\boldsymbol{a}}

\newcommand {\bff}{\boldsymbol{f}}

\newcommand {\bi}{\boldsymbol{i}}

\newcommand {\bn}{\boldsymbol{n}}

\newcommand {\bq}{\boldsymbol{q}}

\newcommand {\bt}{\boldsymbol{t}}
\newcommand {\bu}{\boldsymbol{u}}
\newcommand {\bv}{\boldsymbol{v}}

\newcommand {\bx}{\boldsymbol{x}}

\newcommand {\bnu}{\mbox{\boldmath$\nu$}}

\newcommand {\bvphi}{\mbox{\boldmath$\varphi$}}

\newcommand {\mA}{\mathbf{A}}

\newcommand {\mK}{\mathbf{K}}
\newcommand {\mL}{\mathbf{L}}

\newcommand {\mN}{\mathbf{N}}

\newcommand {\mP}{\mathbf{P}}

\newcommand {\mX}{\mathbf{X}}

\newcommand {\bA}{\boldsymbol{A}}
\newcommand {\bB}{\boldsymbol{B}}
\newcommand {\bC}{\boldsymbol{C}}

\newcommand {\bI}{\boldsymbol{I}}

\newcommand {\bN}{\boldsymbol{N}}

\newcommand {\bT}{\boldsymbol{T}}

\newcommand {\bX}{\boldsymbol{X}}

\newcommand {\sig}{\sigma}

\newcommand {\bsig}{\mbox{\boldmath$\sigma$}}

\newcommand {\bone}{\mathbf{1}}

\newcommand {\bbR}{\mathbb{R}}

\newcommand {\IR}{{\rm\kern.24em
   \vrule width.02em height1.53ex depth-.05ex
   \kern-.3em R}}
\newcommand {\ic}{{\rm\kern.20em
   \vrule width.02em height1.0ex depth-.05ex
   \kern-.22em c}}
\newcommand {\ia}{{\rm\kern.20em
   \vrule width.02em height1.05ex depth-.0ex
   \kern-.25em a}}
\newcommand {\IC}{{\rm\kern.24em
   \vrule width.02em height1.4ex depth-.05ex
   \kern-.26em C}}
\newcommand {\ID}{{\rm\kern.34em
   \vrule width.02em height1.5ex depth-.05ex
   \kern-.36em D}}
\newcommand {\IS}{{\rm\kern.24em
   \vrule width.02em height1.6ex depth.05ex
   \kern-.26em S}}
\newcommand {\IT}{{\rm\kern.50em
   \vrule width.02em height1.55ex depth-.05ex
   \kern-.52em T}}

\newcommand {\IE}{{\rm\kern.24em
   \vrule width.02em height1.55ex depth-.05ex
   \kern-.33em E}}
\newcommand {\IEa}{{\rm\kern.24em
   \vrule width.02em height1.55ex depth-.05ex
   \kern-.33em E}^{1}_{ijkl}}
\newcommand {\IEb}{{\rm\kern.24em
   \vrule width.02em height1.55ex depth-.05ex
   \kern-.33em E}^{2}_{ijkl}}

\newcommand {\sL}{\mathcal{L}}

\newcommand {\sQ}{\mathcal{Q}}

\newcommand {\sS}{\mathcal{S}}

\newcommand {\sV}{\mathcal{V}}

\newcommand {\vaub}{\ba_\beta}

\newcommand {\Ass}[2]{\kern 0.9ex \vrule width0.45em height0.2ex depth0ex \kern -2.1ex \bigwedge_{#1}^{#2}}
\newcommand {\ASS}[2]{\kern 1.45ex \vrule width0.5em height0.2ex depth0ex \kern -2.65ex \bigwedge_{#1}^{#2}}

\newcommand{\ed}{\mathrm{d}}			        		                 

\begin{document}

\begin{center}
\Large{\bf{A stabilized finite element formulation for liquid shells 
and its application to lipid bilayers}}\\

\end{center}

\begin{center}
\large{Roger A. Sauer$^\ast$\footnote{corresponding author, email: sauer@aices.rwth-aachen.de},
Thang X. Duong$^\ast$, Kranthi K. Mandadapu$^{\dag\S}$ and David J. Steigmann$^\ddag$}\\
\vspace{4mm}

\small{\textit{
$^\ast$Aachen Institute for Advanced Study in Computational Engineering Science (AICES), \\ 
RWTH Aachen University, Templergraben 55, 52056 Aachen, Germany \\[1.1mm]
$^\dag$Department of Chemical and Biomolecular Engineering, University of California at Berkeley, \\
110 Gilman Hall, Berkeley, CA 94720-1460, USA \\[1.1mm]
$^\S$Chemical Sciences Division, Lawrence Berkeley National Laboratory, CA 94720, USA \\[1.1mm]
$^\ddag$Department of Mechanical Engineering, University of California at Berkeley, \\
6141 Etcheverry Hall, Berkeley, CA 94720-1740, USA
}}



\end{center}

\vspace{3mm}


\rule{\linewidth}{.15mm}
{\bf Abstract}

This paper presents a new finite element (FE) formulation for liquid shells that is based on an explicit, 3D surface discretization using $C^1$-continuous finite elements constructed from NURBS interpolation. 
Both displacement-based and mixed FE formulations are proposed. 
The latter is needed for area-incompressible material behavior, where penalty-type regularizations can lead to misleading results.
In order to obtain quasi-static solutions, several numerical stabilization schemes are proposed based on either stiffness, viscosity or projection.
Several numerical examples are considered in order to illustrate the accuracy and the capabilities of the proposed formulation, and to compare the different stabilization schemes.
The presented formulation is capable of simulating non-trivial surface shapes associated with tube formation and protein-induced budding of lipid bilayers. 
In the latter case, the presented formulation yields non-axisymmetric solutions, which have not been observed in previous simulations. 
It is shown that those non-axisymmetric shapes are preferred over axisymmetric ones.


\vspace{-4mm}
\rule{\linewidth}{.15mm}

\tableofcontents

\hspace{5mm}

{\Large\bf List of important symbols}

\begin{tabbing}
$\bone$ \qquad~~ \= identity tensor in $\bbR^3$  \\
$\ba_\alpha$ \> co-variant tangent vectors of surface $\sS$ at point $\bx$; $\alpha=1,2$ \\
$\bA_\alpha$ \> co-variant tangent vectors of surface $\sS_0$  at point $\bX$; $\alpha=1,2$ \\
$\ba^\alpha$ \> contra-variant tangent vectors of surface $\sS$  at point $\bx$; $\alpha=1,2$ \\
$\bA^\alpha$  \> contra-variant tangent vectors of surface $\sS_0$  at point $\bX$; $\alpha=1,2$ \\
$\ba_{\alpha,\beta}$ \> parametric derivative of $\ba_\alpha$ w.r.t.~$\xi^\beta$ \\
$\ba_{\alpha;\beta}$ \> co-variant derivative of $\ba_\alpha$ w.r.t.~$\xi^\beta$ \\
$a_{\alpha\beta}$ \> co-variant metric tensor components of surface $\sS$  at point $\bx$ \\
$A_{\alpha\beta}$ \> co-variant metric tensor components of surface $\sS_0$  at point $\bX$ \\
$\ma$ \> class of stabilization methods based on artificial shear viscosity \\
$\mA$ \> class of stabilization methods based on artificial shear stiffness \\
$b_{\alpha\beta}$ \> co-variant curvature tensor components of surface $\sS$  at point $\bx$ \\
$B_{\alpha\beta}$ \> co-variant curvature tensor components of surface $\sS_0$ at point $\bX$ \\
$\bB$ \> left surface Cauchy-Green tensor \\
$c^{\alpha\beta\gamma\delta}$ \> contra-variant components of the material tangent \\
$\bC$ \> right surface Cauchy-Green tensor \\
$\gamma$ \> surface tension of $\sS$ \\
$\Gamma^\gamma_{\alpha\beta}$ \> Christoffel symbols of the second kind \\
$\dif a$ \> differential surface element on $\sS$ \\
$\dif A$ \> differential surface element on $\sS_0$ \\
$\delta...$ \> variation of ...\\
$e$ \> index numbering the finite elements; $e=1,\,...,\,n_\mathrm{el}$ \\
$\epsilon$ \> penalty parameter \\
$\mf^e$ \> finite element force vector of element $\Omega^e$ \\
$g$ \> expression for the area-incompressibility constraint \\
$G$ \> expression for the weak form \\
$G^e$ \> contribution to $G$ from finite element $\Omega^e$ \\
$\mg^e$ \> finite element `force vector' of element $\Omega^e$ due to constraint $g$ \\
$H$ \> mean curvature of $\sS$ at $\bx$ \\
$H_0$ \> spontaneous curvature prescribed at $\bx$ \\
$I$ \> index numbering the finite element nodes \\ 
$I_1$, $I_2$ \> first and second invariants of the surface Cauchy Green tensors \\
$\bi$ \> surface identity tensor on $\sS$ \\
$\bI$ \> surface identity tensor on $\sS_0$ \\ 
$J$ \> surface area change \\
$k$  \> bending modulus \\
$k^\ast$ \> Gaussian modulus \\
$K$ \> initial in-plane membrane bulk modulus \\
$K_\mathrm{eff}$ \> effective in-plane membrane bulk modulus \\
$\mk^e$ \> finite element tangent matrix associated with $\mf^e$ and $\mg^e$ \\
$\kappa$ \> Gaussian curvature of surface $\sS$ at $\bx$ \\
$\kappa_1$, $\kappa_2$ \> principal curvatures of surface $\sS$ at $\bx$ \\
$L_I$ \> pressure shape function of finite element node $I$ \\
$\lambda_1$, $\lambda_2$ \> principal surface stretches of $\sS$ at $\bx$ \\
$m_e$ \> number of pressure nodes of finite element $\Omega^e$ \\
$m_\nu$, $m_\tau$ \> bending moment components acting at $\bx\in\partial\sS$ \\
$\bar m_\nu$, $\bar m_\tau$ \> prescribed bending moment components \\
$M^{\alpha\beta}$ \> contra-variant bending moment components \\
$\mu$ \> initial in-plane membrane shear stiffness \\
$\mu_\mathrm{eff}$ \> effective in-plane membrane shear stiffness \\
$n_\mathrm{no}$ \> total number of finite element nodes used to discretize $\sS$ \\
$n_\mathrm{el}$ \> total number of finite elements used to discretize $\sS$ \\
$n_\mathrm{mo}$ \> total number of finite element nodes used to discretize pressure $q$ \\
$n_e$ \> number of displacement nodes of finite element $\Omega^e$ \\
$N^{\alpha\beta}$ \> total, contra-variant, in-plane membrane stress components \\
$N_I$ \> displacement shape function of finite element node $I$ \\
$\bn$ \> surface normal of $\sS$ at $\bx$ \\
$\bN$ \> surface normal of $\sS_0$ at $\bX$ \\
$\mN$ \> array of the shape functions for element $\Omega^e$ \\
$\nu$ \> in-plane membrane shear viscosity \\
$\bnu$ \> normal vector on $\partial\sS$ \\
$\xi^\alpha$ \> convective surface coordinates; $\alpha=1,2$ \\
$\mP$ \> class of stabilization methods based on normal projection; projection matrix \\
$q$ \> Lagrange multiplier associated with area-incompressibility \\
$\mq$ \> array of all Lagrange multipliers $q_I$ in the system; $I=1,\,...,\,n_\mathrm{mo}$ \\
$\mq_e$ \> array of all Lagrange multipliers $q_I$ for finite element $\Omega^e$; $I=1,\,...,\,m_e$ \\
$S^\alpha$ \> contra-variant, out-of-plane shear stress components \\
$\sS$ \> current configuration of the surface \\
$\sS_0$ \> initial configuration of the surface \\
$\bsig$ \> Cauchy stress tensor of the shell\\
$\sig^{\alpha\beta}$ \> stretch related, contra-variant, in-plane membrane stress components \\
$\bt$ \> effective traction acting on the boundary $\partial\sS$ normal to $\bnu$ \\
$\bar\bt$ \> prescribed boundary tractions on Neumann boundary $\partial_t\sS$ \\
$\bT$ \> traction acting on the boundary $\partial\sS$ normal to $\bnu$ \\
$\bT^\alpha$ \> traction acting on the boundary $\partial\sS$ normal to $\ba^\alpha$ \\
$\sV$, $\sQ$ \> admissible function spaces \\
$\bvphi$ \> deformation map of surface $\sS$ \\
$\bar\bvphi$ \> prescribed boundary deformations on boundary $\partial_x\sS$ \\
$w$ \> hyperelastic stored surface energy density per current surface area \\
$W$ \> hyperelastic stored surface energy density per reference surface area \\
$\bx$ \> current position of a surface point on $\sS$ \\
$\bX$ \> initial position of $\bx$ on the reference surface $\sS_0$ \\
$\bx_I$ \> position vector of finite element node $I$ lying on $\sS$ \\
$\bX_I$ \> initial position of finite element node $I$ on $\sS_0$ \\
$\mx$ \> array of all nodal positions $\bx_I$ of the discretized surface; $I=1,\,...,\,n_\mathrm{no}$ \\
$\mx_e$ \> array of all nodal positions $\bx_I$ for finite element $\Omega^e$; $I=1,\,...,\,n_e$ \\
$\mX_e$ \> array of all nodal positions $\bX_I$ for finite element $\Omega^e_0$; $I=1,\,...,\,n_e$ \\
$\Omega^e$ \> current configuration of finite element $e$ \\
$\Omega_0^e$ \> reference configuration of finite element $e$
\end{tabbing}

\section{Introduction}

Biological membranes form the boundaries of cells and cell-internal organelles such as the endoplasmic reticulum, the golgi complex, mitochondria and endosomes. 
Mechanically they are liquid shells that exhibit fluid-like behavior in-plane and solid-like behavior out-of-plane. 
They mainly consist of self-assembled lipid bilayers and proteins. At the macroscopic level, these membranes exist in different shapes 
such as invaginations, buds and cylindrical tubes \citep{zimmerberg06, shibata09, mcmahon05, shemesh14}. These shapes arise as a result of the lateral loading due to cytoskeletal filaments and protein-driven spontaneous curvature.
Cell membranes undergo many morphological and topological shape transitions to enable important biological processes such as endocytosis \citep{buser13, kukulski12, peter04}, cell motility \citep{keren11} and vesicle formation \citep{gruenberg04,budin09}. The shape transitions occur as a result of lateral loading on the membranes from cytoskeletal filaments, such as actin, from osmotic pressure gradients across the membrane, or from membrane-protein interactions. 
For example, in endocytosis -- a primary mode of transport of cargo between the exterior of the cell and its interior -- proteins bind to the flat membrane and induce 
invaginations or bud shapes, followed by a tube formation by actin-mediated pulling forces \citep{kukulski12,walani15}. 

Most of the computational studies regarding biological membranes are 
restricted to shapes resulting from axisymmetric conditions. 
However, many of the processes in cells are non-axisymmetric in nature. 
To the best of our knowledge, there exist only a few studies that allow for general, non-axisymmetric shapes. 
Therefore, it is important to advance computational methods that can yield solutions for general conditions. 


In the past, several computational models have been proposed for cell membranes. Depending on how the membrane is discretized, two categories can be distinguished: Models based on an explicit surface discretization, and models based on an implicit surface discretization. 
In the second category, the surface is captured by a phase field \citep{du07a} or level set function \citep{salac11} that is defined on the surrounding volume mesh.
In the first category, the surface is captured directly by a surface mesh. The approach is particularly suitable if only surface effects are studied, such that no surrounding volume mesh is needed. This is the approach taken here.
An example is to use Galerkin surface finite elements: 
The first corresponding 3D FE model for lipid bilayer membranes seems to be the formulation of \citet{feng06} and \citet{ma08}. Their FE formulation is based on so-called subdivision surfaces \citep{cirak01}, which provide $C^1$-continuous FE surface discretizations.
Such discretizations are 
advantageous,
since they do not 
require additional degrees of freedom as $C^0$-continuous FE formulations do. 
Still, $C^0$-continuous FEs have been considered to model red blood cell (RBC) membranes and their supporting protein skeleton \citep{dao03,peng10b}, phase changes of lipid bilayers \citep{elliott10}, and viscous cell membranes \citep{tasso13}. 
Subdivision finite elements have been used to study 
confined cells \citep{kahraman12}.
Lipid bilayers can also be modeled with so-called `solid shell' (i.e. classical volume) elements instead of surface shell elements \citep{kloeppel11}. Using solid elements, $C^0$-continuity is sufficient, but the formulation is generally less efficient.
For two-dimensional and axisymmetric problems also $C^1$-continuous B-Spline and Hermite finite elements have been used to study membrane surface flows \citep{arroyo09,rahimi12}, cell invaginations \citep{rim14}, and cell tethering and adhesion \citep{rangarajan15}. The latter work also discusses the generalization to three-dimensional B-spline FE.
For some problems it is also possible to use specific, Monge-patch FE discretizations \citep{rangamani13,rangamani14}. 

There are also several works that do not use finite element approaches. Examples are
numerical ODE integration \citep{agrawal09},
Monte Carlo methods \citep{ramakrishnan10}, 
molecular dynamics \citep{li12}, finite difference methods \citep{lau12,gu14} 
and mesh-free methods \citep{rosolen13}.
There are also non-Galerkin FE approaches that use triangulated 
surfaces, 
e.g.~see \citet{jaric95,jie98}. 

For quasi-static simulations of liquid membranes and shells, the formulation needs to be stabilized.
Therefore, various stabilization methods have been proposed considering artificial viscosity \citep{ma08,droplet}, artificial stiffness \citep{kahraman12} and normal offsets 
-- either as a projection of the solution (with intermediate mesh update steps) \citep{droplet}, or as a restriction of the surface variation \citep{rangarajan15}.
The instability problem is absent, if shear stiffness is present, e.g.~due to an underlying 
cytoskeleton, like in RBCs \citep{dao03,peng10b,kloeppel11}. 
However, this setting does not apply to (purely) liquid membranes and shells. 

Here, a novel FE formulation is presented for liquid shells that is based on an explicit, 3D surface discretization using NURBS \citep{cottrell}. 
The following aspects are new:
\begin{itemize}
\item A theoretical study of the effective bulk and shear stiffness of the Helfrich bending model, 
\item the use of 3D, $C^1$-continuous NURBS-based surface discretizations, considering purely displacement-based and
\item mixed, LBB-conforming\footnote{satisfying the Ladyzhenskaia-Babu\v{s}ka-Brezzi condition, see e.g.~\citet{babuska73,bathe}} finite elements, 
\item a comparison of both these FE formulations, illustrating the limitations of the former,
\item the formulation, application and comparison of various stabilization schemes 
for quasi-static simulations 
\item that allow to accurately compute the in-plane stress state of the liquid, like the surface tension,
\item the verification of the formulation by various analytical solutions, 
\item the possibility to capture complex, non-axisymmetric solutions, and
\item new insight into the budding of cells and vesicles.
\end{itemize}
To the best of our knowledge, non-axisymmetric shapes have not been simulated to the same detail as is presented here.

The remainder of this paper is organized as follows:
Sec.~\ref{s:theo} presents an overview of the underlying theory of thin liquid shells. 
The governing weak form is presented and the model properties are discussed.
Stabilization methods are then addressed in Sec.~\ref{s:stab}. 
The finite element formulation follows in Sec.~\ref{s:FE}.
The formulation is verified and illustrated by several 3D numerical examples in Sec.~\ref{s:ex}. 
In particular, tethering and budding are studied.
The paper then concludes with Sec.~\ref{s:concl}.

\section{Summary of thin liquid shell theory}\label{s:theo}

This section gives a brief overview of the governing equations of thin shell theory considering liquid material behavior.
A much more detailed discussion can be found in \citet{shelltheo} and in the references therein.

\subsection{Thin shell kinematics}\label{s:kine}


The surface of the shell, denoted by $\sS$, is described by the parameterization
\eqb{l}
\bx = \bx(\xi^\alpha)~,\quad\alpha=1,2,
\label{e:x}\eqe
where $\xi^1$ and $\xi^2$ are coordinates that can be associated with a flat 2D domain that is then mapped to $\sS$ by function \eqref{e:x}. The surface is considered deformable. The deformation of $\sS$ is assessed in relation to a reference configuration $\sS_0$, described by the parameterization $\bX = \bX(\xi^\alpha)$. $\sS$ is considered here to coincide with $\sS_0$ initially (e.g.~at time $t=0$), such that $\bx=\bX$ for $t=0$.
From these mappings, the tangent vectors $\bA_\alpha=\bX_{,\alpha}$ and $\ba_\alpha=\bx_{,\alpha}$ can be determined. Here $..._{,\alpha}:=\partial.../\partial\xi^\alpha$ denotes the parametric derivative. From the tangent vectors, the surface normals
\eqb{l}
\bN = \ds\frac{\bA_1\times\bA_2}{\norm{\bA_1\times\bA_2}}~,
\label{e:N}\eqe
\eqb{l}
\bn = \ds\frac{\ba_1\times\ba_2}{\norm{\ba_1\times\ba_2}}~,
\label{e:n}\eqe
and the metric tensor components
\eqb{l}
A_{\alpha\beta} = \bA_\alpha\cdot\bA_\beta~,
\eqe
\eqb{l}
a_{\alpha\beta} = \ba_\alpha\cdot\ba_\beta
\eqe
can be defined. From the inverses $[A^{\alpha\beta}]:=[A_{\alpha\beta}]^{-1}$ and $[a^{\alpha\beta}]:=[a_{\alpha\beta}]^{-1}$, the dual tangent vectors
\eqb{l}
\bA^\alpha = A^{\alpha\beta}\,\bA_\beta~,
\eqe
\eqb{l}
\ba^\alpha = a^{\alpha\beta}\,\ba_\beta
\eqe
(with summation implied on repeated indices) can be introduced, such that $\bA^\alpha\cdot\bA_\beta=\delta^\alpha_\beta$ and $\ba^\alpha\cdot\ba_\beta=\delta^\alpha_\beta$, where $\delta^\alpha_\beta$ is the Kronecker symbol. Further, we can define the surface identity tensors
\eqb{l}
\bI = \bA^\alpha\otimes\bA_\alpha~,
\eqe
\eqb{l}
\bi = \ba^\alpha\otimes\ba_\alpha~,
\eqe
such that the usual, 3D identity becomes 
\eqb{l}
\bone=\bI+\bN\otimes\bN=\bi+\bn\otimes\bn~.
\label{e:1}\eqe
From the second derivatives of $\bA_\alpha$ and $\ba_\alpha$ one can define the curvature tensor components
\eqb{l}
B_{\alpha\beta} = \bA_{\alpha,\beta}\cdot\bN~,
\eqe
\eqb{l}
b_{\alpha\beta} = \ba_{\alpha,\beta}\cdot\bn~,
\eqe
allowing us to compute the mean, Gaussian and principal curvatures
\eqb{l}
H = \frac{1}{2}b^{\alpha\beta}\,a_{\alpha\beta}~,\quad
\kappa = \ds\frac{\det[b_{\alpha\beta}]}{\det[a_{\alpha\beta}]}~,\quad
\kappa_{1/2} = H \pm\sqrt{H^2-\kappa}
\eqe
of surface $\sS$, and likewise for $\sS_0$. 
Here, $b^{\alpha\beta}=a^{\alpha\gamma}\,b_{\gamma\delta}\,a^{\beta\delta}$.

For thin shells, the deformation between $\sS_0$ and $\sS$ is fully characterized by $a_{\alpha\beta}$ and $b_{\alpha\beta}$, and their relation to $A_{\alpha\beta}$ and $B_{\alpha\beta}$. Two important quantities that characterize the in-plane part of the deformation are
\eqb{l}
I_1 = A^{\alpha\beta}\,a_{\alpha\beta}
\eqe
and
\eqb{l}
I_2 = \det{[A^{\alpha\beta}]}\cdot\det{[a_{\alpha\beta}]}~.
\eqe
They define the invariants of the surface Cauchy Green tensors $\bC=a_{\alpha\beta}\,\bA^\alpha\otimes\bA^\beta$ and $\bB=A^{\alpha\beta}\,\ba_\alpha\otimes\ba_\beta$.
The quantity $J=\sqrt{I_2}$ determines the change of surface area between $\sS_0$ and $\sS$.
\\
Given the parametric derivative of $\ba_\alpha$, the so-called co-variant derivative of $\ba_\alpha$ is given by
\eqb{l}
\ba_{\alpha;\beta} = (\bn\otimes\bn)\,\ba_{\alpha,\beta}~,
\eqe
i.e.~as the change of $\ba_{\alpha}$ along $\bn$.
Introducing the Christoffel symbol $\Gamma^\gamma_{\alpha\beta}:=\ba^\gamma\cdot\ba_{\alpha,\beta}$, we can also write
\eqb{l}
\ba_{\alpha;\beta} = \ba_{\alpha,\beta}-\Gamma^\gamma_{\alpha\beta}\,\ba_\gamma~.
\eqe
For general vectors, $\bv=v^\alpha\,\ba_\alpha + v\,\bn$, the parametric and co-variant derivative are considered to agree, i.e.~$\bv_{,\alpha}=\bv_{;\alpha}$. Therefore, $v_{;\alpha}=v_{,\alpha}$, $\bn_{;\alpha}=\bn_{,\alpha}$, $(v^\alpha\,\ba_\alpha)_{;\beta}=(v^\alpha\,\ba_\alpha)_{,\beta}$ and 
\eqb{l}
v^\alpha_{;\beta} = v^\alpha_{,\beta} + \Gamma^\alpha_{\beta\gamma}\,v^\gamma~,
\eqe
due to \eqref{e:1}. Likewise definitions follow for the reference surface.

The variation and linearization of the above quantities can be found in \citet{shelltheo}.

\subsection{Quasi-static equilibrium}

For negligible inertia, equilibrium of the thin shell is governed by the field equation \citep{steigmann99b}
\eqb{l}
\bT^{\alpha}_{;\alpha} + \bff = \mathbf{0}~,
\label{e:PDE}\eqe
where $\bff$ is a source term and 
\eqb{l}
\bT^\alpha = N^{\alpha\beta} \, \ba_{\beta} + S^{\alpha} \, \bn
\label{e:Ta}
\eqe
is the traction vector acting on the surface normal to $\ba^\alpha$. It can be related to the stress tensor
\eqb{l}
\bsig = N^{\alpha\beta}\,\ba_\alpha\otimes\ba_\beta + S^\alpha\,\ba_\alpha\otimes\bn
\label{e:bsig}\eqe
via Cauchy's formula $\bT^\alpha = \bsig^\mrT\ba^\alpha$, or equivalently $\bT=\bsig^\mrT\bnu$ with $\bnu = \nu_\alpha\ba^\alpha$ and $\bT = \nu_\alpha\bT^\alpha$.
In Eqs.~(\ref{e:Ta}) and (\ref{e:bsig}),
\eqb{lll}
N^{\alpha\beta} \is \sigma^{\alpha\beta} + b^\alpha_\gamma \, M^{\gamma\beta}~, \\[2mm]
S^\alpha \is - M^{\beta\alpha}_{~~;\beta}
\label{e:NS}\eqe
are the in-plane and shear stress components acting on the cross section.
The stress and bending moment components $\sigma^{\alpha\beta}$ and $M^{\alpha\beta}$ are given through constitution. 
Eqs.~\eqref{e:PDE} and \eqref{e:NS} are a consequence of momentum balance \citep{shelltheo}.
At the boundary of the surface, $\partial\sS$, the boundary conditions
\begin{equation}\begin{array}{llll}
\bx \is \bar\bvphi ~~ & $on$~\partial_x\sS~,\\[1mm]
\bt \is \bar\bt & $on$~\partial_t\sS~, \\[1mm]
m_\tau \is \bar m_\tau & $on$~\partial_m\sS
\end{array}\end{equation}
can be prescribed.
Here, $m_\tau$ is the bending moment component parallel to boundary $\partial\sS$.  For Kirchhoff-Love shells, bending moments perpendicular to boundary $\partial\sS$, denoted $m_\nu$, affect the boundary traction. Therefore the effective traction
\eqb{l}
\bt := \bT - (m_\nu\bn)'
\eqe
is introduced \citep{shelltheo}. We will consider $m_\nu=0$ in the following.

\subsection{Constitution}\label{s:consti}

The focus here is on the quasi-static behavior of lipid bilayers, which can be described in the framework of hyperelasticity. For thin shells the stored energy function (per reference surface area) takes the form $W=W(a_{\alpha\beta},b_{\alpha\beta})$, such that \citep{steigmann99b,shelltheo}
\eqb{lll}
\sigma^{\alpha\beta} \is \ds\frac{2}{J}\pa{W}{a_{\alpha\beta}}~, \\[4mm]
M^{\alpha\beta} \is \ds\frac{1}{J}\pa{W}{b_{\alpha\beta}}~.
\label{e:sig_def}
\eqe
From this we can then evaluate the in-plane stress $N^{\alpha\beta}$ and the shear $S^\alpha$ according to \eqref{e:NS}.
For convenience, we further define
$\tau^{\alpha\beta}:=J\sigma^{\alpha\beta}$,
$M^{\alpha\beta}_0:=JM^{\alpha\beta}$ and
$N^{\alpha\beta}_0:=JN^{\alpha\beta}$.
The bending behavior of lipid bilayers is commonly described by the bending model of \citet{helfrich73}
\eqb{l}
w = k\,(H-H_0)^2+k^\ast\kappa~,
\label{e:Helf}\eqe
which is an energy density per current surface area. Here $k$ is the bending modulus, $k^\ast$ is the Gaussian modulus and $H_0$ denotes the so-called spontaneous curvature caused by the presence of certain proteins.
Based on \eqref{e:Helf}, we consider the following two constitutive models:

\subsubsection{Area-compressible lipid bilayer}

Combining the Helfrich energy with an energy resulting from the surface area change, we write
\eqb{l}
W = J\,w + \ds\frac{K}{2}(J-1)^2~.
\label{e:W_c}\eqe
Here a simple quadratic term for the compressible part is considered, since the area change of lipid bilayers is very small before rupture occurs (typically $|J-1|<4\%$).
According to \eqref{e:sig_def} and \eqref{e:NS}, the stress and moment components then become
\eqb{lll}
\sigma^{\alpha\beta} \is \big(K\,(J-1) + k\,\Delta H^2-k^\ast\kappa\big)\,a^{\alpha\beta} - 2\,k\,\Delta H\,b^{\alpha\beta}~, \\[2mm]
M^{\alpha\beta} \is \big(k\,\Delta H+2\,k^\ast H\big)\,a^{\alpha\beta} - k^\ast b^{\alpha\beta}~, \\[2mm]
N^{\alpha\beta} \is \big( K(J-1) + k\,\Delta H^2\big)\,a^{\alpha\beta} - k\,\Delta H\,b^{\alpha\beta}~,
\label{e:sig_c}\eqe
where $\Delta H:=H-H_0$.

{\bf Remark}: Here, $k$ and $k^\ast$ are material constants that have the units \textit{strain energy per current surface area}.
In principle, also $Jk$ and $Jk^\ast$ could be regarded as material constants that now have the units \textit{strain energy per reference surface area}. 
This alternative would lead to different expressions for $\sigma^{\alpha\beta}$ and $N^{\alpha\beta}$.

\subsubsection{Area-incompressible lipid bilayer}

Since $K$ is usually very large for lipid-bilayers, one may as well consider the surface to be fully area-incompressible.
Using the Lagrange multiplier approach, we now have
\eqb{l}
W = J\,w + q\,g~,
\label{e:W_i}\eqe
where the incompressibility constraint 
\eqb{l}
g := J-1 = 0 
\label{e:g}\eqe
is enforced by the Lagrange multiplier $q$, which is an independent variable that needs to be accounted for. 
Physically, it corresponds to a surface tension.
The stress and moment components now become
\eqb{lll}
\sigma^{\alpha\beta} \is \big(q+k\,\Delta H^2-k^\ast\kappa\big)\,a^{\alpha\beta} - 2\,k\,\Delta H\,b^{\alpha\beta}~, \\[2mm]
M^{\alpha\beta} \is \big(k\,\Delta H+2\,k^\ast H\big)\,a^{\alpha\beta} - k^\ast b^{\alpha\beta}~, \\[2mm]
N^{\alpha\beta} \is \big( q + k\,\Delta H^2\big)\,a^{\alpha\beta} - k\,\Delta H\,b^{\alpha\beta}~,
\label{e:sig_i}\eqe
which is identical to \eqref{e:sig_c} for $q=Kg$.


As $K$ becomes larger and larger both models approach the same solution. So from a physical point of view it may not make a big difference which model is used. Computationally, model \eqref{e:W_c} is easier to handle but can become inaccurate for large $K$ as is shown in Sec.~\ref{s:ex}. In analytical approaches, often \eqref{e:W_i} is preferred as it usually simplifies the solution. Examples for the latter case are found in \citet{baesu04} and \citet{agrawal09}; the former case is considered in the original work of \citet{helfrich73}.

\subsubsection{Model properties}

In both models, the membrane part only provides bulk stiffness, but lacks shear stiffness. For quasi-static computations the model can thus become unstable and should be stabilized, as is discussed in Sec.~\ref{s:stab}. Interestingly, the bending part of the Helfrich model can contribute an in-plane shear stiffness, which is shown in the following.

We first introduce the surface tension $\gamma$ of the surface as the average trace of the stress tensor, giving
\eqb{l}
\gamma:=\frac{1}{2}\,\bsig:\bi = \frac{1}{2} N^\alpha_\alpha~.
\label{e:gamma}\eqe
For both \eqref{e:sig_c} and \eqref{e:sig_i} we find
\eqb{l}
\gamma = q - k\,H_0\,\Delta H~,
\label{e:gamma_i}\eqe
where $q=Kg$ in the former case. 
It can be seen that for $H_0\neq0$, the bending part contributes to the surface tension. The surface tension is therefore not given by the membrane part alone
\citep{rangamani14}.
For the compressible case, the effective bulk modulus can then be determined from
\eqb{l}
K_\mathrm{eff} := \ds\pa{\gamma}{J}~,
\eqe
i.e.~as the change of $\gamma$ w.r.t $J$. 
We find 
\eqb{l}
K_\mathrm{eff} = K + k\,H_0\,H/J~,
\label{e:Keff}\eqe
since $\partial H/\partial J = -H/J$.
Likewise we can define the effective shear modulus from
\eqb{l}
\mu_\mathrm{eff} := J\,a_{\alpha\gamma}\,\ds\frac{\partial N^{\alpha\beta}_\mathrm{dev}}{2\,\partial a_{\gamma\delta}}\,a_{\beta\delta}\,
,
\label{e:mueffdef}\eqe
i.e.~as the change of the deviatoric stress w.r.t~to the deviatoric deformation (characterized by $a_{\gamma\delta}/J$). The deviatoric in-plane stress is given by
\eqb{l}
N^{\alpha\beta}_\mathrm{dev} :=  N^{\alpha\beta} - \gamma\,a^{\alpha\beta}~.
\eqe
We find
\eqb{l}
N^{\alpha\beta}_\mathrm{dev} = k\,\Delta H\,\big(H\,a^{\alpha\beta}-b^{\alpha\beta}\big)
\eqe
for both \eqref{e:sig_c} and \eqref{e:sig_i}. 
Evaluating \eqref{e:mueffdef} 
thus gives
\eqb{l}
\mu_\mathrm{eff} = Jk\,\big(3H^2 - 2HH_0 - \kappa \big)/2~.
\label{e:mueff}\eqe
The model therefore provides stabilizing shear stiffness if $3H^2>2HH_0+\kappa$. 
Since this is not always the case 
(e.g. for flat surface regions), 
additional shear stabilization should be provided in general. This is discussed in Sec.~\ref{s:stab}.
The value of $\mu_\mathrm{eff}$ is discussed in detail in the examples of Sec.~\ref{s:ex}. It is shown that $\mu_\mathrm{eff}$ can sufficiently stabilize the problem such that no additional shear stabilization is needed. It is also shown that $\mu_\mathrm{eff}$ does not necessarily need to be positive to avoid instabilities. Geometric stiffening, arising in large deformations, can also stabilize the shell.

\subsection{Weak form}\label{s:wf}

The computational solution technique proposed here is based on the weak form governing the mechanics of the lipid bilayer.
Consider a kinematically admissible variation of the surface denoted by $\delta\bx\in\sV$. Such a variation of $\sS$ causes variations of $\ba_\alpha$, $\bn$, $a_{\alpha\beta}$ and $b_{\alpha\beta}$. Contracting field equation \eqref{e:PDE} with $\delta\bx$, integrating over $\sS$ and applying Stokes' theorem leads to the weak form \citep{shelltheo}
\begin{equation}
G = G_\mathrm{int} - G_\mathrm{ext} = 0 \quad\forall\,\delta\bx\in\sV~,
\label{e:wf}\end{equation}
with
\eqb{lll}
G_\mathrm{int} 
\is \ds\int_{\sS} \frac{1}{2}\,\delta a_{\alpha\beta} \, \sigma^{\alpha\beta} \, \dif a 
+ \int_{\sS} \delta b_{\alpha\beta} \, M^{\alpha\beta} \, \dif a~, \\[4mm]
G_\mathrm{ext} 
\is \ds\int_{\sS}\delta\bx\cdot\bff\,\dif a 
+ \int_{\partial_t\sS}\delta\bx\cdot\bt\,\dif s + \int_{\partial_m\sS}\delta\bn\cdot m_\tau\,\bnu\,\dif s~.
\label{e:Giie}\eqe
Denoting the in-plane and out-of-plane components of $\delta\bx$ by $w_\alpha$ and $w$, such that $\delta\bx := w_\alpha\,\ba^\alpha + w\,\bn$,
we find that
\eqb{l}
\delta a_{\alpha\beta} = w_{\alpha;\beta} + w_{\beta;\alpha} - 2w\,b_{\alpha\beta} ~.
\eqe
Thus, the first part of $G_\mathrm{int}$ can be split into in-plane and out-of-plane contributions as
\eqb{l}
\ds\int_{\sS} \frac{1}{2}\,\delta a_{\alpha\beta} \, \sigma^{\alpha\beta} \, \dif a = G_{\sigma}^\mathrm{in} + G_{\sigma}^\mathrm{out}~,
\label{e:split}\eqe
with
\eqb{l}
G_{\sigma}^\mathrm{in} = \ds\int_{\sS} w_{\alpha;\beta} \, \sigma^{\alpha\beta} \, \dif a
\label{e:Gsigi}\eqe
and
\eqb{l}
G_{\sigma}^\mathrm{out} = - \ds\int_{\sS} w \, b_{\alpha\beta}\,\sigma^{\alpha\beta} \, \dif a~.
\label{e:Gsigo}\eqe
In principle -- although not needed here -- the second part of $G_\mathrm{int}$ can also be split into in-plane and out-of-plane contributions \citep{shelltheo}.
In the area-incompressible case, we additionally have to satisfy the weak form of constraint \eqref{e:g},
\eqb{l}
G_g = \ds\int_{\mathrm{\sS_0}}\delta q\,g\,\dif A = 0\quad\forall\,\delta q\in\sQ~,
\label{e:G_g}\eqe
where $\sQ$ is an admissible space for the variation of Lagrange multiplier $q$.

\section{Liquid shell stabilization}\label{s:stab}

As noted above, the system is unstable for quasi-static computations. There are two principal ways to stabilize the system without modifying/affecting the original problem. They are discussed in the following two sections and then summarized in Sec.~\ref{s:stabsum}. 

\subsection{Adding stiffness}

Firstly, the system can be stabilized by adding a stabilization stress $\sig^{\alpha\beta}_\mathrm{sta}$ to 
$\sigma^{\alpha\beta}$ in order to provide additional stiffness. This stress can be defined from a (convex) shear energy or from numerical viscosity.
An elegant and accurate way to stabilize the system is to add the stabilization stress only to the in-plane contribution \eqref{e:Gsigi} while leaving the out-of-plane contribution \eqref{e:Gsigo} unchanged.
The advantage of this approach is that the out-of-plane part, responsible for the shape of the bilayer, is not affected by the stabilization, at least not in the continuum limit of the 
surface discretization. 
There are several different ways to define the stabilization stress, which we will group into two categories.
An overview of all the options is then summarized in Tab.~\ref{t:stab}.

\subsubsection{In-plane shear and bulk stabilization}\label{s:stab_orig}

The first category goes back to \citet{droplet}, who used it
to stabilize liquid membranes governed by constant surface tension. 
The stabilization stress for such membranes requires shear and bulk contributions. Those are given for example by the stabilization stress
\eqb{l}
\sigma_\mathrm{sta}^{\alpha\beta} = \mu/J\big(A^{\alpha\beta}-a^{\alpha\beta}\big)~,
\label{e:A}\eqe
based on numerical stiffness, and
\eqb{l}
\sigma_\mathrm{sta}^{\alpha\beta} = \mu/J\big(a_\mathrm{pre}^{\alpha\beta}-a^{\alpha\beta}\big)~,
\label{e:a}\eqe
based on numerical viscosity. Here $a_\mathrm{pre}^{\alpha\beta}$ denotes the value of $a^{\alpha\beta}$ at the preceding computational step. 
These stabilization stresses are then only included within 
Eq.~\eqref{e:Gsigi} and not in Eq.~\eqref{e:Gsigo},
and the resulting two stabilization schemes are denoted `A' (for \eqref{e:A}) and `a' (for \eqref{e:a}) following \citet{droplet}. 
This reference shows that scheme `a' is highly accurate and performs much better than scheme `A'. 
It is also shown that applying the stabilization stresses \eqref{e:A} and \eqref{e:a} only to the in-plane part is much more accurate than applying it throughout the system (i.e. in both 
Eqs.~\eqref{e:Gsigi} and \eqref{e:Gsigo}), which we denote as schemes `A-t' and `a-t'. 

\subsubsection{Sole in-plane shear stabilization}\label{s:stab_shear}

If the surface tension is not constant, as in the lipid bilayer models introduced above, only shear stabilization is required. Therefore, the following new category of stabilization schemes is defined. 
Consider a split of the surface 
Cauchy-Green tensor into dilatational and deviatoric parts, such that $\bC=J\,\widehat\bC$, where $\widehat\bC:=J^{-1}\bC$ describes only the deviatoric deformation (since $\det{\widehat\bC}=1$).   
The stored energy function of an elastic membrane can then be defined for example by
\eqb{l}
W = \ds\frac{K}{4}\big(J^2 - 1 - 2\ln J\big) + \ds\frac{\mu}{2}\big(\widehat I_1 - 2\big)~,
\label{e:Wsplit}\eqe
where $\widehat I_1=I_1/J$ is the first invariant of $\widehat\bC$. The first term in \eqref{e:Wsplit} captures purely dilatoric deformations, while the second part captures purely deviatoric deformations. The formulation is analogous to the 3D case described for example in \citet{wriggers-fee}.
Since the bilayer energy introduced in Sec.~\ref{s:consti} already contains a bulk part, we now only need to consider the contribution
\eqb{l}
W_\mathrm{sta} = \ds\frac{\mu}{2}\big(\widehat I_1 - 2\big)
\label{e:W_sta}
\eqe
to derive the new stabilization scheme. From (\ref{e:sig_def}) we then find the stabilization stress
\eqb{l}
\sigab_\mathrm{sta} =\ds\frac{\mu}{J^2}\Big(\Aab - \frac{I_1}{2}\,\aab\Big)~,
\eqe
or
\eqb{l}
\tauab_\mathrm{sta} =\ds\frac{\mu}{J}\Big(\Aab - \frac{I_1}{2}\,\aab\Big)~.
\label{e:As}\eqe
As before, this stress will only be applied to 
Eq.~\eqref{e:Gsigi} and not in Eq.~\eqref{e:Gsigo}, even though it has been derived from a potential and should theoretically apply to both terms. 
Following earlier nomenclature we denote this scheme by `A-s'. Replacing $A^{\alpha\beta}$ by $a_\mathrm{pre}^{\alpha\beta}$ in (\ref{e:As}) gives
\eqb{l}
\tauab_\mathrm{sta} =\ds\frac{\mu}{J^*}\Big(a^{\alpha\beta}_{\mathrm{pre}} - \frac{I_1^*}
{2}\,\aab\Big)~,
\label{e:as}\eqe
with 
$J^*:= \sqrt{\det a_{\alpha\beta}\big/\det a_{\alpha\beta}^{\mathrm{pre}}}$ and 
$I_1^*:= a^{\alpha\beta}_{\mathrm{pre}}\,a_{\alpha\beta}$, which is an alternative shear-stabilization scheme based on numerical viscosity. We denote it `a-s'.
If stresses \eqref{e:As} and \eqref{e:as} are applied throughout the system (i.e. to both 
\eqref{e:Gsigi} and \eqref{e:Gsigo}), we denote the corresponding schemes `A-st' and `a-st'. 

%
If the shell is (nearly) area-incompressible the two stabilization methods of Sec.~\ref{s:stab_orig} and \ref{s:stab_shear} can behave identical, as can be seen by example~\ref{s:strip}.

\subsubsection{Relation to physical viscosity}

Schemes \eqref{e:a} and \eqref{e:as} are related to physical viscosity. 
Considering (near) area-incompressibility ($J^\ast=J=1$), the viscous stress for a Newtonian fluid is given by
\eqb{l}
\sigma^{\alpha\beta}_\mathrm{visc} = -\nu\,\dot a^{\alpha\beta}
\eqe
\citep{aris,rangamani13,rangamani14}, where $\dot a^{\alpha\beta}=-a^{\alpha\gamma}\,\dot a_{\gamma\delta}\,a^{\delta\beta}$.
Considering the first order rate approximation
\eqb{l}
\dot a^{\alpha\beta} \approx \ds\frac{1}{\Delta t}\Big(a^{\alpha\beta}-a^{\alpha\beta}_\mathrm{pre}\Big)~,
\eqe
and a small time step ($I_1^\ast\approx2$), immediately leads to expressions \eqref{e:a} and \eqref{e:as} with
\eqb{l}
\nu=\mu\,\Delta t~.
\label{e:nu}\eqe


\subsection{
Normal projection}\label{s:stab_P}

The second principal way to stabilize the system consists of a simple projection of the formulation onto the solution space defined by the normal surface direction. We apply this step directly to the discretized formulation as was proposed by \citet{droplet}. According to this, for the given discrete system of linear equations for displacement increment $\Delta\muu$ given by $\mK\,\Delta\muu=-\mf$, the reduced system for increment $\Delta\muu_\mathrm{red}=\mP\,\Delta\muu$ is simply obtained as
\eqb{l}
\mK_\mathrm{red}\,\Delta\muu_\mathrm{red} = -\mf_\mathrm{red}~, 
\mK_\mathrm{red} := \mP\,\mK\,\mP^T~,~~
\mf_\mathrm{red} := \mP\,\mf~,
\label{e:P}\eqe
where  
\eqb{l}
\mP := \left[
         \begin{array}{cccc}
           \bn_1^T & \mathbf{0}^T & \cdots & \mathbf{0}^T \\[2mm]
           \mathbf{0}^T & \bn_2^T & \cdots & \mathbf{0}^T \\[1mm]
	     \vdots & \vdots & \ddots & \vdots \\[2mm]
           \mathbf{0}^T & \mathbf{0}^T & \cdots & \bn_{n_\mathrm{no}}^T
         \end{array}
       \right] 
\label{e:Pmat}\eqe
is a projection matrix defined by the nodal normal vectors $\bn_I$. Since this method can lead to distorted FE meshes, a mesh update can be performed by applying any of the stabilization techniques discussed above. If this is followed by a projection step at the same load level, a dependency on parameter $\mu$ is avoided. 

\begin{figure}[h]
\begin{center} \unitlength1cm
\begin{picture}(0,5.3)
\put(-8,-.4){\includegraphics[height=59mm]{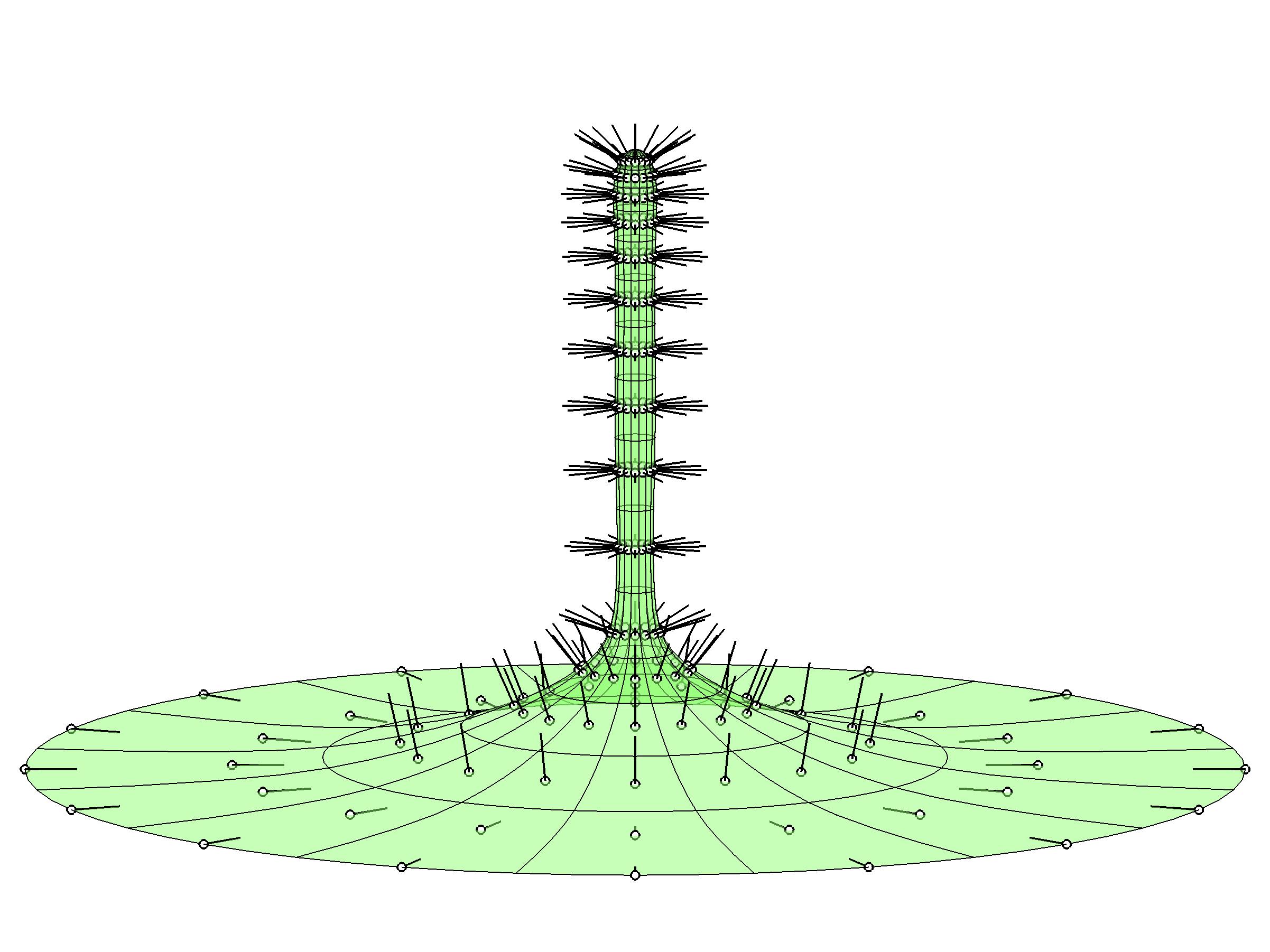}}
\put(-3,2.3){\includegraphics[height=24mm]{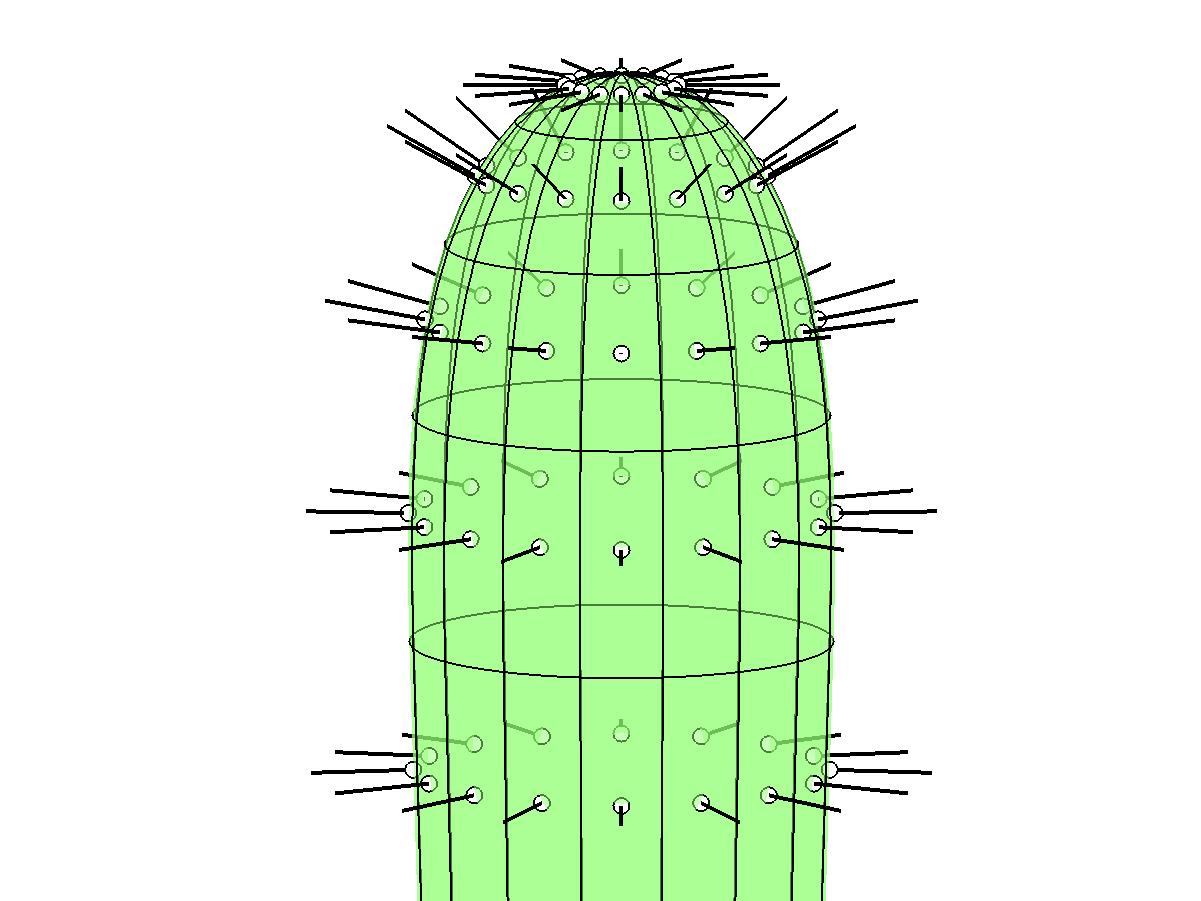}}
\put(0.2,-.4){\includegraphics[height=59mm]{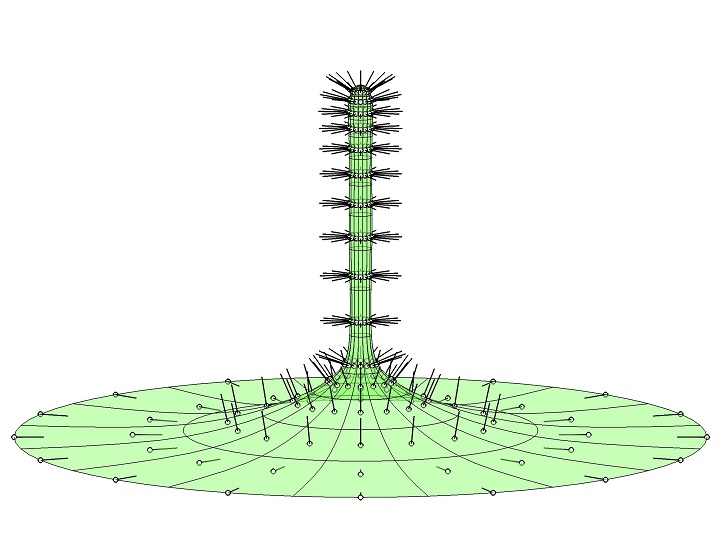}}
\put(5.2,2.3){\includegraphics[height=24mm]{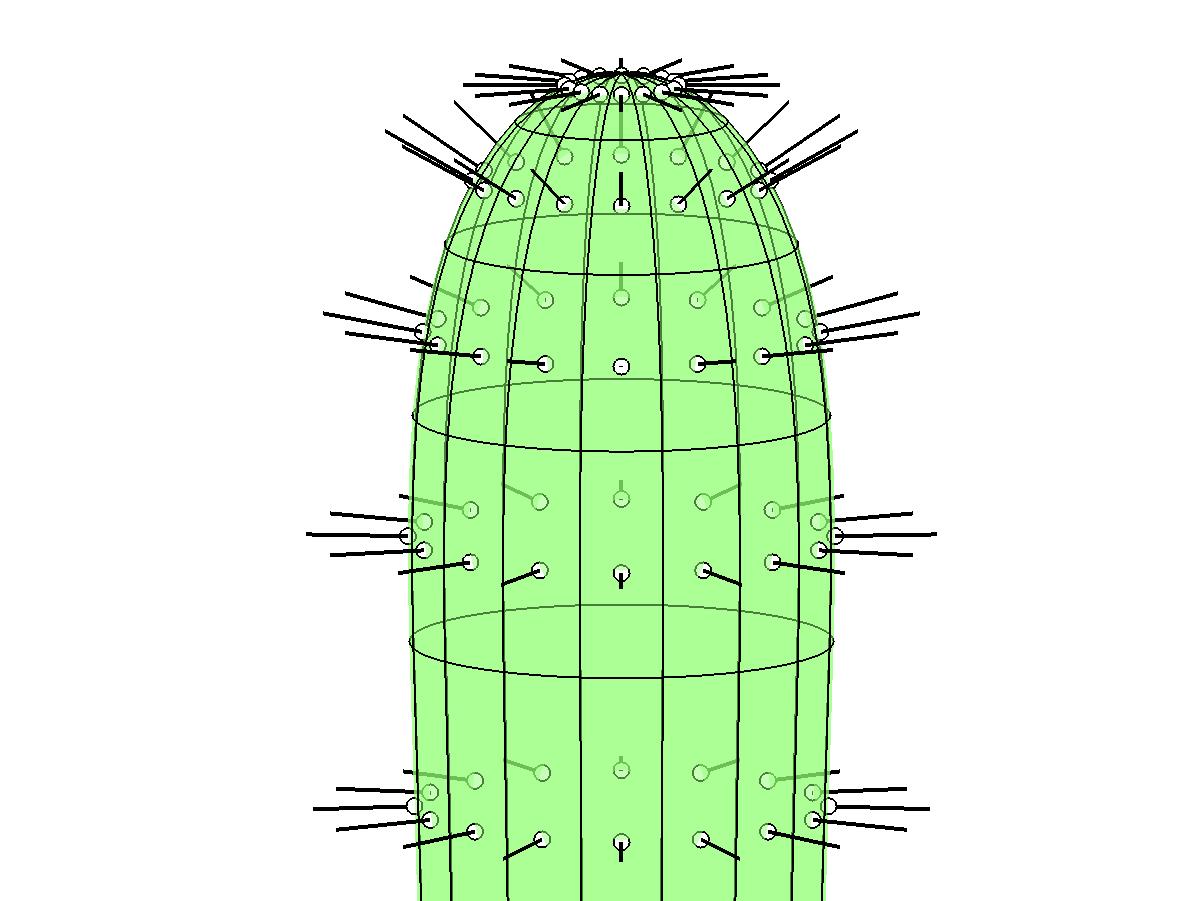}}
\put(-7.9,0){a.}
\put(.3,0){b.}
\end{picture}
\caption{Search direction $\bn_I$ for stabilization scheme `P': a.~Considering the normals at the projected control points; b.~Considering the normals at the current location of the initially projected control points.}
\label{f:normal}
\end{center}
\end{figure}

For NURBS discretizations, the computation of the normal $\bn_I$ corresponding to control point $I$ is not trivial, since the control points generally do not lie on the surface. The control points can be projected onto the actual surface in order to evaluate the normal at the projection point. As a simplification of this approach one can also work with the approximate normal obtained at the current location of the initial projection point. The difference between the two approaches is very small as Fig.~\ref{f:normal} shows.
There are minor differences in the locations of the projection points where $\bn_I$ is evaluated. But $\bn_I$ itself does not change much. In the example considered here and in \citet{droplet}, no siginificant difference was therefore found in the numerical results.
In order to solve the reduced problem \eqref{e:P} appropriate boundary conditions are needed. Boundary conditions can be applied before reduction or they can be suitably adapted to the reduced system. This is also shown in Fig.~\ref{f:normal}. In this example (see Sec.~\ref{s:tube}) the vertical displacement at the outer two and inner two rings of nodes is prescribed. After a mesh update at a given solution step no further vertical displacement can therefore occur on those nodes. But the nodes can still move radially. Therefore one can simply replace $\bn_I$ at those nodes by radial unit vectors.



\textbf{Remark}: For liquid droplets, stabilization scheme `P' shows superior accuracy over all other schemes \citep{droplet}. However this is not the case for area-incompressible surfaces:
For a curved surface, non-zero values of $\Delta\muu_\mathrm{red}$ lead to a change in surface area. In regions of non-zero curvature, $\Delta\muu_\mathrm{red}$ will therefore tend to be zero, yielding scheme `P' ineffective. This is confirmed by the results in Sec.~\ref{s:tube}.

\subsection{Summary of the stabilization schemes}\label{s:stabsum}

The nine stabilization schemes presented above are summarized in Tab.~\ref{t:stab}.
\begin{table}[h]
\centering
\begin{tabular}{|l|l|l|l|l|}
  \hline
  & & & & \\[-4mm]
  class & scheme & stab. stress $\sigma^{\alpha\beta}_\mathrm{sta}/\mu$ & application of $\sigma^{\alpha\beta}_\mathrm{sta}$ & dependence \\[0.5mm] \hline 
   & & & & \\[-4mm]
   \textbf{A} & A & $\big(A^{\alpha\beta}-a^{\alpha\beta}\big)/J$ & only in \eqref{e:Gsigi} & only on $\mu$ \\[1mm]
    & A-t & $\big(A^{\alpha\beta}-a^{\alpha\beta}\big)/J$ & both in \eqref{e:Gsigi} \& \eqref{e:Gsigo} & only on $\mu$ \\[1mm]
   & A-s & $\big(A^{\alpha\beta}-\frac{1}{2}I_1\,a^{\alpha\beta}\big)/J^2$ & only in \eqref{e:Gsigi} & only on $\mu$ \\ [1mm]
   & A-st & $\big(A^{\alpha\beta}-\frac{1}{2}I_1\,a^{\alpha\beta}\big)/J^2$ & both in \eqref{e:Gsigi} \& \eqref{e:Gsigo} & only on $\mu$ \\ [0.5mm] \hline
   & & & & \\[-4mm]   
   \textbf{a} & a   & $\big(a^{\alpha\beta}_\mathrm{pre}-a^{\alpha\beta}\big)/J$ & only in \eqref{e:Gsigi} & on $\mu$ and $n_t$ \\[1mm] 
   & a-t & $\big(a^{\alpha\beta}_\mathrm{pre}-a^{\alpha\beta}\big)/J$ & both in \eqref{e:Gsigi} \& \eqref{e:Gsigo} & on $\mu$ and $n_t$ \\[1mm]
   & a-s & $\big(a^{\alpha\beta}_\mathrm{pre}-\frac{1}{2}I_1^\ast\,a^{\alpha\beta}\big)/{J^\ast}^2$ & only in \eqref{e:Gsigi} & on $\mu$ and $n_t$ \\[1mm] 
   & a-st & $\big(a^{\alpha\beta}_\mathrm{pre}-\frac{1}{2}I_1^\ast\,a^{\alpha\beta}\big)/{J^\ast}^2$ & both in \eqref{e:Gsigi} \& \eqref{e:Gsigo} & on $\mu$ and $n_t$ \\[0.5mm] \hline
   & & & & \\[-4mm]   
   \textbf{P} & P & $0$ & -- & on nodal $\bn_I$ \\[0.5mm] \hline
\end{tabular}
\caption{Summary of the stabilization schemes presented in Sec.~\ref{s:stab_orig} and \ref{s:stab_shear}}
\label{t:stab}
\end{table}
They can be grouped into three classes: \textbf{A}, \textbf{a} and \textbf{P}.
The schemes of class \textbf{A} depend only on $\mu$ but require this value to be quite low. The schemes of class \textbf{a} also depend on the number of computational steps, $n_t$. If this number is high the schemes provide stiffness without adding much stress. The shell is then stabilized without modifying the solution much, even when $\mu$ is high. 
Scheme `P' depends on the nodal projection vector $\bn_I$, which is usually taken as the surface normal. 
The performance of the different stabilization schemes is investigated in the examples of Sec.~\ref{s:ex}.

\section{FE formulation}\label{s:FE}

This section presents a general finite element formulation 
for lipid bilayers using both material models of Sec.~\ref{s:consti}. The formulation applies to general 3D surfaces.
It is based on the solid shell formulation of \citet{solidshell}.

\subsection{FE approximation}

The geometry of the reference surface and the current surface are approximated by the 
finite element interpolations
\begin{equation}
\bX \approx \mN\,\mX_e
\label{e:Xapprox}\end{equation}
and
\begin{equation}
\bx \approx \mN\,\mx_e~,
\label{e:xapprox}\end{equation}
where $\mN := [N_1\bone,~...,~N_{n_e}\bone]$ is a $(3\times3n_e)$ array containing the $n_e$ nodal shape functions
$N_I=N_I(\xi^1,\xi^2)$ of element $\Omega^e$ defined in parameter space. $\mX_e := [\bX_1^\mrT,~....,~\bX^\mrT_{n_e}]^\mrT$ and $\mx_e := [\bx_1^\mrT,~....,~\bx^\mrT_{n_e}]^\mrT$ contain the $n_e$ nodal position vectors of $\Omega^e$.
In order to ensure $C^1$-continuity of the surface, NURBS-based shape functions are used \citep{borden11}.
The tangent vectors of the surface are thus approximated by
\begin{equation}
\ds\bA_{\alpha}=\pa{\bX}{\xi^\alpha} \approx 
\mN_{,\alpha}\,\mX_e
\label{e:aapprox}\end{equation}
and
\begin{equation}
\ds\ba_{\alpha}=\pa{\bx}{\xi^\alpha} \approx 
\mN_{,\alpha}\,\mx_e~,
\label{e:aapprox}\end{equation}
From these, the normal vectors of the surface are then determined from \eqref{e:N} and \eqref{e:n}.
The variation of $\bx$ and $\ba_\alpha$ are approximated in the same manner, i.e.
\begin{equation}
\ds\delta\bx \approx 
\mN\,\delta\mX_e
\label{e:dx_approx}\end{equation}
and
\begin{equation}
\ds\delta\ba_\alpha \approx 
\mN_{,\alpha}\,\delta\mx_e~.
\label{e:da_approx}\end{equation}
Based on these expressions, all the kinematical quantities discussed in Sec.~\ref{s:kine} as well as their variation \citep{shelltheo} can be evaluated.

\subsection{Discretized weak form}\label{s:dwf}

Inserting the above interpolations into the weak form of Sec.~\ref{s:wf} leads to the approximated weak form
\eqb{l}
G \approx 
\ds\sum_{e=1}^{n_\mathrm{el}}G^e
= \ds\sum_{e=1}^{n_\mathrm{el}}(G^e_\mathrm{int}-G^e_\mathrm{ext})~.
\eqe
The internal virtual work of each finite element is given by \citep{solidshell}
\eqb{l}
G^e_\mathrm{int} = \delta\mx_e^\mrT\,\big(\mf^e_\mathrm{int\sig} + \mf^e_{\mathrm{int}M}\big)~,
\label{e:Pinth}
\eqe
with the FE force vectors due to the membrane stress $\sig^{\alpha\beta}$ and the bending moment $M^{\alpha\beta}$
\eqb{l}
\mf^e_\mathrm{int\sig} 
:=  \ds\int_{\Omega^e} \sig^{\alpha\beta} \,\mN_{,\alpha}^\mrT\,\vaub\,\dif a~,
\label{e:fint}
\eqe
and
\eqb{l}
\mf^e_{\mathrm{int}M} 
:=  \ds\int_{\Omega^e} \, M^{\alpha\beta}\, \big( \mN_{,\alpha\beta}^\mrT - \Gamma^\gamma_{\alpha\beta}\,\mN_{,\gamma}^\mrT\big)\,\bn\, \dif a~.
\label{e:Mint}
\eqe
Following decomposition \eqref{e:split}, $\mf^e_{\mathrm{int}\sig}$ can be split into the in-plane and out-of-plane contributions \citep{membrane}
\eqb{lll}
\mf^e_\mathrm{inti} \dis  \mf^e_\mathrm{int\sig} - \mf^e_\mathrm{into}~, \\[2mm]
\mf^e_\mathrm{into} \dis -\ds\int_{\Omega^e} \sig^{\alpha\beta}\,b_{\alpha\beta} \,\mN^\mrT\,\bn\,\dif a~.
\label{e:fintio}
\eqe
The external virtual work for each FE due to the external loads $\bff$, $\bt$ and $m_\tau$ is given by \citep{solidshell}
\eqb{l}
G^e_\mathrm{ext} = \delta\mx_e^\mrT\,\big(\mf^e_{\mathrm{ext}f} 
+\mf^e_{\mathrm{ext}t}+\mf^e_{\mathrm{ext}m}\big)~,
\label{e:Pexth}
\eqe
with
\eqb{lll}
\mf^e_{\mathrm{ext}f} \dis \ds\int_{\Omega^e}\mN^\mrT\,\bff\,\dif a~, \\[4mm]
\mf^e_{\mathrm{ext}t} \dis \ds\int_{\partial_t\Omega^e}\mN^\mrT\,\bt\,\dif s~, \\[4mm]
\mf^e_{\mathrm{ext}m} \dis \ds\int_{\partial_m\Omega^e}\mN_{,\alpha}^\mrT\,\nu^\alpha\,m_\tau\,\bn\,\dif s~.
\label{e:fext}\eqe
For the complete linearization of these terms, see \citet{membrane}, \citet{solidshell} and Appendix~\ref{s:mfe}. 
The linearization requires the material tangents of $\sig^{\alpha\beta}$ and $M^{\alpha\beta}$. 
For the material models in \eqref{e:sig_c} and \eqref{e:sig_i} these are given in \citet{shelltheo}. 
For the stabilization schemes in Sec.~\ref{s:stab_orig} the tangent is given in \citet{droplet}, while the tangent for schemes `A-s' and `A-st' (see \eqref{e:As}) in Sec.~\ref{s:stab_shear} is
\eqb{lll}
c^{\alpha\beta\gamma\delta} := 2\ds\pa{\tau^{\alpha\beta}}{a_{\gamma\delta}}
= \ds\frac{\mu}{J}\left(\frac{I_1}{2}\,\aab\,\agd -I_1\,a^{\alpha\beta\gamma\delta} - \aab\, A^{\gamma\delta} - \Aab\,\agd \right)~.
\eqe
For schemes `a-s' and `a-st', the terms $I_1$ and $J$ are simply substituted by $I_1^*$ and $J^*$, respectively. Here, $a^{\alpha\beta\gamma\delta}$ is given in \citet{shelltheo}.

\subsection{Area constraint} \label{s:areac}

For the area-incompressible case, a further step is to discretize the constraint and the corresponding Lagrange multiplier. For the latter, we write
\eqb{l}
q\approx 
\mL\,\mq_e~,
\eqe
as in Eq.~\eqref{e:xapprox}. Here $\mL := [L_1,~...,~L_{m_e}]$ is a $(1\times m_e)$ array containing the $m_e$ nodal shape functions $L_I=L_I(\xi^1,\xi^2)$ of surface element $\Omega^e$, and $\mq_e := [q_1,~....,~q_{m_e}]^\mrT$ contains the $m_e$ nodal Lagrange multipliers of the element.
It follows that
\eqb{l}
\delta q\approx 
\mL\,\delta\mq_e~,
\eqe
such that weak constraint \eqref{e:G_g} becomes
\eqb{l}
G_g \approx 
\ds\sum_{e=1}^{n_\mathrm{el}}G_g^e~,
\eqe
where
\eqb{l}
G^e_g = \delta\bq_e^\mrT\,\mg^e~,
\eqe
with
\eqb{l}
\mg^e := \ds\int_{\Omega_0^e}\mL^\mrT\,g\,\dif A~.
\eqe
The linearization of $\mg^e$, needed for the following solution procedure, is provided in Appendix \ref{s:mge}.

\subsection{Solution procedure}

The elemental vectors $\mf^e_\mathrm{int\sig}$, $\mf^e_{\mathrm{int}M}$, $\mf^e_{\mathrm{ext}f}$, $\mf^e_{\mathrm{ext}t}$, $\mf^e_{\mathrm{ext}m}$ and $\mg^e$ are assembled into the global vectors $\mf$ and $\mg$ by adding corresponding entries. The discretized weak form then reads
\eqb{l}
\delta\mx^\mrT\,\mf(\mx,\mq) + \delta\mq^\mrT\,\mg(\mx) = 0~,
\quad\forall\,\delta\mx\in\sV^h~\&~\delta\mq\in\sQ^h~,
\label{e:wf_xq}\eqe
where $\mx$, $\mq$, $\delta\mx$ and $\delta\mq$ are global vectors containing all nodal deformations, Lagrange multipliers and their variations.
$\sV^h$ and $\sQ^h$ are corresponding discrete spaces.
Eq.~\eqref{e:wf_xq} is satisfied if $\mf=\mathbf{0}$ and $\mg=\mathbf{0}$ at nodes where no Dirichlet BC apply. 
These two nonlinear equations are then solved with Newton's method for the unknowns $\mx$ and $\mq$.
We note that the discretization of $\bx$ and $q$ should satisfy the LBB-condition \citep{babuska73,bathe}.
For that, we consider here $C^1$-continuous, bi-quadratic NURBS interpolation for $\bx$ and $C^0$-continuous, bi-linear Lagrange interpolation for $q$.
If no constraint is present (like in model \eqref{e:W_c}), the parts containing $\mq$ and $\mg$ are simply skipped.
A comparison between the different models (\eqref{e:W_c} and \eqref{e:W_i}) is presented in Sec.~\ref{s:ex}.

\subsection{Rotational constraints}

To constrain rotations, we add the constraint potential
\eqb{l}
\Pi_\mathrm{n} =  \ds \int_{\sL_0} \frac{\epsilon}{2}\, (\bn - \bar\bn)\cdot (\bn - \bar\bn)\, \ed S
\label{e:Pin}
\eqe
to the shell formulation.
This approach can be used to apply rotations at boundaries, to fix rotations at symmetry boundaries, and to equalize normals at patch boundaries. The particular boundary under consideration is denoted as $\sL_0$ in the reference configuration. $\epsilon$ is a penalty parameter.
The variation, linearization and FE discretization of (\ref{e:Pin}) is discussed in \citet{solidshell}. 

\subsection{Normalization}\label{s:norm}

For a numerical implementation, the above expressions need to be normalized.
We normalize the geometry and deformation by some length scale $L$, i.e.~$\bar\bX=\bX/L$ and $\bar\bx=\bx/L$, where the bar indicates non-dimensional quantities, and non-dimensionalize all the kinematics based on this.
The material parameters 
are chosen to be normalized by parameter $k$, which has the unit [force $\times$ length]. 
The non-dimensional material parameters thus are 
\eqb{lll}
\bar k \is 1~, \\[1mm]
\bar k^\ast \is k^\ast/k~, \\[1mm]
\bar K \is K\,L^2/k~, \\[1mm]
\bar\mu \is \mu\,L^2/k~, \\[1mm]
\bar\epsilon \is \epsilon\,L/k~.
\eqe
With the chosen normalization parameters $k$ and $L$, the normalization of stress and moment components become\footnote{supposing that the parameters $\xi^\alpha$ carry units of length, so that $a_{\alpha\beta}$ and $a^{\alpha\beta}$ become dimensionless}
\eqb{lll}
\bar q \is q\,L^2/k ~,\\[1mm]
\bar\sigma^{\alpha\beta} \is \sigma^{\alpha\beta}\,L^2/k ~,\\[1mm] 
\bar M^{\alpha\beta} \is M^{\alpha\beta}\,L/k~,
\eqe
while the normalization of the loading follows as
\eqb{lll}
\bar\bff \is \bff\,L^3/k~,\\[1mm]
\bar\bt \is \bt\,L^2/k~,\\[1mm]
\bar m_\tau \is m_\tau\,L/k~.
\eqe

\section{Numerical examples}\label{s:ex}

To illustrate the performance of the proposed finite element model, four examples -- marked by increasing computational complexity -- are presented here.
The first three examples have analytical solutions that are used for model verification.

\subsection{Pure bending and stretching of a flat strip}\label{s:strip}

\begin{figure}[h]
\begin{center} \unitlength1cm
\begin{picture}(0,5.8)
\put(-8.2,-.2){\includegraphics[height=60mm]{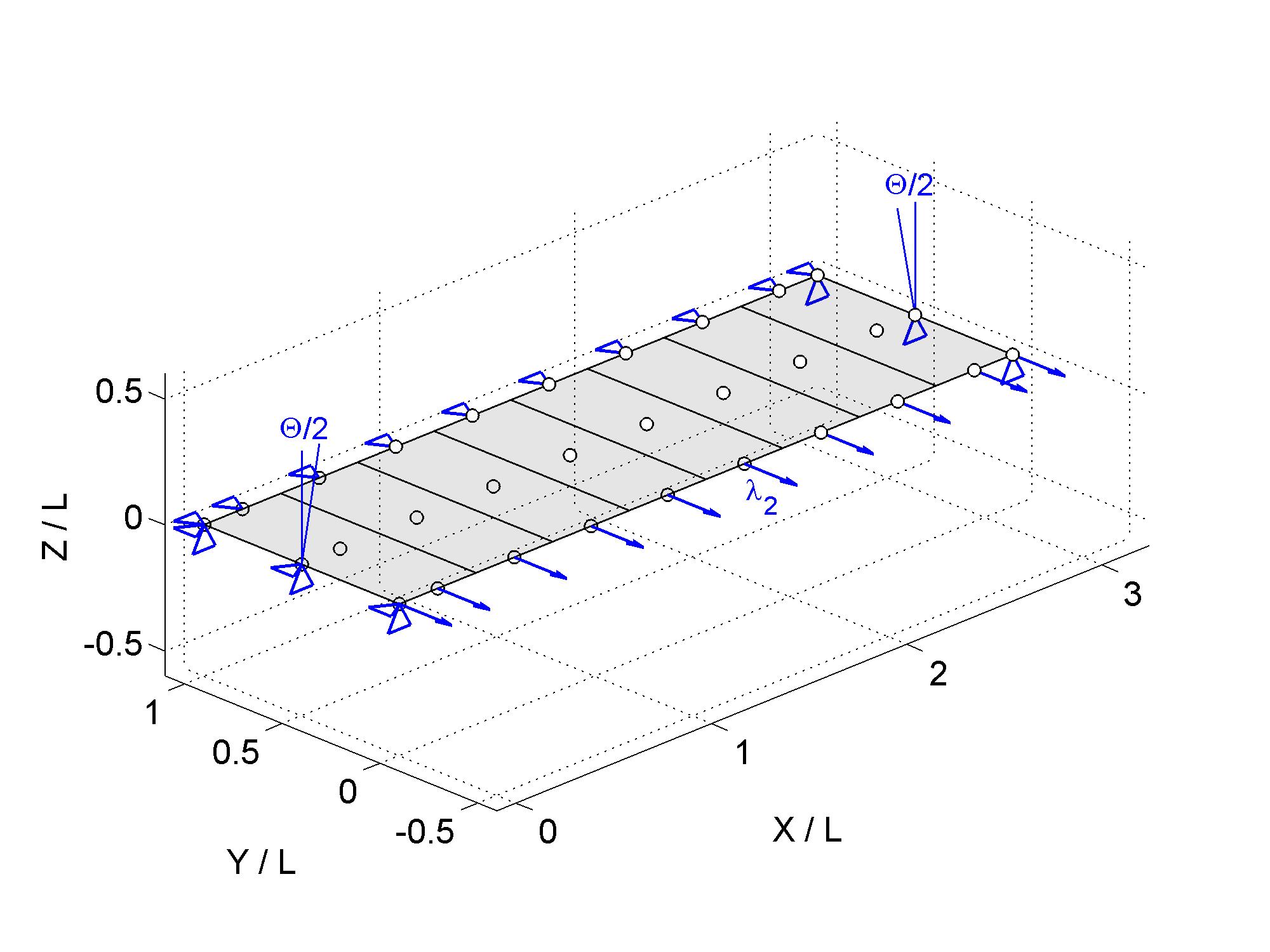}}
\put(-0.8,-.2){\includegraphics[height=60mm]{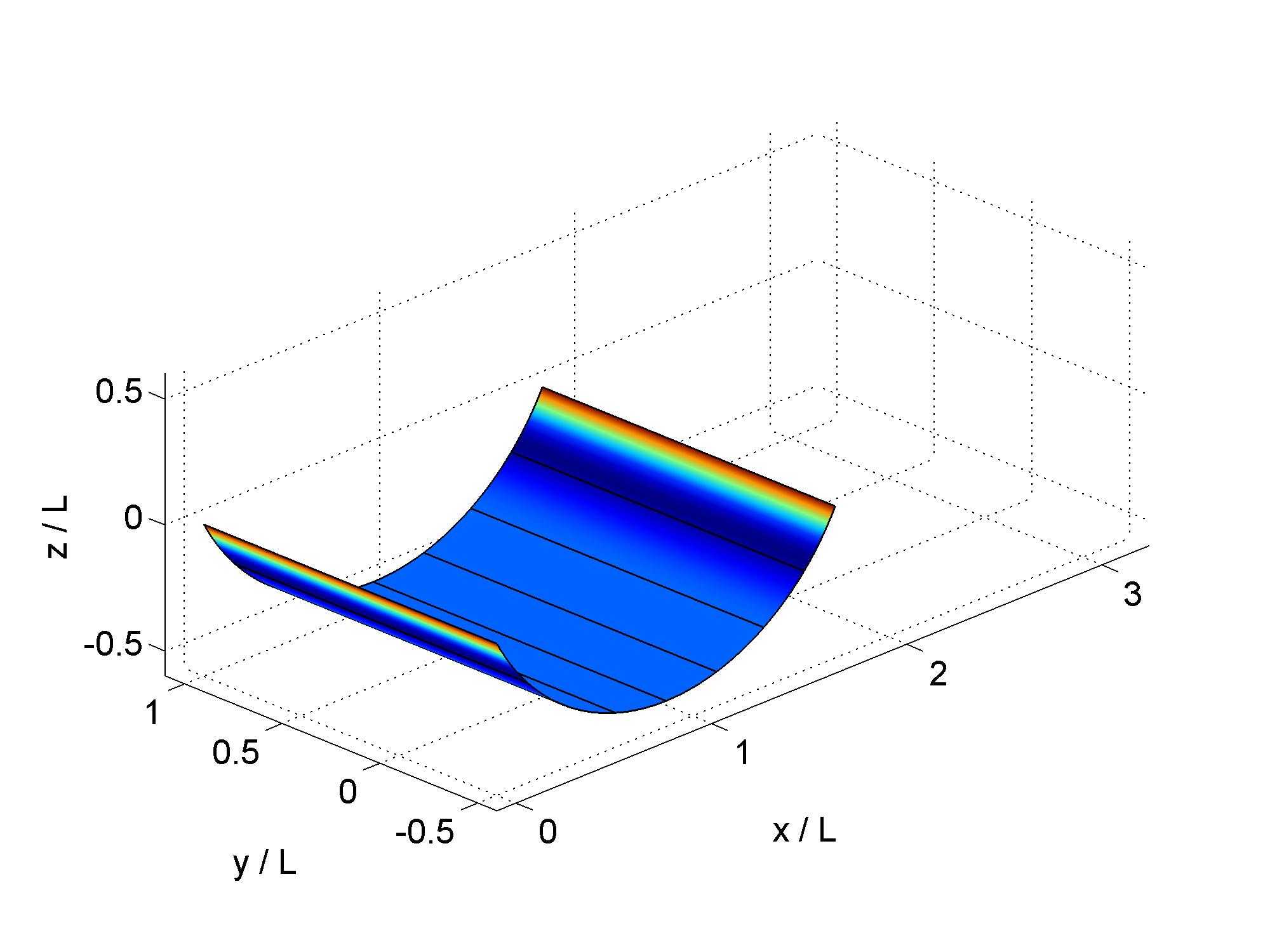}}
\put(7,-.2){\includegraphics[height=55mm]{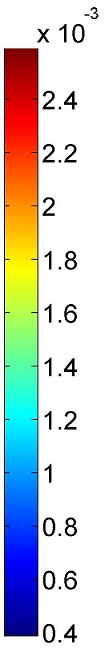}}
\put(-7.8,0){a.}
\put(-.3,0){b.}
\end{picture}
\caption{Pure bending: a.~initial FE configuration and boundary conditions (for $S=\pi L$ discretized with $m=8$ elements); b.~current FE configuration for an imposed rotation and stretch of $\Theta/2=\pi/3$ and $\lambda_2=1.5$. The color shows the relative error in the mean curvature, defined as $e_H := 1 - H_\mathrm{FE}/H_\mathrm{exact}$, considering model \eqref{e:W_i} without any stabilization.}
\label{f:bend_0}
\end{center}
\end{figure}

The first example considers the bending of a flat strip, with dimension $S\times L$, by applying the rotation $\Theta/2$ at the ends of the strip as is shown in
Fig.~\ref{f:bend_0}a. 
Further, a uniform stretch, with magnitude $\lambda_2$, is applied as shown in the figure.
The remaining boundaries are supported as shown. In particular, the rotation along the boundaries at $Y=0$ and $Y=L$ is not constrained. 
We will study the problem by examining the support reactions $M=M(\Theta,\lambda_2)$ (the distributed moment along $X=0$) and $N=N(\Theta,\lambda_2)$ (the traction along $Y=0$). 
The analytical solution for this problem is given by \citet{shelltheo}.
According to this, the strip deforms into a curved sheet with dimension 
$s\times\ell=\lambda_1S\times\lambda_2L$ and constant mean curvature 
\eqb{l}
H = \ds\frac{\kappa_1}{2}~,
\eqe
where $\lambda_1=s/S$, $\lambda_2=\ell/L$ and $\kappa_1=\Theta/s$ are the in-plane stretches and the out-of-plane curvature of the strip.
Further,
\eqb{l}
[a^{\alpha\beta}] = \left[\begin{array}{cc}
\lambda_1^{-2} & 0 \\ 
0 & \lambda^{-2}_2
\end{array} \right]
\eqe
and
\eqb{l}
[b^{\alpha\beta}] = \left[\begin{array}{cc}
\kappa_1\lambda_1^{-2} & 0 \\ 
0 & 0
\end{array} \right].
\eqe
With this, the in-plane stress components become
\eqb{llll}
N^1_1 \is q - k\,H^2~,\\[1mm]
N^2_2 \is q + k\,H^2 
\label{e:N_pb}
\eqe
both for the area-incompressible model of \eqref{e:W_i} and the compressible model of \eqref{e:W_c} with $q=K(J-1)$.
For the considered boundary conditions, $N^1_1=0$, so that
\eqb{llll}
q = k\,H^2,
\label{e:q_pb}\eqe
and we thus have the support reaction (per current length) $N:=N_2^2 = 2k\,H^2$ along $Y=0$ and $Y=\ell$. 
Per reference length this becomes $N_0=\lambda_1N$. 
The bending moment necessary to support the applied rotation (along $X=0$ and $X=\pi R$) becomes \citep{shelltheo}
\eqb{l}
M = k\,H
\eqe
per current length of the support (or $M_0=\lambda_2\,M$ per reference length).
If the special case $k^\ast=-k/2$ \citep{canham70} is considered, there is no bending in the $Y$-direction.

For the area-incompressible model of \eqref{e:W_i}, we have $\lambda_1=1/\lambda_2$. 
For the area-compressible case according to model \eqref{e:W_c}, we can determine $\lambda_1$ from \eqref{e:q_pb} with $J=\lambda_1\lambda_2$, giving
\eqb{l}
\lambda_1 = \ds\frac{1}{\lambda_2}\Big[\frac{k}{K}H^2+1\Big]~.
\eqe

\begin{figure}[h]
\begin{center} \unitlength1cm
\begin{picture}(0,5.8)
\put(-7.9,-.1){\includegraphics[height=60mm]{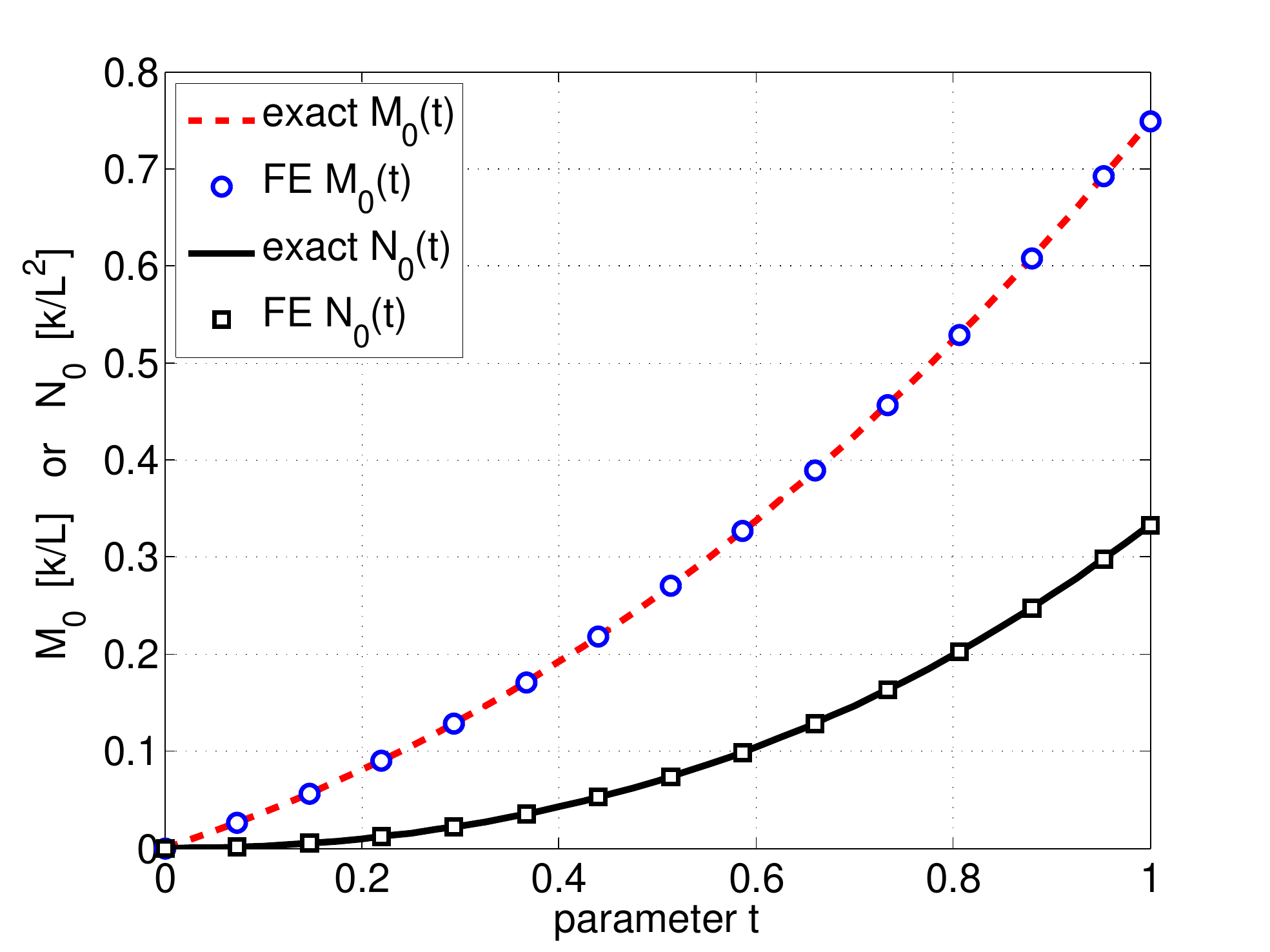}}
\put(0.4,-.1){\includegraphics[height=60mm]{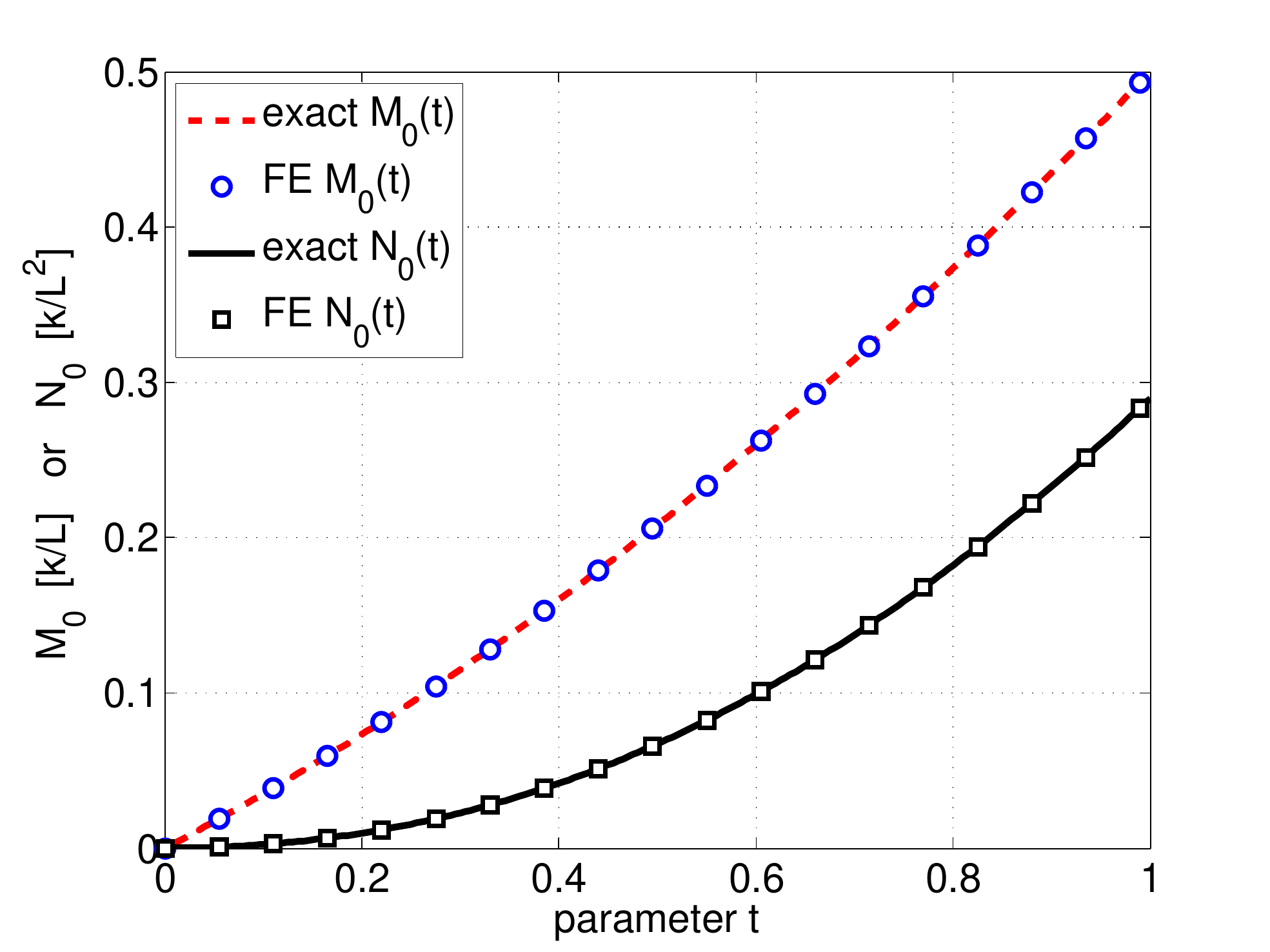}}
\put(-7.7,0){a.}
\put(.6,0){b.}
\end{picture}
\caption{Pure bending: distributed boundary moment $M_0(t)$ and normal traction $N_0(t)$ as obtained analytically and computationally for a.~the area-incompressible model \eqref{e:W_i} and b.~the area-compressible model \eqref{e:W_c} for $\bar K=2.5$.}
\label{f:bend_MN}
\end{center}
\end{figure}

The two cases are solved numerically using the computational setup shown in Fig.~\ref{f:bend_0}. The FE mesh is discretized by $m$ elements along $X$.
The parameter $t$ is introduced to apply the rotation $\Theta = t\pi/6$ and stretch $\lambda_2 = 1 + t/2$ by increasing $t$ linearly from 0 to 1 in $n_t$ steps, where $n_t$ is chosen as a multiple of $m$. The mean curvature then follows as $H = \Theta/(2\lambda_1S)$. 
For the unconstrained case, $\lambda_1$ is then the solution of the cubic equation 
\eqb{l}
\lambda_1^3\lambda_2 - \lambda_1^2 - \ds\frac{\Theta^2}{4\bar K} = 0~,
\eqe
with $\bar K=KS^2/k$.
Numerically, the rotation is applied according to \eqref{e:Pin} considering the penalty parameter $\epsilon=100\,n_x\,k/L$.
Fig.~\ref{f:bend_MN} shows the FE solution and analytical solution for $M_0(t)$ and $N_0(t)$, normalizing $M$ by $k/L$ and $N$ by $k/L^2$.

\begin{figure}[h]
\begin{center} \unitlength1cm
\begin{picture}(0,12.2)
\put(-7.9,6.3){\includegraphics[height=60mm]{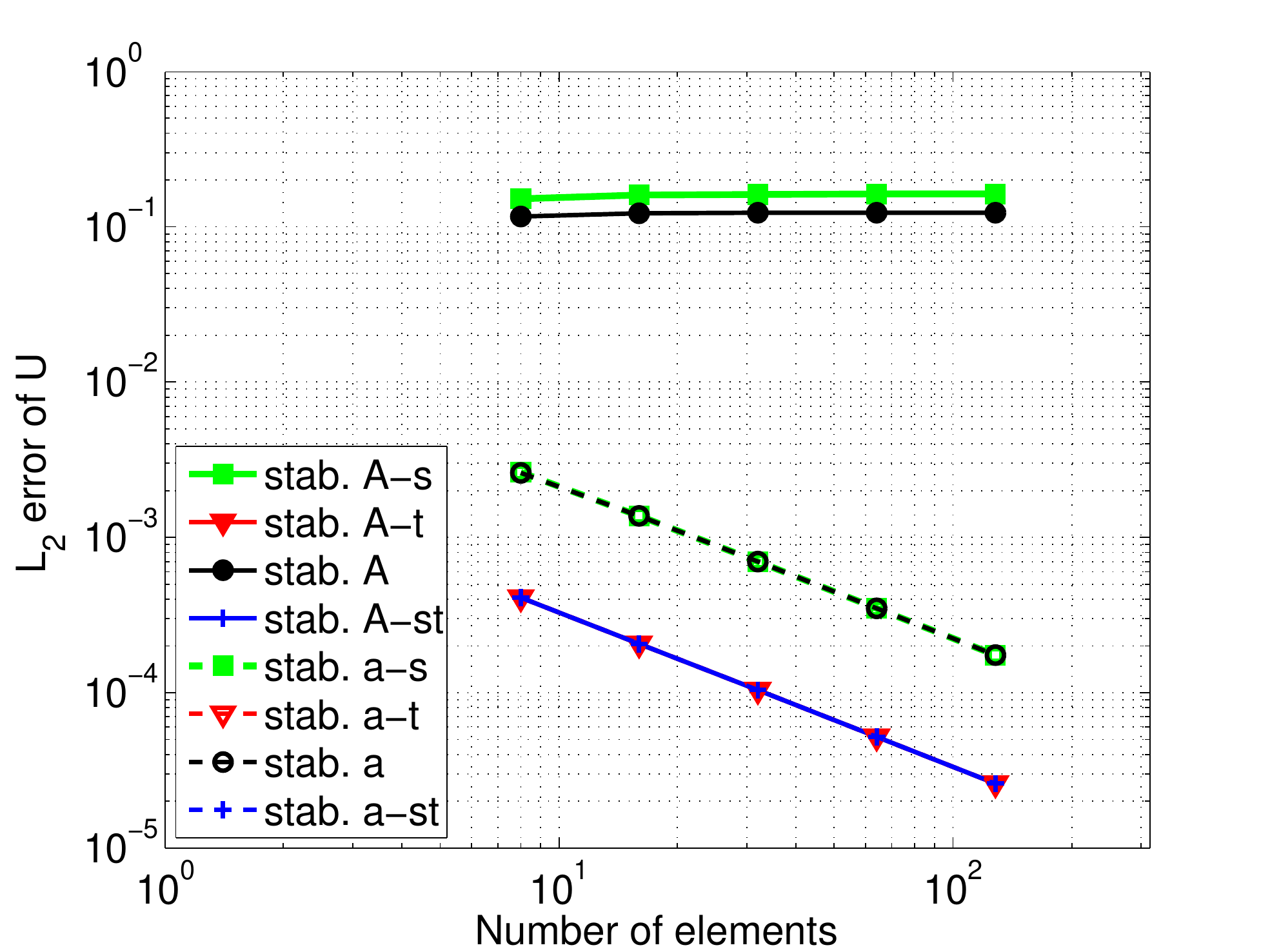}}
\put(0.3,6.2){\includegraphics[height=61mm]{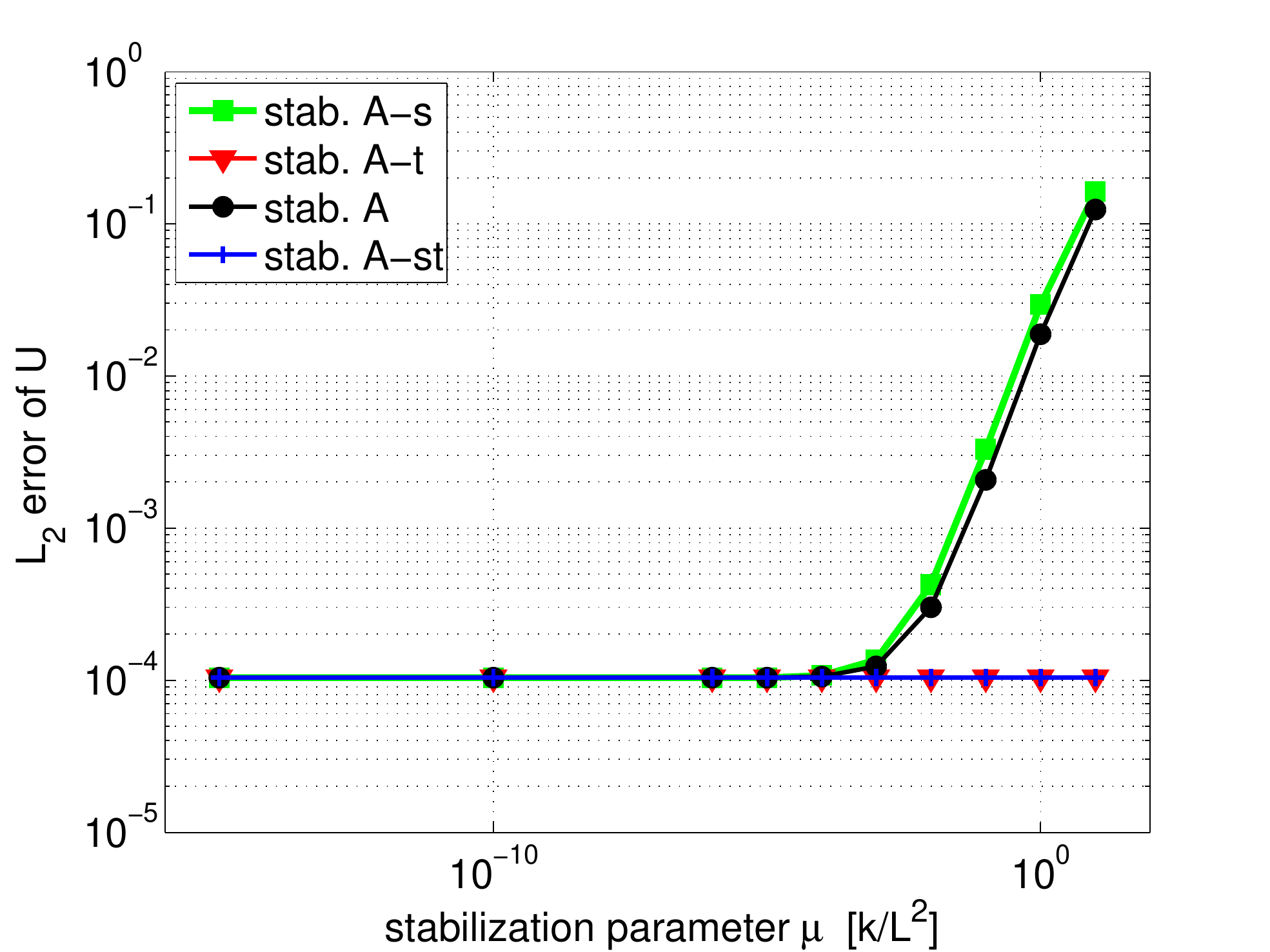}}
\put(-7.8,6.3){a.}
\put(.5,6.3){b.}
\put(-7.9,0){\includegraphics[height=60mm]{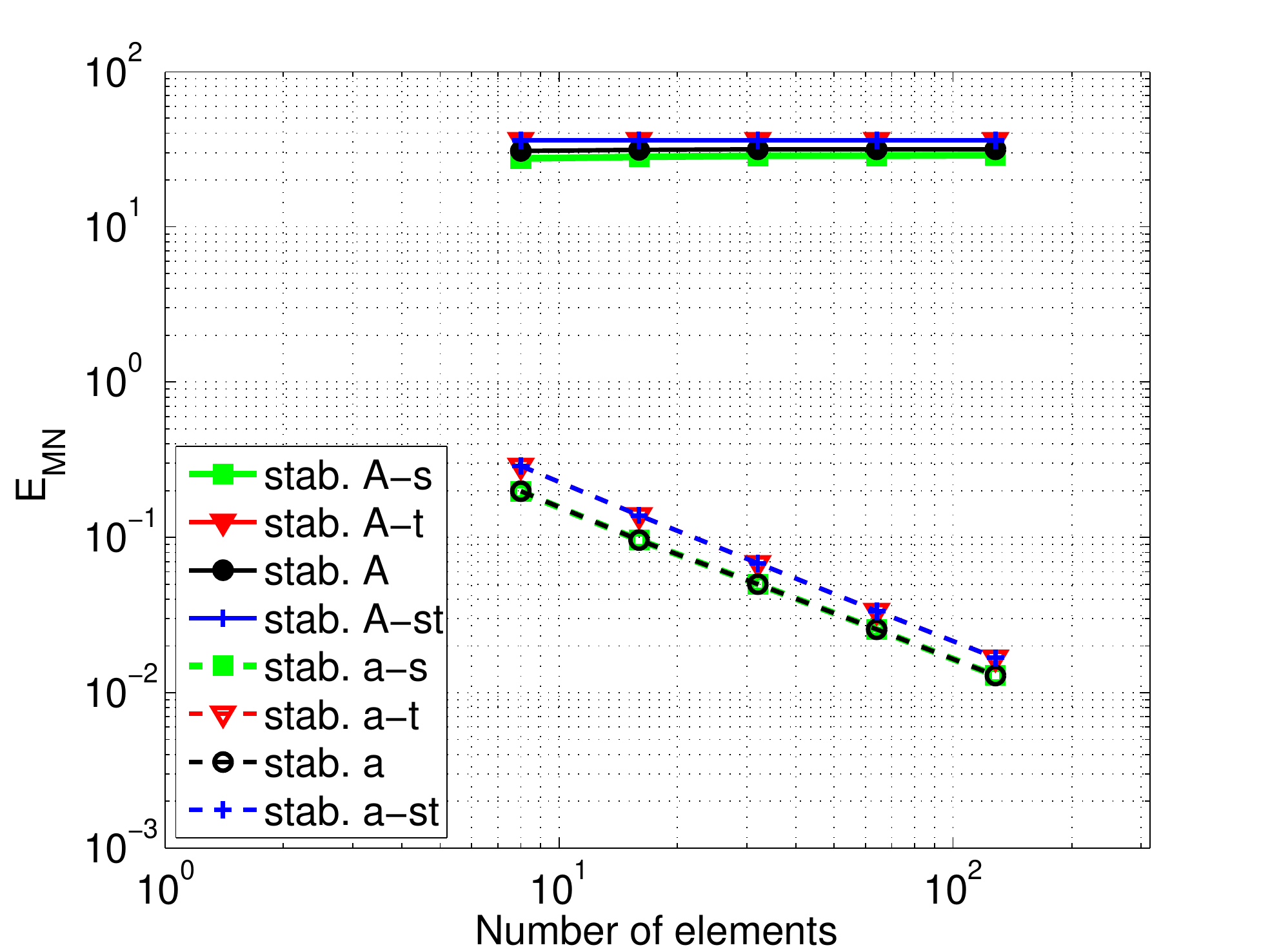}}
\put(0.3,-.1){\includegraphics[height=61mm]{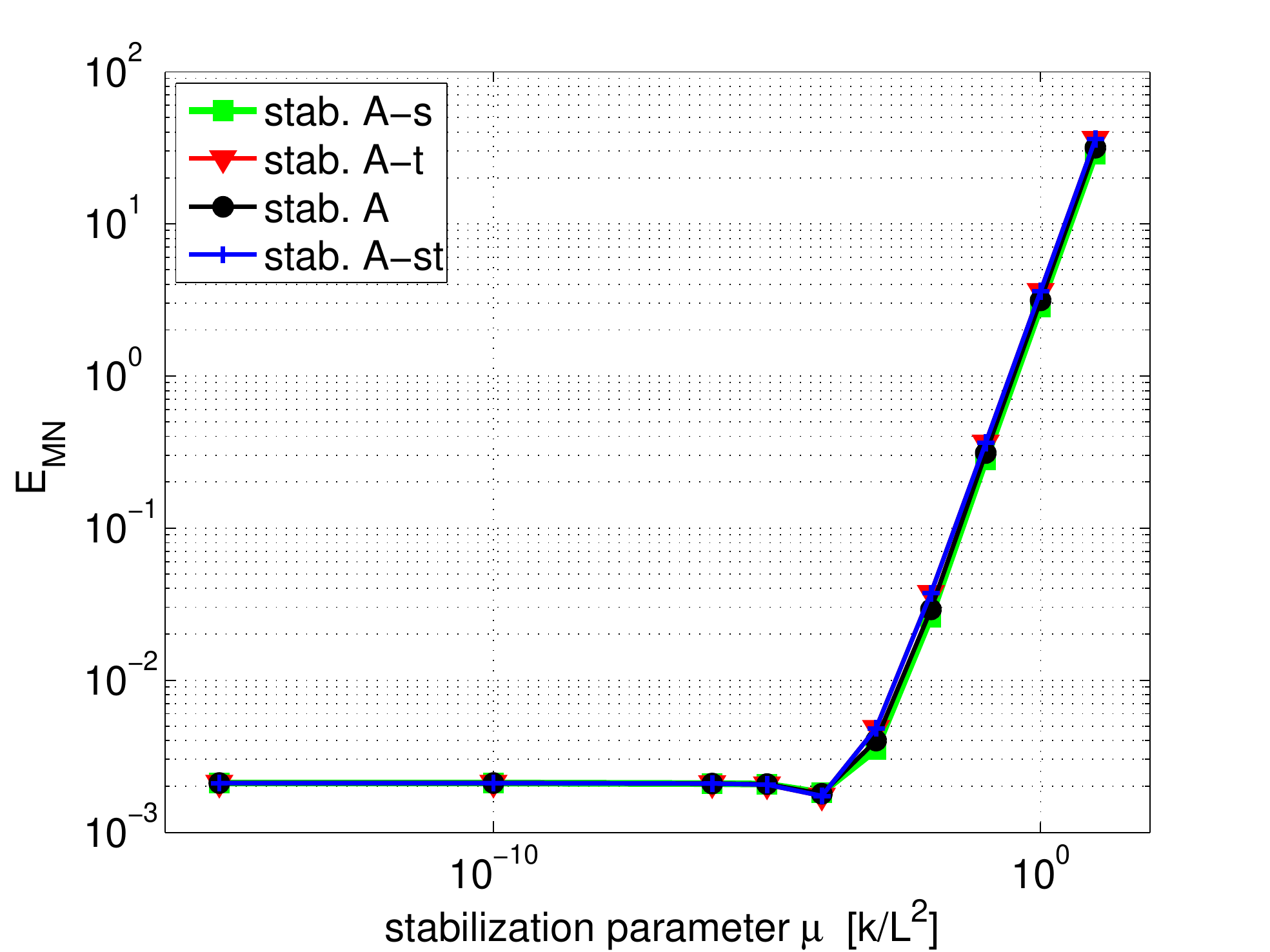}}
\put(-7.8,0){c.}
\put(.5,0){d.}
\end{picture}
\caption{Pure bending: accuracy for the area-incompressibile model \eqref{e:W_i}: a.~$L_2$-error vs.~$m$ considering stabilization classes $\mA$ and $\ma$ with $\bar\mu=10$ and $n_t = 12.5\,m$; b.~$L_2$-error vs.~$\mu$ considering stabilization class $\mA$ with $m=32$; c.--d.~same as a.--b.,~but now for error $E_{MN}$. Considered is $\Theta/2=\pi/3$ and $\lambda_2=1.5$.}
\label{f:bend_ie}
\end{center}
\end{figure}

Next, the accuracy of the different stability schemes is studied in detail by examining the $L_2$-error of the solution, defined by
\eqb{l}
L_2 := \ds \sqrt{  \frac{1}{SL}  \int_{\sS_0} \norm{ \bu_\mathrm{exact} - \bu_\mathrm{FE}}^2 \,\dif A}~,
\eqe
and the error in $M$ and $N$, defined by
\eqb{l}
E_{MN}:= \ds\frac{|M_{\mathrm{exact}} - M_{\mathrm{FE}}|}{M_{\mathrm{exact}}} + \ds\frac{|N_{\mathrm{exact}} - N_{\mathrm{FE}}|}{N_{\mathrm{exact}}}~,
\eqe
where $M_{\mathrm{FE}}$ and  $N_{\mathrm{FE}}$ are the computed mean values along the respective boundaries.
The first error is a measure of the kinematic accuracy, while the second is a measure of the kinetic accuracy.
Fig.~\ref{f:bend_ie} shows the two errors for the area-incompressible model of Eq.~\eqref{e:W_i}.
Looking at the $L_2$-error, schemes `A-t', `A-st', `a-t' and `a-st' perform best.
In case of error $E_{NM}$, schemes `a' and `a-s' perform best.
Class $\mA$ generally converges with $\mu$, but it may not converge with the number of elements for high values of $\mu$. 
Interestingly, the $L_2$-error of scheme `A-t' and `A-st' is not affected by $\mu$, as schemes `A' and `A-s' are. For sufficiently low $\mu$ (for $m=32$ about $\bar\mu<10^{-3}$), the accuracy of class $\mA$ (both in $L_2$ and $E_{MN}$) reaches that of class $\ma$ and then only improves with mesh refinement. Class $\mA$ with low $\mu$ may even surpass class $\ma$ with high $\mu$.
But generally, class $\ma$ is more accurate and robust (as $\mu$ does not need to be very small). There is no clear favourite in class $\ma$ for this test case.

\begin{figure}[h]
\begin{center} \unitlength1cm
\begin{picture}(0,12.2)
\put(-7.9,6.3){\includegraphics[height=60mm]{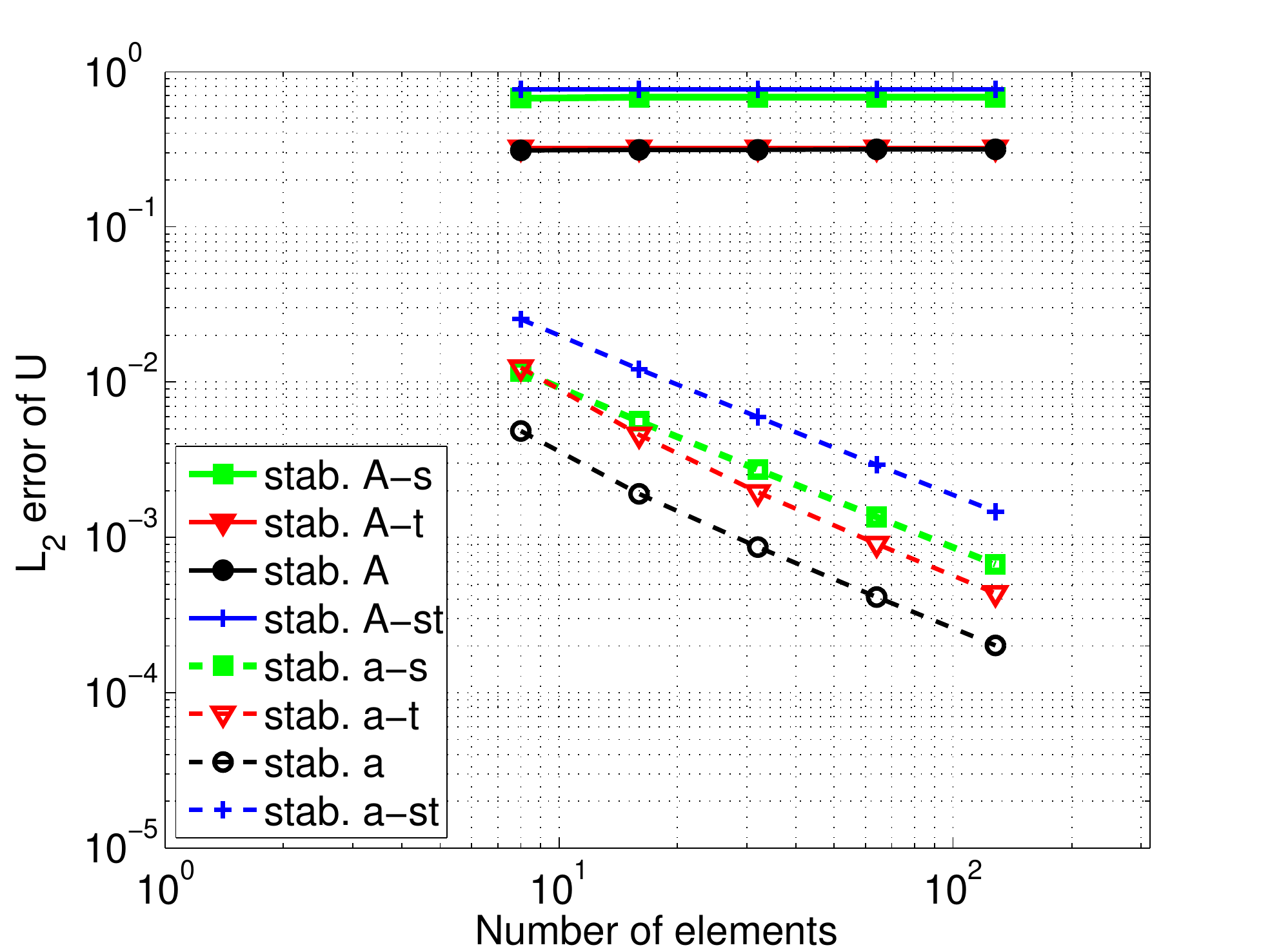}}
\put(0.3,6.2){\includegraphics[height=61mm]{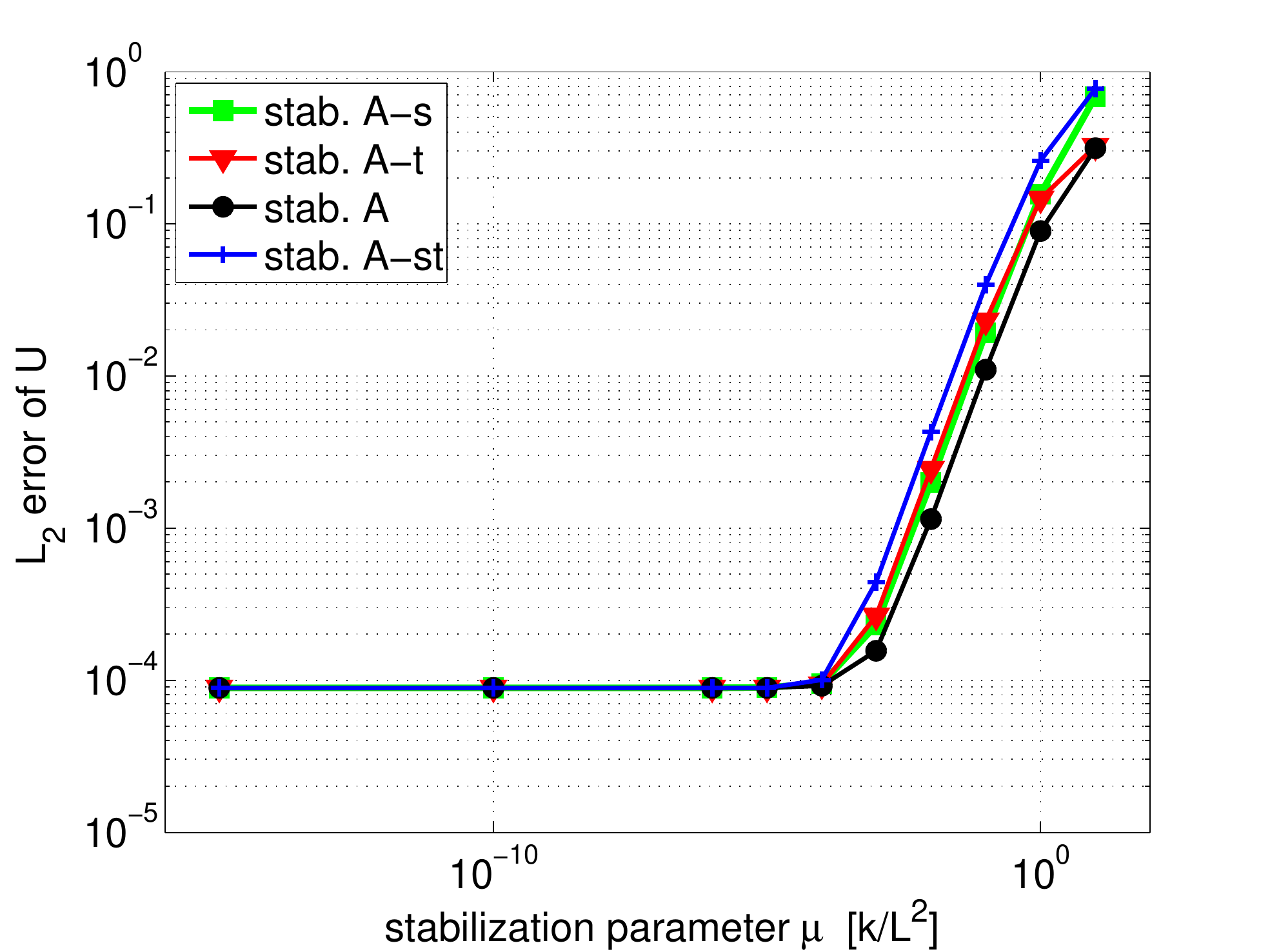}}
\put(-7.8,6.3){a.}
\put(.5,6.3){b.}
\put(-7.9,-.0){\includegraphics[height=60mm]{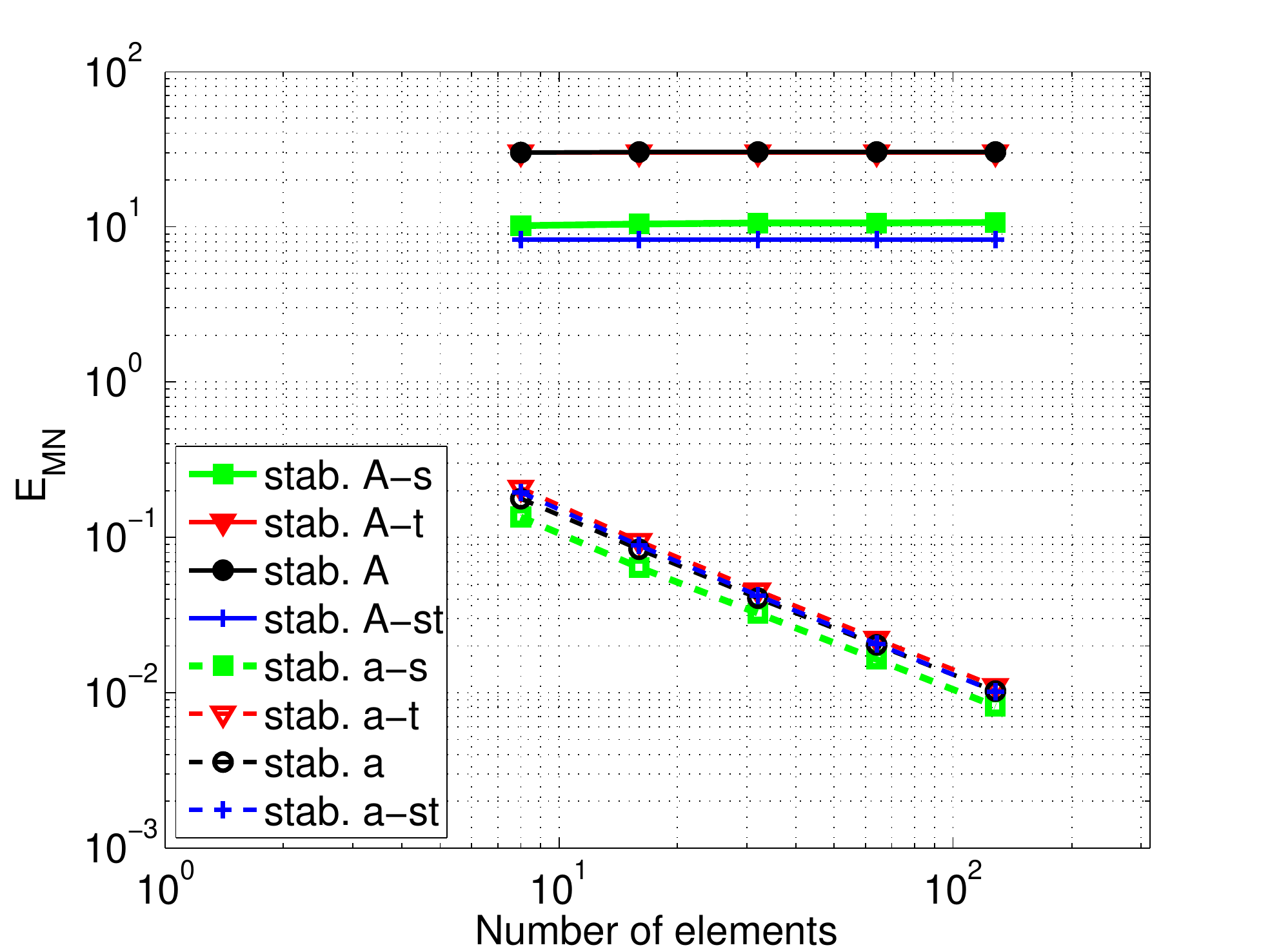}}
\put(0.3,-.1){\includegraphics[height=61mm]{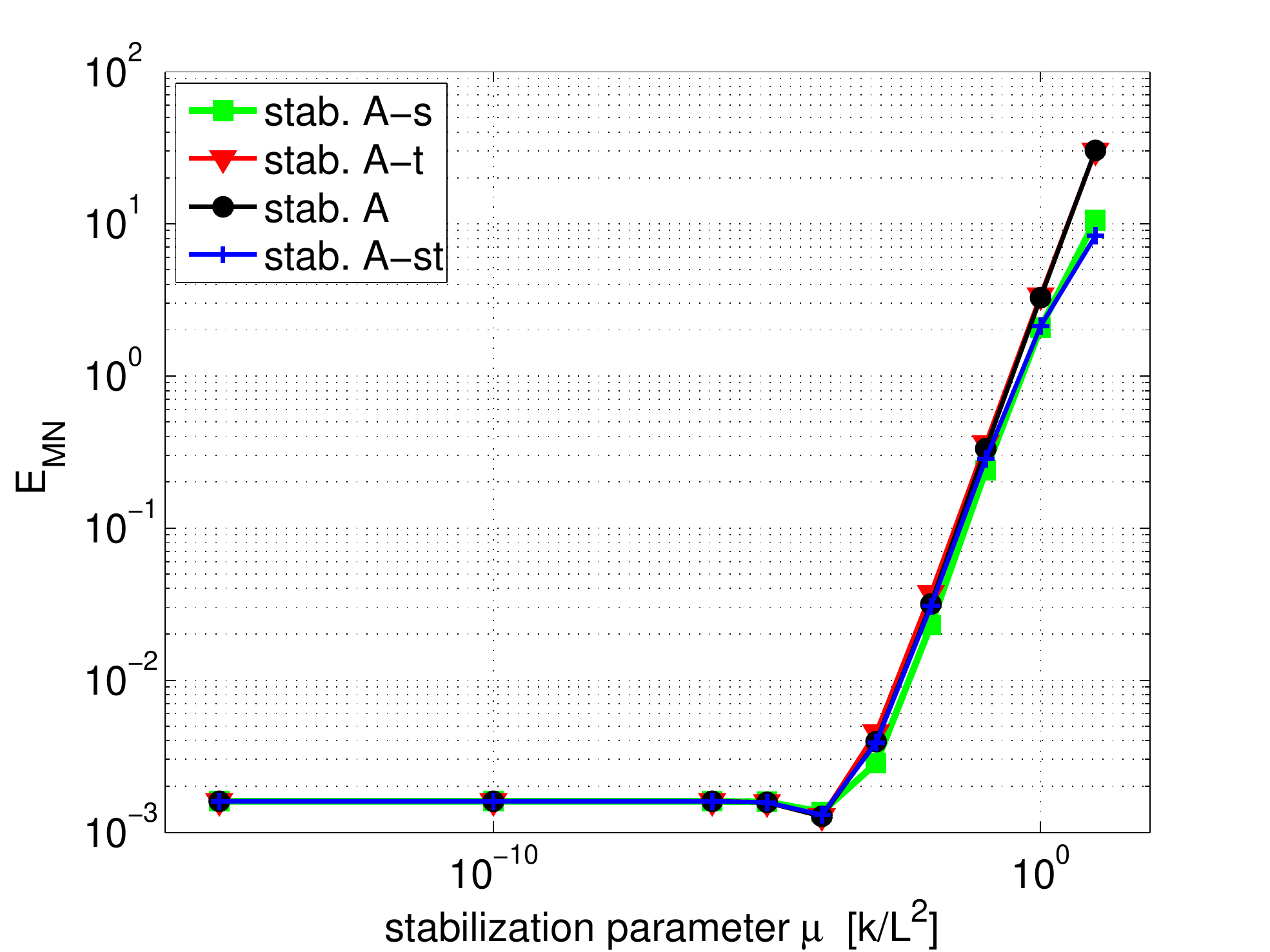}}
\put(-7.8,0){c.}
\put(.5,0){d.}
\end{picture}
\caption{Pure bending: accuracy of the area-compressible model \eqref{e:W_c}: a.~$L_2$-error vs.~$m$ considering stabilization classes $\mA$ and $\ma$ with $\bar\mu=10$ and $n_t=12.5\,m$; b.~$L_2$-error vs.~$\mu$ considering stabilization class $\mA$ with $m=32$;  c.--d.~same as a.--b.,~but now for error $E_{MN}$. Considered is $\bar K=2.5$, $\Theta/2=\pi/3$ and $\lambda_2=1.5$.} 
\label{f:bend_ce}
\end{center}
\end{figure}
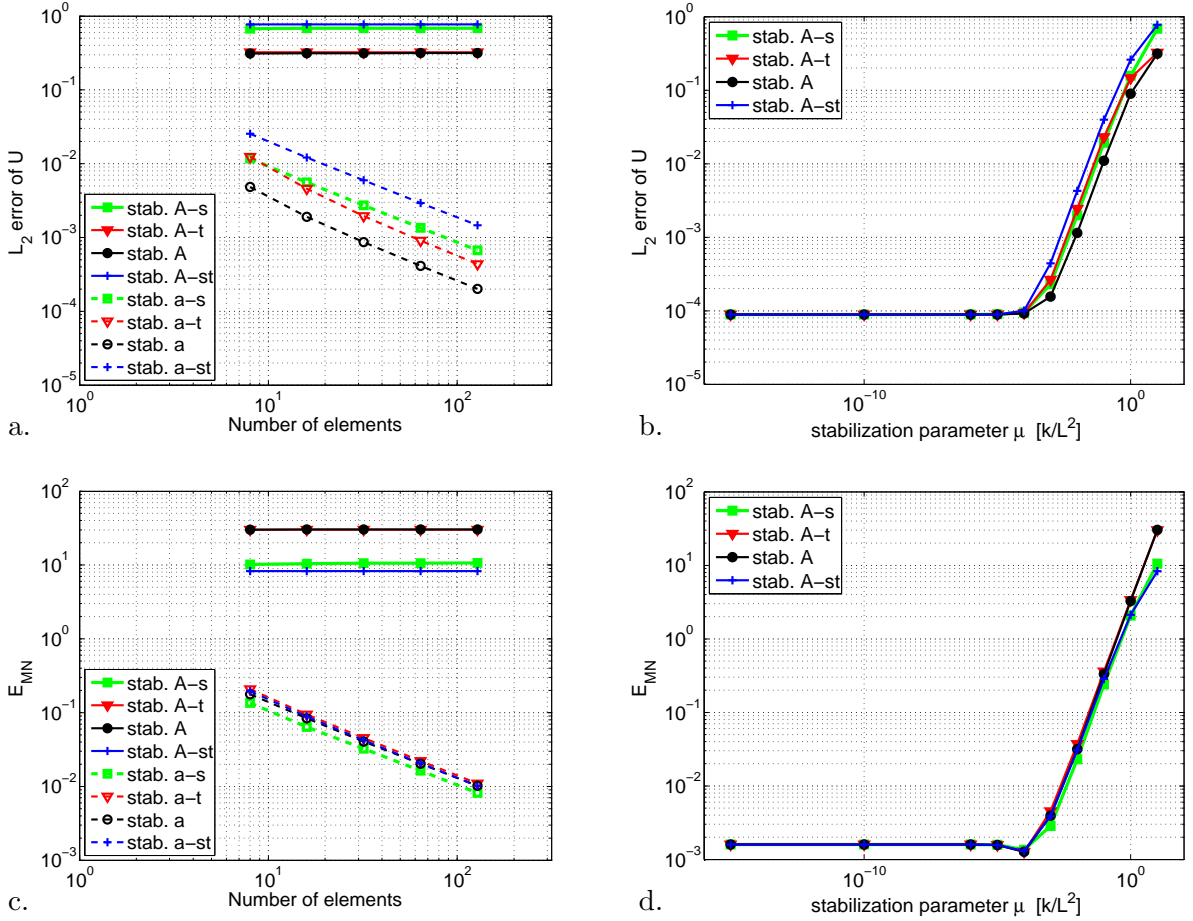

Fig.~\ref{f:bend_ce} shows the two errors for the area-compressible model of Eq.~\eqref{e:W_c} considering $\bar K=2.5$.
In case of the $L_2$-error, scheme `a' performs best, while for error $E_{MN}$, scheme `a-s' is best.
As before, class $\mA$ is poor for large $\mu$, but reaches (and may surpass) the accuracy of class $\ma$ at some $\mu$ depending on $m$.

As the plots show, not a single stabilization scheme stands out here and the accuracy depends both on the model and the error measure. In general, all schemes are suitable to solve the problem. If class $\mA$ is used, the value of $\mu$ needs to be suitably low. For class $\ma$ even large values for $\mu$ can be used.
In this example it is even possible to set $\mu=0$ in the code. 
This works since the effective shear stiffness according to \eqref{e:mueff} is positive here, i.e.~$\mu_\mathrm{eff} = 3JkH^2/2>0$.
For other problems $\mu_\mathrm{eff}$ can be negative, and stabilization is required.

\subsection{Inflation of a sphere}\label{s:sphere}

The second example considers the inflation of a spherical cell. Contrary to the previous example, the FE mesh now also contains interfaces between NURBS patches. Since the surface area increases during inflation, potential~(\ref{e:W_c}) is considered. For this model, the in-plane traction component, given in (\ref{e:sig_c}), is
\eqb{l}
N^{\alpha\beta} = N_a\,a^{\alpha\beta} + N_b\,b^{\alpha\beta}~,
\eqe
with
\eqb{lll}
N_a \dis k\,\Delta H^2 + K\,(J-1)~, \\[1mm]
N_b \dis -k\,\Delta H~. 
\eqe
The initial radius of the sphere is denoted by $R$, the initial volume is denoted by $V_0=4\pi R^3/3$. The cell remains spherical during inflation, so that we can obtain an analytical reference solution. The current radius during inflation shall be denoted by $r$, the current volume by $V=4\pi r^3/3$. 
Considering the surface parameterization
\eqb{l}
\bx(\phi,\theta) = \left[\begin{array}{c}
r\,\cos\phi\,\sin\theta \\ 
r\,\sin\phi\,\sin\theta \\ 
-r\,\cos\theta
\end{array} \right],
\eqe
we find
\eqb{l}
[a^{\alpha\beta}] = \ds\frac{1}{r^2}\left[\begin{array}{cc}
1/\sin^{2}\theta & 0 \\[1mm] 
0 & 1 
\end{array} \right],
\eqe
$b^{\alpha\beta}=-a^{\alpha\beta}/r$ and $H=-1/r$ for this example. The traction vector 
$\bT=\nu_\alpha\bT^\alpha$ on a cut $\perp\bnu$ thus becomes
\eqb{l}
\bT = (N_a-N_b/r)\,\bnu + S^\alpha\nu_\alpha\bn
\eqe
according to (\ref{e:Ta}). The in-plane component $T_\nu:=N_a-N_b/r$ must equilibrate the current pressure according to the well-known relation
\eqb{l}
p = \ds\frac{2T_\nu}{r}~.
\eqe
We can thus establish the analytical pressure-volume relation
\eqb{l}
\bar p(\bar V) 
= 2\bar H_0\,\bar V^{-\frac{2}{3}}
- 2\bar H_0^2\,\bar V^{-\frac{1}{3}}
+ 2\bar K\,\Big(\bar V^{\frac{1}{3}}-\bar V^{-\frac{1}{3}}\Big)~,
\label{e:pV}\eqe
normalized according to the definitions $\bar p := pR^3/k$, $\bar V := V/V_0$, $\bar H_0 := H_0R$ and $\bar K := KR^2/k$.

\begin{figure}[h]
\begin{center} \unitlength1cm
\begin{picture}(0,5.5)
\put(-7.8,4){\includegraphics[height=54mm,angle=-90]{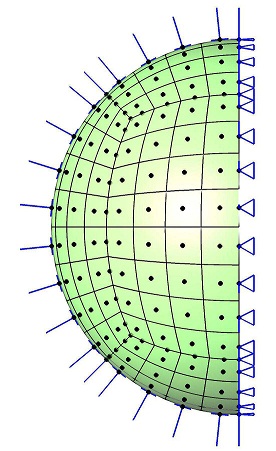}}
\put(-1.8,-.3){\includegraphics[height=55mm]{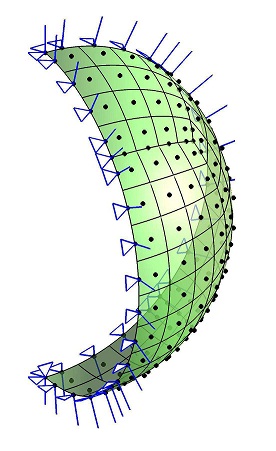}}
\put(3,-.5){\includegraphics[height=60mm]{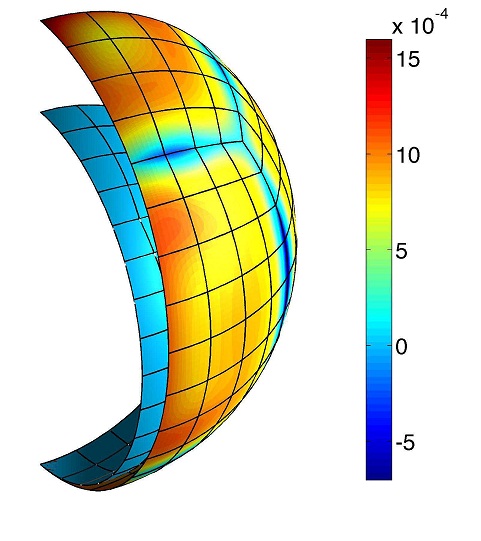}}
\put(-7.7,0){a.}
\put(2.8,0){b.}
\end{picture}
\caption{Sphere inflation: a.~initial FE configuration and boundary conditions (for mesh $m=8$); b.~current FE configuration for an imposed volume of $\bar V = 2$ compared to the initial configuration; the colors show the relative error in the surface tension $T_\nu$.}
\label{f:inflate_0}
\end{center}
\end{figure}

\begin{figure}[h]
\begin{center} \unitlength1cm
\begin{picture}(0,5.9)
\put(-8.2,-.1){\includegraphics[height=60mm]{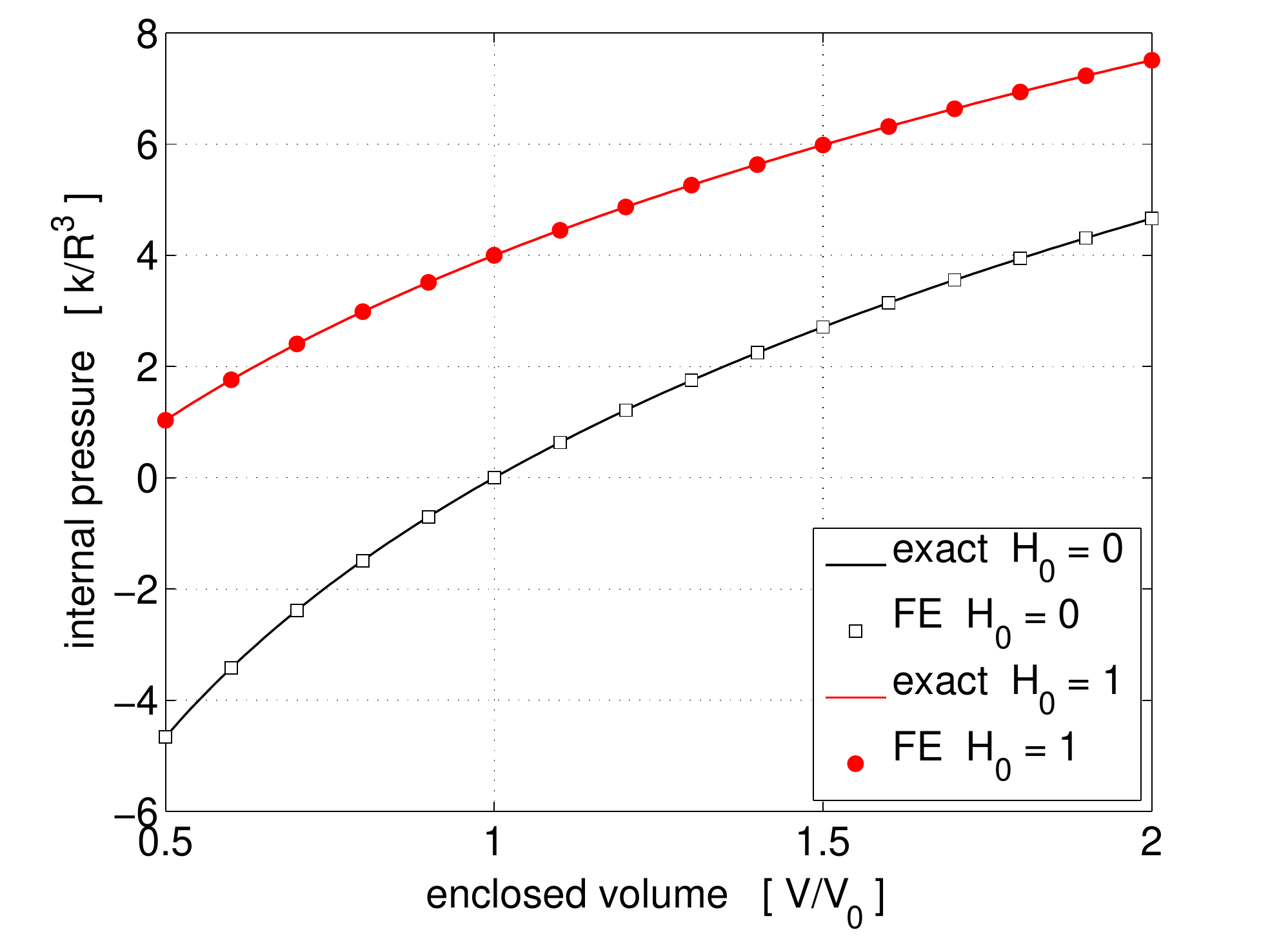}}
\put(0.1,0){\includegraphics[height=59mm]{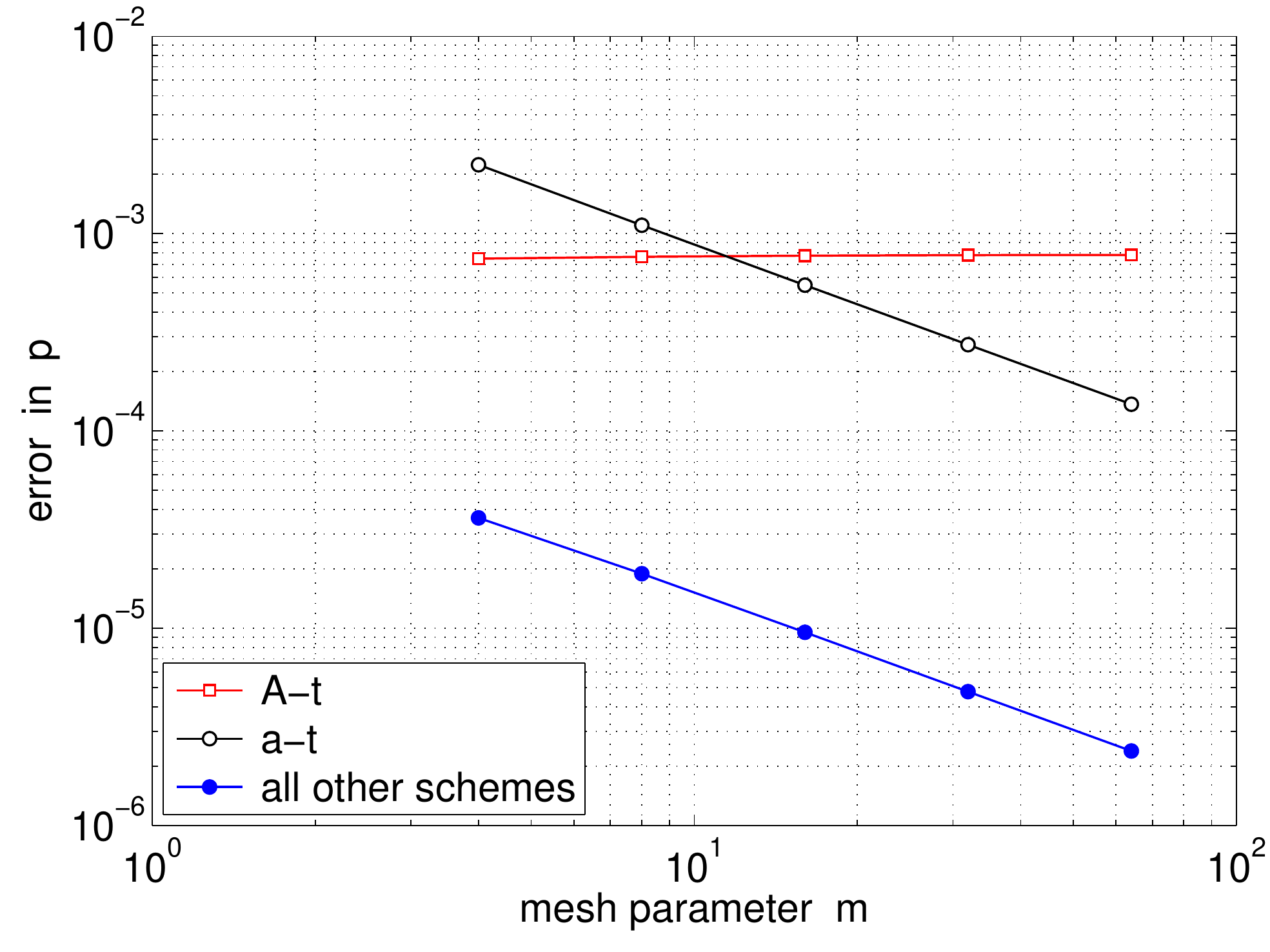}}
\put(-7.9,0){a.}
\put(.2,0){b.}
\end{picture}
\caption{Sphere inflation: a.~pressure-volume relation; b.~FE convergence for the different stabilization schemes.}
\label{f:inflate_P}
\end{center}
\end{figure}

Fig.~\ref{f:inflate_0} shows the computational setup of the problem.
The computational domain consists of a quarter sphere discretized with four NURBS patches. 
The quarter sphere contains $3m^2/2$ elements where $m$ is the number of elements along the equator of the quarter sphere. At the boundaries and at the patch interfaces $C^1$-continuity is enforced using 
\eqref{e:Pin} with $\epsilon=4000 mk/R$.
The area bulk modulus is taken as $K = 5 k/R^2$, while $k^\ast$ is taken as zero. Two cases are considered: $H_0=0$ and $H_0=1/R$.
Fig.~\ref{f:inflate_P} shows that the computational $p(V)$-data converge to the exact analytical result of (\ref{e:pV}).
Here the pressure error 
\eqb{l}
e_\mrp = \ds\frac{|p_\mathrm{exact}-p_\mathrm{FE}|}{p_\mathrm{exact}}
\eqe
is examined for $H_0=1/R$ and $\bar V=2$ considering the 9 stabilization schemes of Tab.~\ref{t:stab} with $\bar\mu=0.01$ for class $\mA$ and $\bar\mu=1$ and $n_t=5m$ for class $\ma$. For schemes `A', `A-s', `A-st', `a', `a-s', `a-st' and `P' this error converges nicely (and is indistinguishable in the figure). Only schemes `A-t' and `a-t' behave significantly different. They introduce further errors that only converge if $\mu$ is decreased or $n_t$ is increased.
The reason why all other schemes have equal error, is that here the error is actually determined by the penalty parameter $\epsilon$ used within patch constraint \eqref{e:Pin}. The error stemming from the stabilization methods (apart from `A-t' and `a-t') is insignificant compared to that.
It is interesting to note that `A-st' and `a-st' perform much better than `A-t' and `a-t', even though no shear is present in the analytical solution. `A-st' and `a-st' can therefore be considered as the best choices here, since they are the most efficient schemes to implement.

We finally note that for a sphere $\mu_\mathrm{eff}=JkH(H-H_0)$, where $H=-1/r$. Thus $\mu_\mathrm{eff}>0$ for $H<H_0$, 
which is the case here.

\subsection{Drawing of a tube}\label{s:tube}

Tube-like shapes are one of the most common non-trivial shapes in biological cell membranes. They can be observed in the endoplasmic reticulum \citep{shibata06, shibata09, hu09} and mitochondria \citep{fawcett, griparic01, shibata09}, where individual tubules or a complex network of tubules co-exist. The tubes can also be formed when a membrane is pulled by actin or microtubule polymerization, or by molecular motors attached to membranes but walking on cytoskeletal filaments \citep{terasaki86, waterman98, koster03, itoh05}. These situations can be modeled by means of a lateral force that acts on a membrane. An analytical solution for this problem has been obtained by \citet{derenyi02} by assuming axisymmetry and infinitely long tubes, which we use to verify our 
finite element formulation.
The dynamics of tether formation and relaxation have been also studied by \citet{rahimi12}.

\begin{figure}[h]
\begin{center} \unitlength1cm
\begin{picture}(0,5.0)
\put(-8,-.1){\includegraphics[height=53mm]{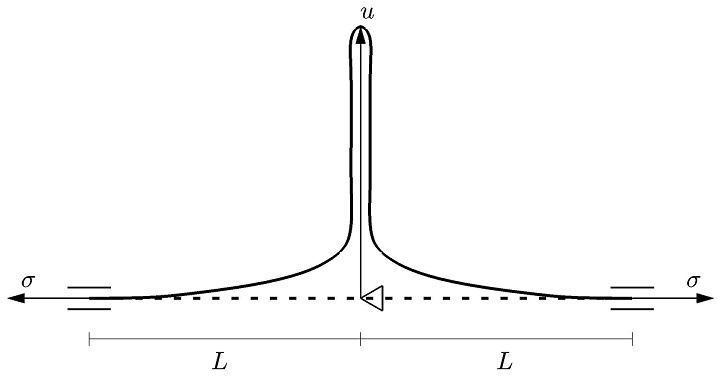}}
\put(2.6,-.1){\includegraphics[height=52mm]{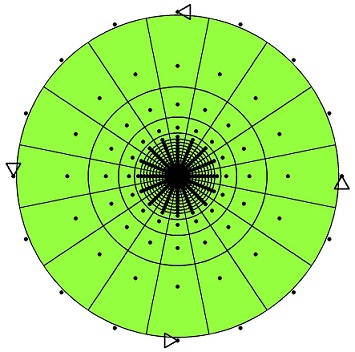}}
\put(-7.8,0){a.}
\put(2.8,0){b.}
\end{picture}
\caption{Tube drawing: a.~boundary conditions and b.~coarse FE mesh of the initial configuration. The dots show the control points of the NURBS discretization.}
\label{f:tube_0}
\end{center}
\end{figure}

To simulate the tube drawing process we consider the setup shown in Fig.~\ref{f:tube_0}.
The cell membrane is modeled as a circular, initially flat disc with initial radius $L$. 
The effect of the surrounding membrane is captured by the boundary tension $\sigma$ (w.r.t.~the current boundary length).
The surface is described by material model (\ref{e:W_c}). $L$ and $k$ are used for normalization. The remaining material parameters are chosen as $\bar k=-0.7\,k$ and $K=20,\!000\,k/L^2$. $H_0$ is taken as zero. We consider $\sigma \in\{100,\,200,\,400,\,800\}k/L^2$. 
Stabilization scheme `A-s' is considered with $\mu=0.1\,k/L^2$. 
The shell is clamped at the boundary, but free to move in the in-plane direction. The traction $\bt=\sigma\bnu$ is imposed and applied numerically via (\ref{e:fext}.2). Even though $\bt$ is constant during deformation, the boundary length $\dif s$ appearing in $\mf^e_{\mathrm{ext}t}$ changes and has to be linearized. This is, for example, discussed in \citet{droplet}. 
At the center, the displacement $u$ is imposed on the initially flat, circular surface. To prevent rigid rotations, selected nodes are supported as shown in Fig.~\ref{f:tube_0}b. 

Fig.~\ref{f:tube_0}b also shows one of the chosen finite element discretizations of the initial configuration. Quadratic, NURBS-based, $C^1$-continuous finite elements are used. For those, the outer ring of control points lies outside of the mesh; their distance to the center is therefore slightly larger than $L$. A finer discretization is chosen at the center, where the tube is going to form. The chosen NURBS discretization degenerates at the center, such that the $C^1$-continuity is lost there. It is regained if displacement $u$ is applied not only to the central control point but also to the first ring of control points around the center. This ensures that the tangent plane remains horizontal at the tip. Likewise, a horizontal tangent is enforced at the outer boundary by fixing the height of the outer two rings of control points.

\begin{figure}[h]
\begin{center} \unitlength1cm
\begin{picture}(0,15.4)
\put(3,-.2){\includegraphics[width=48mm]{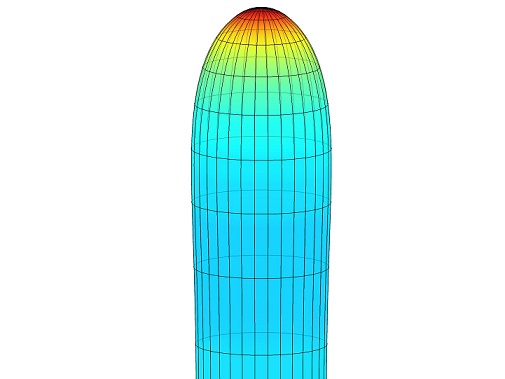}}
\put(-2,-.5){\includegraphics[width=58mm]{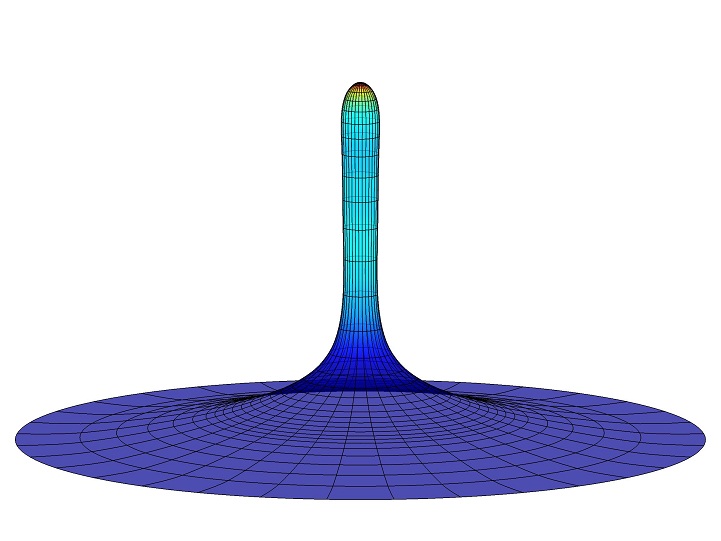}}
\put(-8,-.5){\includegraphics[width=58mm]{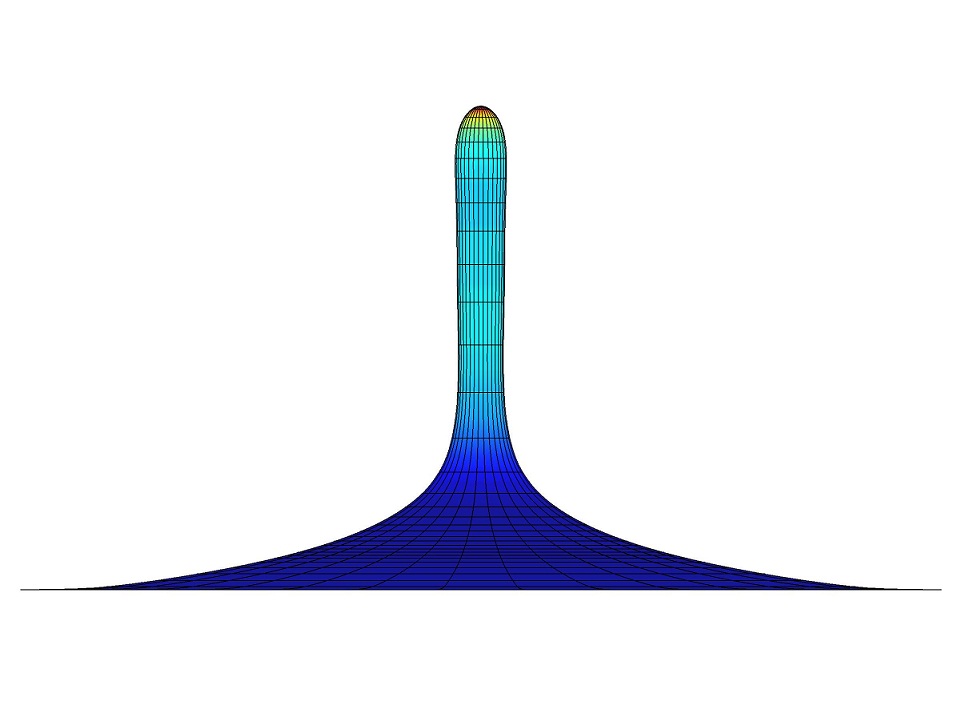}}
\put(3,3.8){\includegraphics[width=48mm]{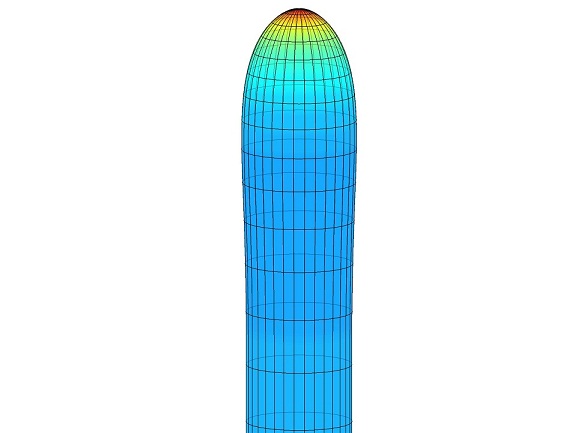}}
\put(-2,3.5){\includegraphics[width=58mm]{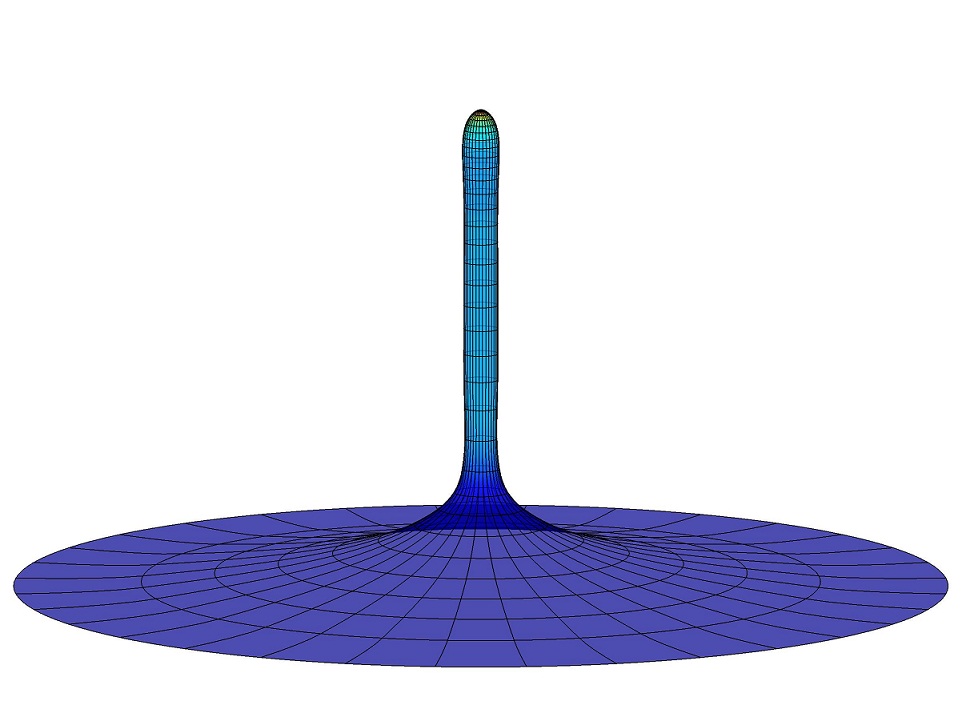}}
\put(-8,3.5){\includegraphics[width=58mm]{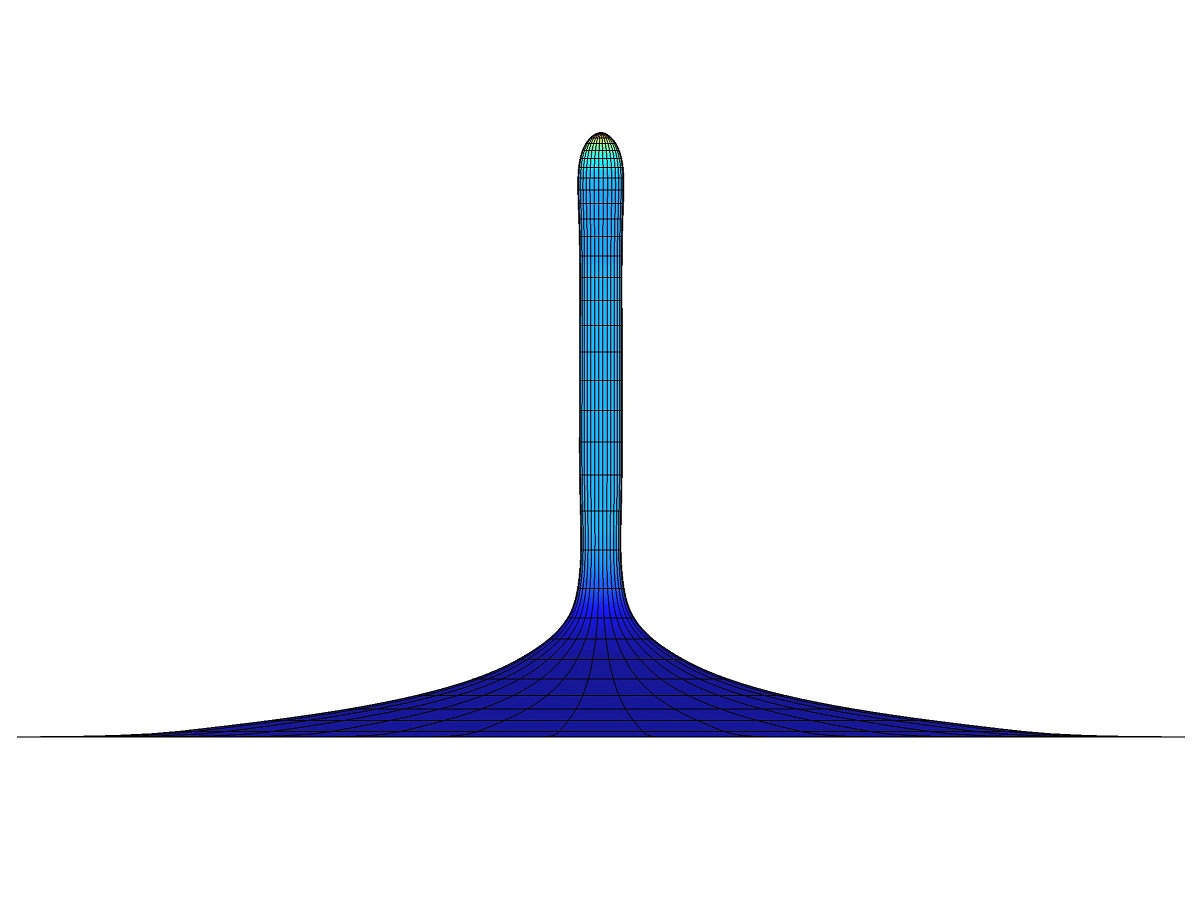}}
\put(3,7.8){\includegraphics[width=48mm]{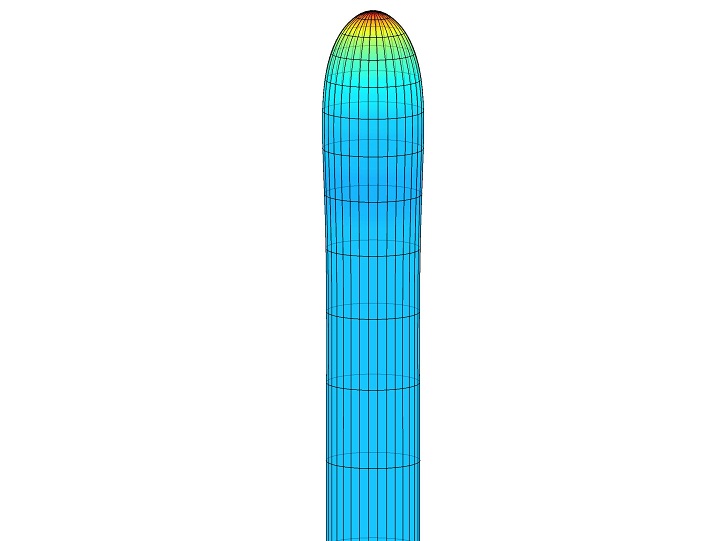}}
\put(-2,7.5){\includegraphics[width=58mm]{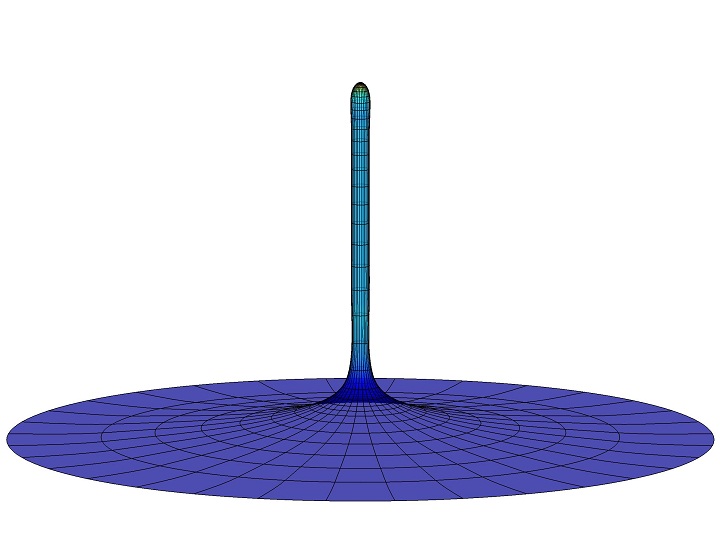}}
\put(-8,7.5){\includegraphics[width=58mm]{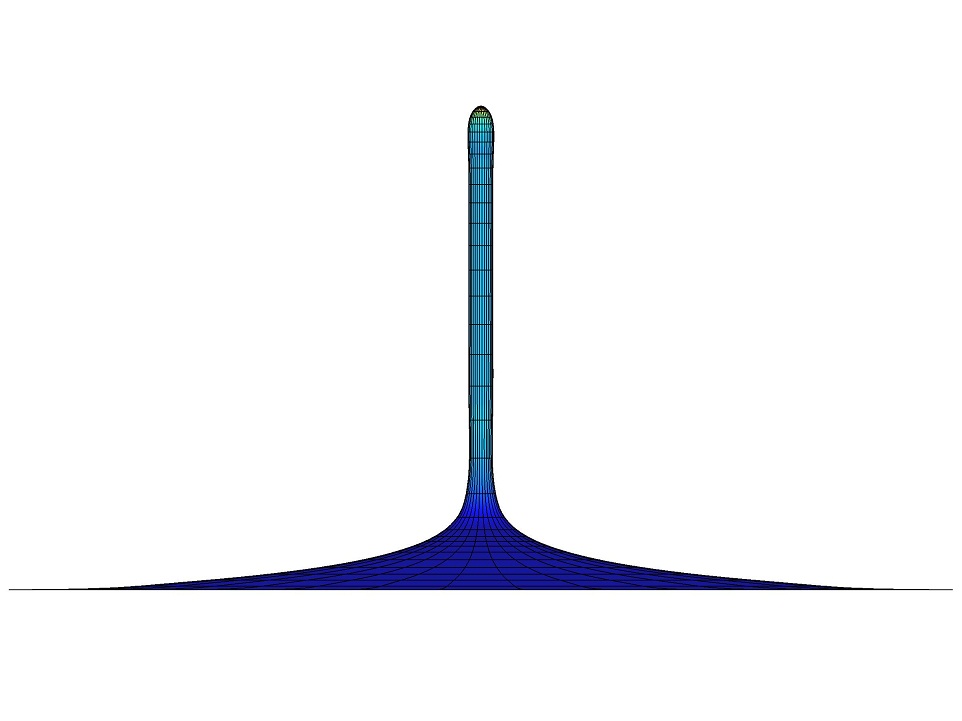}}
\put(3,11.8){\includegraphics[width=48mm]{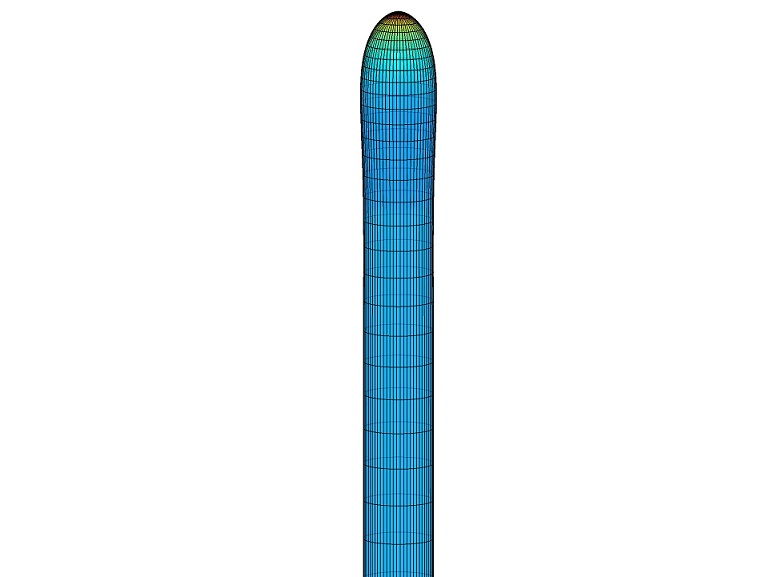}}
\put(-2,11.5){\includegraphics[width=58mm]{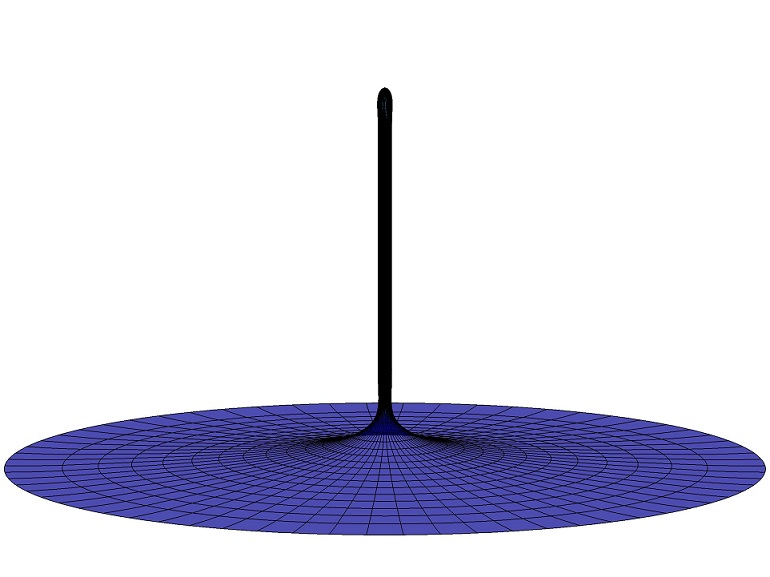}}
\put(-8,11.5){\includegraphics[width=58mm]{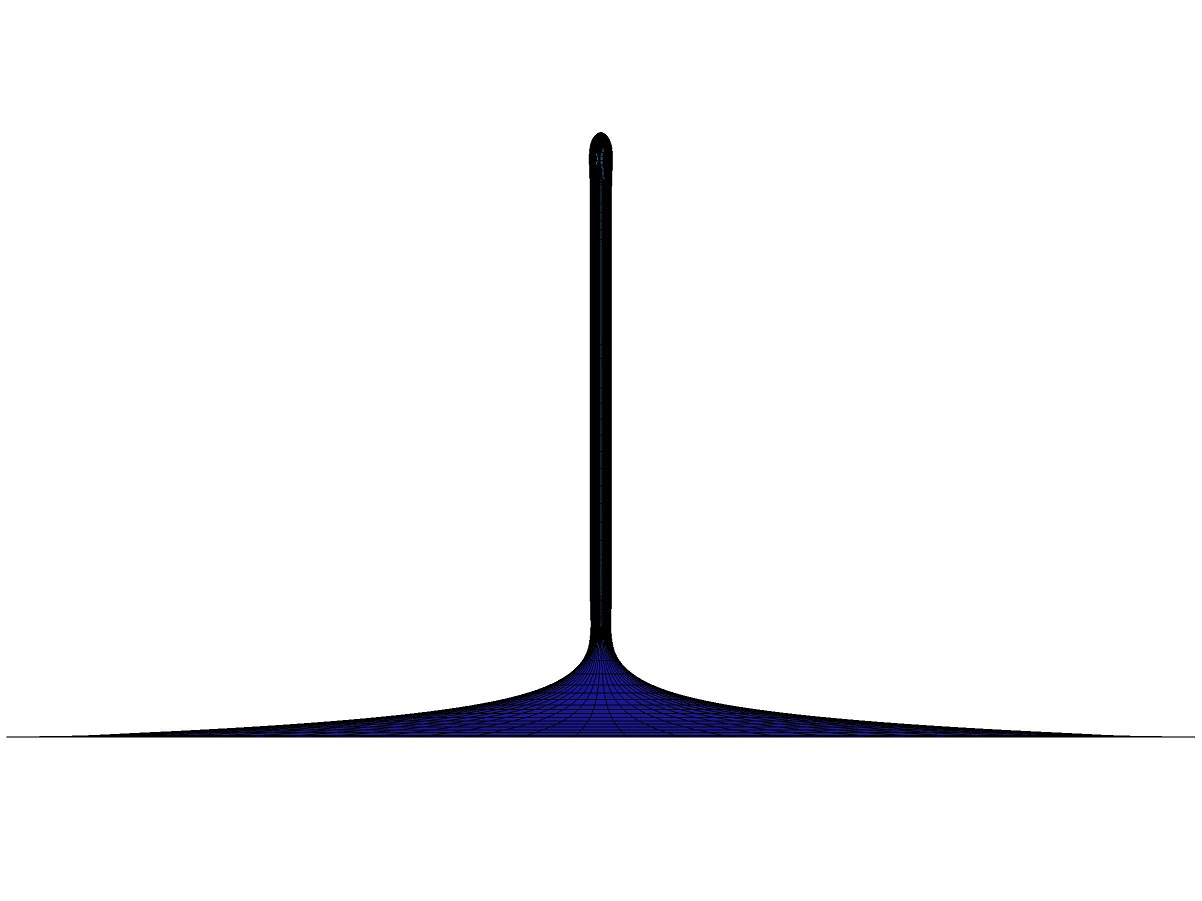}}
\put(7.4,11.25){\includegraphics[height=43mm]{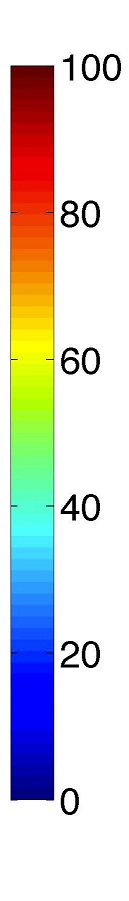}}
\put(7.4,7.25){\includegraphics[height=43mm]{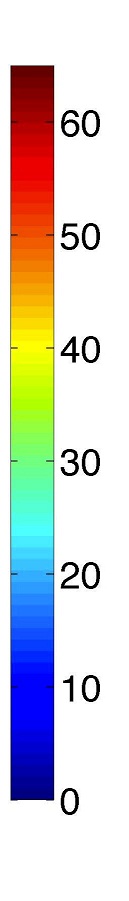}}
\put(7.4,3.25){\includegraphics[height=43mm]{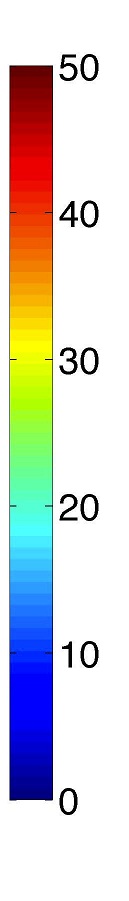}}
\put(7.4,-.75){\includegraphics[height=43mm]{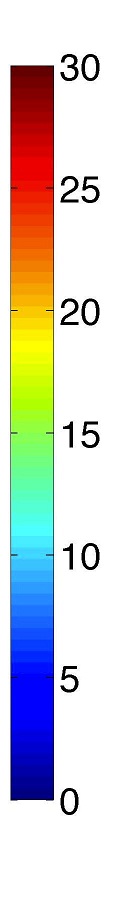}}
\end{picture}
\caption{Tube drawing: Results for $\sigma\in\{100,\,200,\,400,\,800\}\,k/L^2$ (bottom to top); colorscale shows the mean curvature $H$ normalized by $L^{-1}$. 
}
\label{f:tube_x}
\end{center}
\end{figure}

\begin{figure}[h]
\begin{center} \unitlength1cm
\begin{picture}(0,5.6)
\put(-8.1,0.1){\includegraphics[height=57mm]{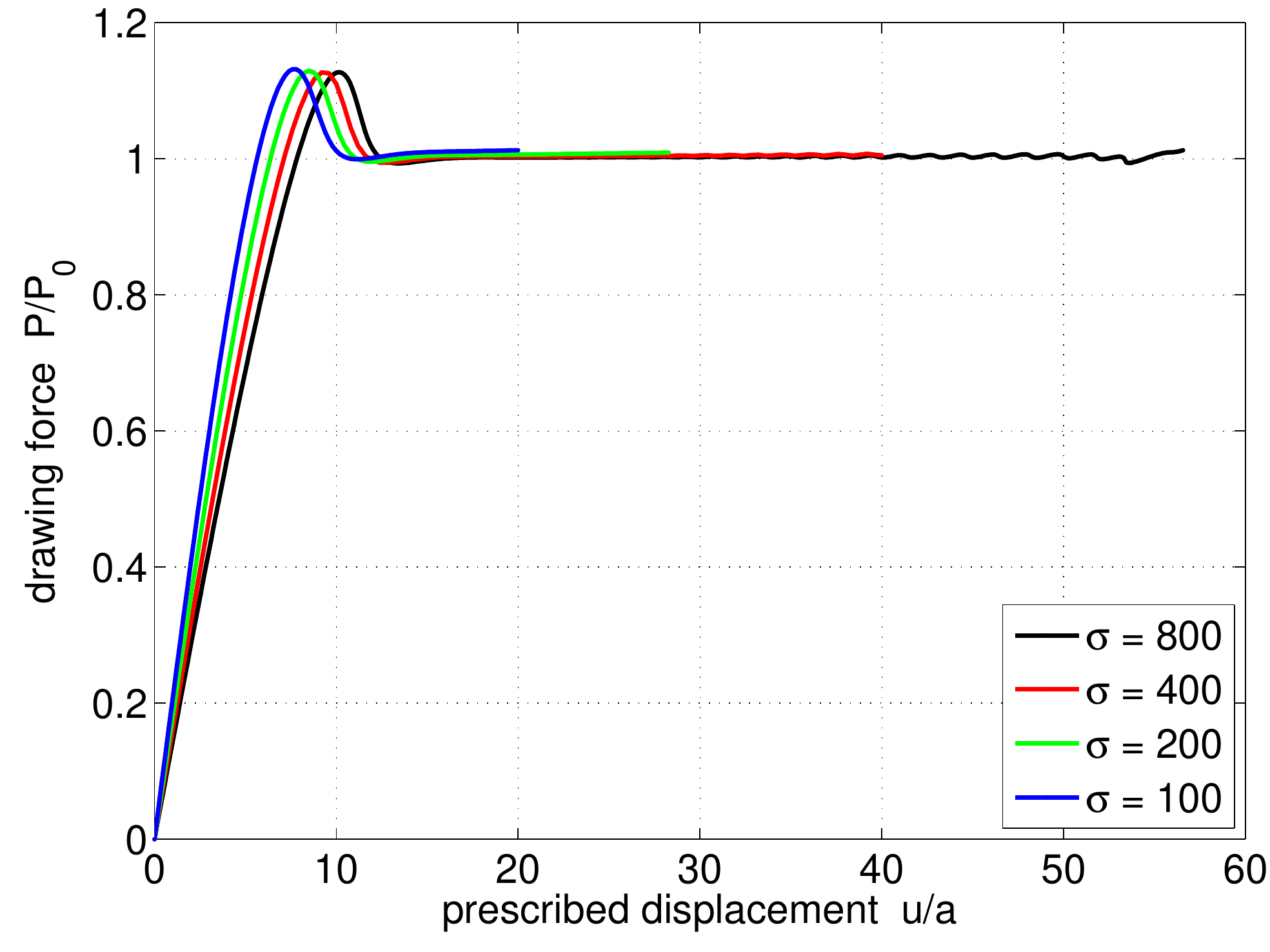}}
\put(0.2,0){\includegraphics[height=58.5mm]{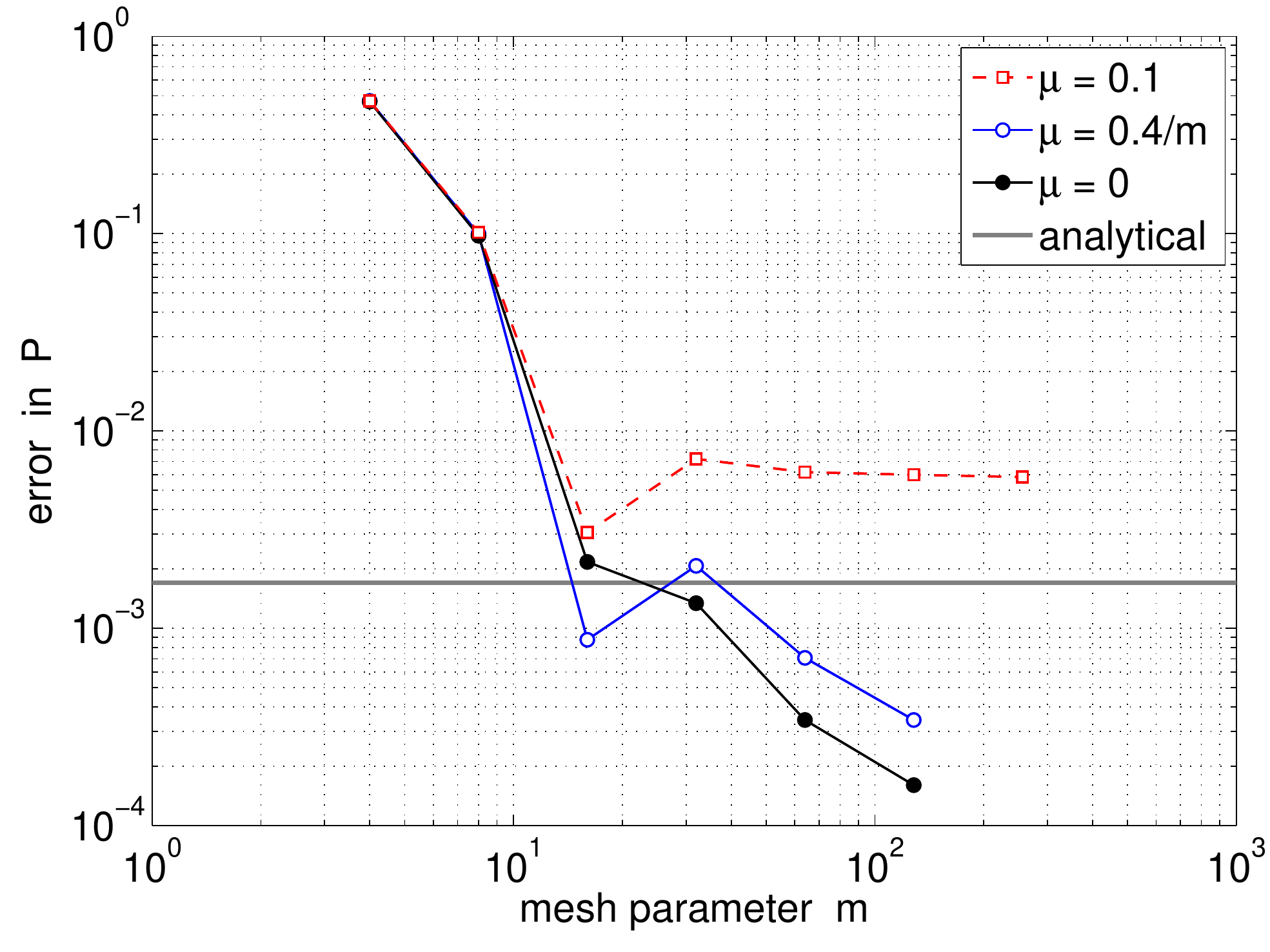}}
\put(-7.9,0){a.}
\put(.4,0){b.}
\end{picture}
\caption{Tube drawing: a.~load-displacement curve; b.~FE convergence.} 
\label{f:tube_P}
\end{center}
\end{figure}

Fig.~\ref{f:tube_x} shows the deformed surface at $u=L$, considering different values of the far-field surface tension $\sigma$.
The surface tension affects the slenderness of the appearing tube.
\citet{derenyi02} showed from theoretical considerations\footnote{Assuming that the tube is sufficiently long and can be idealized by a perfect cylinder.} that the tube radius is
\eqb{l}
a = \ds\frac{1}{2}\sqrt{\frac{k}{\sigma}}~,
\eqe
while the steady force during tube drawing is
\eqb{l}
P_0 = 2\pi\sqrt{\sigma\,k}~.
\eqe
These values are confirmed by our computations, which is illustrated in Fig.~\ref{f:tube_P}. \clearpage
The left inset shows the force-displacement relation during drawing.
Oscillations appear in the numerical solution due to the mesh discretization error. They are more pronounced for more slender tubes, as the black curve in Fig.~\ref{f:tube_P}a shows. They disappear upon mesh refinement, as the solution converges.
The convergence of $P_0$ for $\sigma=200 k/L^2$ and $u=L$ ($=28.28a$) is shown in Fig.~\ref{f:tube_P}b by examining the error 
\eqb{l}
e(P_0^\mathrm{FE}) := \ds\frac{|P^\mathrm{ref}_0-P_0^\mathrm{FE}|}{P^\mathrm{ref}_0}~,
\eqe
where $P^\mathrm{ref}_0$ is the FE solution for $m=256$ and $\mu=0$. 
The mesh sequence given in Tab.~\ref{t:tube_m} is used for the convergence study.
\begin{table}[h]
\centering
\begin{tabular}{|r|r|r|r|}
  \hline
  $m$ & $n_\mathrm{el}=4\,m^2$ & $n_\mathrm{no}=4m(m+1)$ & $\bar\mu$ \\ \hline
     4 &   64 &   80 & 1/10 \\
     8 & 256 & 288 & 1/20 \\
   16 & 1024 & 1088 & 1/40 \\ 
   32 & 4096 & 4224 & 1/80 \\ 
   64 & 16384 & 16640 & 1/160 \\
 128 & 65336 & 66048 & 1/320 \\ 
 256 & 262144 & 263168 & 1/640 \\ 
\hline
\end{tabular}
\caption{tube drawing: computational parameters (number of surface elements $n_\mathrm{el}$, number of control points $n_\mathrm{no}$ and stability parameter $\bar\mu$) for the convergence study of Fig.~\ref{f:tube_P}b.}
\label{t:tube_m}
\end{table}
For the convergence study, the following stabilization cases are considered: 
\begin{enumerate}
\item scheme `A-s' with fixed $\bar\mu=0.1$,
\item scheme `A-s' with varying $\bar\mu$, as specified in Tab.~\ref{t:tube_m},
\item scheme `P' using the solution of case 1 as initial guess,
\item scheme `P' using the solution of case 2 as initial guess,
\item no stabilization ($\mu=0$), using the solution of case 2 as initial guess.
\end{enumerate}
It turns out that scheme `P' does not at all improve the results obtained by scheme `A-s', probably due to the issue noted in the remark of Sec.~\ref{s:stab_P}.
Case 1 does not converge below an error of about $0.6\%$, which reflects the error caused by $\bar\mu=0.1$.
Case 5 works due to the inherent shear stiffness of the Helfrich model given in \eqref{e:mueff}.
For the cylindrical part, $\bar\mu_\mathrm{eff}=3H^2/2>0$. With $H=-1/(2a)$ follows $\bar\mu_\mathrm{eff}=3/(2a^2)$.
At the tip, the principal curvatures are equal ($\kappa_1=\kappa_2$), so that 
$\bar\mu_\mathrm{eff}=\kappa_1^2>0$. At the tip the curvature is almost twice as large as the cylinder curvature (i.e.~$\kappa_1\approx-2/a$), so that $\bar\mu_\mathrm{eff}\approx4k/a^2$.
In the cap, $\bar\mu_\mathrm{eff}$ ranges in between these two extreme values, and so $\mu_\mathrm{eff}$ is always positive.
The initial flat disk has $\mu_\mathrm{eff}=0$, so that stabilization is needed for the initial step. 
In all cases, the error reported in Fig.~\ref{f:tube_P}b is assessed by comparison to the finest mesh of case 2. 
From this we can find that the analytical solution itself has an error of about $0.2\%$, due to its underlying assumptions.




\subsection{Cell budding}\label{s:bud}

The last example considers the budding of spherical cells. The example is used to demonstrate the capabilities of the proposed computational model, in particular in the context of non-trivial and non-axisymmetric deformations.

\subsubsection{Motivation}

It is known that protein adsorption can mediate shape changes in biological membranes \citep{zimmerberg06, mcmahon05, kozlov14, shi15}. The lipid membrane deforms whenever its curvature is incompatible with the inherent structure of a protein, giving rise to a spontaneous curvature. Such protein-induced spontaneous curvature is common in many important biological phenomena such as endocytosis \citep{buser13, kukulski12, peter04}, cell motility \citep{keren11} and vesicle formation \citep{gruenberg04}. For example, in endocytosis, a clathrin protein coat is formed on the membrane which is incompatible with the membrane curvature, thus driving the out-of-plane shape changes, finally leading to fission. Moreover, another set of curvature generating proteins, the so-called BAR proteins \citep{peter04, kukulski12}, play an important role in modulating shape changes leading to fission in the later stages of endocytosis.

In this example, we use the proposed finite element formulation to study the shape changes in membranes due to protein-induced spontaneous curvature. This spontaneous curvature usually depends on the concentration of proteins adsorbed onto the membrane. In obtaining the shapes, we assume that any proteins that are bound to the membrane do not diffuse and are concentrated in a specific region \citep{walther06, agrawal09, karotki11, kishimoto11, rangamani14}. Such processes are common in all endocytosis related phenomena, where it is known that the clathrin protein coat does not diffuse in the membrane once adsorbed. Therefore, our example aims at helping to understand the shapes arising in the early stages of endocytosis.

\subsubsection{Computational setup}

\begin{figure}[h]
\begin{center} \unitlength1cm
\begin{picture}(0,4.7)
\put(-8.7,-.3){\includegraphics[height=50mm]{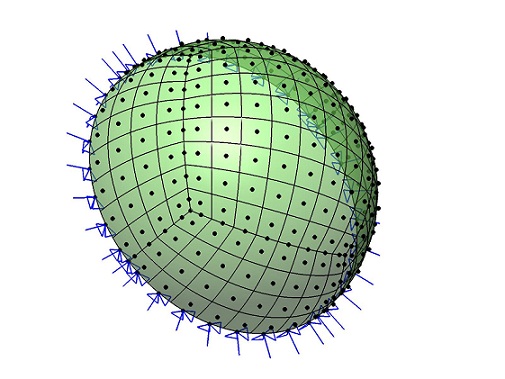}}
\put(-3.2,-.3){\includegraphics[height=49mm]{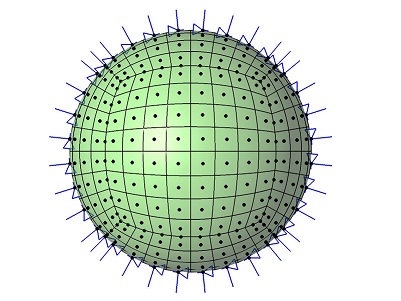}}
\put(2.4,-.3){\includegraphics[height=49mm]{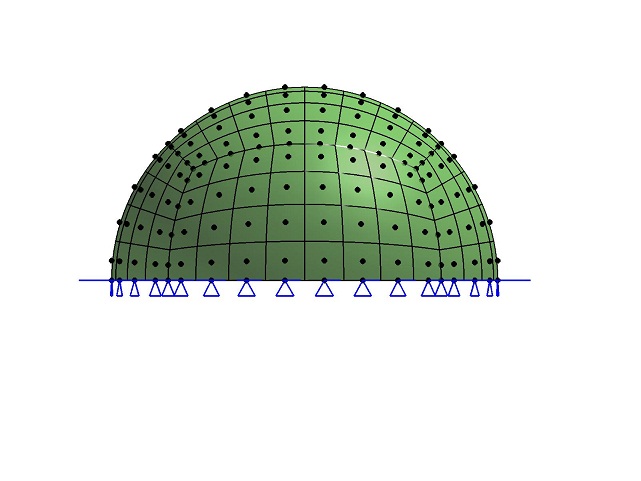}}
\end{picture}
\caption{Cell budding: initial configuration, FE discretization and boundary conditions. The boundary normal is fixed and the boundary nodes are only free to move in the radial direction.}
\label{f:bud_0}
\end{center}
\end{figure}

For our example, we consider a hemi-spherical cell with initial radius $R$; the cell surface is clamped at the boundary, but free to expand radially as is shown in Fig.~\ref{f:bud_0}.
On the tip of the cell, within the circular region of radius $0.2R$, a constant spontaneous curvature $\bar H_0$ is prescribed.
Unless otherwise specified, model (\ref{e:W_c}) is used with the material parameters 
$\bar k^\ast=-0.7$ and $\bar K=10,\!000$, while $k$ and $R$ are used for normalization according to Sec.~\ref{s:norm} and remain unspecified.
Further, stabilization scheme `A-s' is used with $\bar\mu=0.01$.
The FE discretization shown in Fig.~\ref{f:bud_0}, consisting of five NURBS patches, is used. Where the patches meet, constraint \eqref{e:Pin} is added to ensure rotational continuity and moment transfer. Constraint \eqref{e:Pin} is also used to fix the boundary normal.
The actual FE mesh is much finer than in Fig.~\ref{f:bud_0} and uses 12228 elements (64 times more than in the figure).  
The penalty parameter of the rotational constraint is $\bar\epsilon=6,\!400$.
Gaussian quadrature is considered, using $3\times3$ points for surface integrals and 4 points for line integrals.

\subsubsection{Bud shapes}

In past numerical studies, axisymmetric bud shapes have been reported, e.g.~\citet{walani15}. These shapes should be a natural solution due to the axisymmetry of the problem. However, as is shown below, non-axisymmetric solutions are also possible, and in fact energetically favorable, indicating that axisymmetric solutions can become unfavored.
\begin{table}[h]
\centering
\begin{tabular}{|l|l|l|l|l|l|}
  \hline
  case & bud shape & $H_0$ region & stabilization & $\bar\mu$ & in-plane stress state \\ \hline
     1 & axisymmetric & circle & A-s & 0.01 & hydro-static \\
     2 & general & circle & A-s & 0.01 & hydro-static \\
     3 & general & ellipse & A-s & 0.01 & hydro-static \\ 
     4 & general & ellipse & A-st & 10 & elastic shear \\ 
     5 & general & ellipse & a-st & 1250 & viscous shear \\
\hline
\end{tabular}
\caption{Cell budding: different physical test cases considered.}
\label{t:bud_cases}
\end{table}
This is illustrated by considering the five different physical test cases shown in Tab.~\ref{t:bud_cases} and discussed in the following:

\textbf{Case 1}: The deformation is constrained to remain axisymmetric (the FE nodes are only allowed to move in radial direction\footnote{For the given mesh, this does not enforce axisymmetry exactly as a close inspection of the results shows.}). The resulting deformation for $\bar H_0=-25$ is shown in Fig.~\ref{f:bud_A} and in the supplemental movie file \verb"bud1.mpg".
\begin{figure}[h]
\begin{center} \unitlength1cm
\begin{picture}(0,5.6)
\put(-8.9,-.4){\includegraphics[height=31mm]{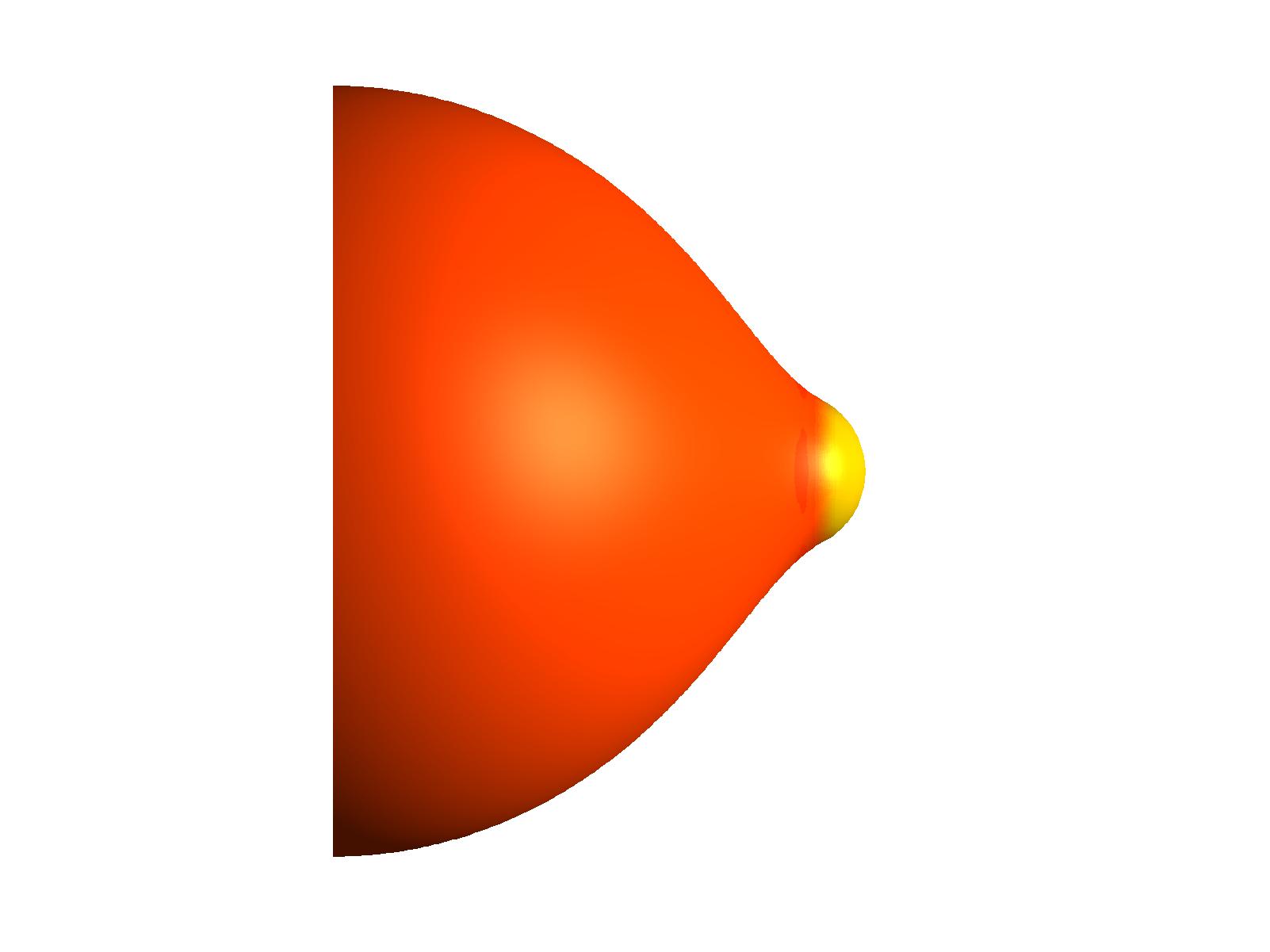}}
\put(-5.65,-.4){\includegraphics[height=31mm]{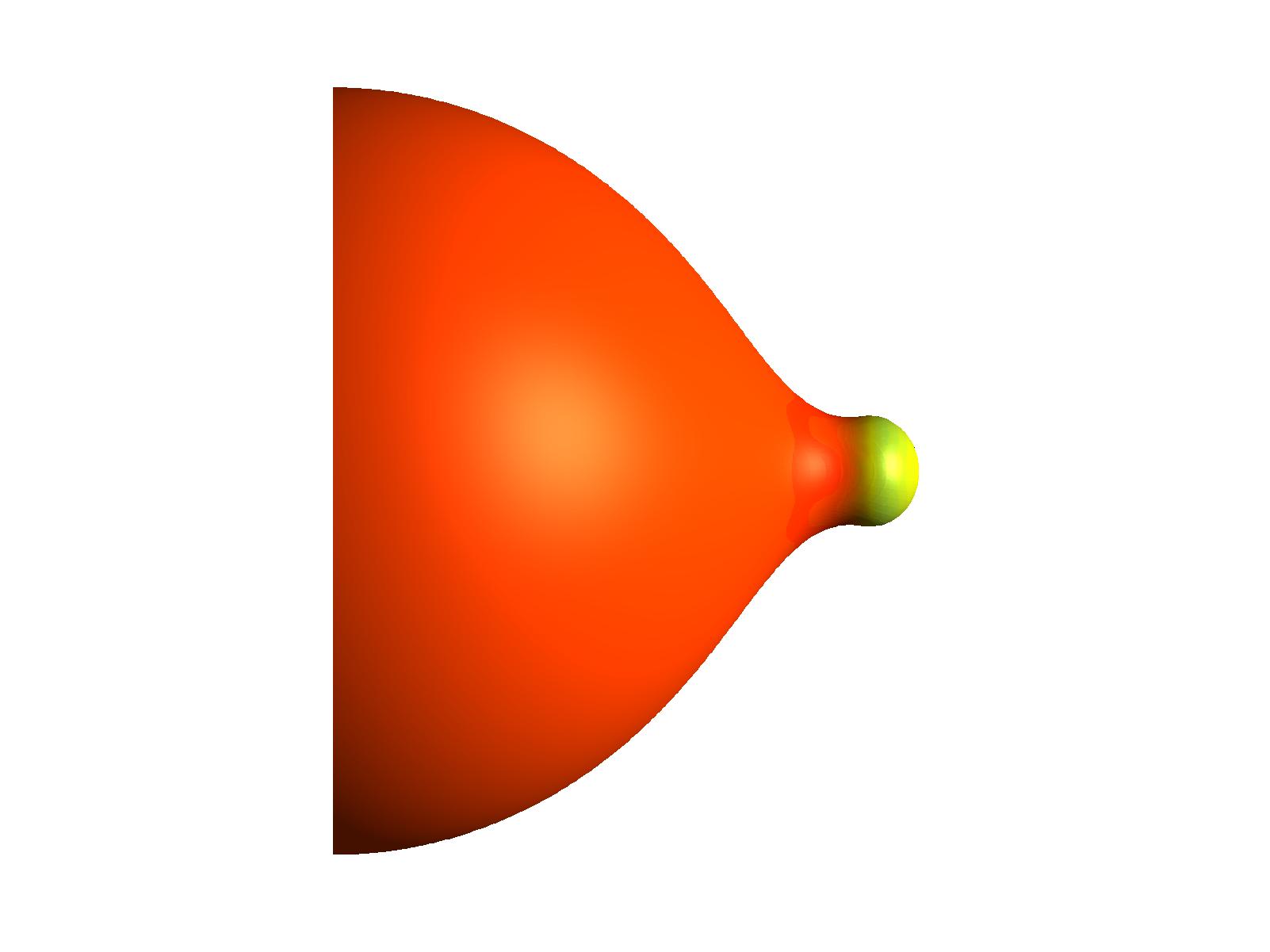}}
\put(-2.4,-.4){\includegraphics[height=31mm]{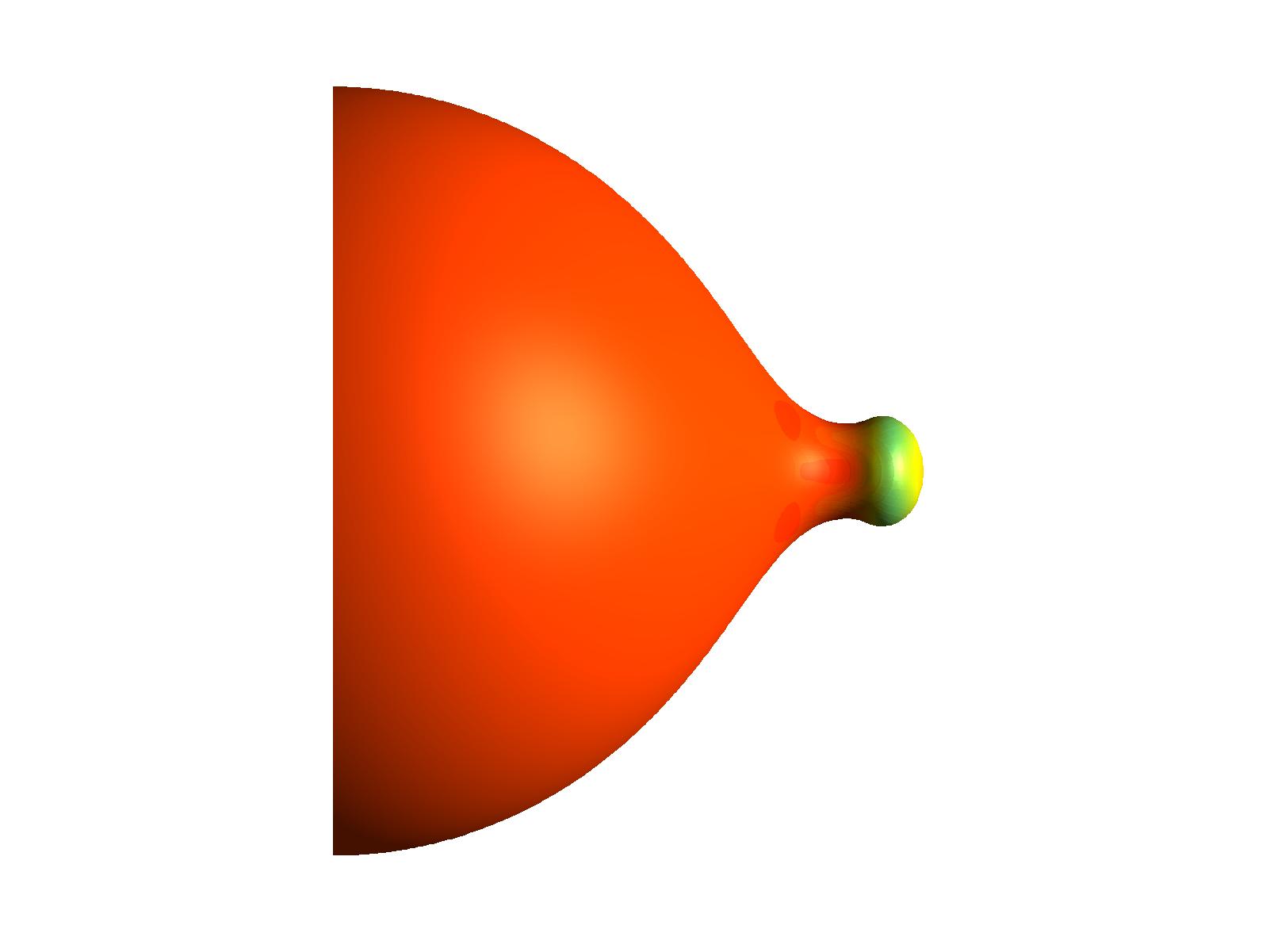}}
\put(0.85,-.4){\includegraphics[height=31mm]{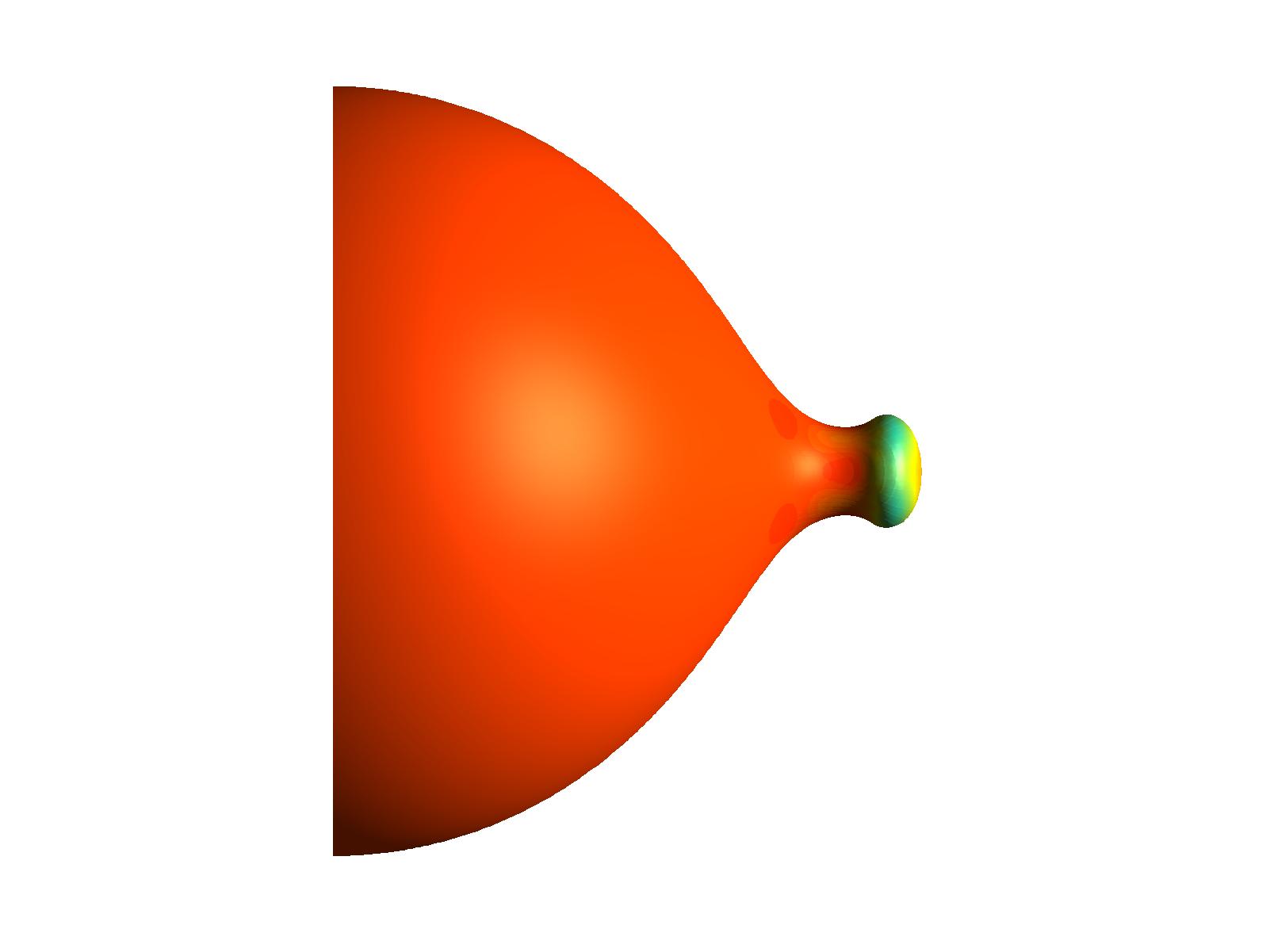}}
\put(4.05,-.4){\includegraphics[height=31mm]{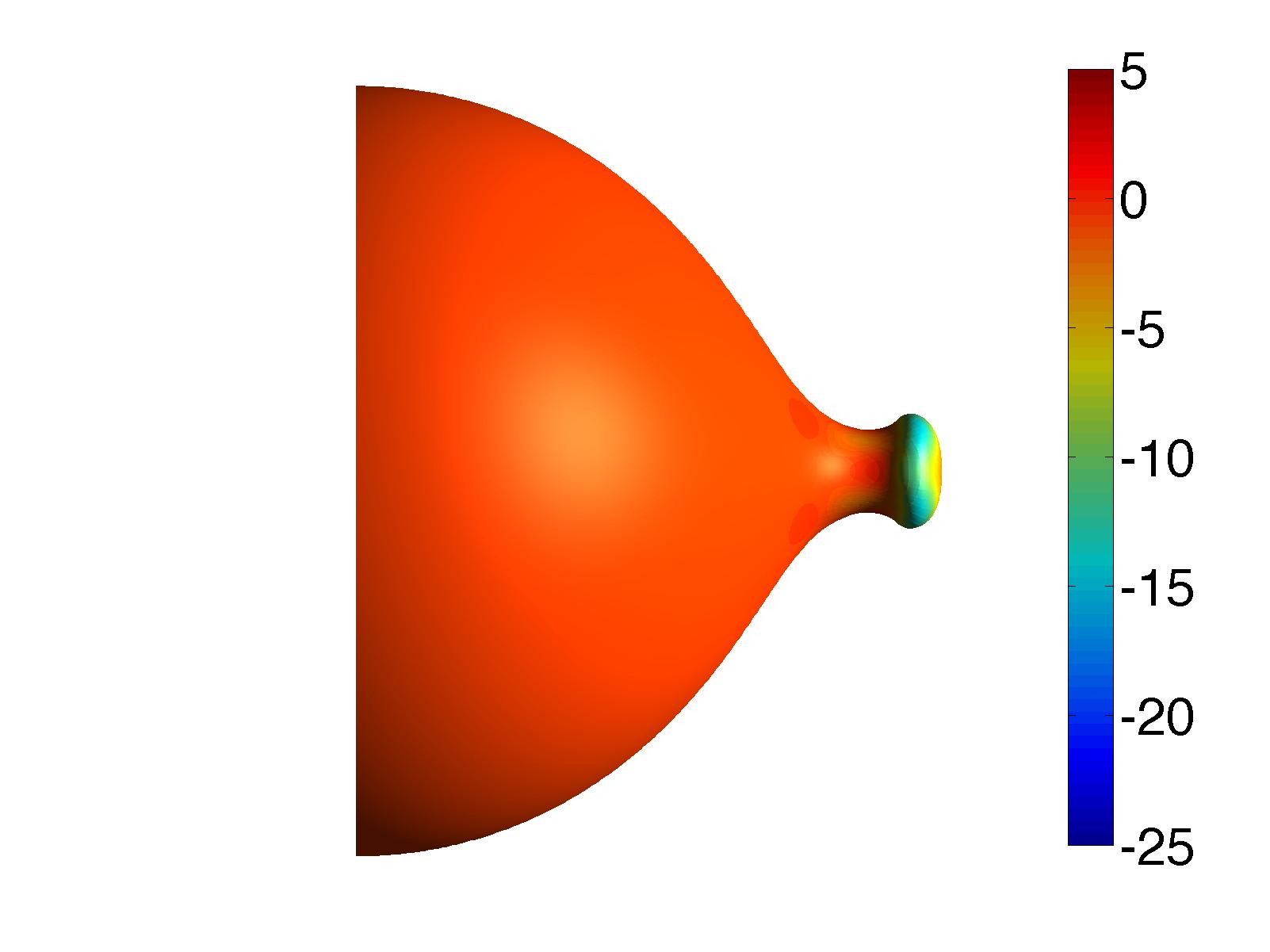}}
\put(-8.5,2.6){\includegraphics[height=31mm]{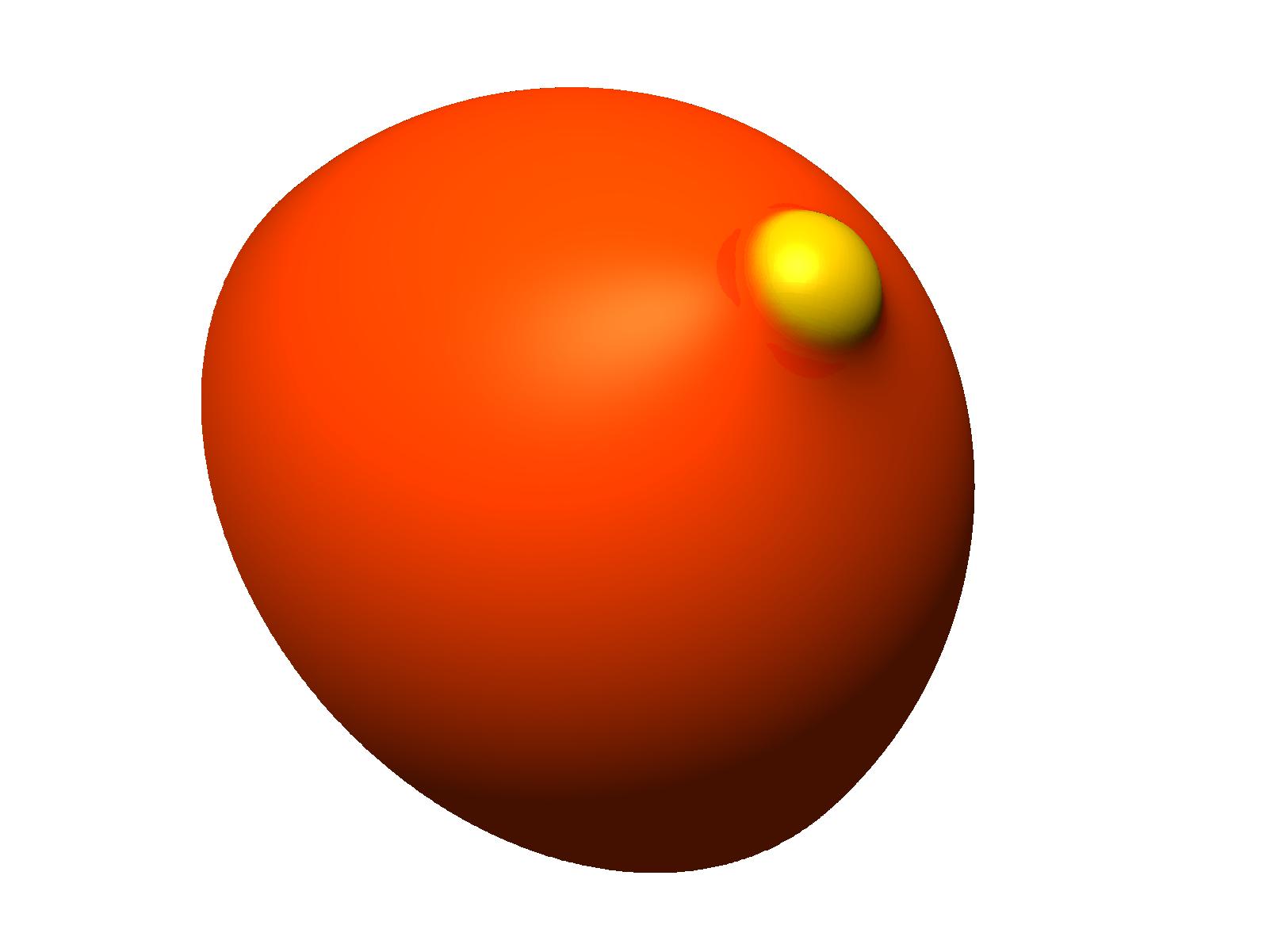}}
\put(-5.25,2.6){\includegraphics[height=31mm]{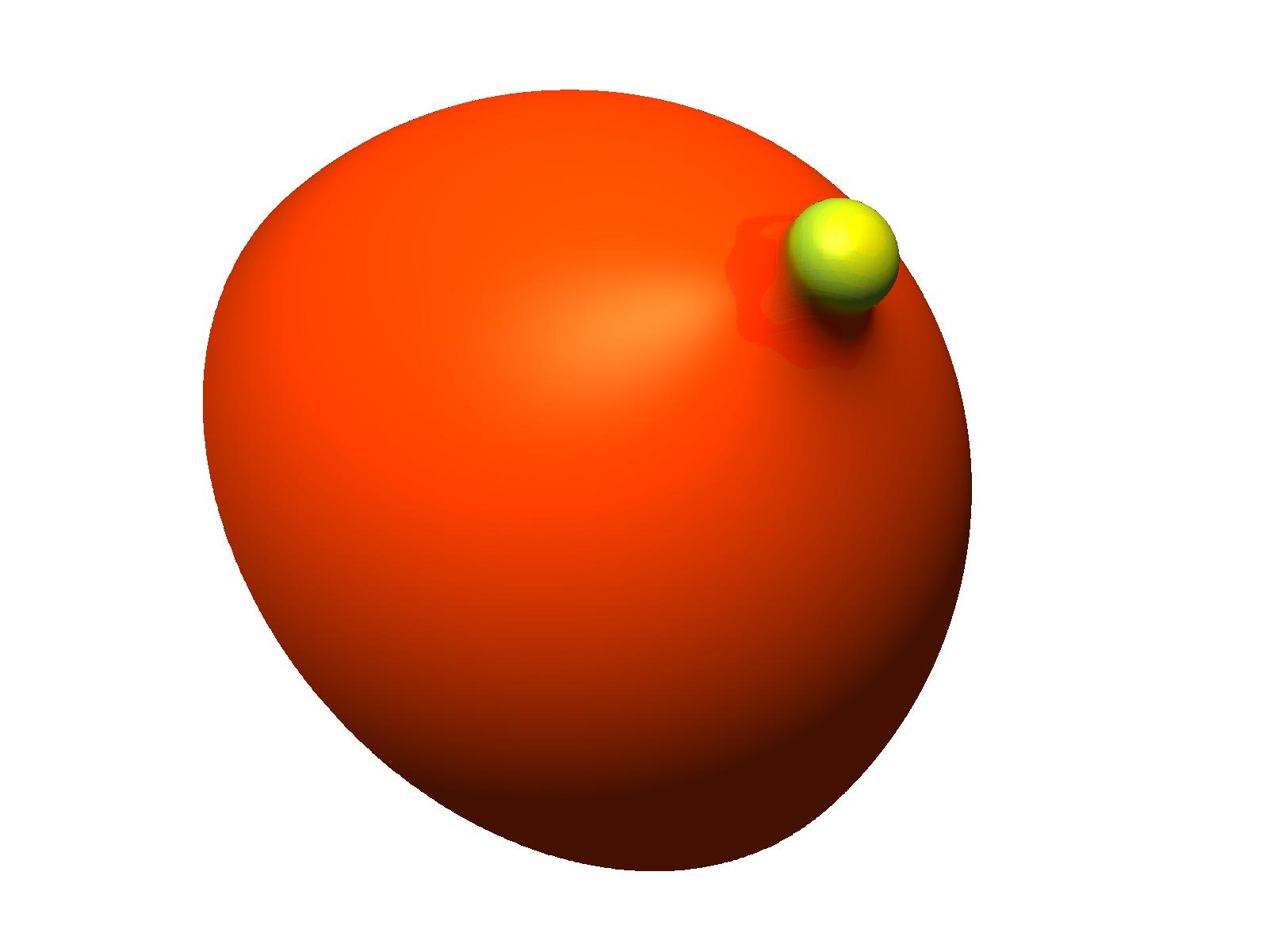}}
\put(-2,2.6){\includegraphics[height=31mm]{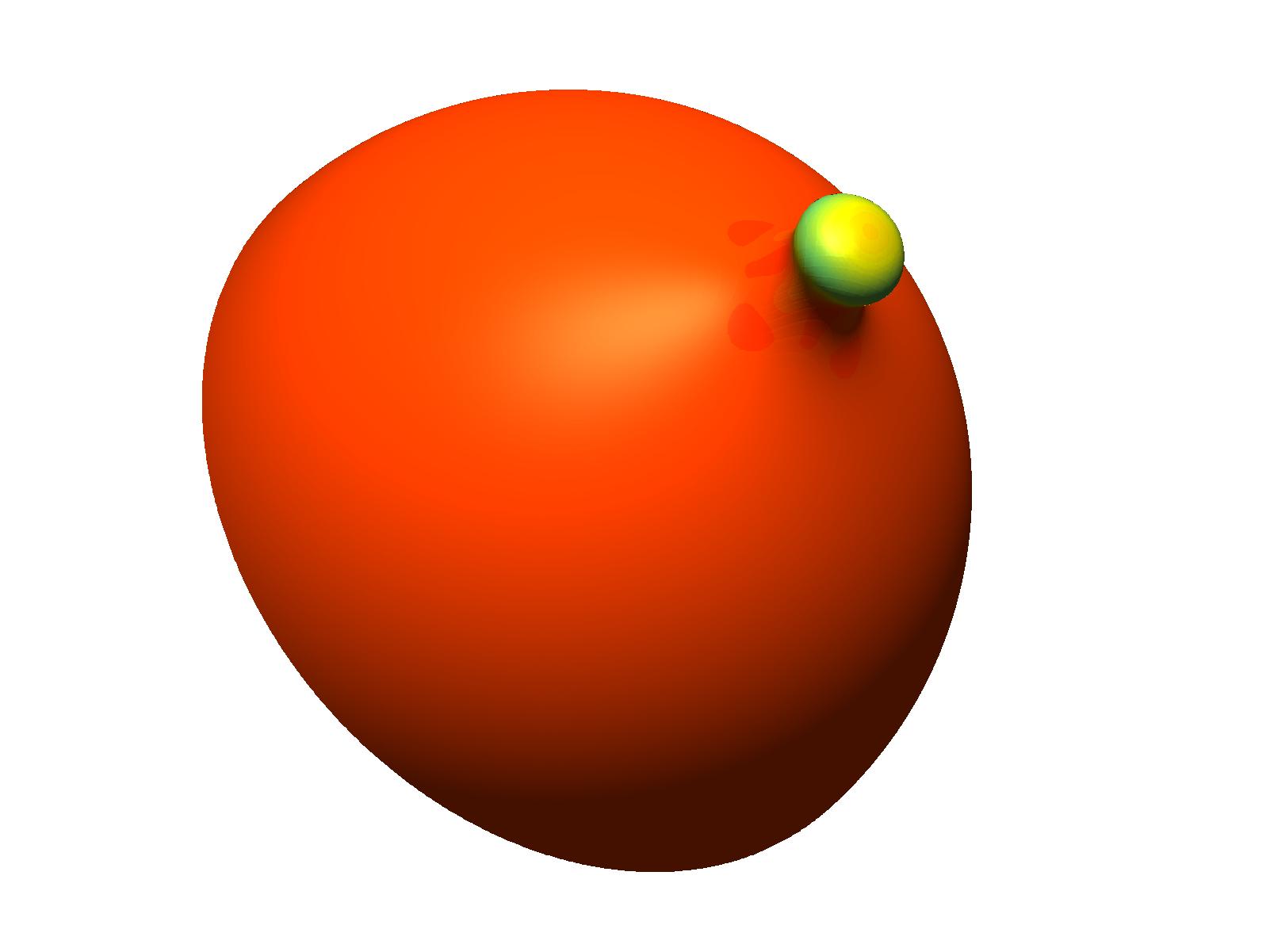}}
\put(1.25,2.6){\includegraphics[height=31mm]{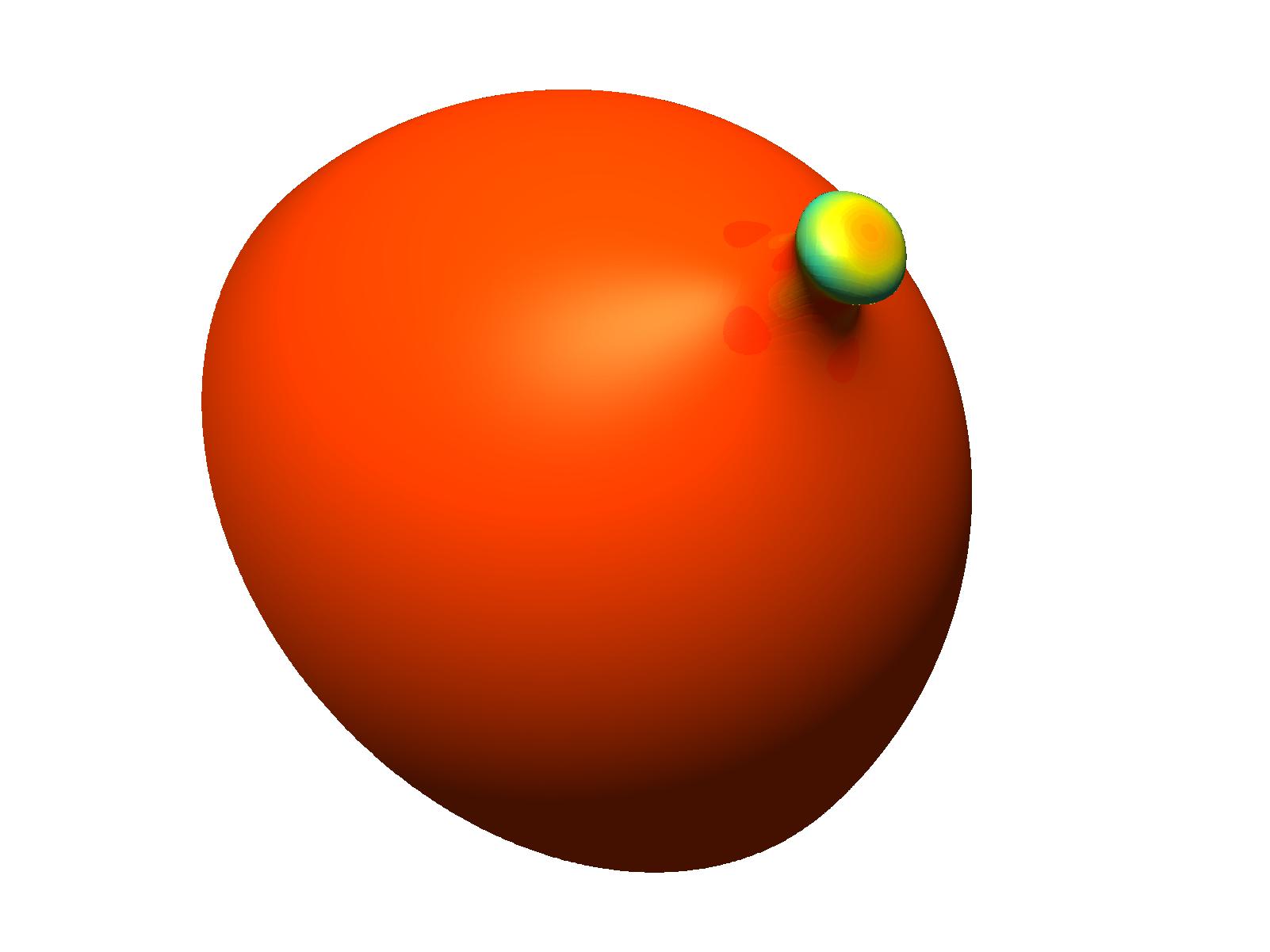}}
\put(4.5,2.6){\includegraphics[height=31mm]{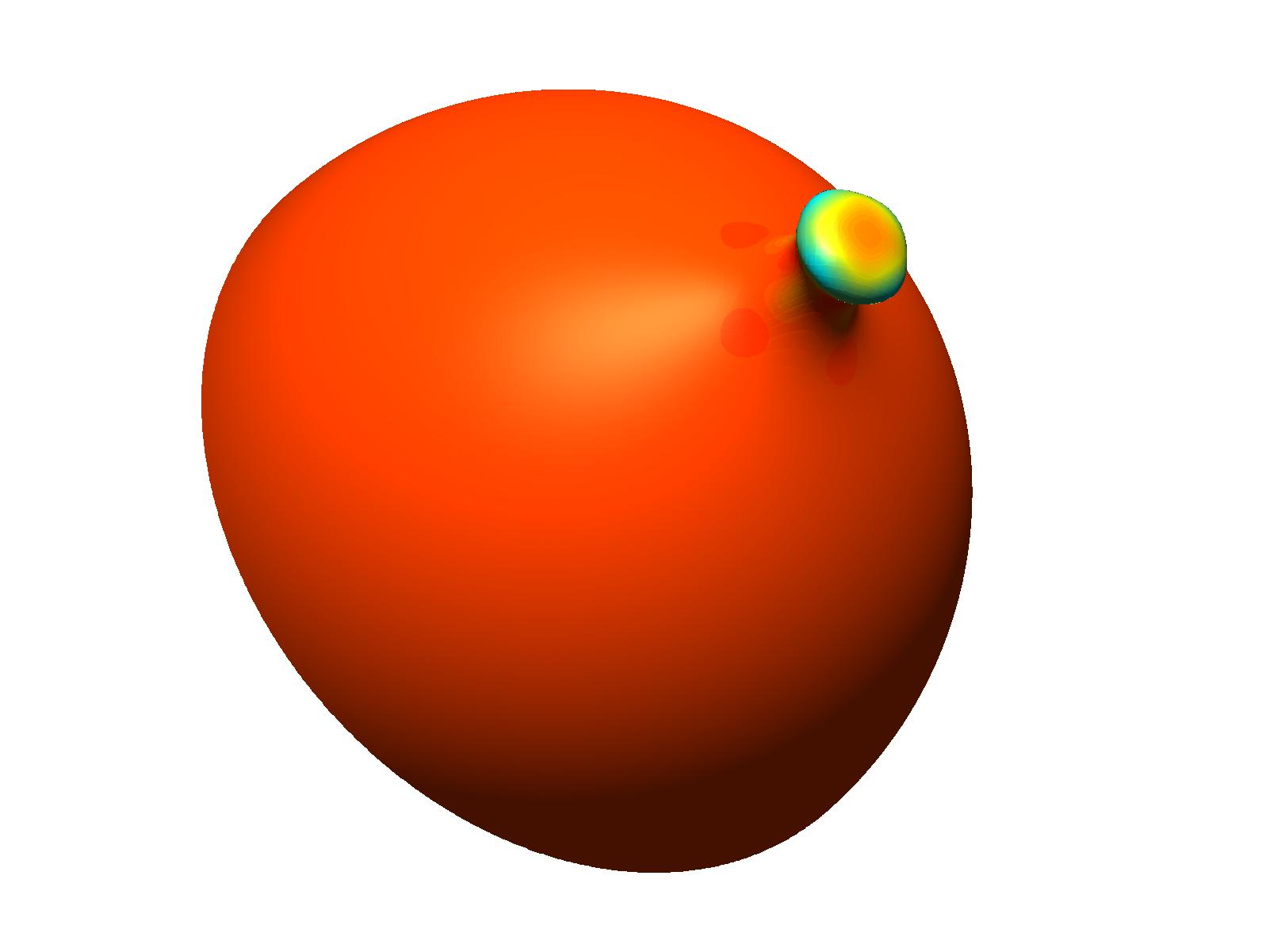}}
\end{picture}
\caption{Cell budding: Axisymmetric case at $\bar H_0=-5,-10,-15,-20,-25$ (left to right): 3D and side view of deformation and curvature $\bar H$. Here $\bar H\in[-15.0,\,0.31]$.}
\label{f:bud_A}
\end{center}
\end{figure}

\begin{figure}[h]
\begin{center} \unitlength1cm
\begin{picture}(0,5.6)
\put(-8.9,-.4){\includegraphics[height=31mm]{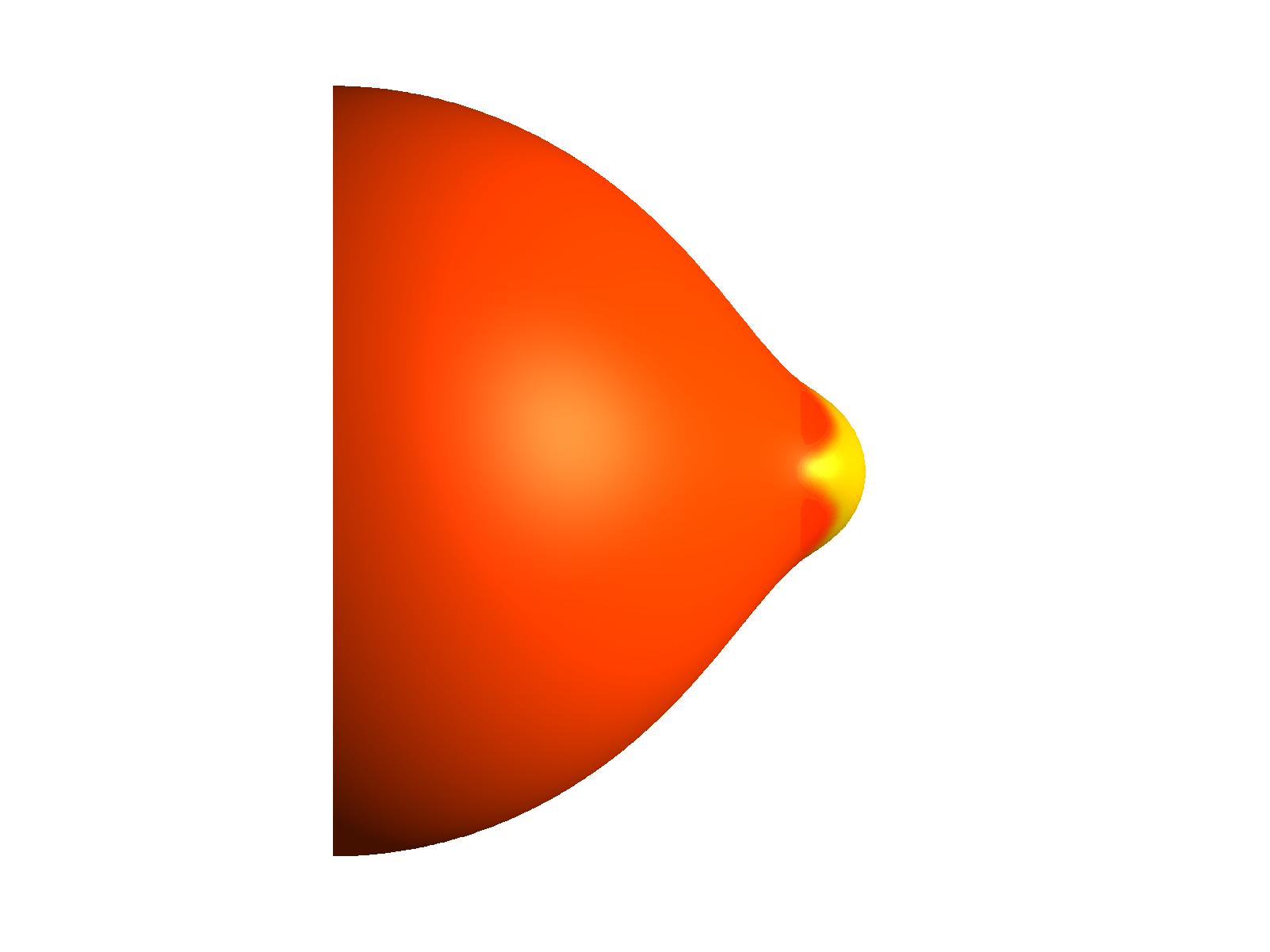}}
\put(-5.65,-.4){\includegraphics[height=31mm]{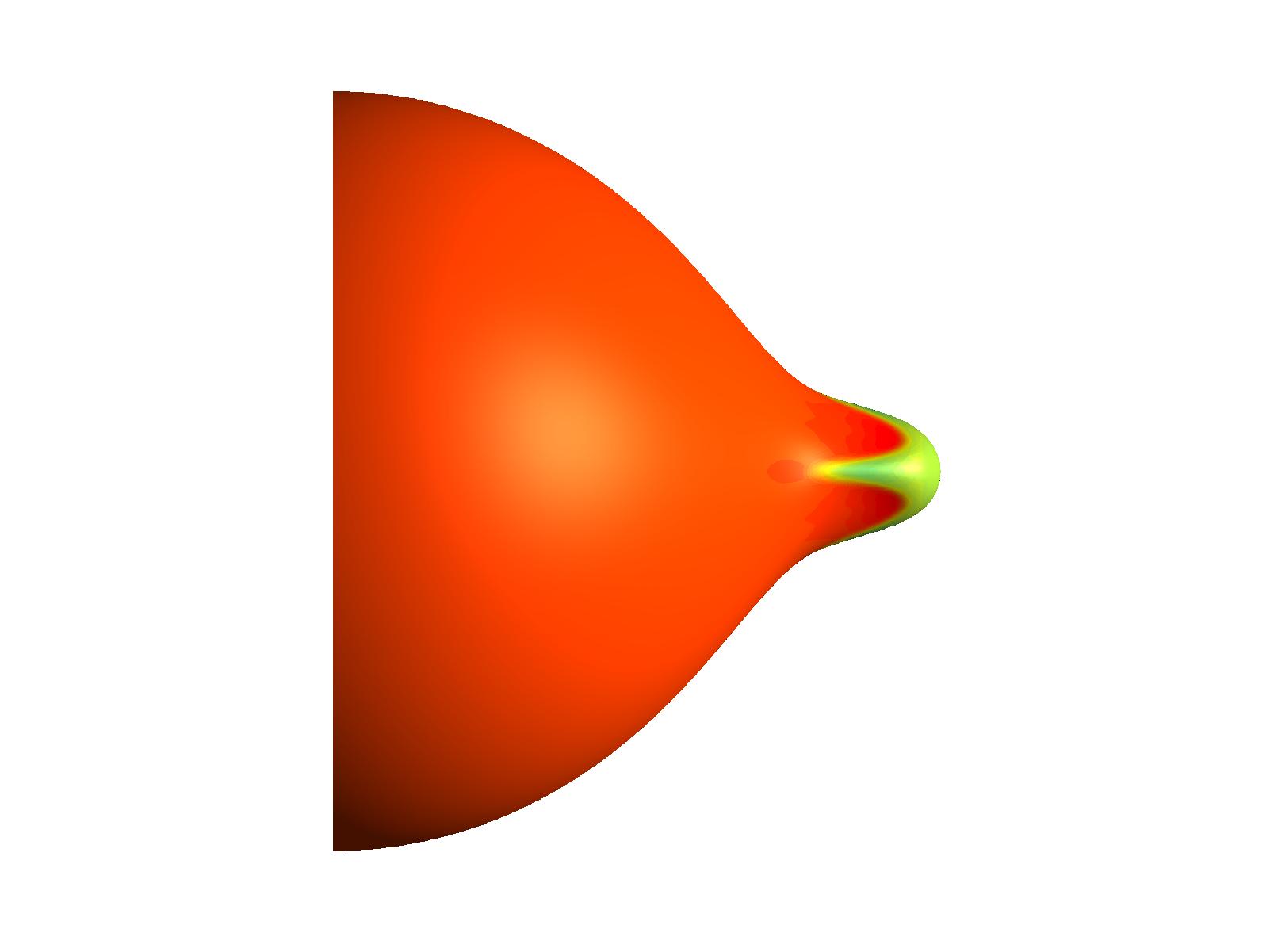}}
\put(-2.4,-.4){\includegraphics[height=31mm]{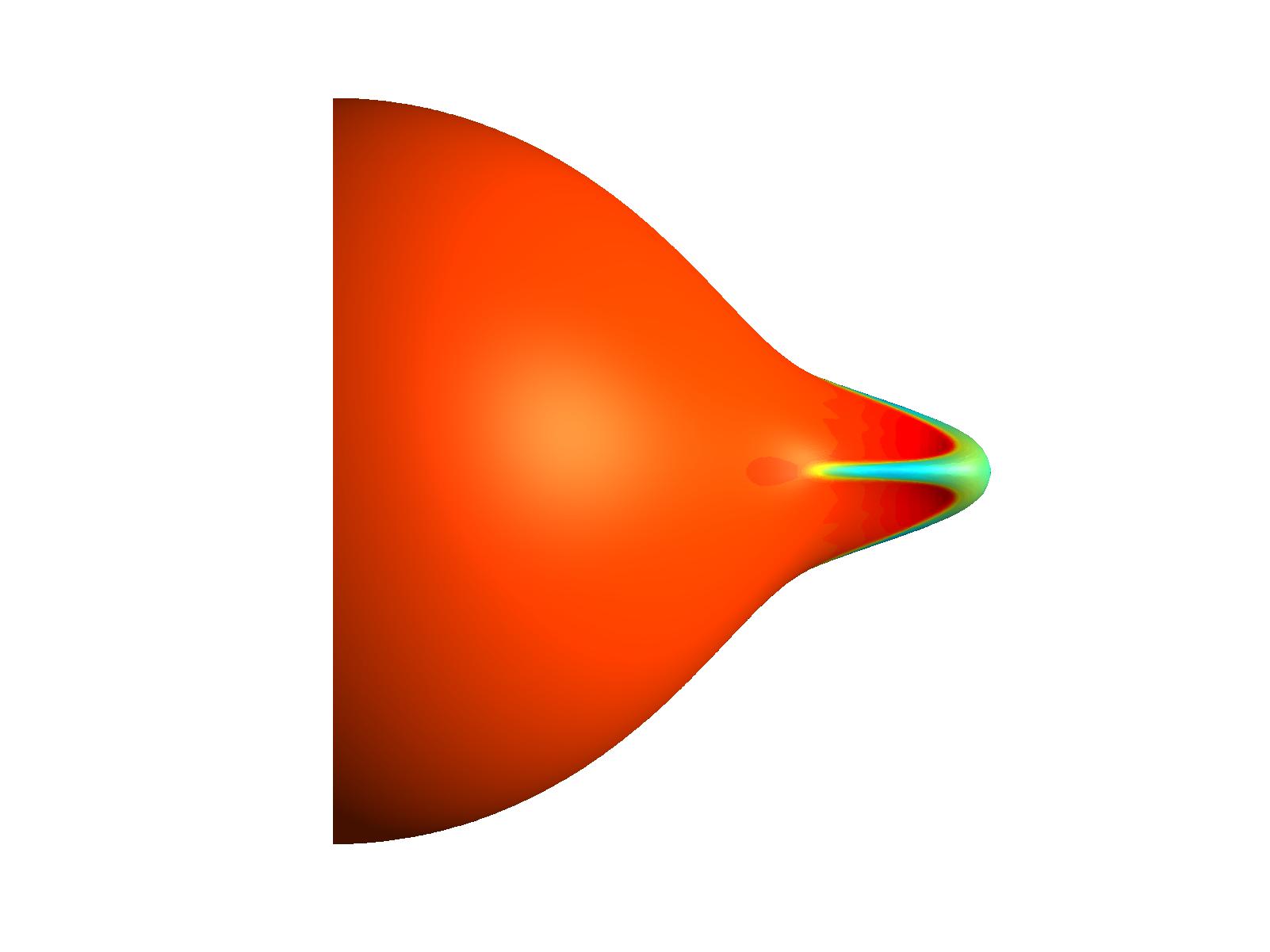}}
\put(4.05,-.4){\includegraphics[height=31mm]{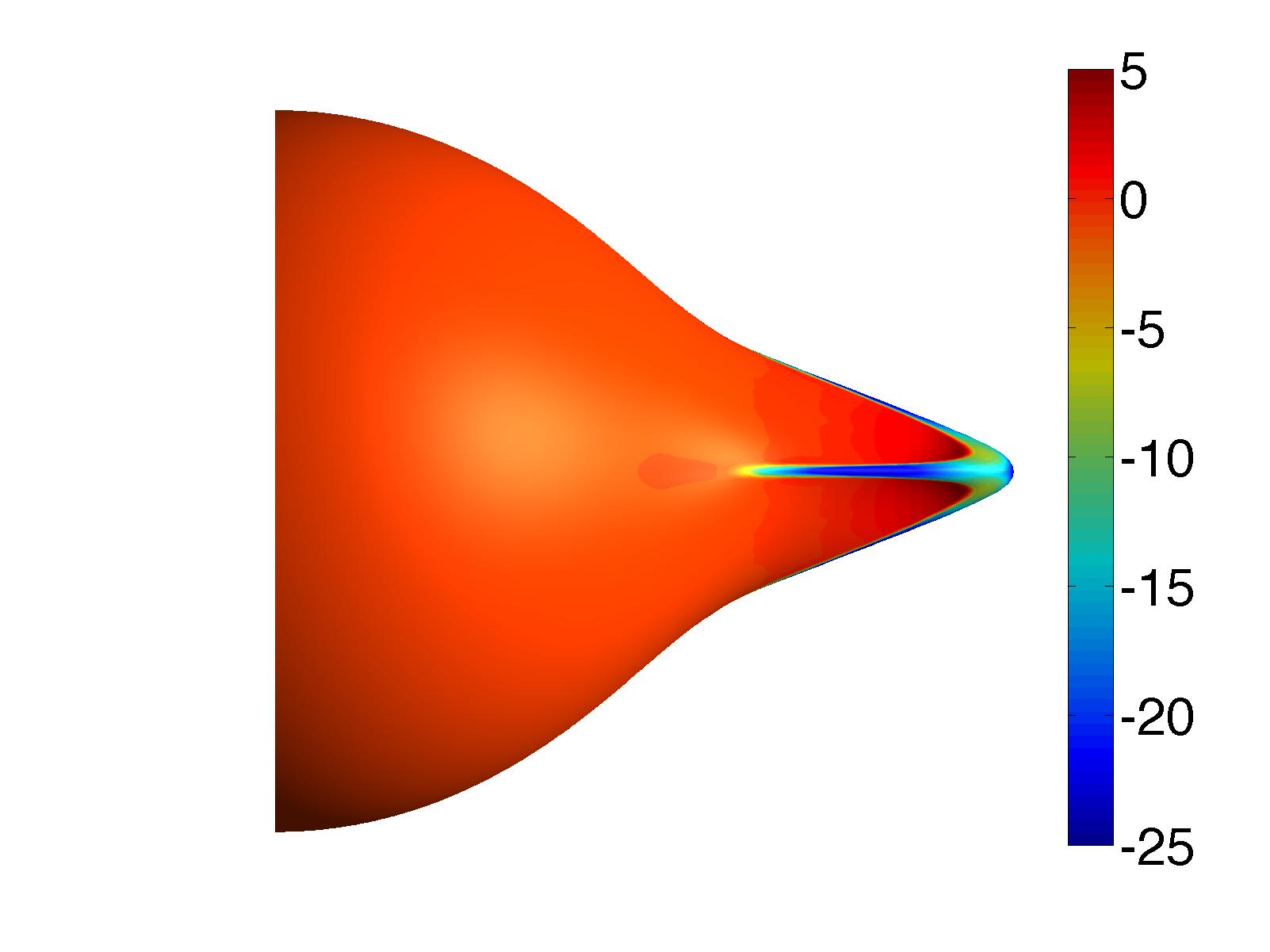}}
\put(0.85,-.4){\includegraphics[height=31mm]{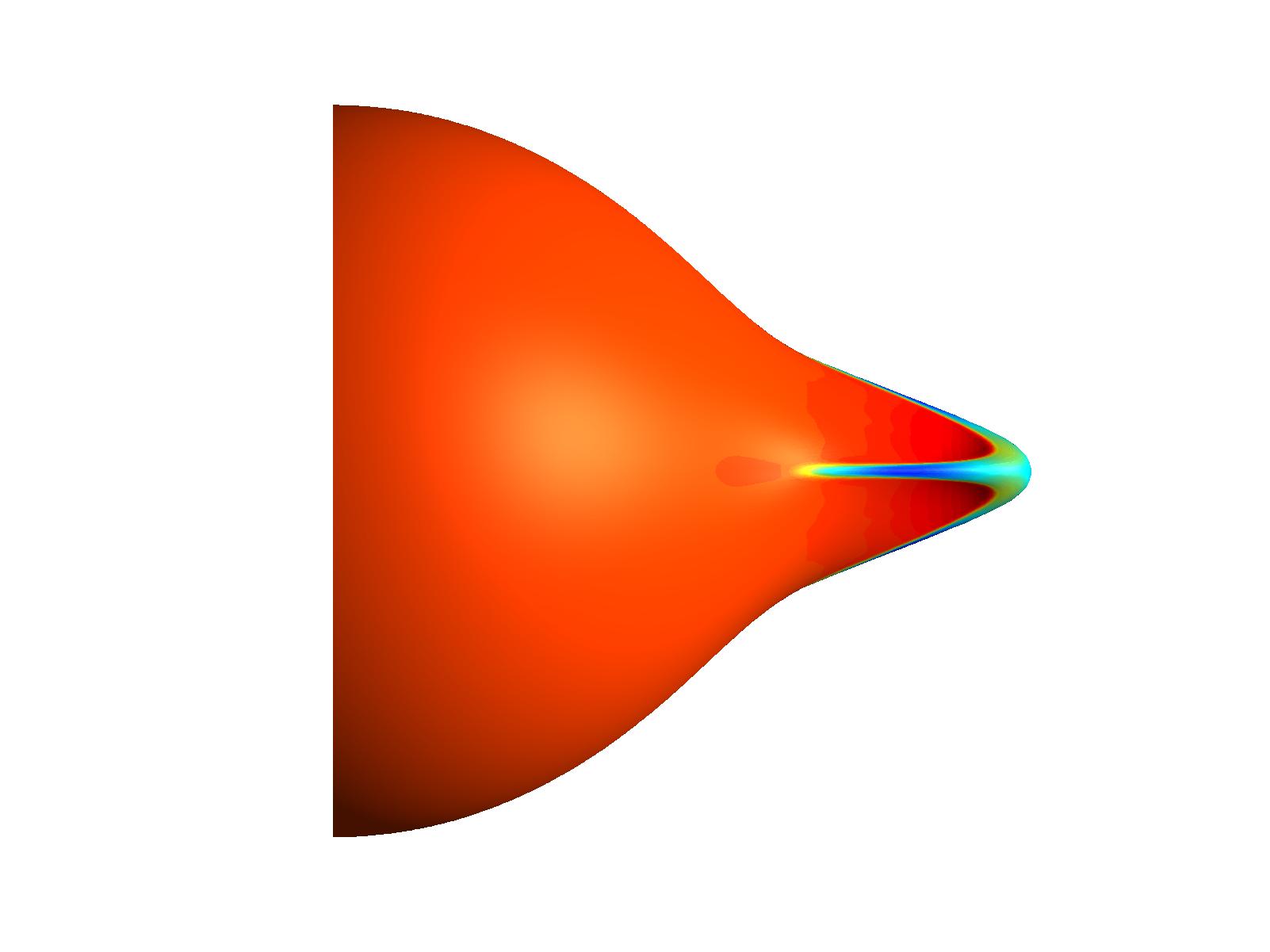}}
\put(-8.5,2.6){\includegraphics[height=31mm]{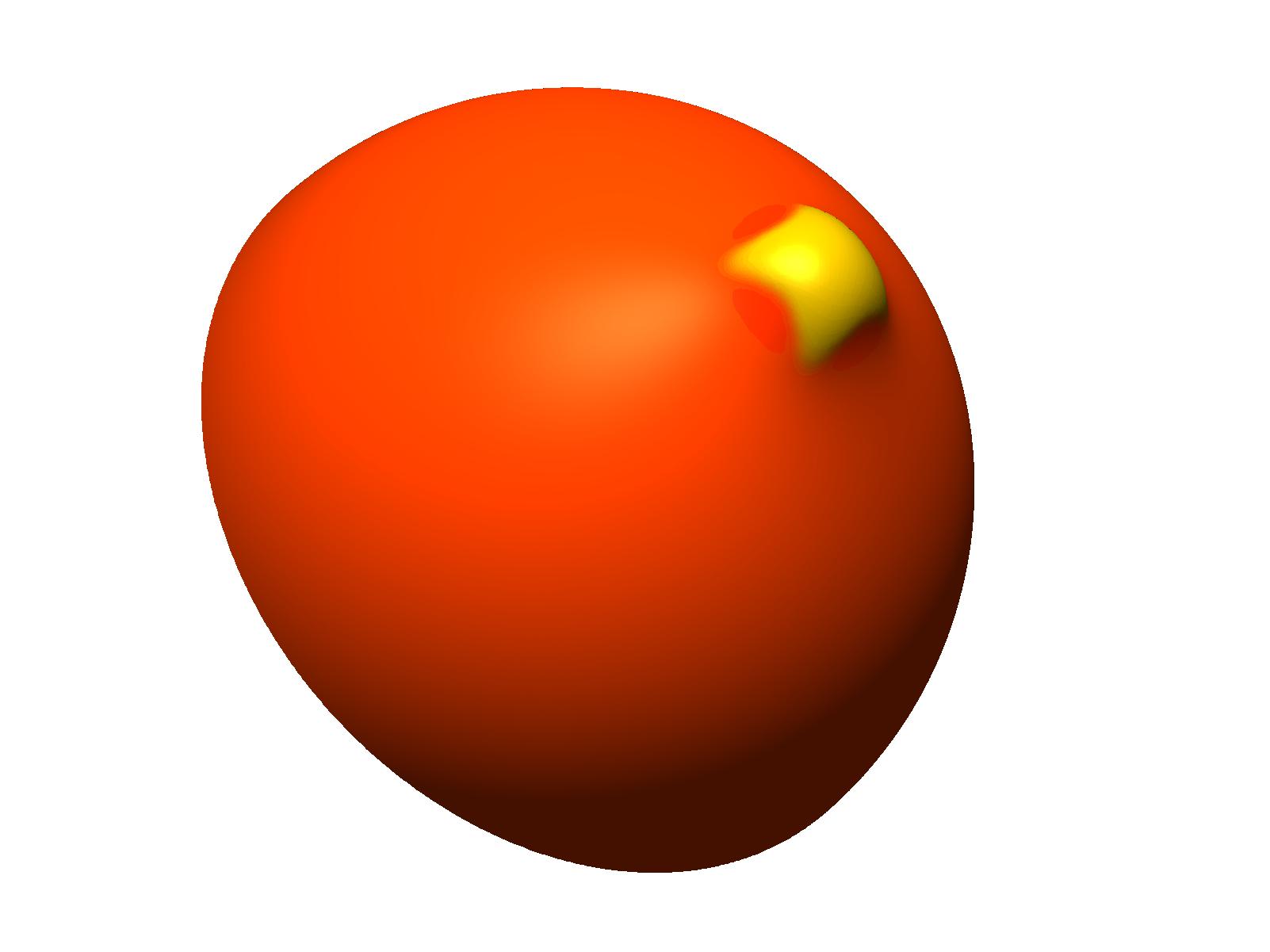}}
\put(-5.25,2.6){\includegraphics[height=31mm]{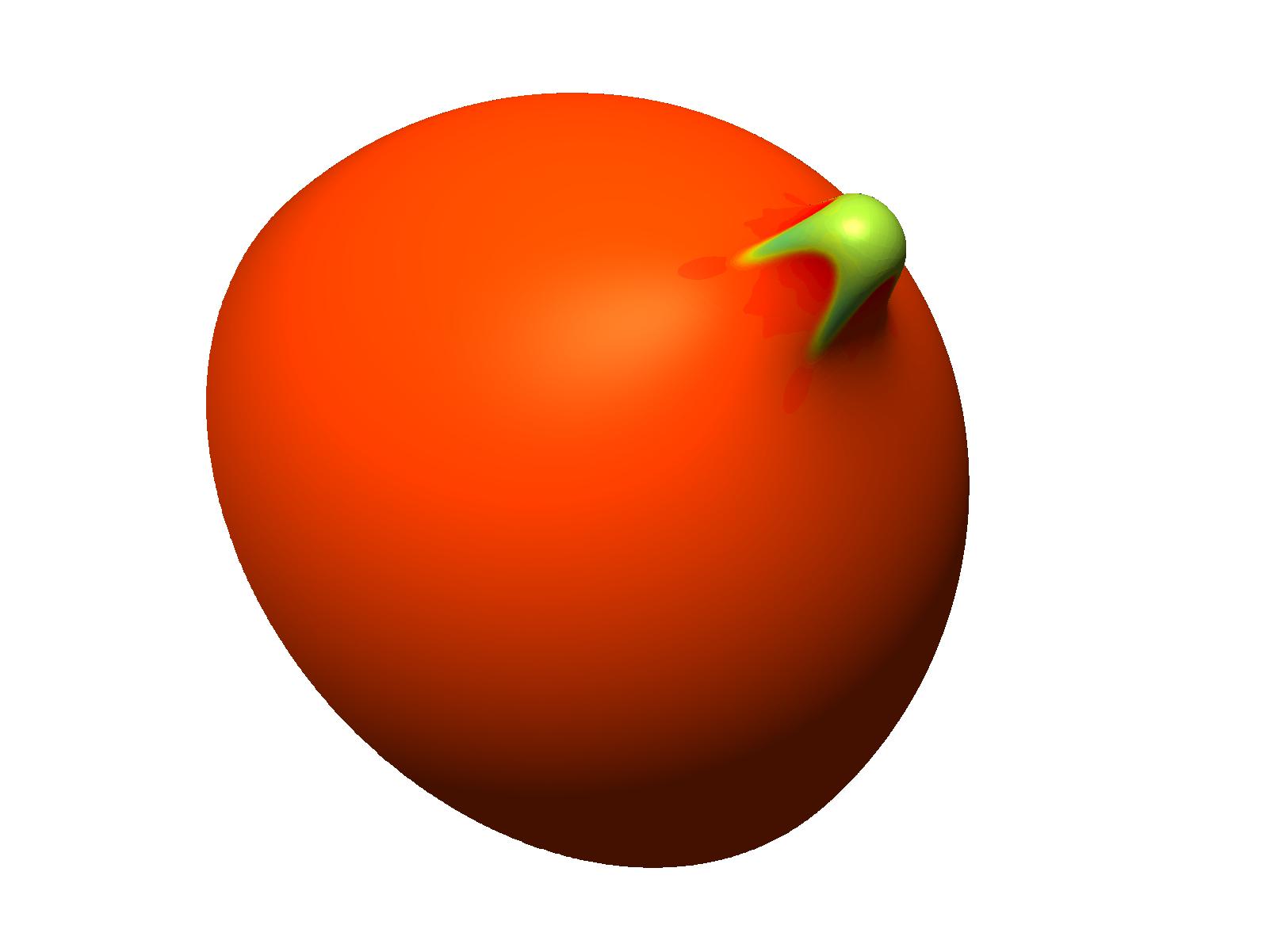}}
\put(-2,2.6){\includegraphics[height=31mm]{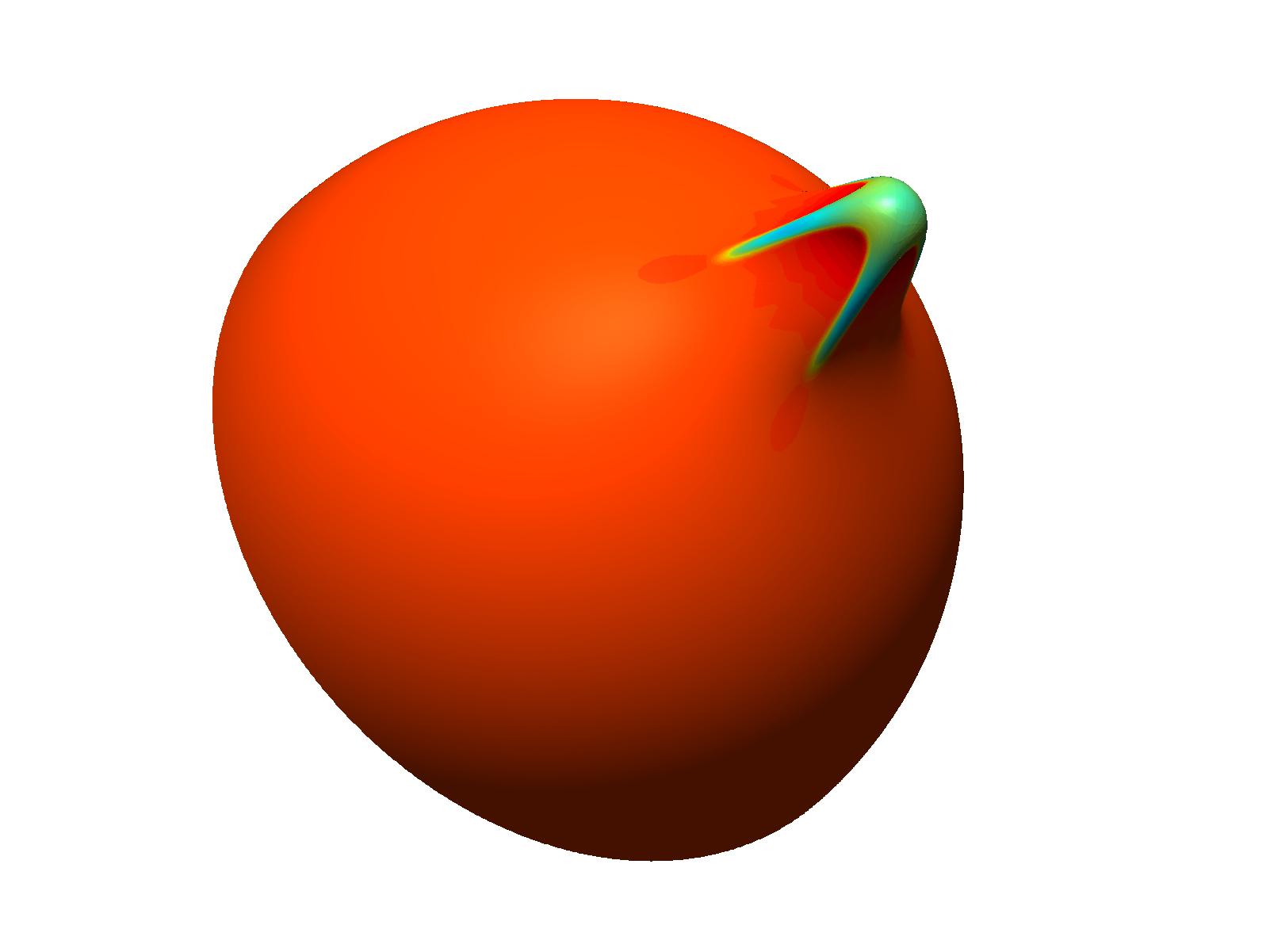}}
\put(1.25,2.6){\includegraphics[height=31mm]{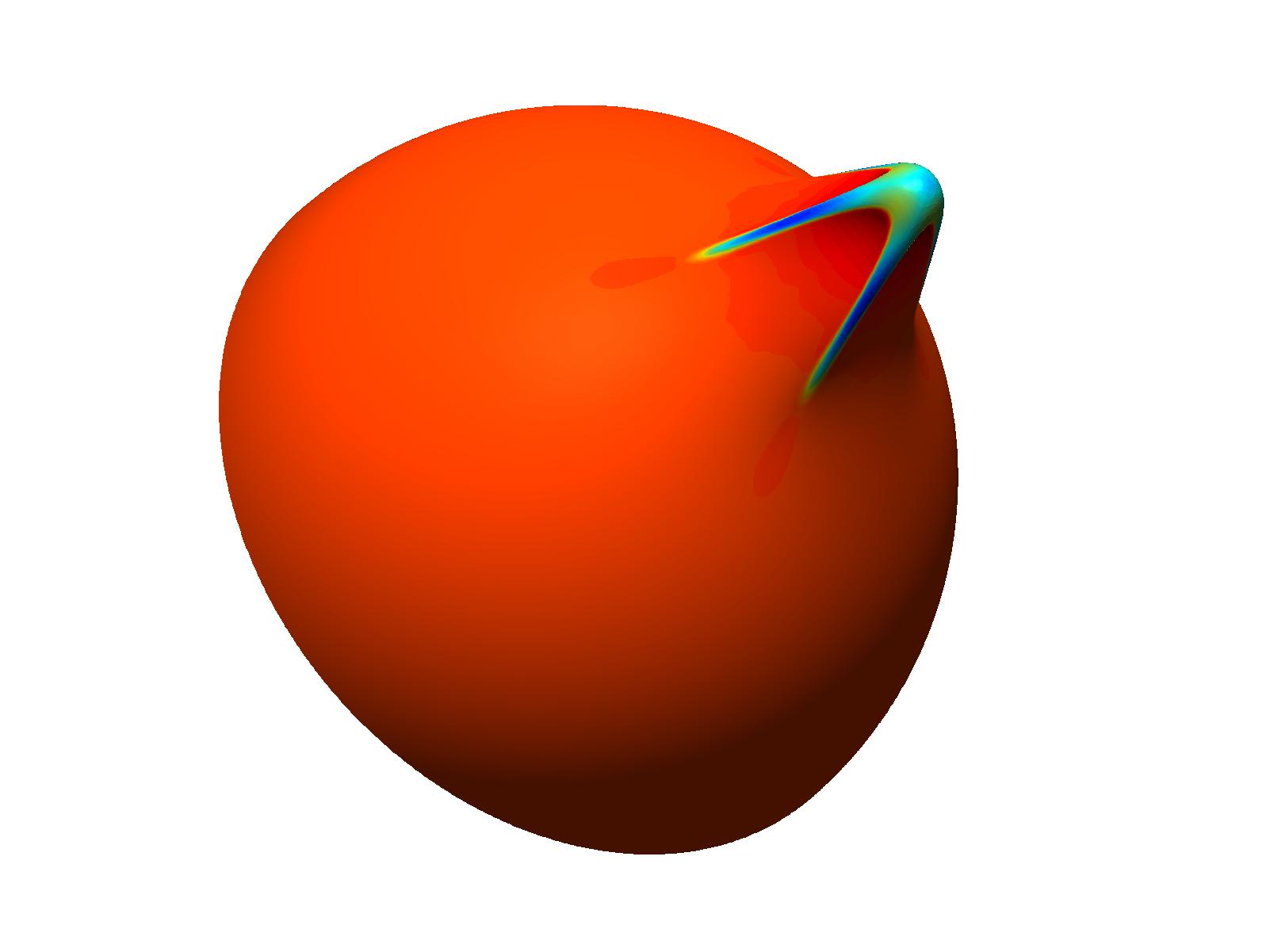}}
\put(4.5,2.6){\includegraphics[height=31mm]{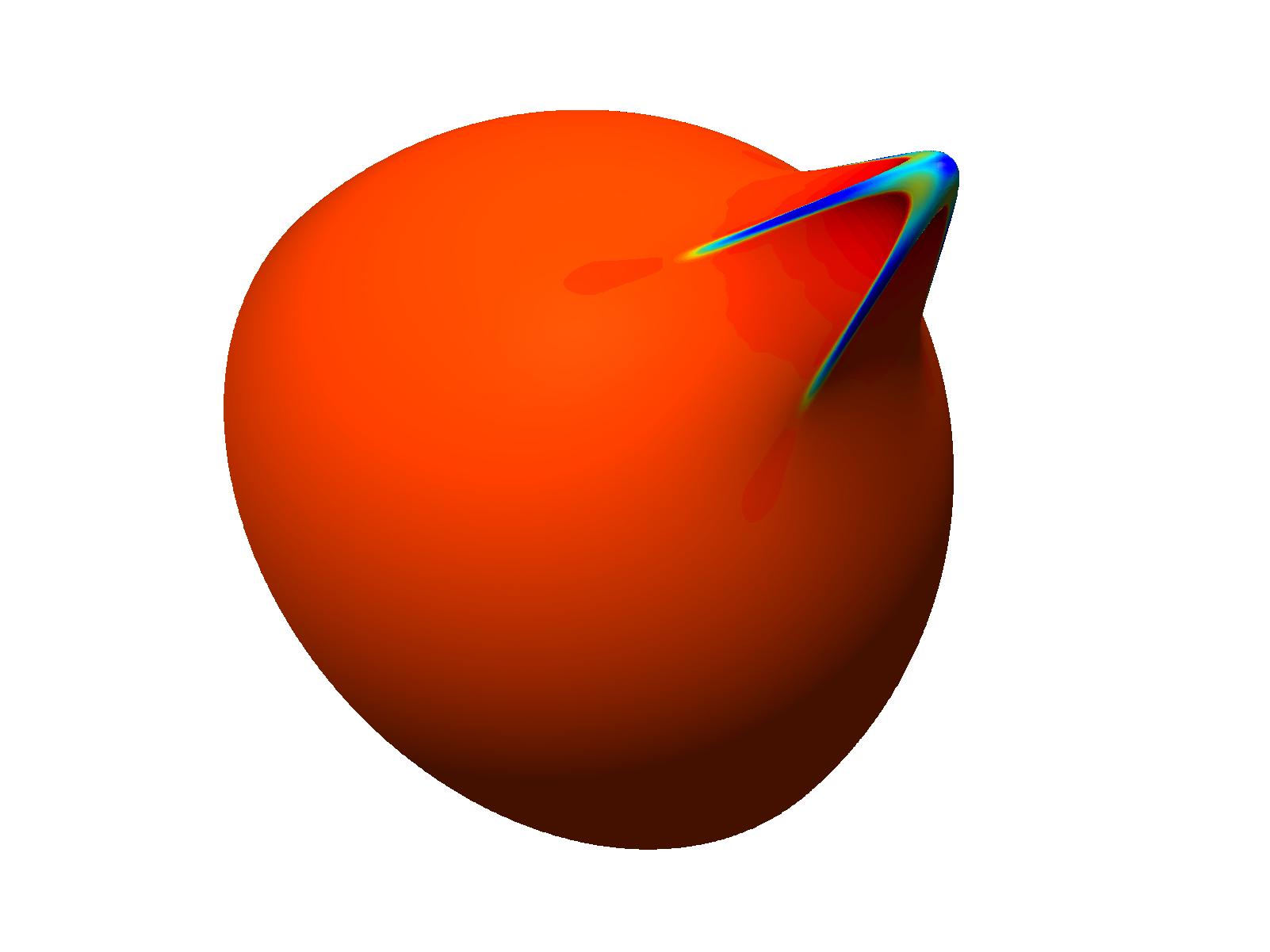}}
\end{picture}
\caption{Cell budding: Unconstrained, perfect case at $\bar H_0=-5,-10,-15,-20,-23.17$ (left to right): 3D and side view of deformation and curvature $\bar H$. Here $\bar H\in[-24.5,\,6.12]$.}
\label{f:bud_C}
\end{center}
\end{figure}

\begin{figure}[h]
\begin{center} \unitlength1cm
\begin{picture}(0,5.6)
\put(0.9,-.6){\includegraphics[height=64mm]{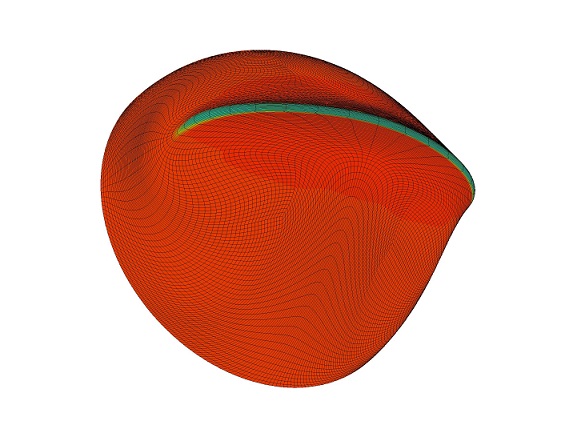}}
\put(-8.9,-.4){\includegraphics[height=31mm]{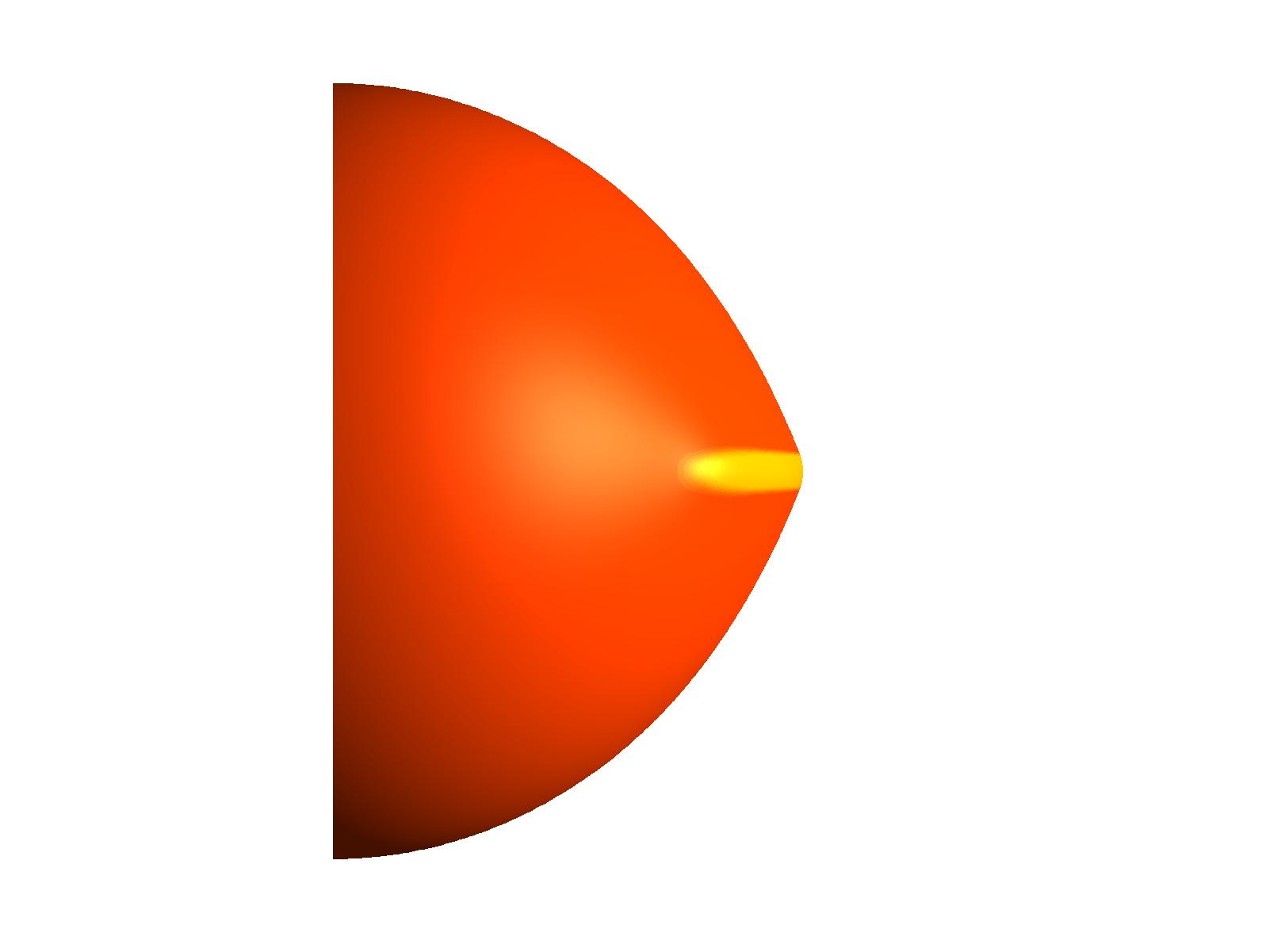}}
\put(-5.65,-.4){\includegraphics[height=31mm]{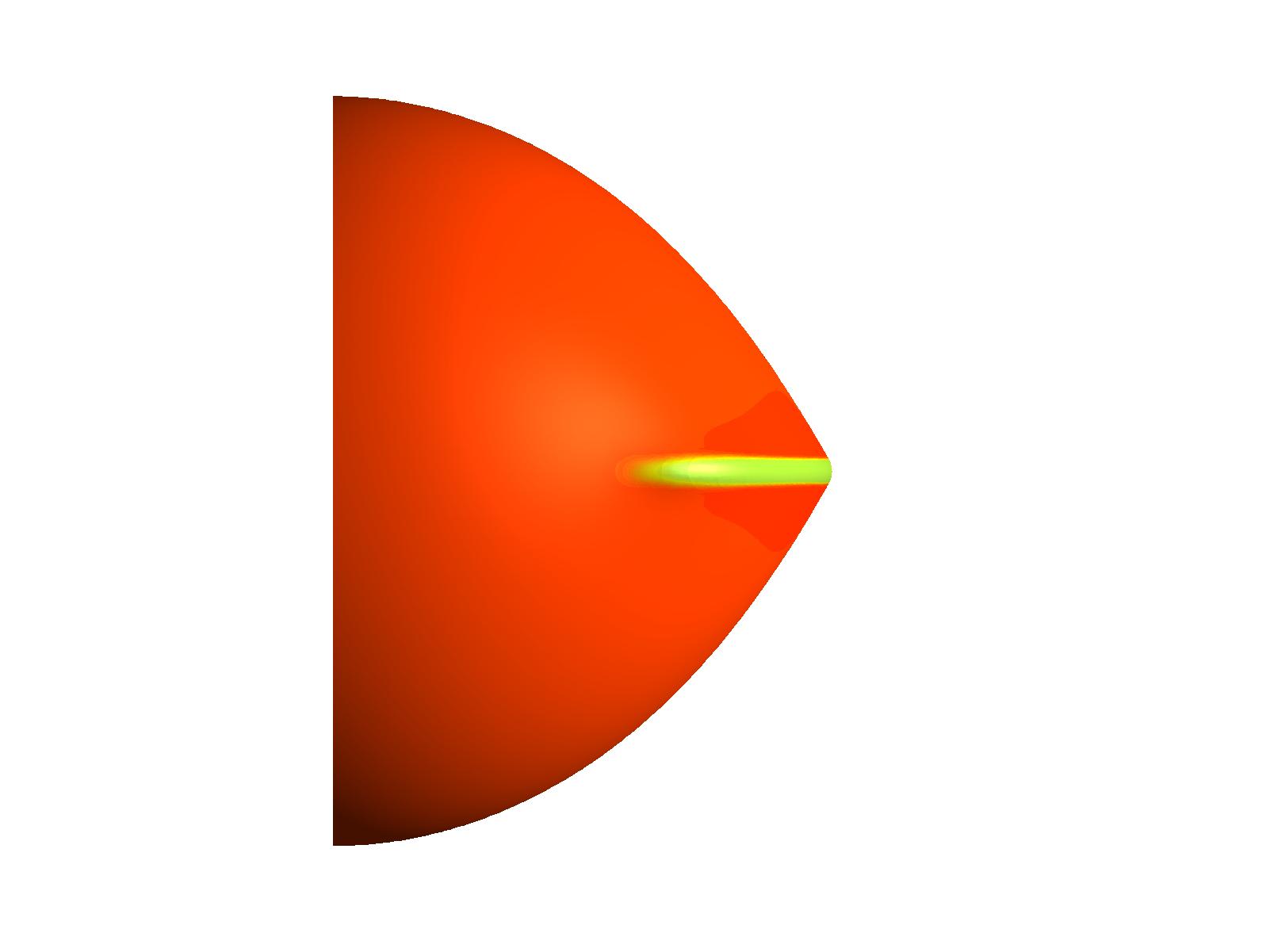}}
\put(-2.3,-.4){\includegraphics[height=31mm]{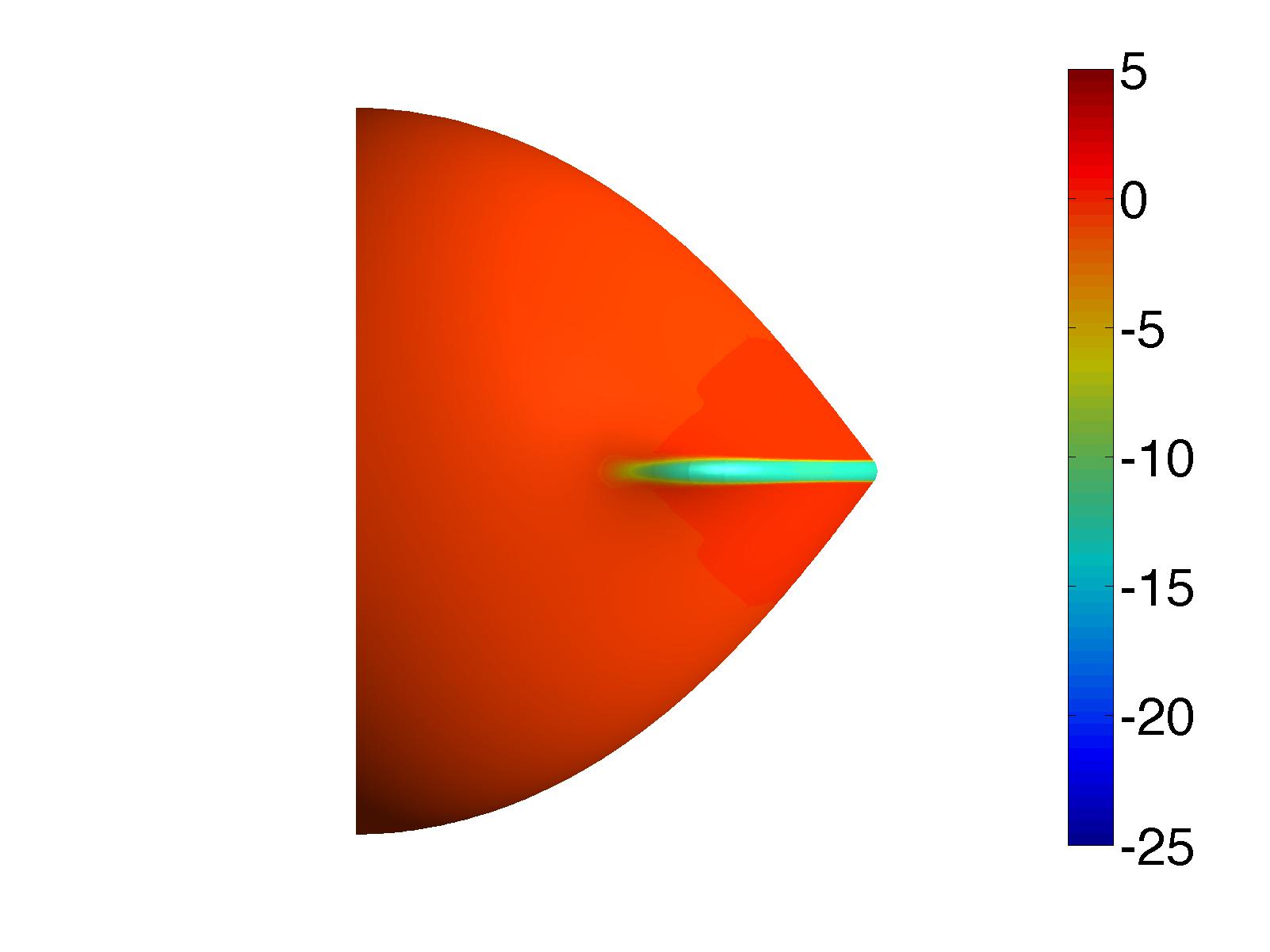}}
\put(-8.5,2.6){\includegraphics[height=31mm]{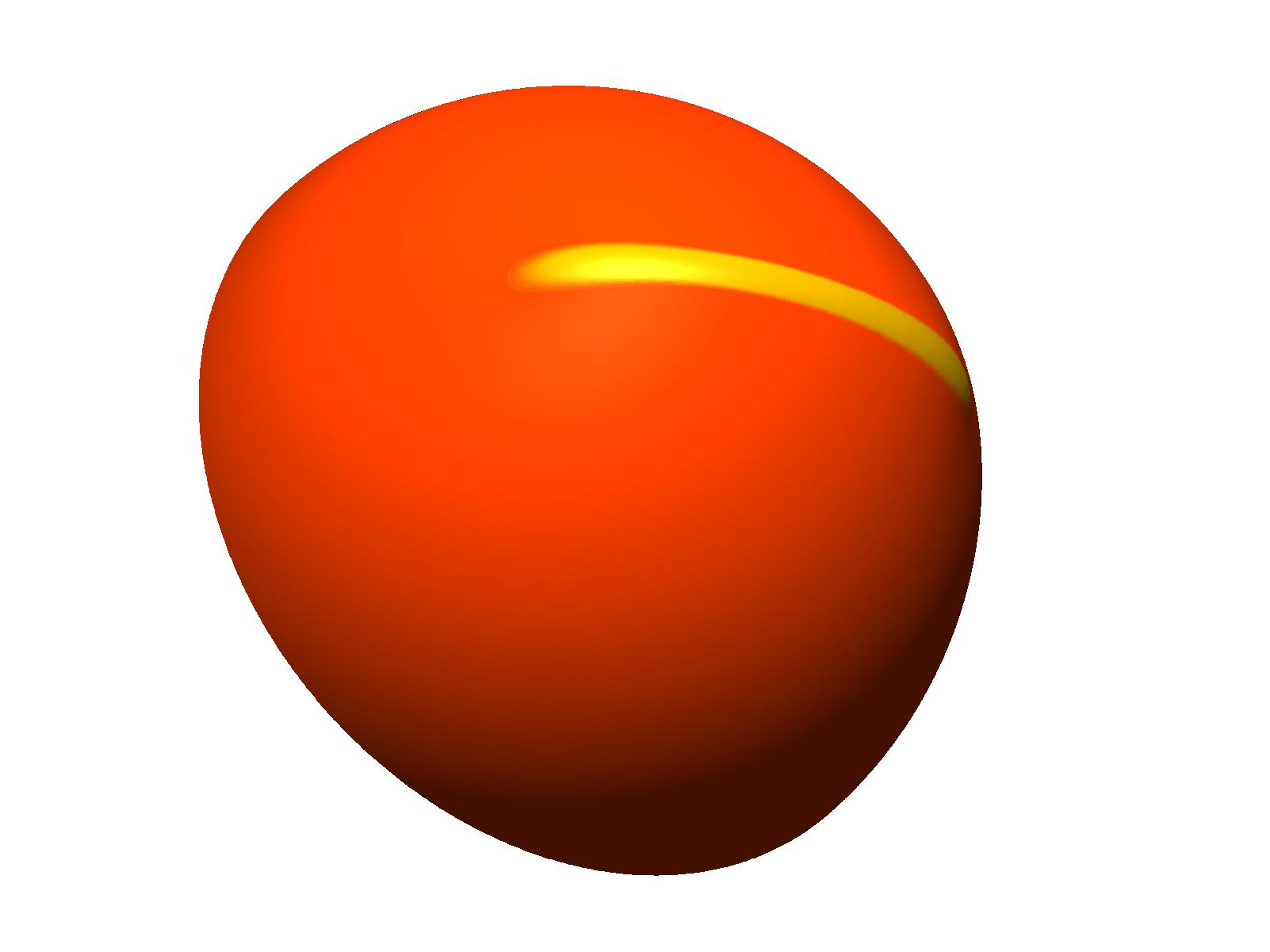}}
\put(-5.25,2.6){\includegraphics[height=31mm]{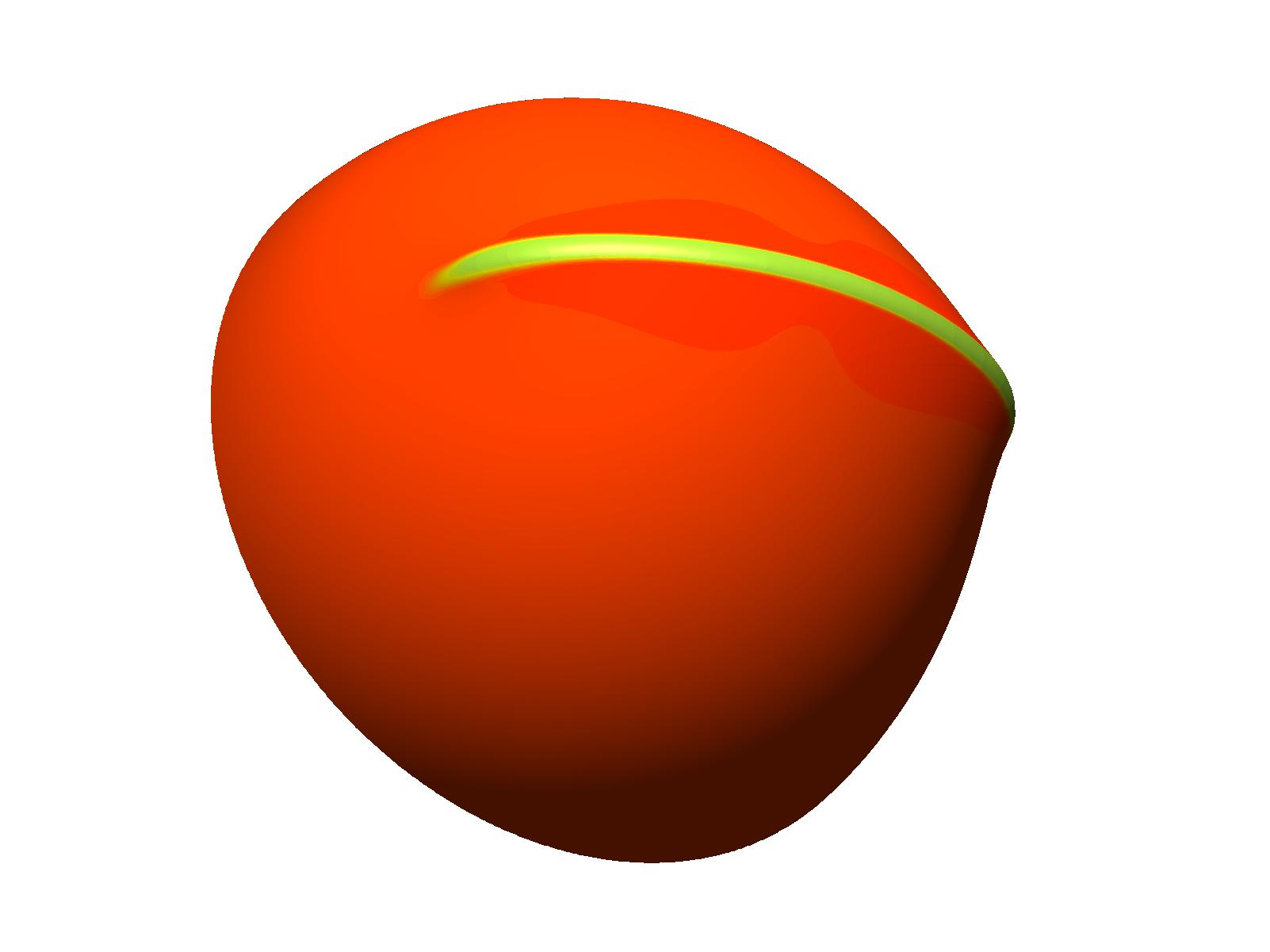}}
\put(-1.9,2.6){\includegraphics[height=31mm]{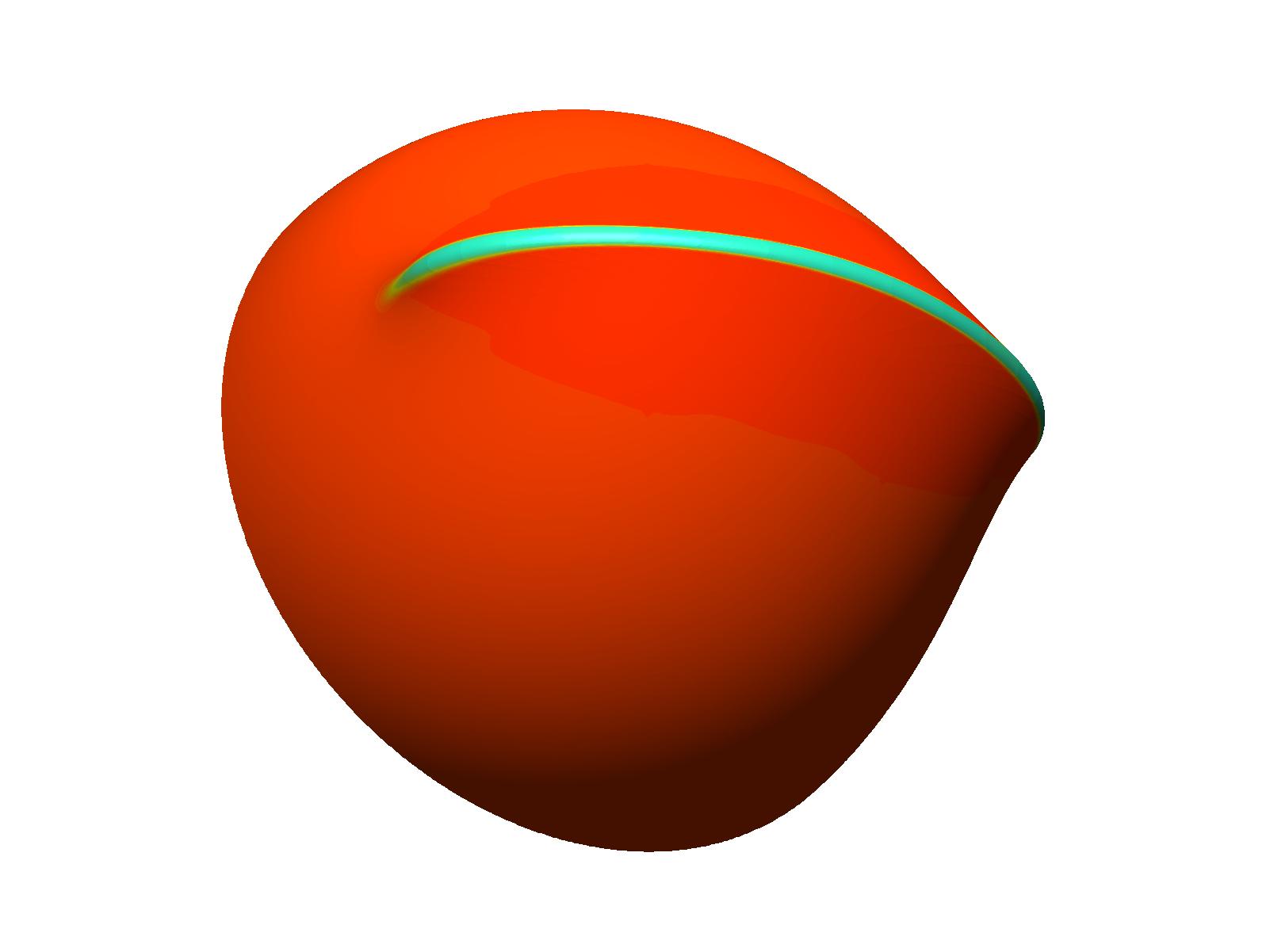}}
\end{picture}
\caption{Cell budding: Unconstrained, imperfect case at $\bar H_0=-4,-8,-11.98$ (left to right): 3D and side view of deformation and curvature $\bar H$. Here $\bar H\in[-12.5,\,-0.27]$.}
\label{f:bud_I}
\end{center}
\end{figure}

\textbf{Case 2}: The deformation is not constrained. Consequently, the non-axisymmetric bud shape shown in Fig.~\ref{f:bud_C} appears. 
Viewed from the top, the bud takes the shape of a `$+$'. Essentially, the initially circular region, where $H_0$ is prescribed, tries to flow away in order to lower the elastic energy. The flow leads to large mesh distortions and consequently the simulation crashes at $\bar H_0=-23.17$.
It is reasonable to suspect that the `$+$' shape is a consequence of the particular discretization. To confirm this, the following case is considered:

\textbf{Case 3}: Instead of a circle, $H_0$ is prescribed within an imperfect circle, i.e.~an ellipse with half-axes $a=0.22R$ and $b=0.18R$. The bud now flows into a `$-$' shape, as is shown in Fig.~\ref{f:bud_I}.
The distortion of the mesh seen on the right clearly shows how the material flows outward in horizontal direction and back inward in vertical direction. 
The mesh distortion becomes so large that the simulation crashes at $\bar H_0=-11.98$.
In the current case the flow is not resisted mechanically (since $\mu$ is so low). Resistance is provided either by shear stiffness (e.g. due to an underlying cytoskeleton) or by viscosity. This is considered in the remaining two cases. 

\begin{figure}[h]
\begin{center} \unitlength1cm
\begin{picture}(0,5.6)
\put(-8.9,-.4){\includegraphics[height=31mm]{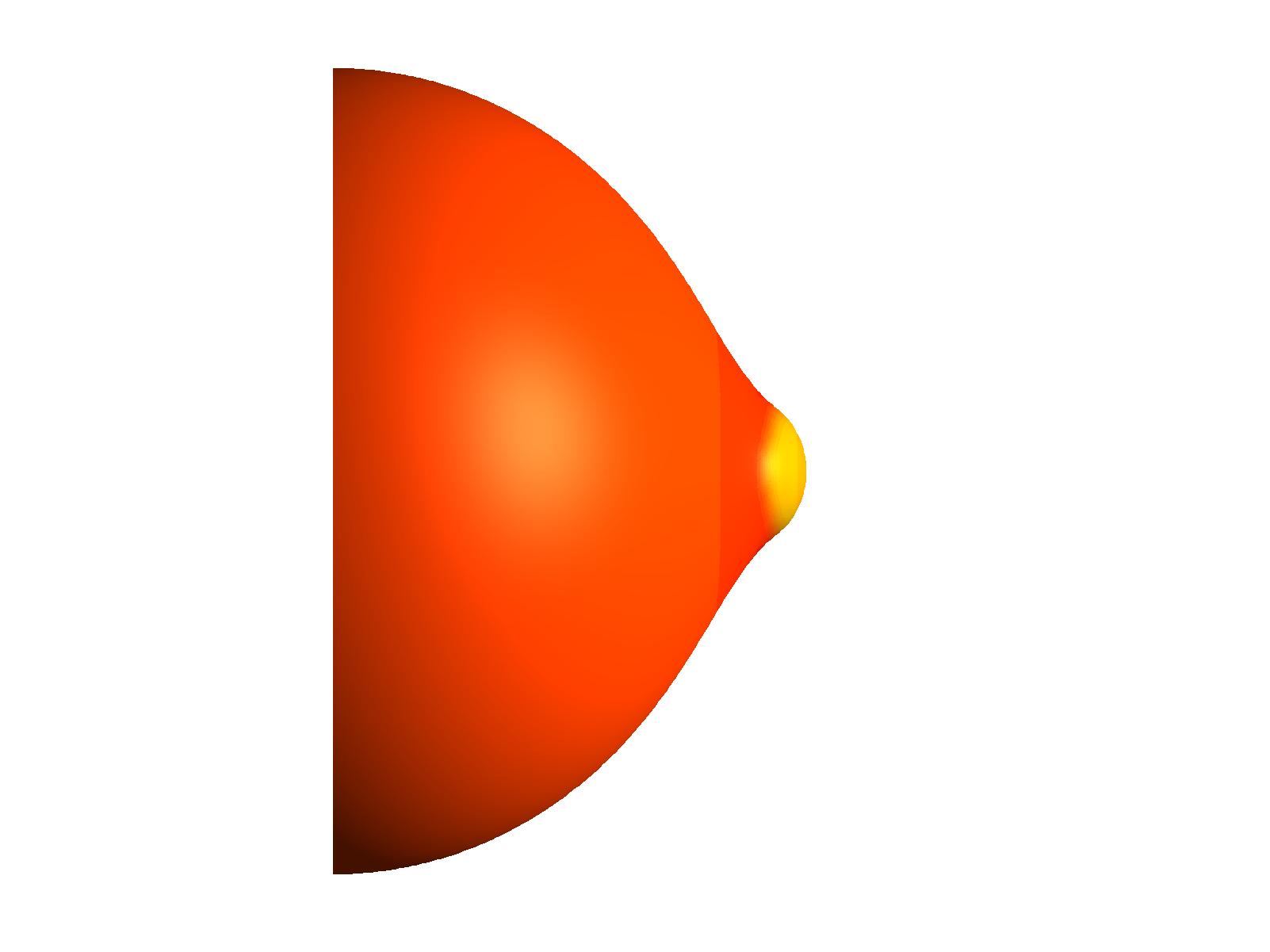}}
\put(-5.65,-.4){\includegraphics[height=31mm]{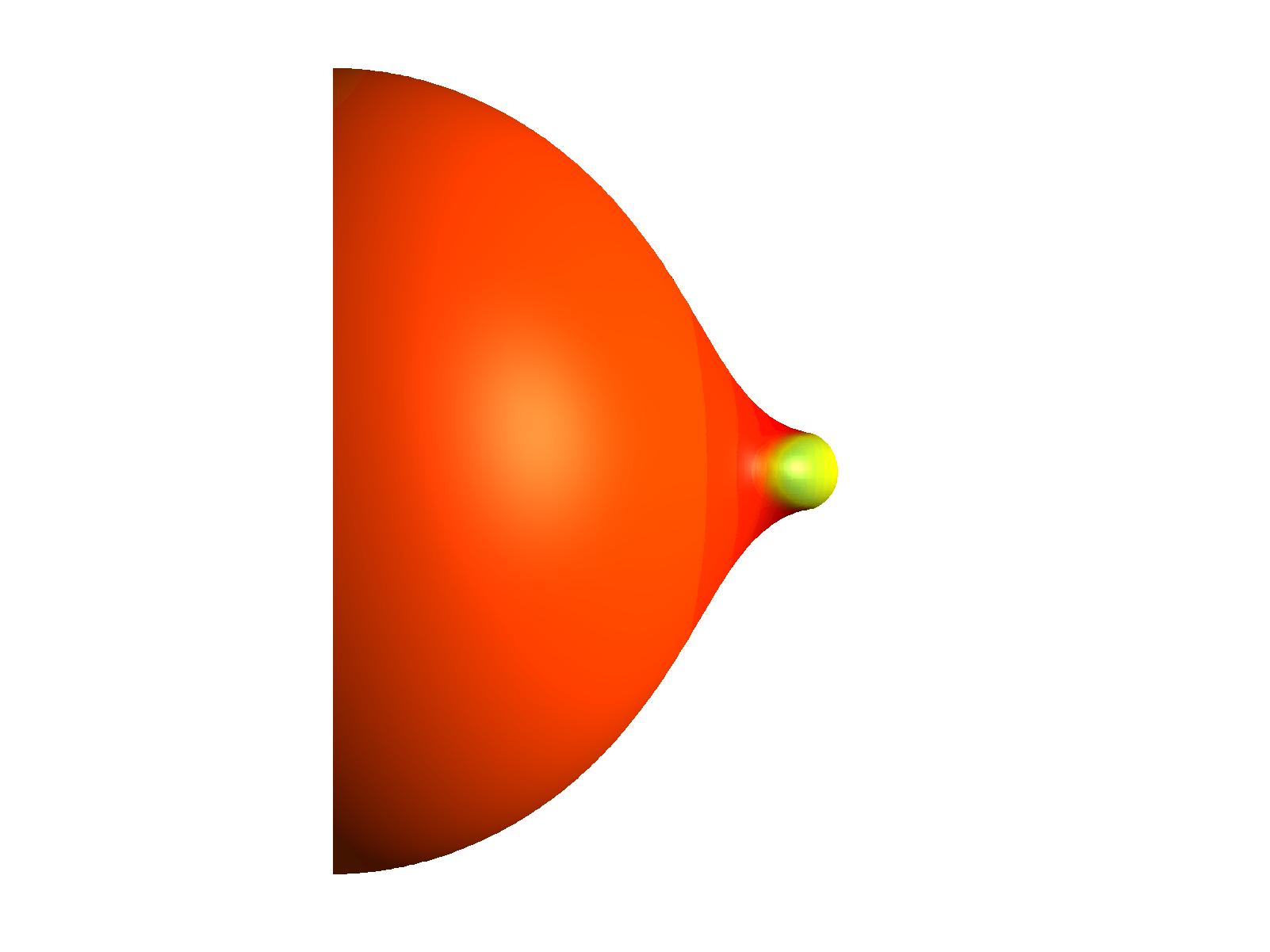}}
\put(-2.4,-.4){\includegraphics[height=31mm]{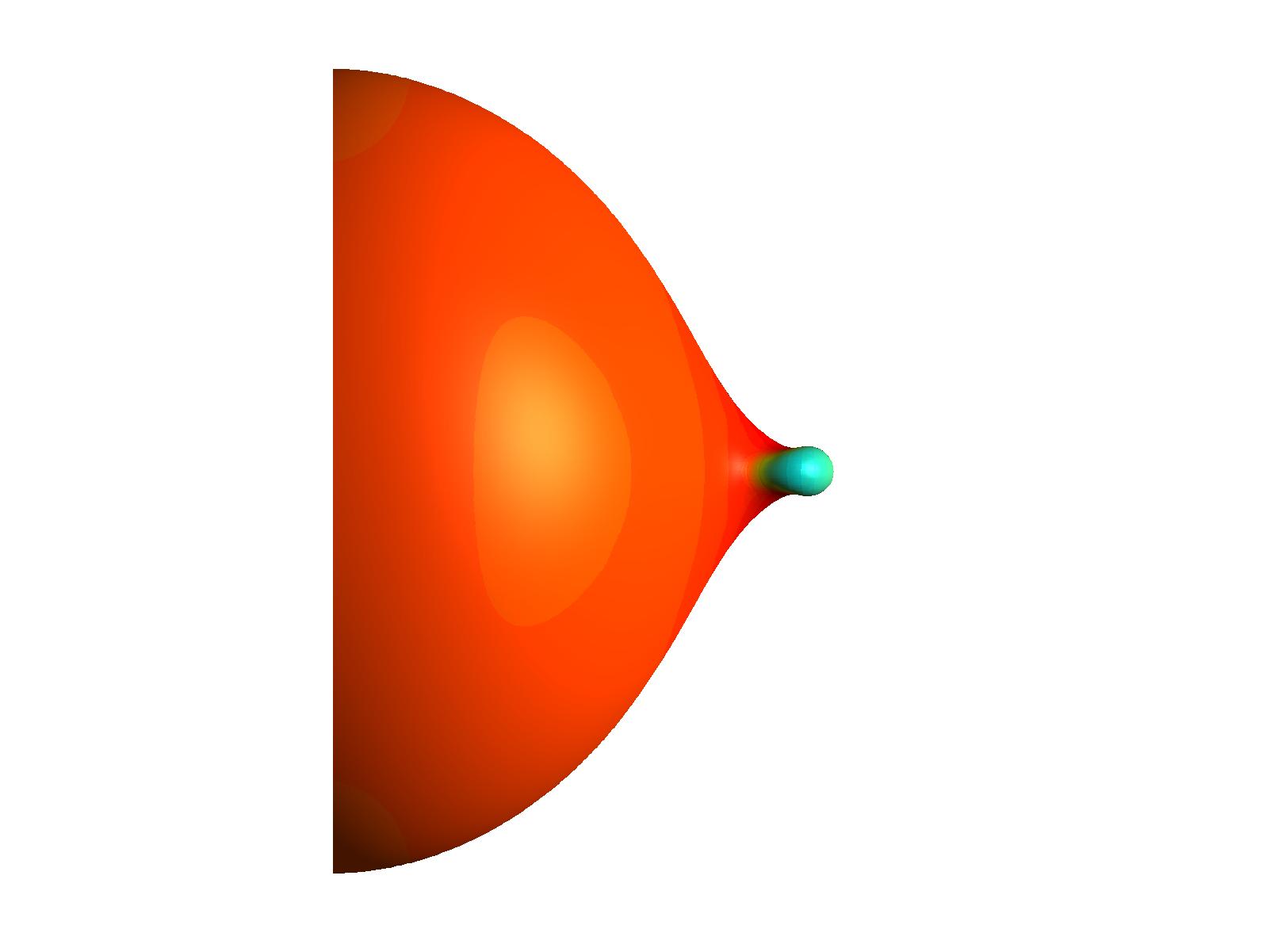}}
\put(4.05,-.4){\includegraphics[height=31mm]{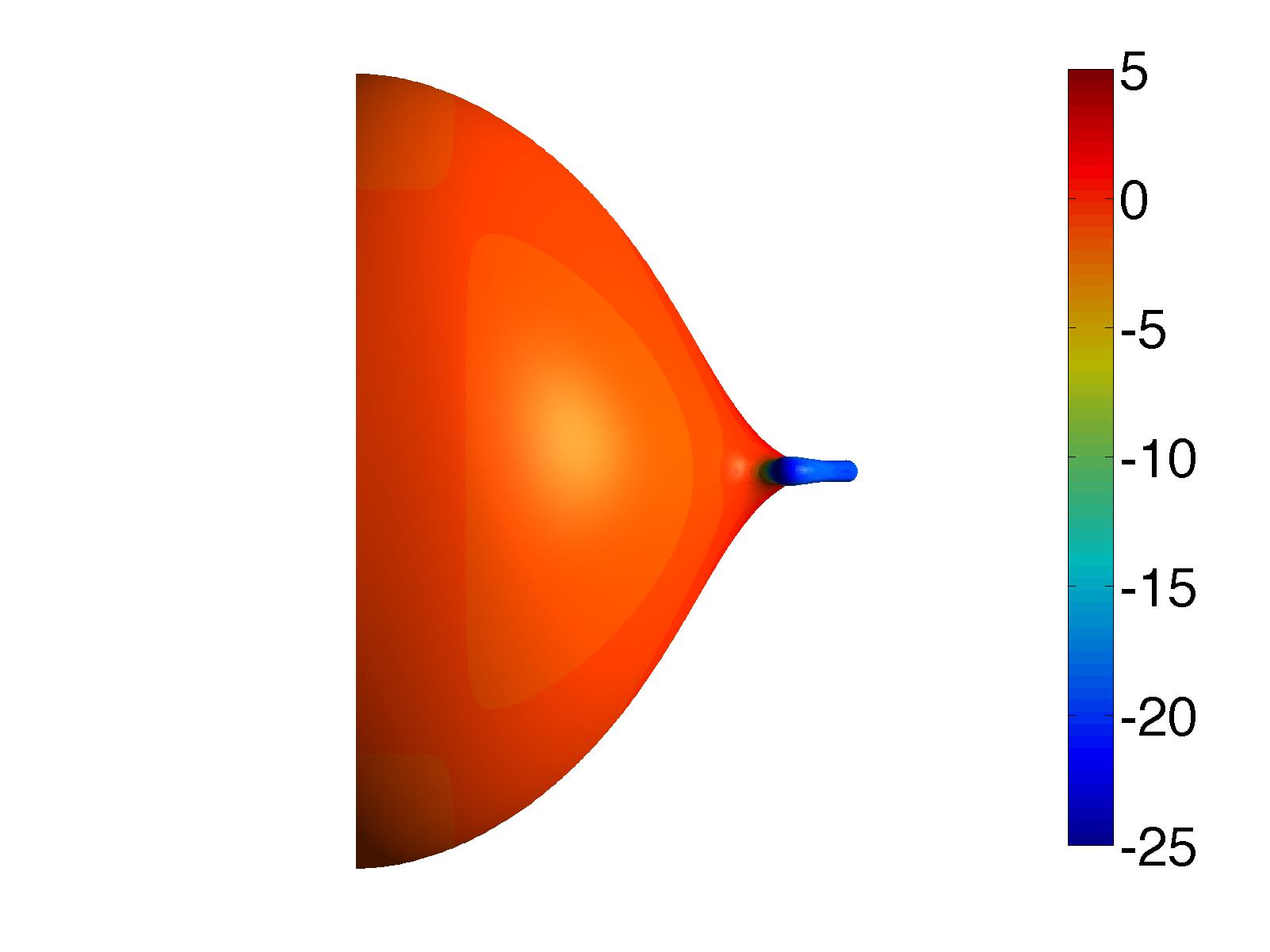}}
\put(0.85,-.4){\includegraphics[height=31mm]{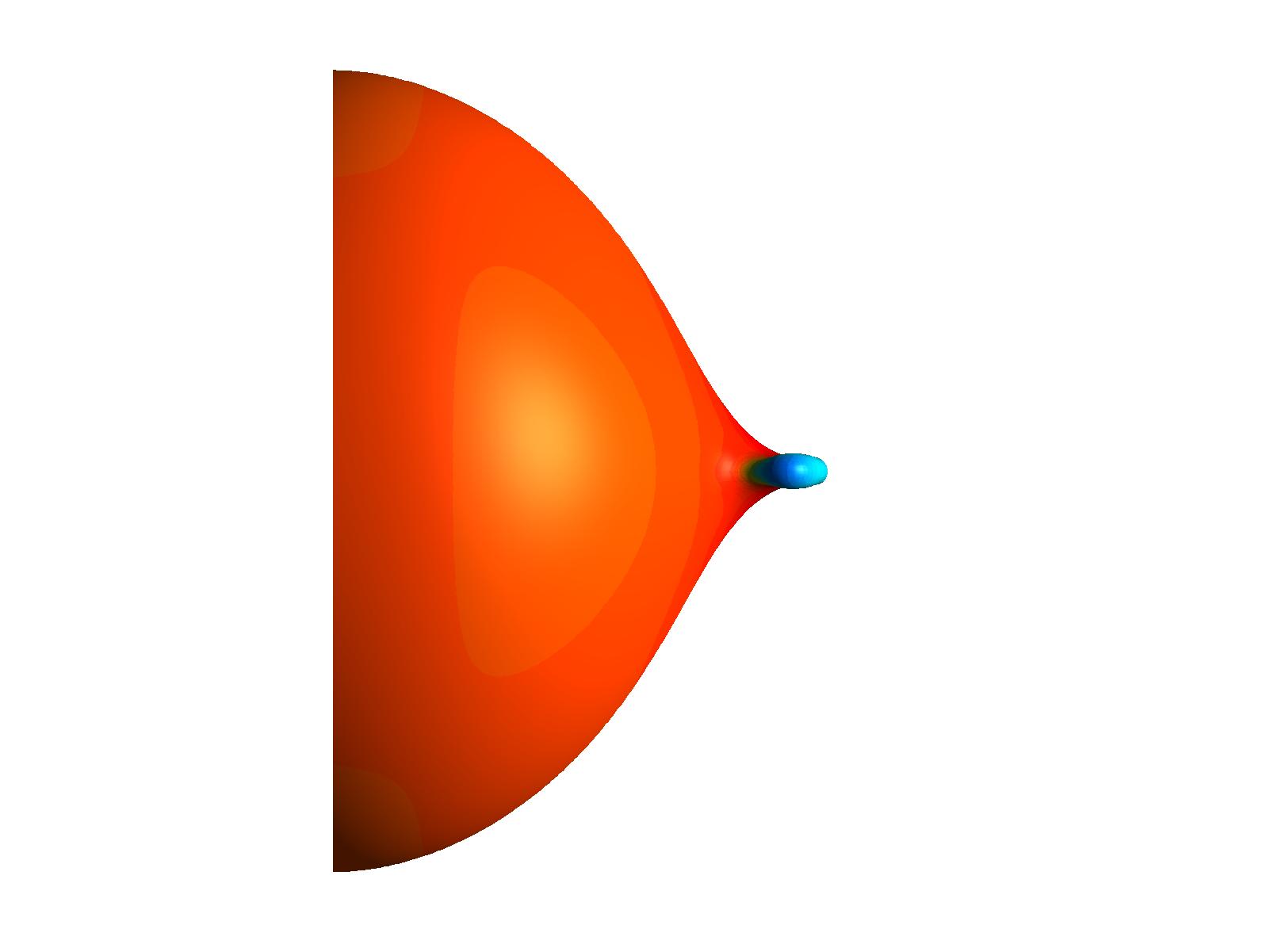}}
\put(-8.5,2.6){\includegraphics[height=31mm]{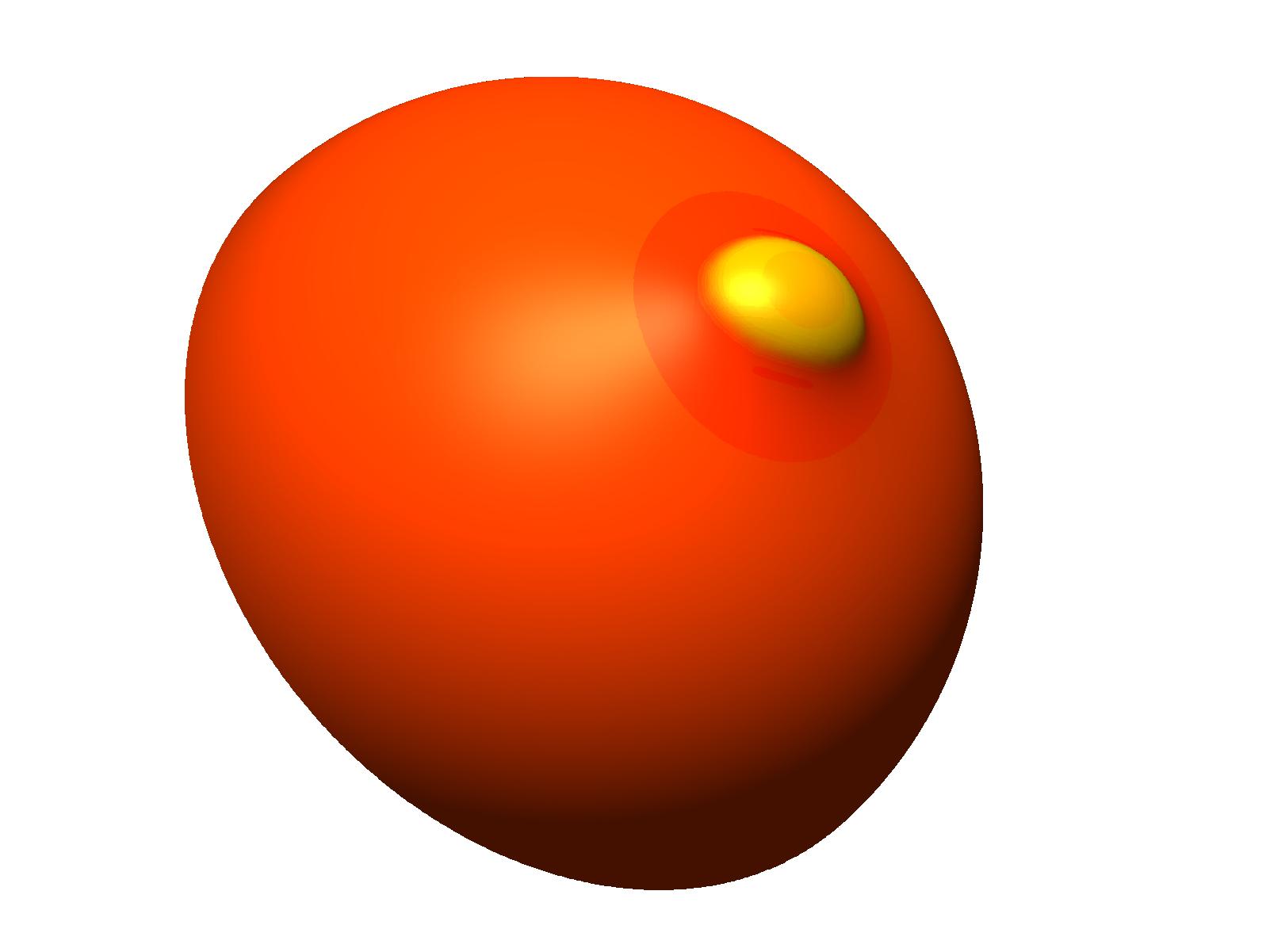}}
\put(-5.25,2.6){\includegraphics[height=31mm]{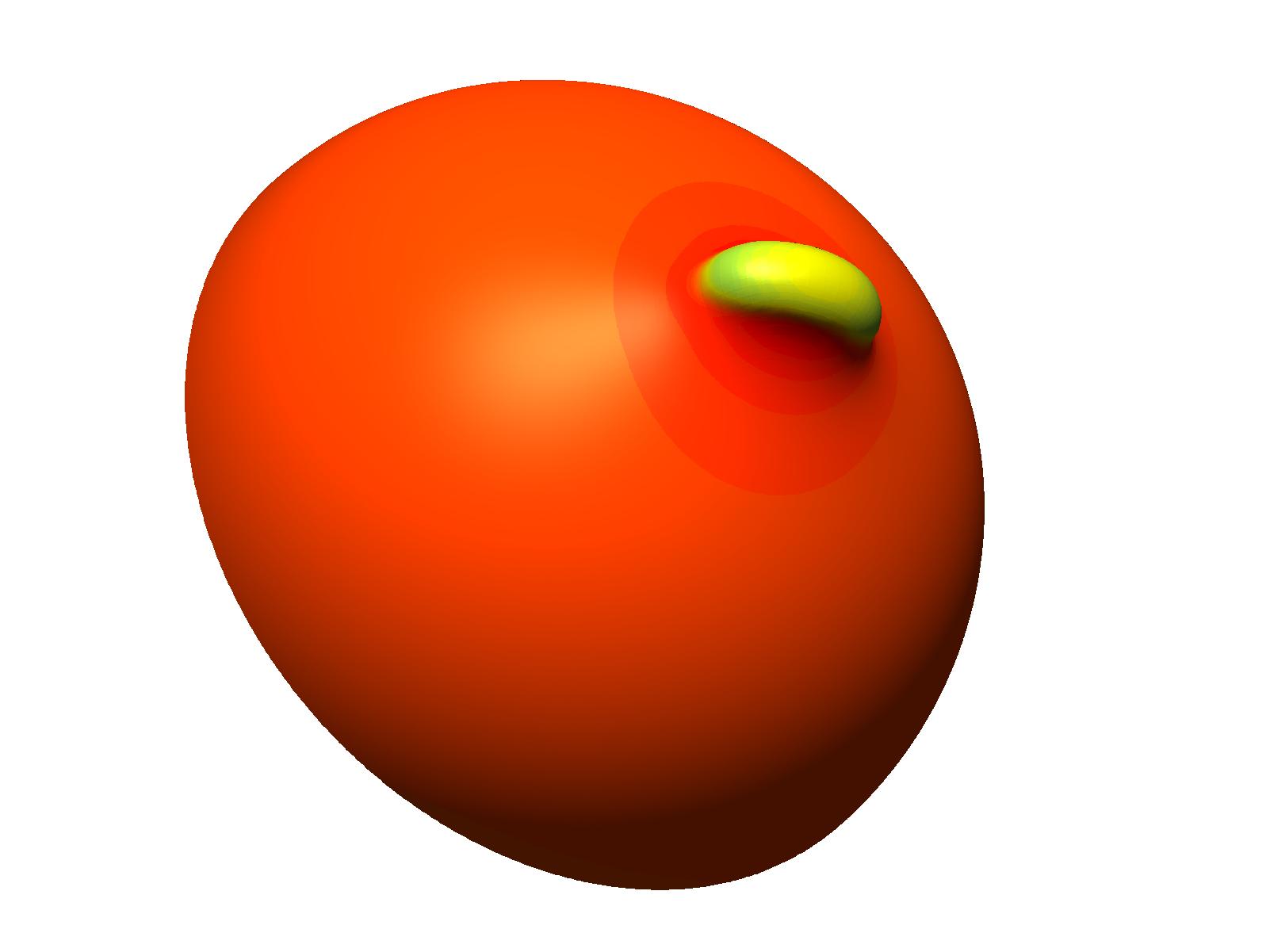}}
\put(-2,2.6){\includegraphics[height=31mm]{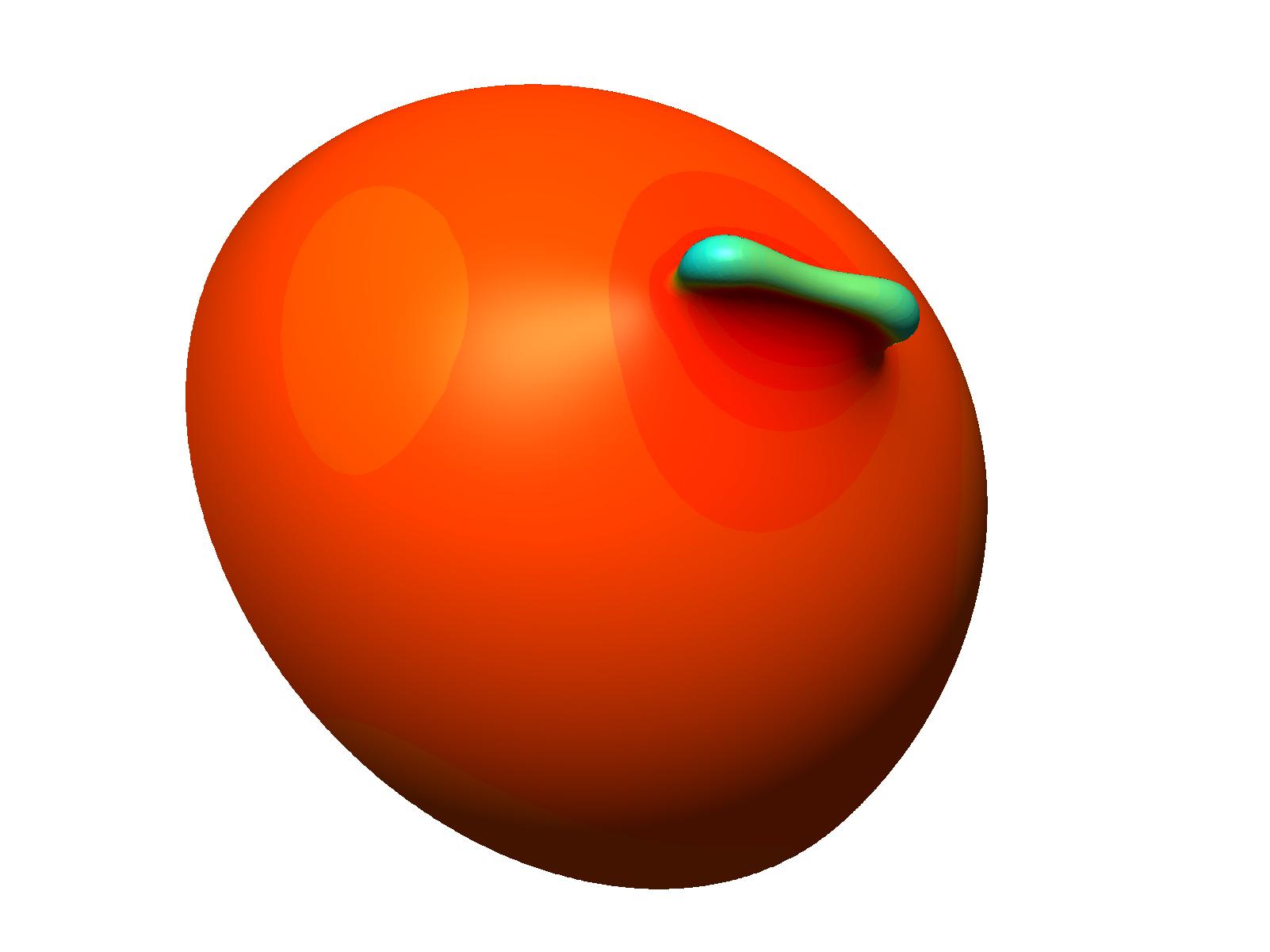}}
\put(1.25,2.6){\includegraphics[height=31mm]{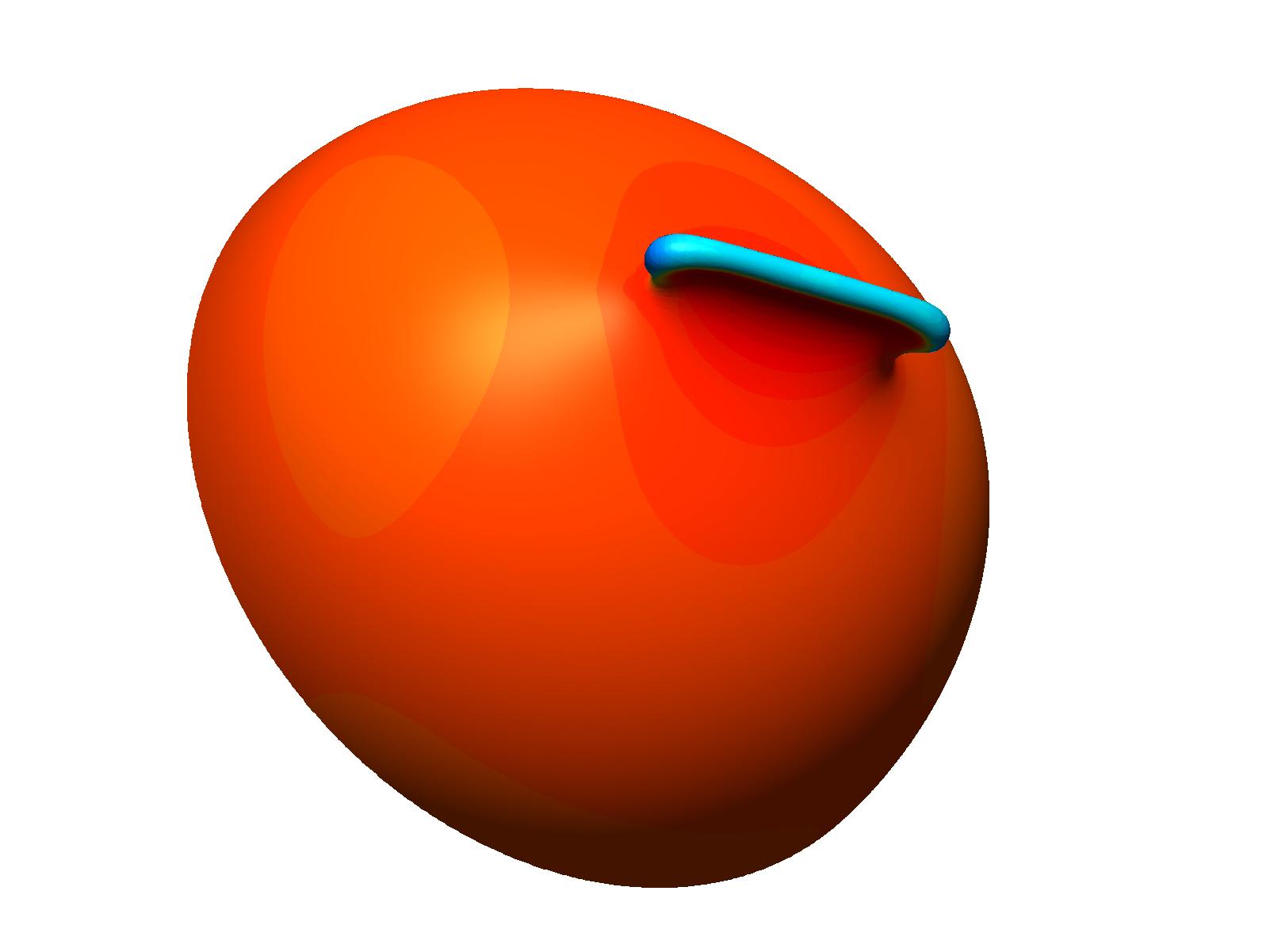}}
\put(4.5,2.6){\includegraphics[height=31mm]{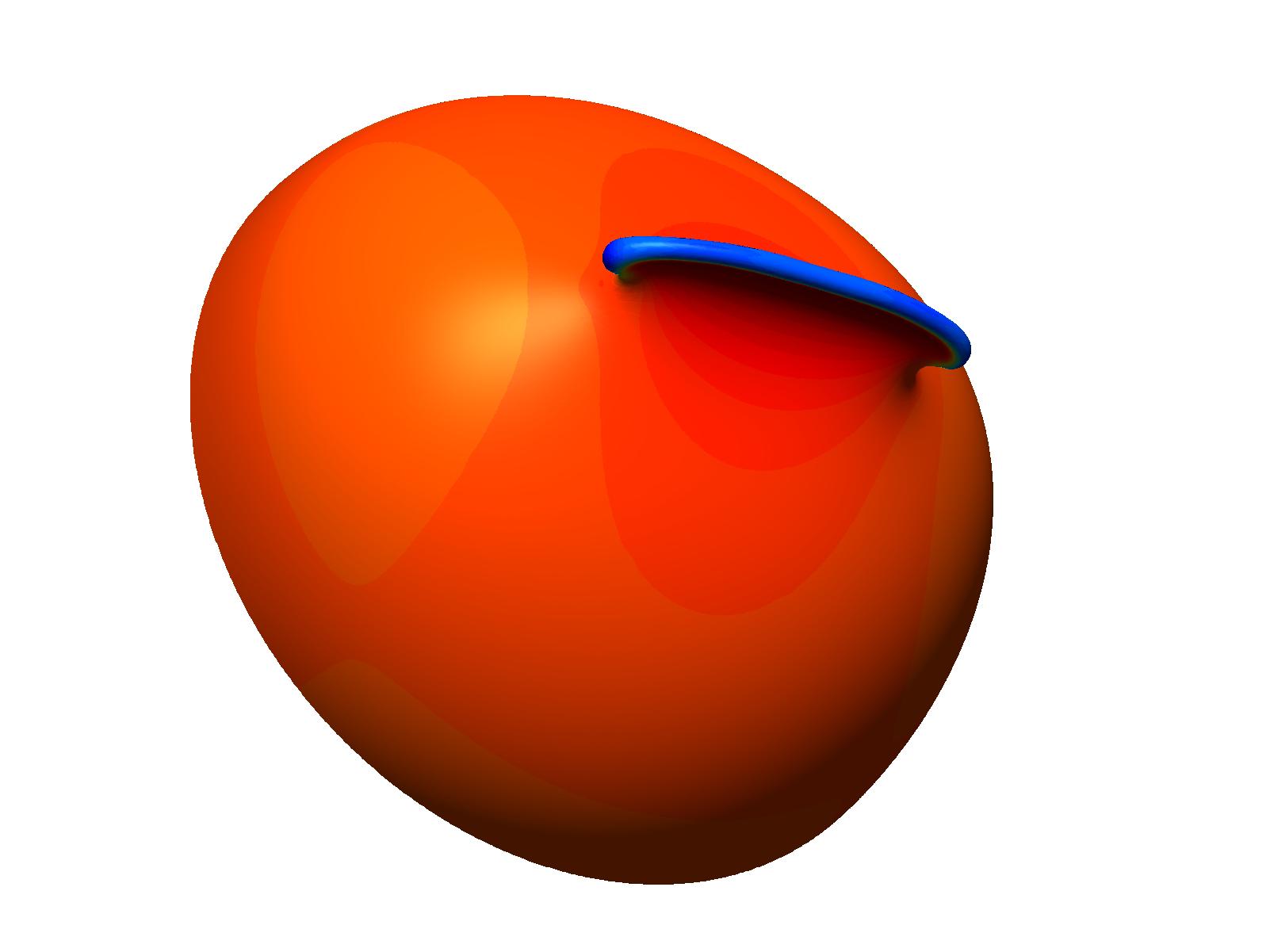}}
\end{picture}
\caption{Cell budding: Shear resistant, imperfect case at $\bar H_0=-5,-10,-15,-20,-25$ (left to right): 3D and side view of deformation and curvature $\bar H$. Here $\bar H\in[-22.8,\,4.61]$.}
\label{f:bud_Imu}
\end{center}
\end{figure}

\textbf{Case 4}: Fig.~\ref{f:bud_Imu} shows the deformation for the same case as before ($a=0.22R$ and $b=0.18R$) considering now a shear resisting membrane.
Model `A-st'\footnote{The shear stresses are now physical and need to be applied both in-plane and out-of-plane.} 
is now considered with $\bar\mu=10$. 
The shear resistance prevents the unbounded spreading of the bud observed earlier. Now the bud remains localized in the center. The bud starts growing in an almost circular fashion, but then degenerates into the shape shown in Fig.~\ref{f:bud_Imu}. The process suggests that the initial imperfection is only a trigger for the final shape, but does not affect it in magnitude. The evolving deformation for case 4 is also shown in the supplemental movie file \verb"bud4.mpg".

\textbf{Case 5}: The final case also considers the imperfect circle from before ($a=0.22R$ and $b=0.18R$), but now provides significant shear stress through physical viscosity. 
\begin{figure}[h]
\begin{center} \unitlength1cm
\begin{picture}(0,5.6)
\put(-8.9,-.4){\includegraphics[height=31mm]{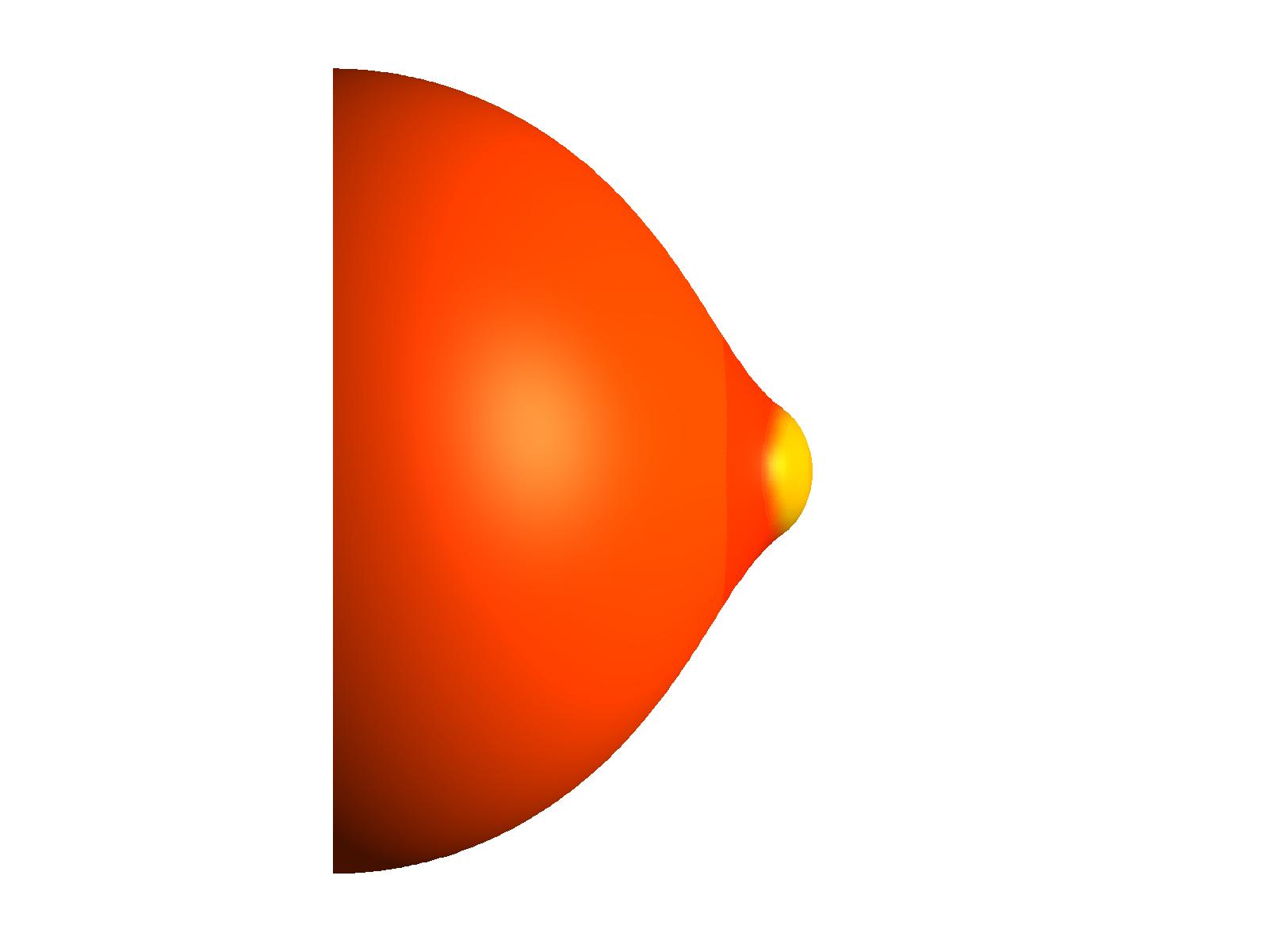}}
\put(-5.65,-.4){\includegraphics[height=31mm]{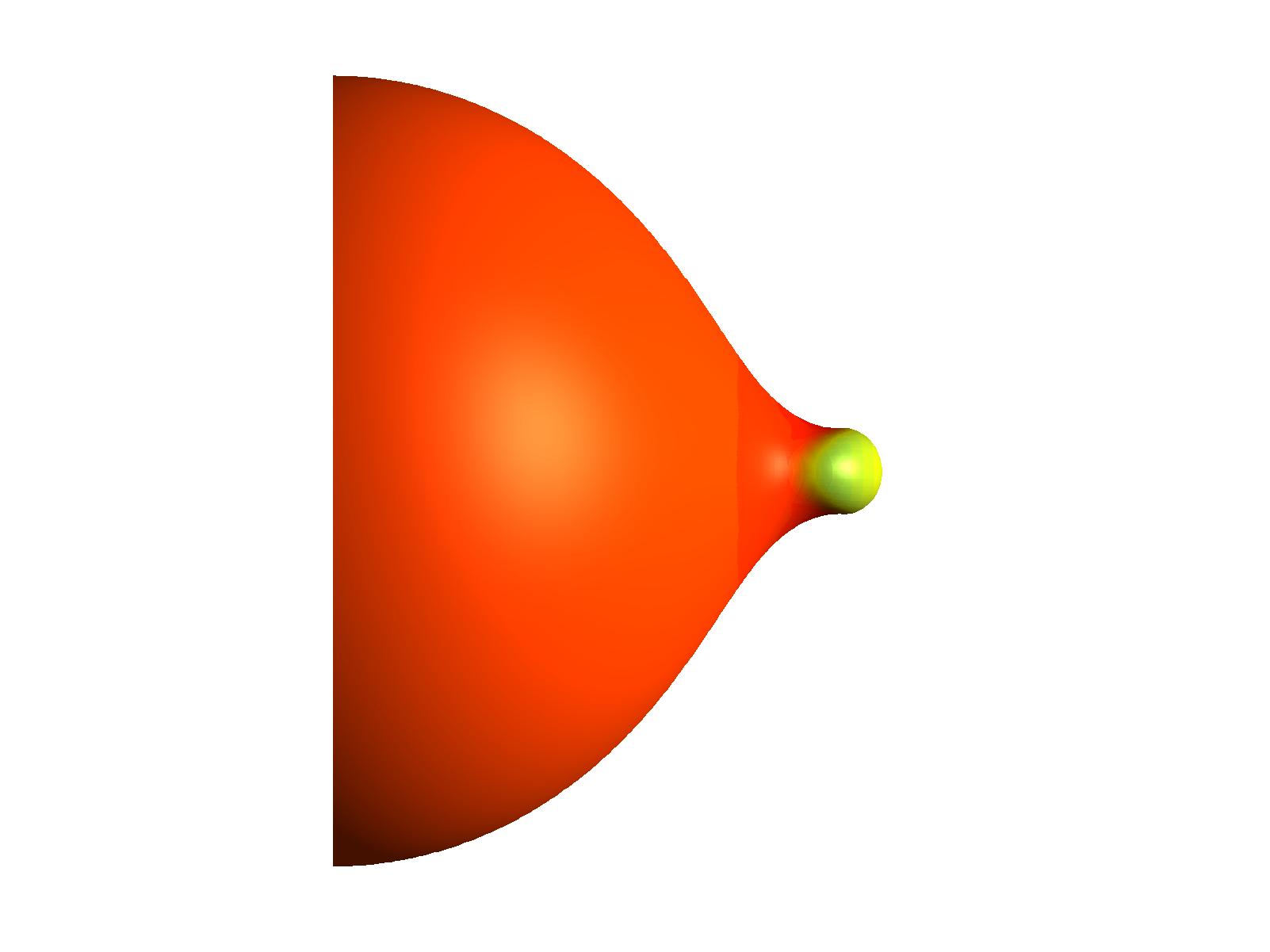}}
\put(-2.4,-.4){\includegraphics[height=31mm]{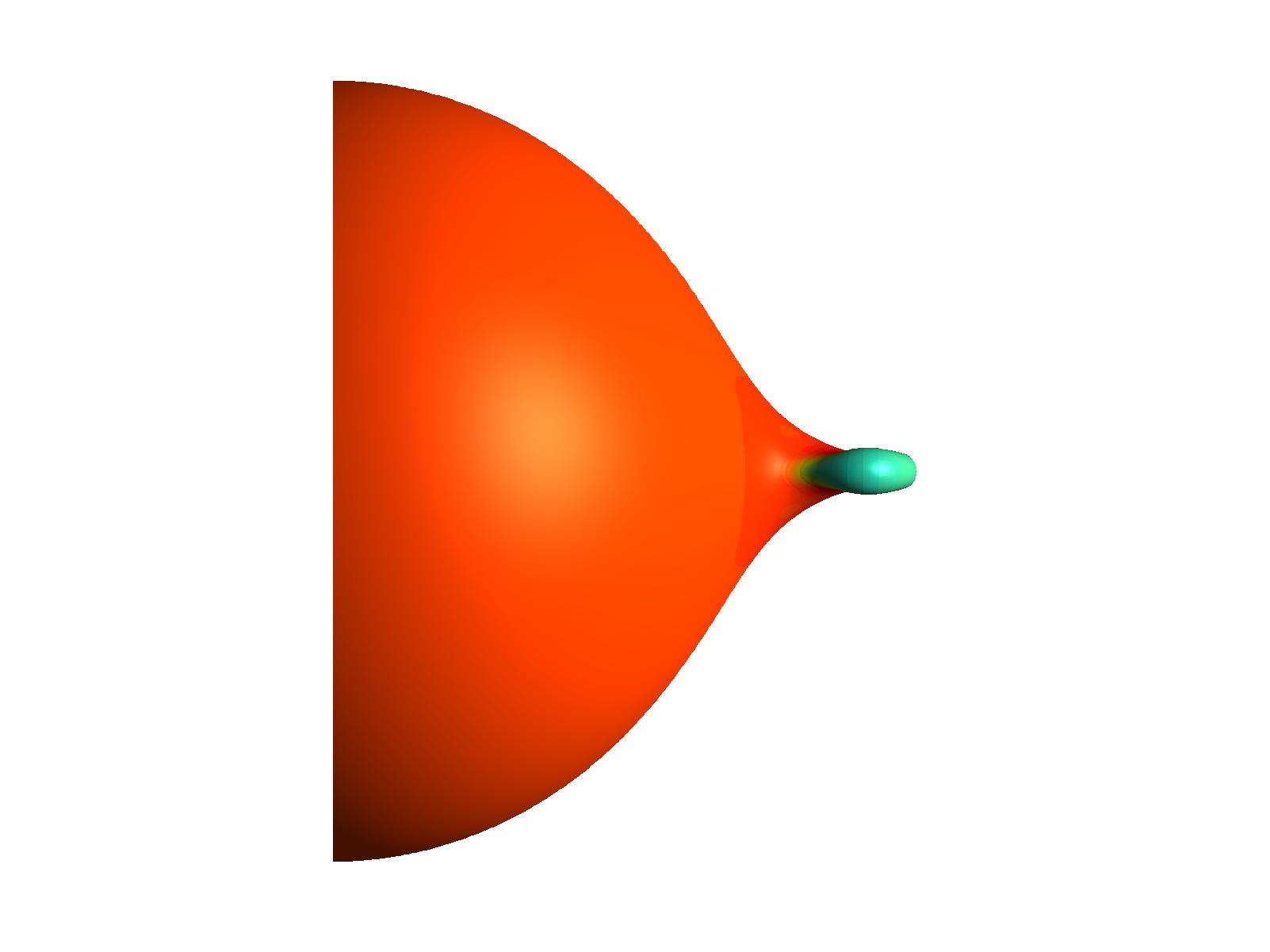}}
\put(0.85,-.4){\includegraphics[height=31mm]{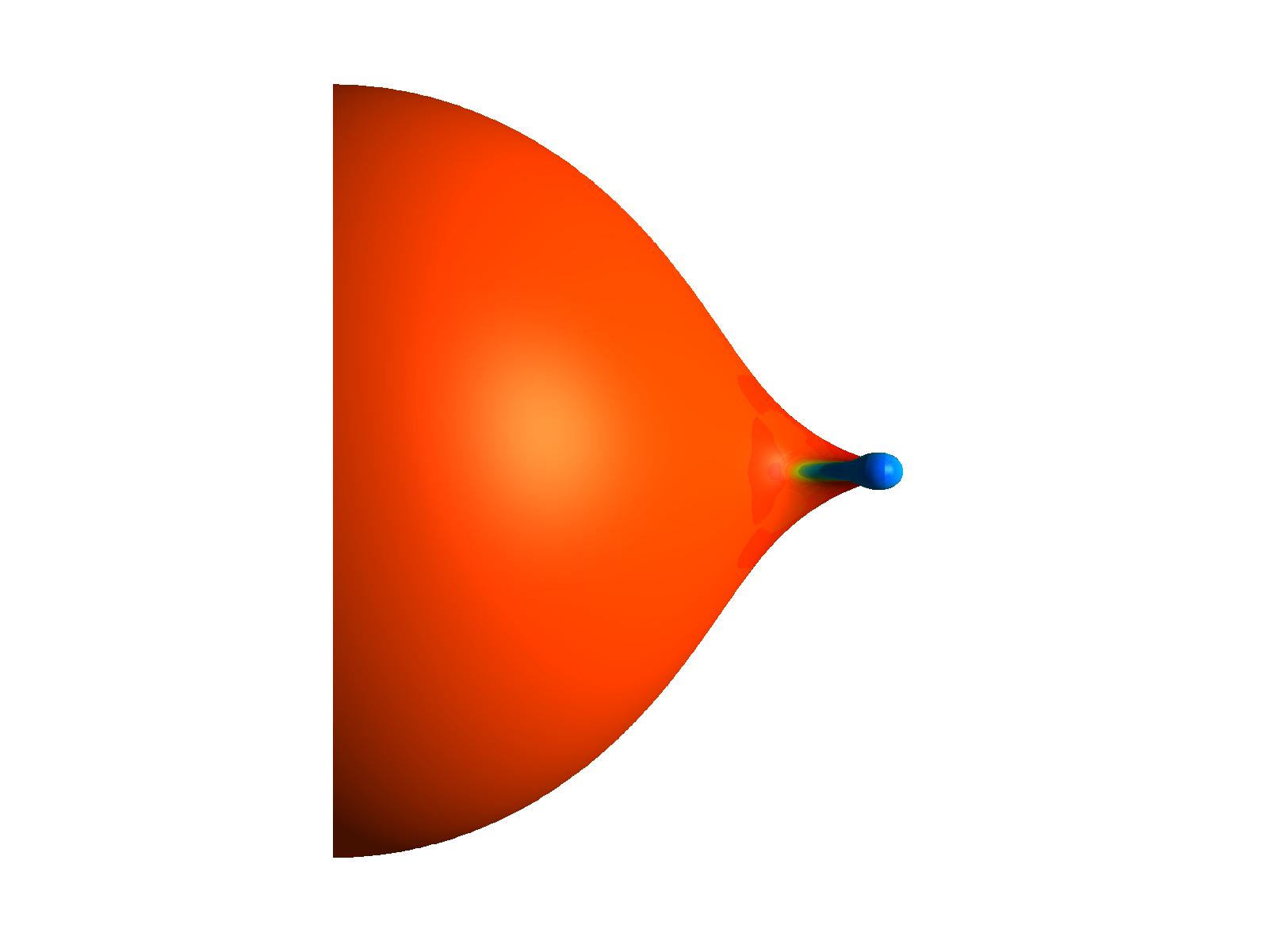}}
\put(4.05,-.4){\includegraphics[height=31mm]{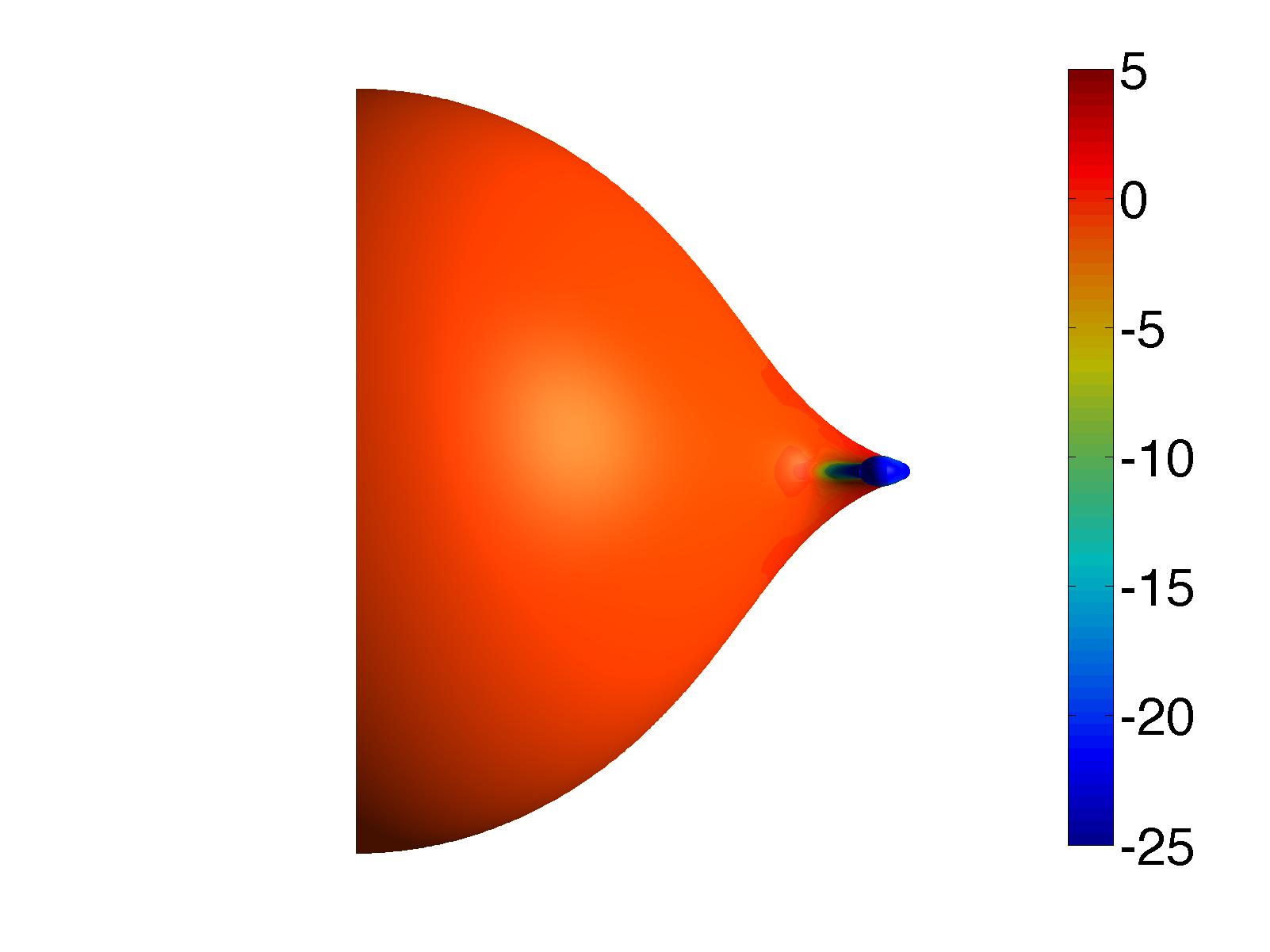}}
\put(-8.5,2.6){\includegraphics[height=31mm]{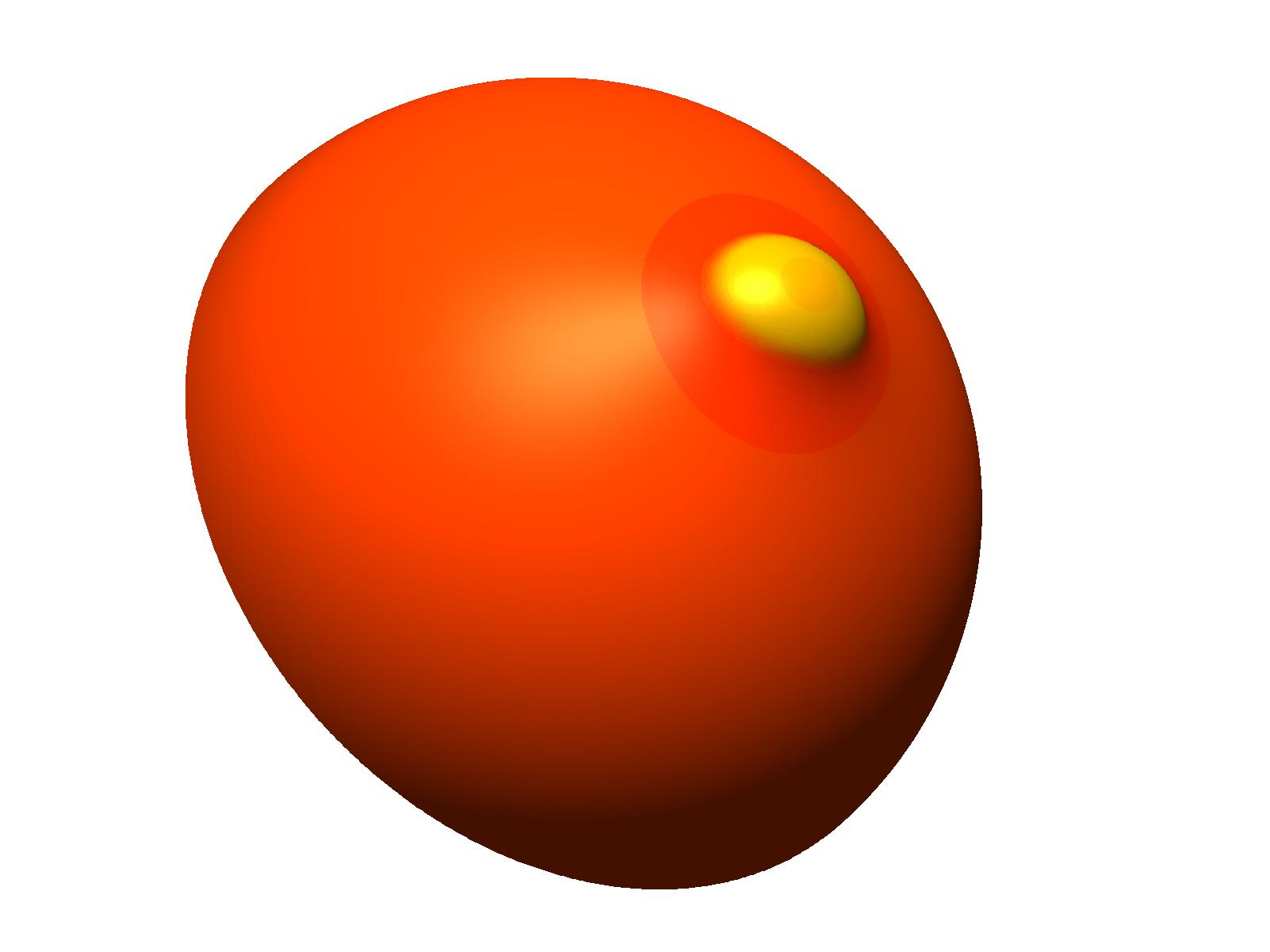}}
\put(-5.25,2.6){\includegraphics[height=31mm]{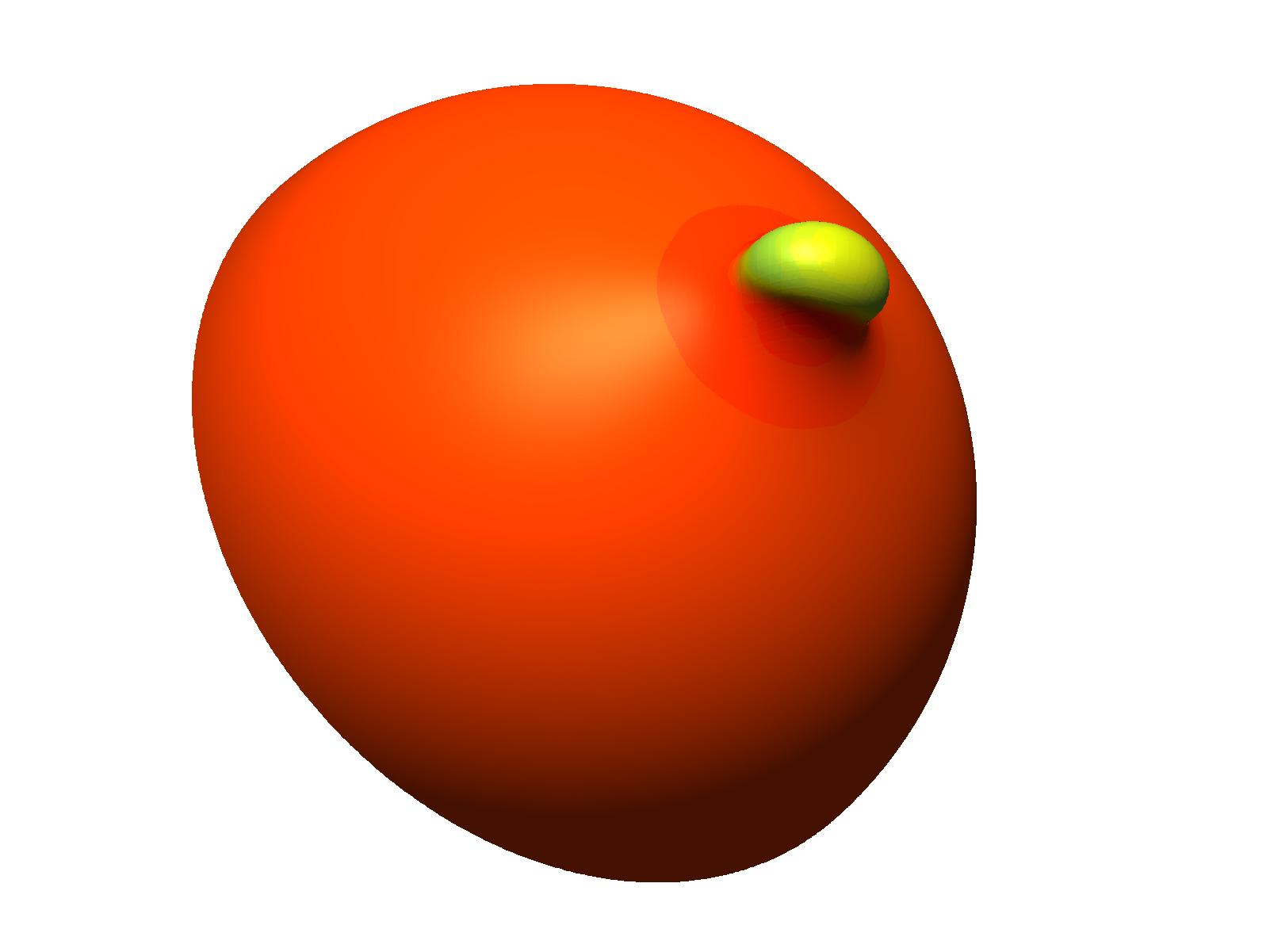}}
\put(-2,2.6){\includegraphics[height=31mm]{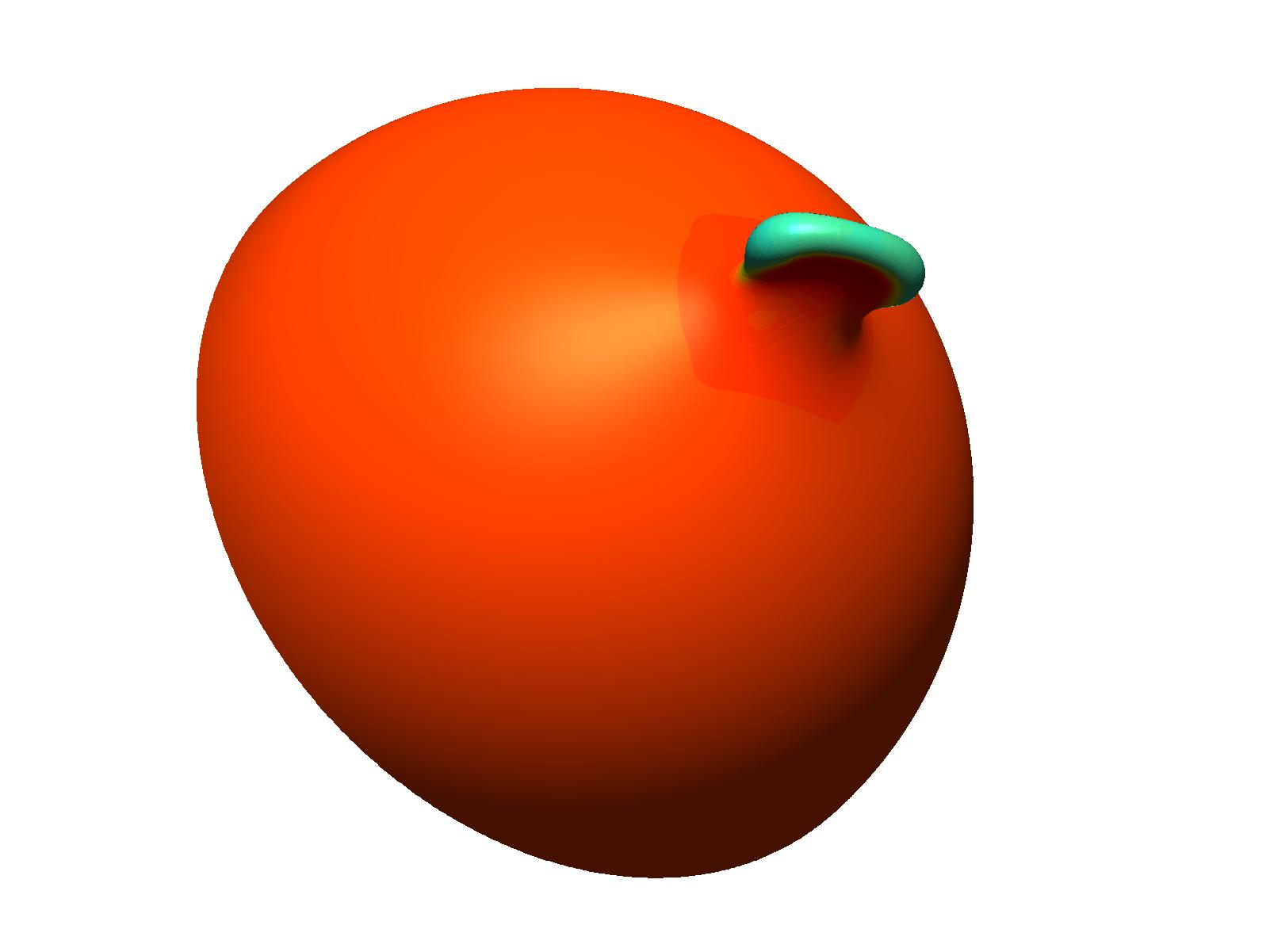}}
\put(1.25,2.6){\includegraphics[height=31mm]{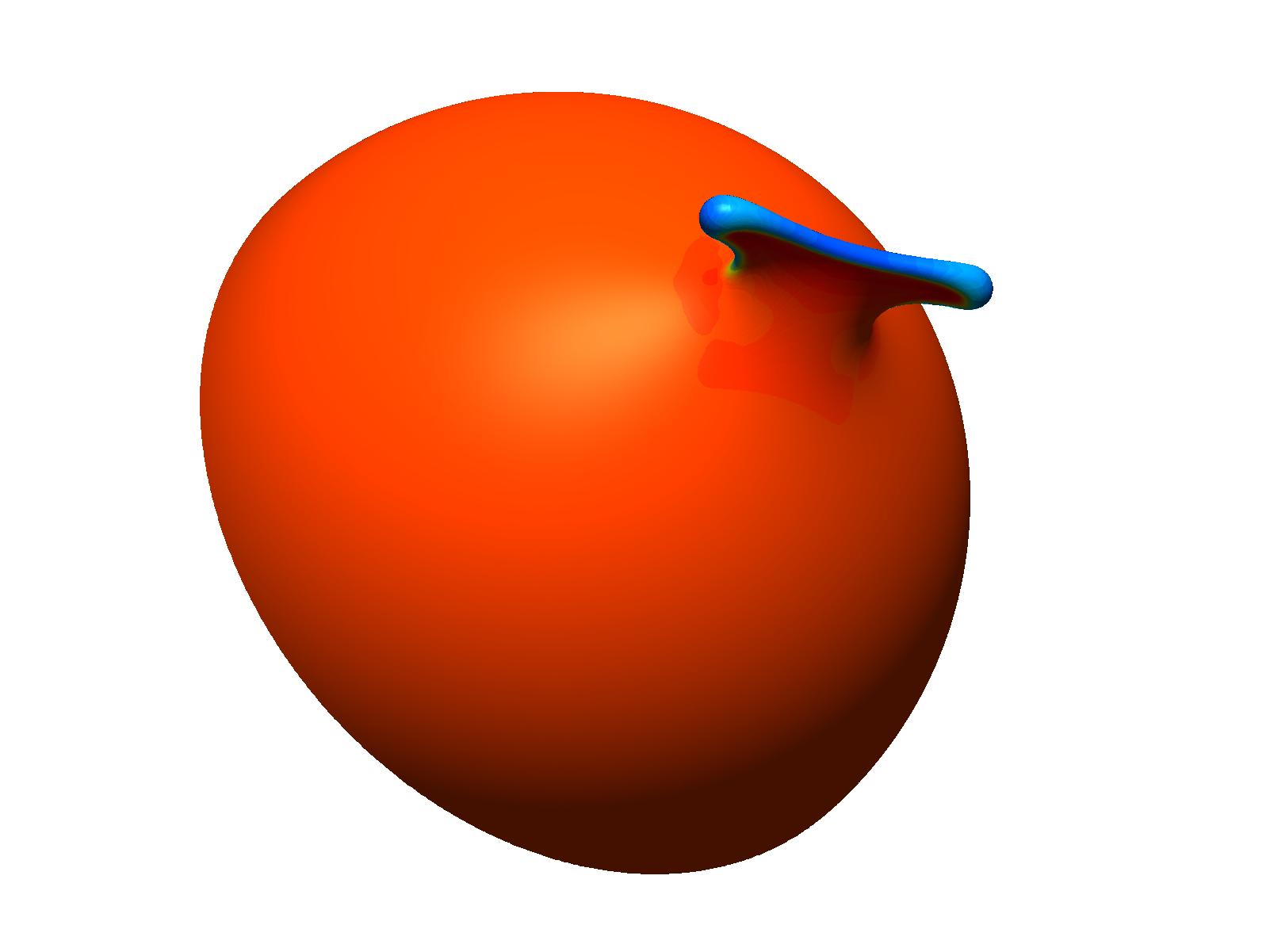}}
\put(4.5,2.6){\includegraphics[height=31mm]{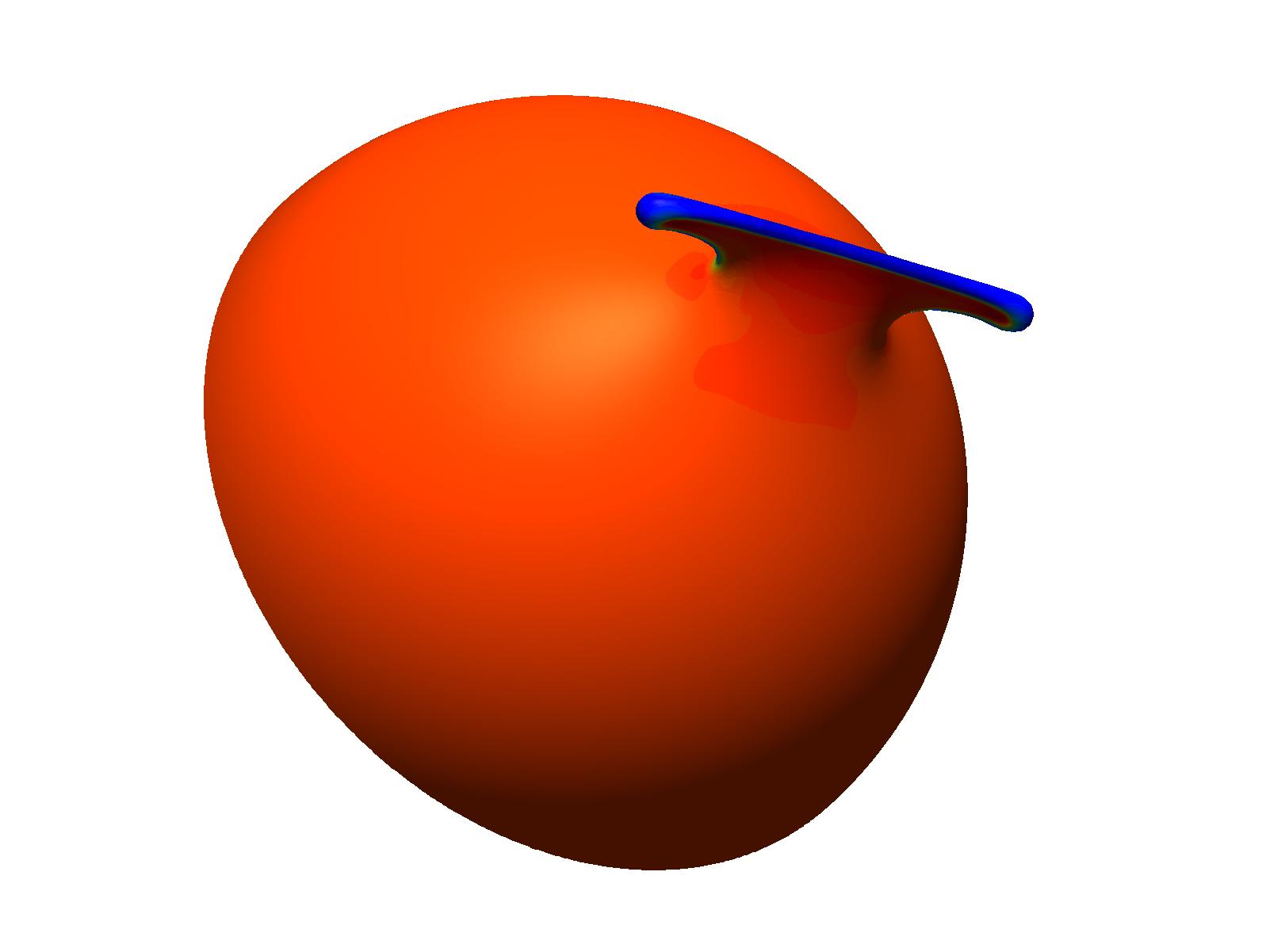}}
\end{picture}
\caption{Cell budding: Viscous, imperfect case at $\bar H_0=-5,-10,-15,-20,-25$ (left to right): 3D and side view of deformation and curvature $\bar H$. Here $\bar H\in[-24.4,\,4.29]$.}
\label{f:bud_Inu}
\end{center}
\end{figure}
This is captured through model `a-st' using relation 
\eqref{e:nu} with $\bar\mu=1250$ and a load stepping increment for $H_0$ of $\Delta \bar H_0=0.02$
(such that $\nu = 25k/L^3/\dot{H}_0$, where $\dot H_0$ is the rate with which the spontaneous curvature is prescribed).
The evolution of the bud with $H_0$ is shown in Fig.~\ref{f:bud_Inu} and in the supplemental movie file \verb"bud5.mpg".
Again, the bud starts growing in an almost circular fashion, but then degenerates into a plate-like shape.
If $H_0$ is kept fixed over time, the solution of case 5 will relax to the solution of case 3.



\subsubsection{Surface energy}

\begin{figure}[h]
\begin{center} \unitlength1cm
\begin{picture}(0,5.7)
\put(-8.05,0){\includegraphics[height=58mm]{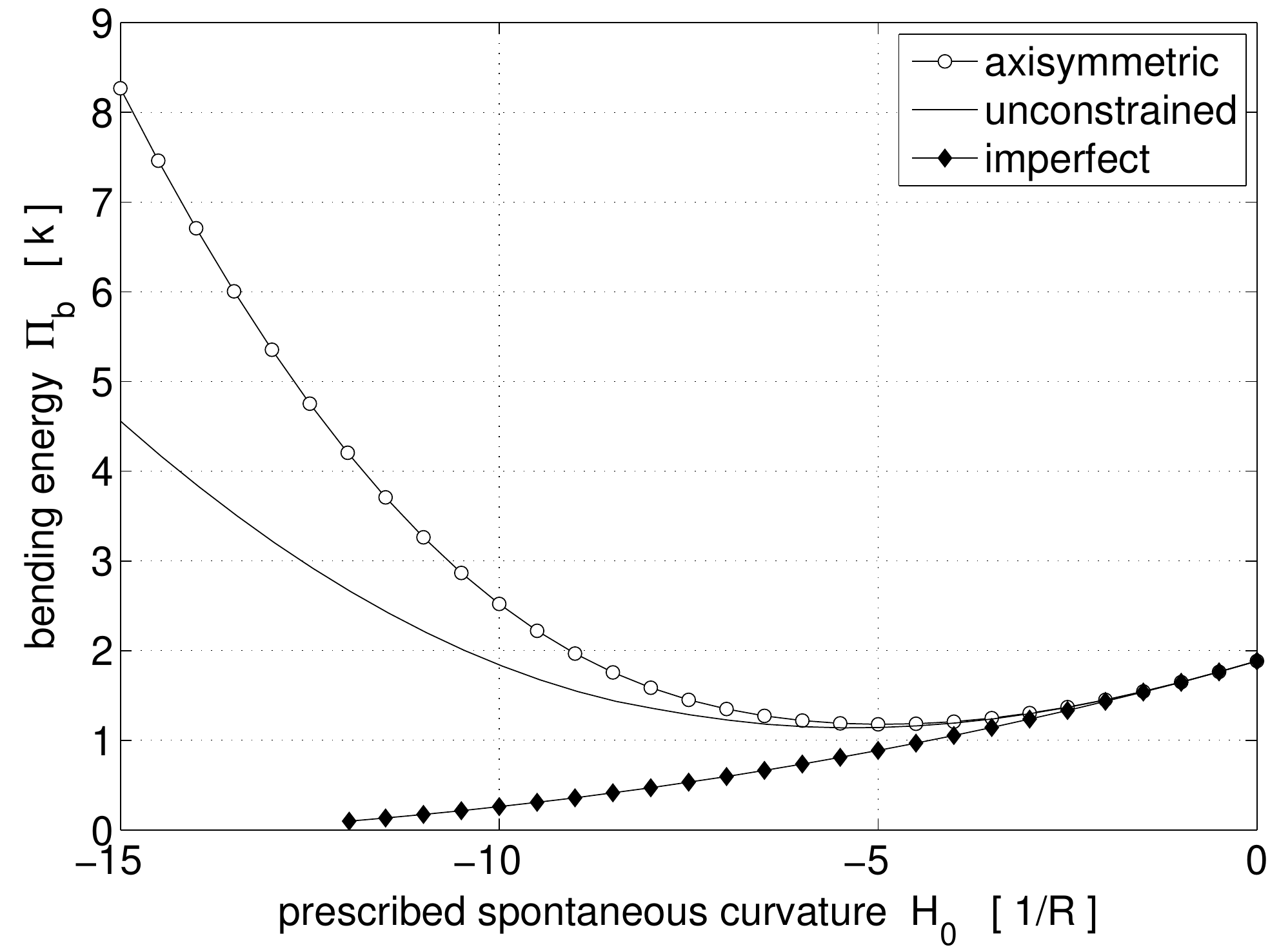}}
\put(0.25,0){\includegraphics[height=58mm]{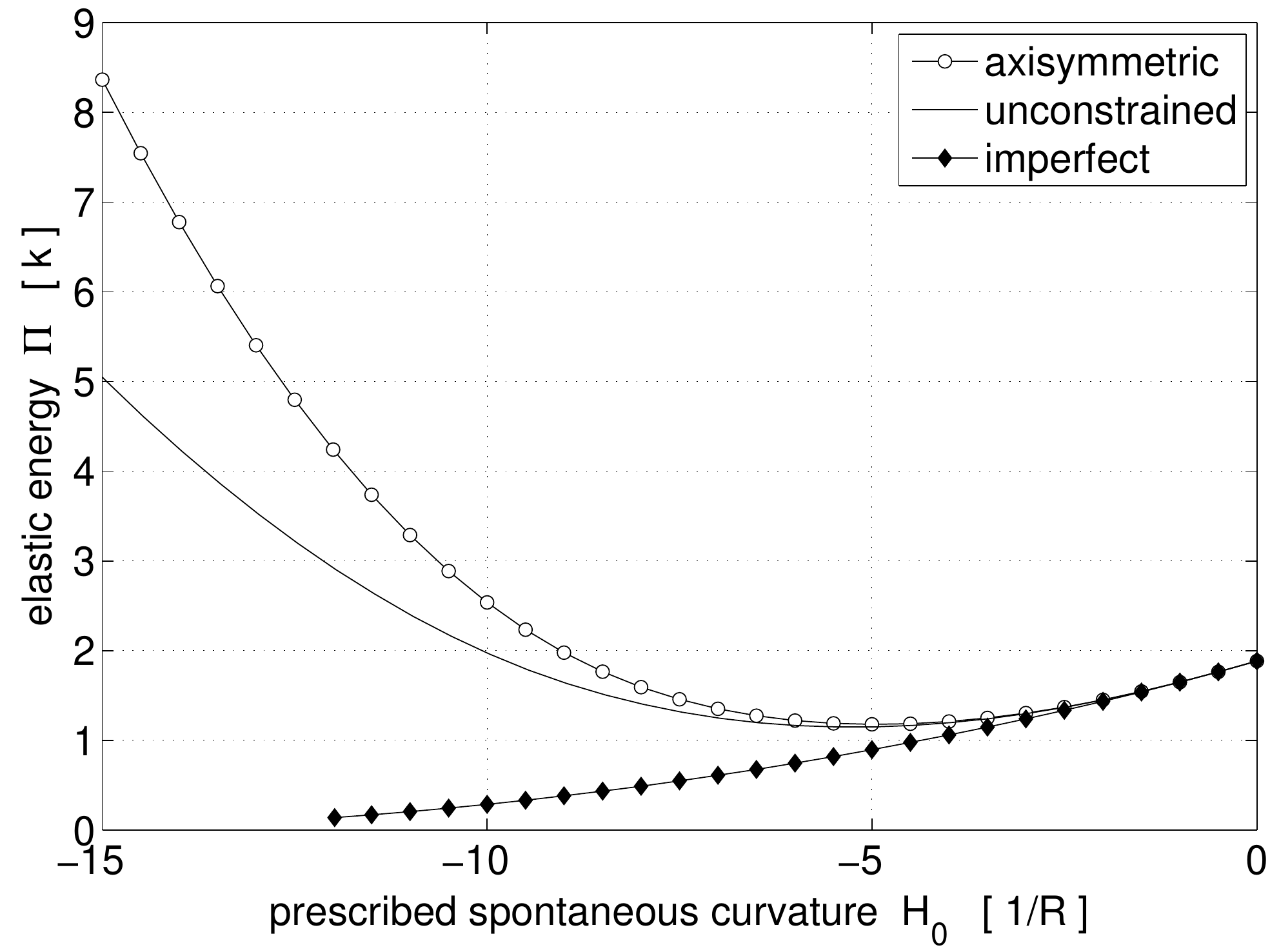}}
\put(-7.9,-.1){a.}
\put(0.4,-.1){b.}
\end{picture}
\caption{Cell budding: elastic surface energy vs.~spontaneous curvature for the three cases shown in Figs.~\ref{f:bud_A}, \ref{f:bud_C} and \ref{f:bud_I}: a.~bending energy; b.~total energy (bending $+$ areal part).}
\label{f:bud_Pu}
\end{center}
\end{figure}

By examining the surface energy
\eqb{l}
\Pi := \ds\int_{\sS_0}W\,\dif A~, 
\eqe
it can be seen that the non-axisymmetric shapes are preferred. As Fig.~\ref{f:bud_Pu} shows case 2 and 3 have much lower surface energy than case 1. 
The difference becomes especially large below $H_0=4/R$, when the deformations become large. 
As the system tries to minimize energy, this shows that the axisymmetric bud shape of case 1 is not a preferred solution.
As the figure shows, almost all of the energy goes into bending (contribution $Jw$ in \eqref{e:W_c}), as the areal part (contribution $Kg^2/2$ in \eqref{e:W_c}) becomes negligible for near incompressibility.


\subsubsection{Surface tension}

One of the advantages of the proposed finite element formulation is that
the surface tension $\gamma$ can be studied. 
This is done here for cases 1, 4 and 5 listed above.
It was shown before that the surface tension is not uniform under protein-induced spontaneous curvature through FE simulations utilizing Monge patches \citep{rangamani14}. The current simulations confirm this result and in addition yield further understanding for large deformations.

\begin{figure}[h]
\begin{center} \unitlength1cm
\begin{picture}(0,5.6)
\put(3.95,-.4){\includegraphics[height=31mm]{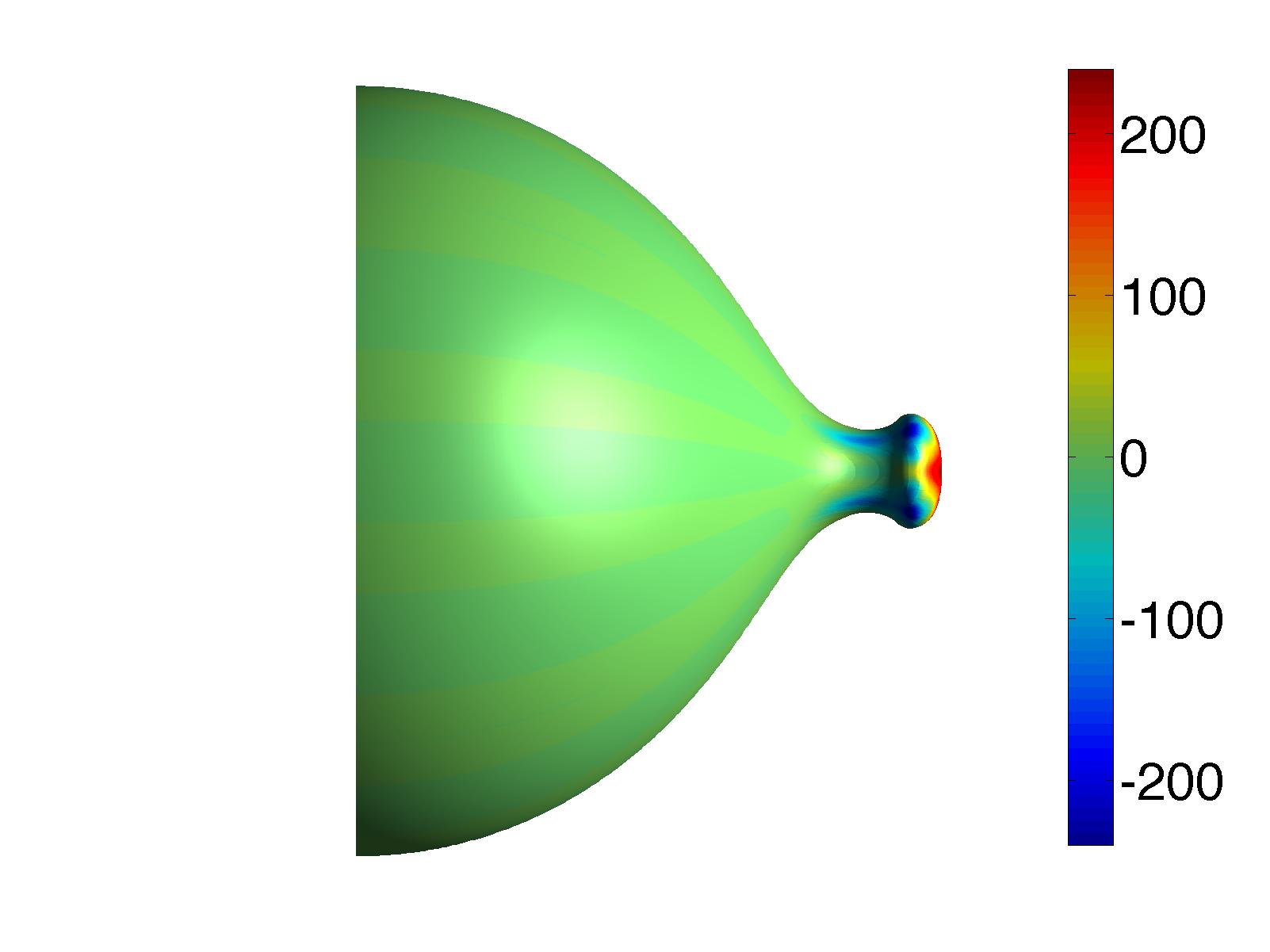}}
\put(0.75,-.4){\includegraphics[height=31mm]{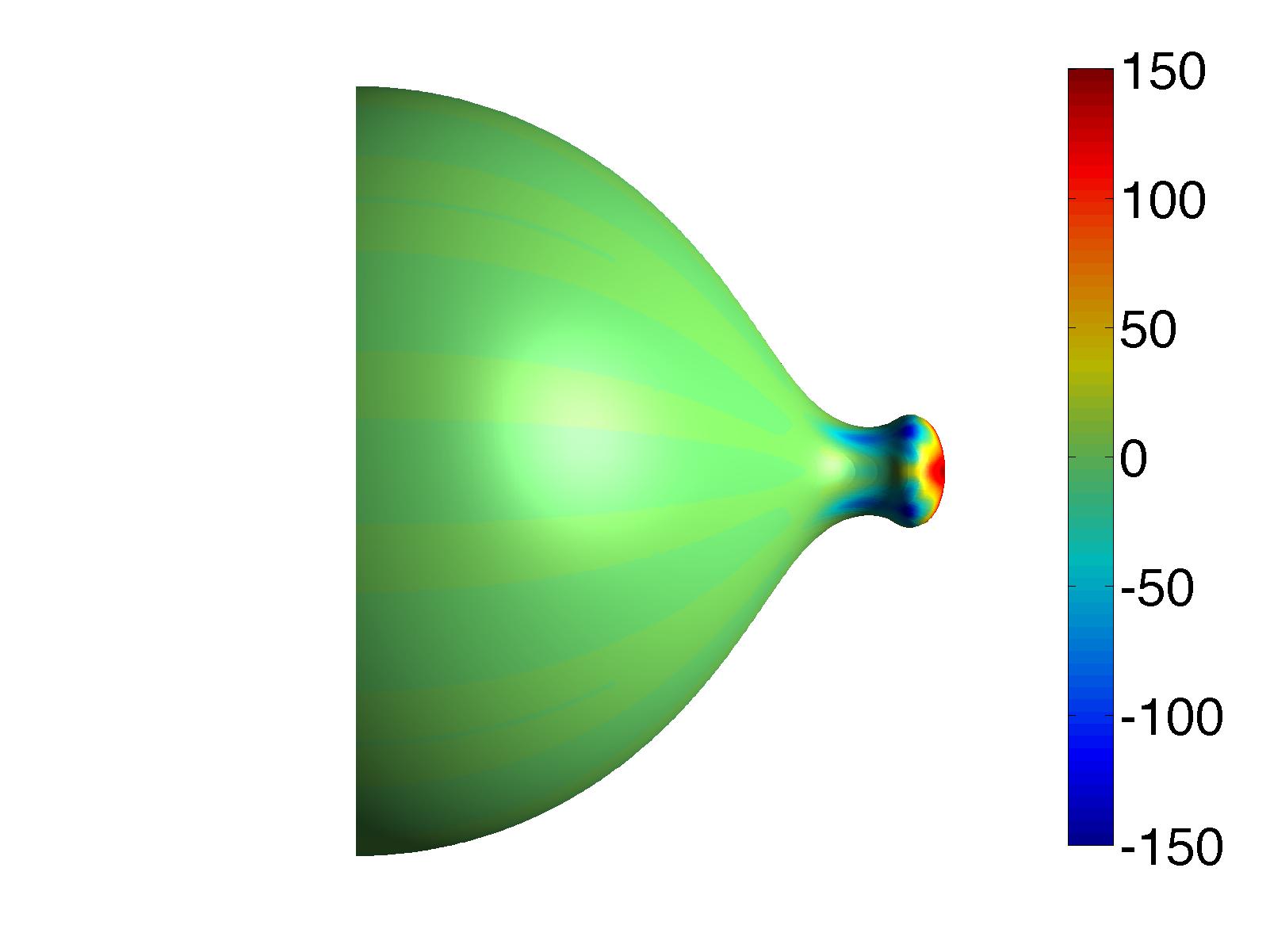}}
\put(-2.5,-.4){\includegraphics[height=31mm]{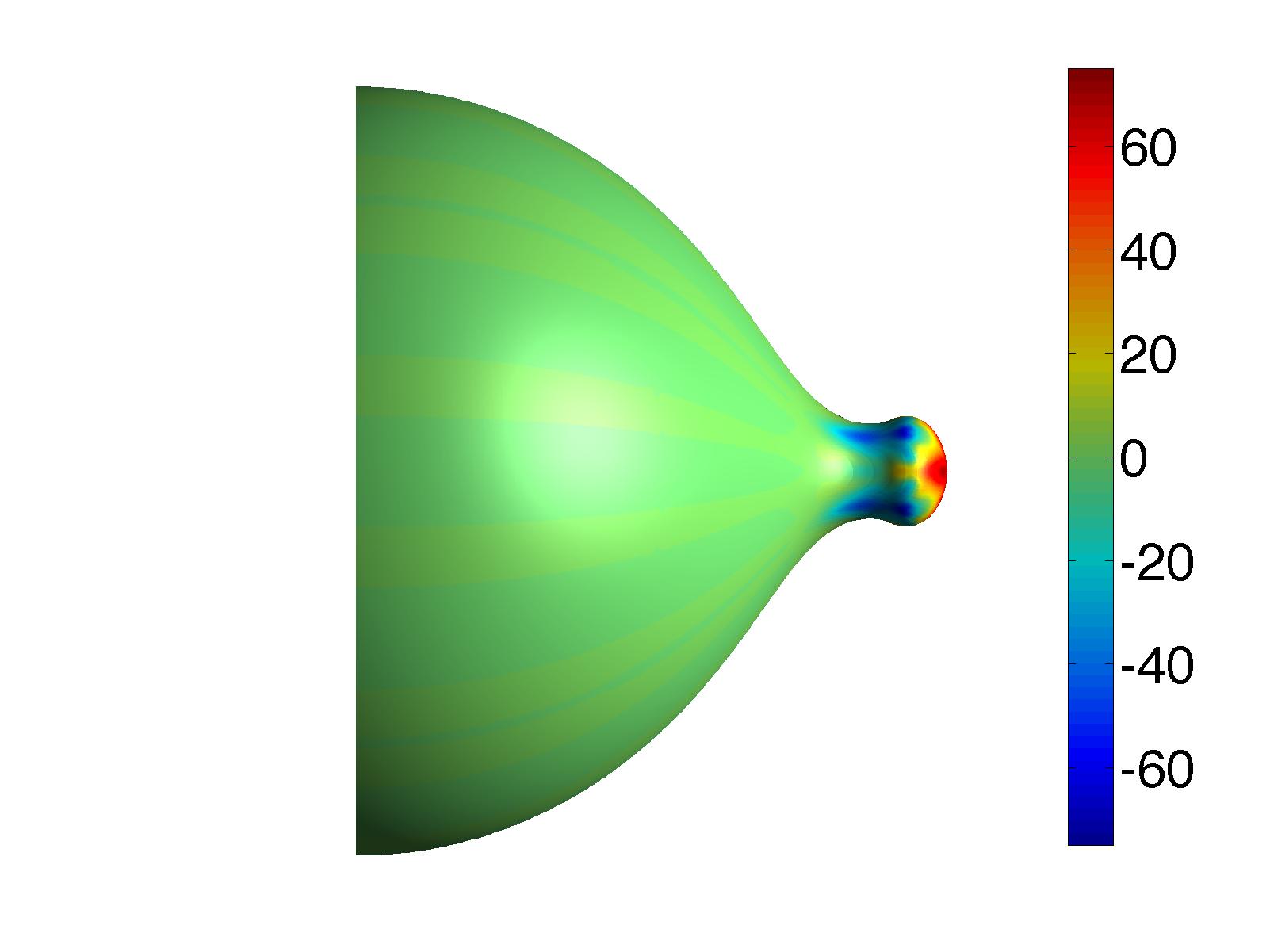}}
\put(-5.75,-.4){\includegraphics[height=31mm]{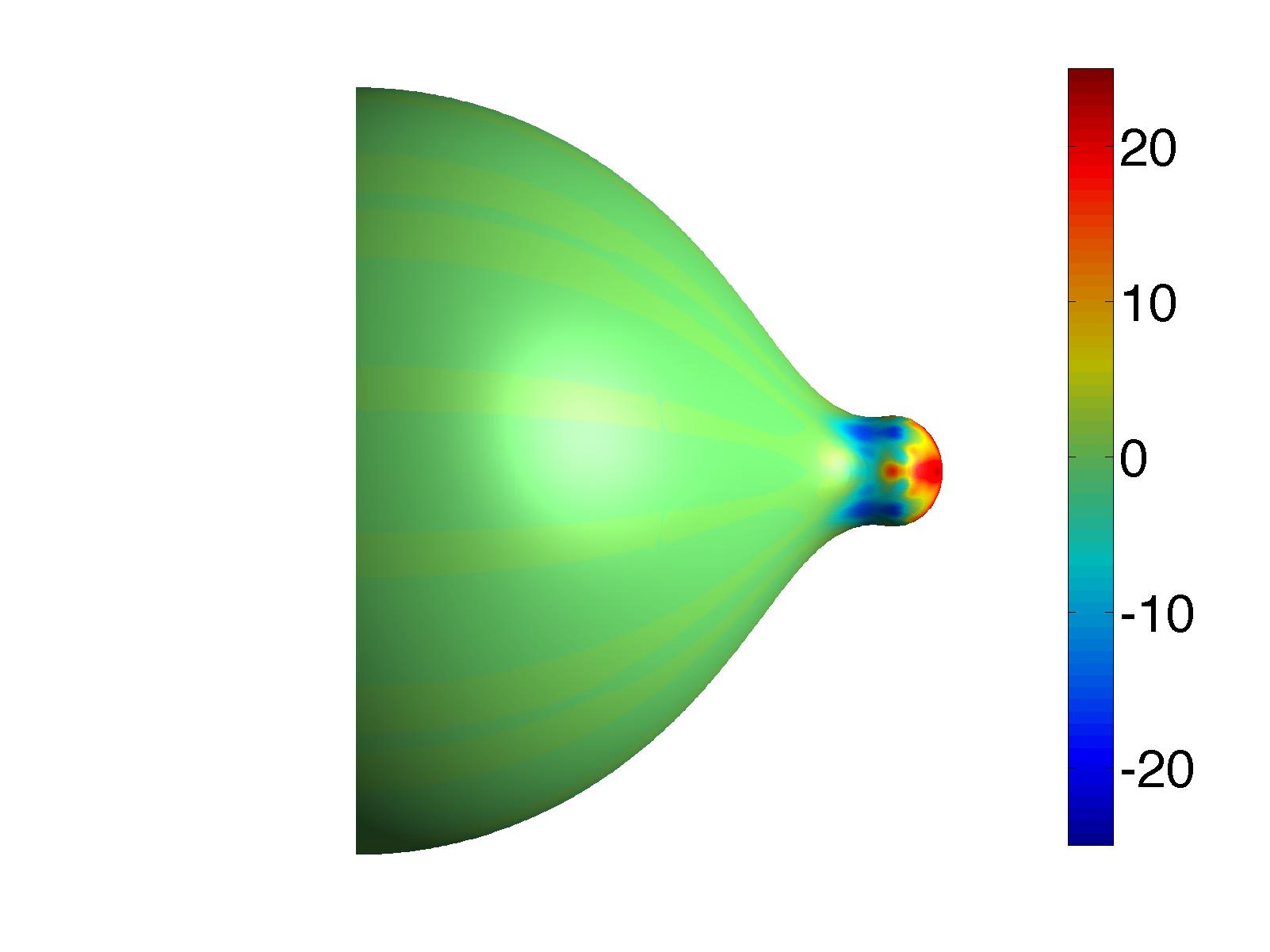}}
\put(-9.0,-.4){\includegraphics[height=31mm]{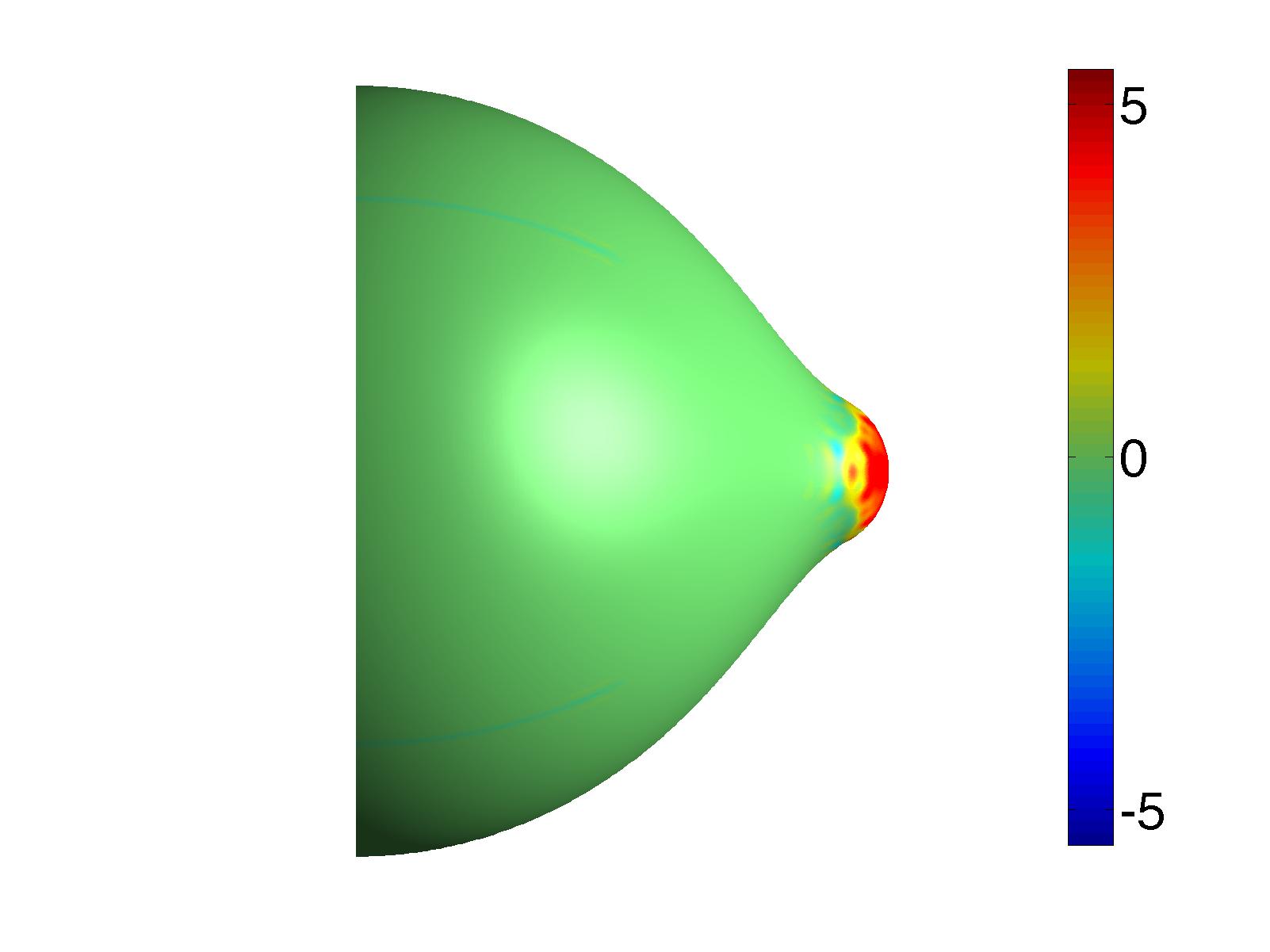}}
\put(-8.6,2.6){\includegraphics[height=31mm]{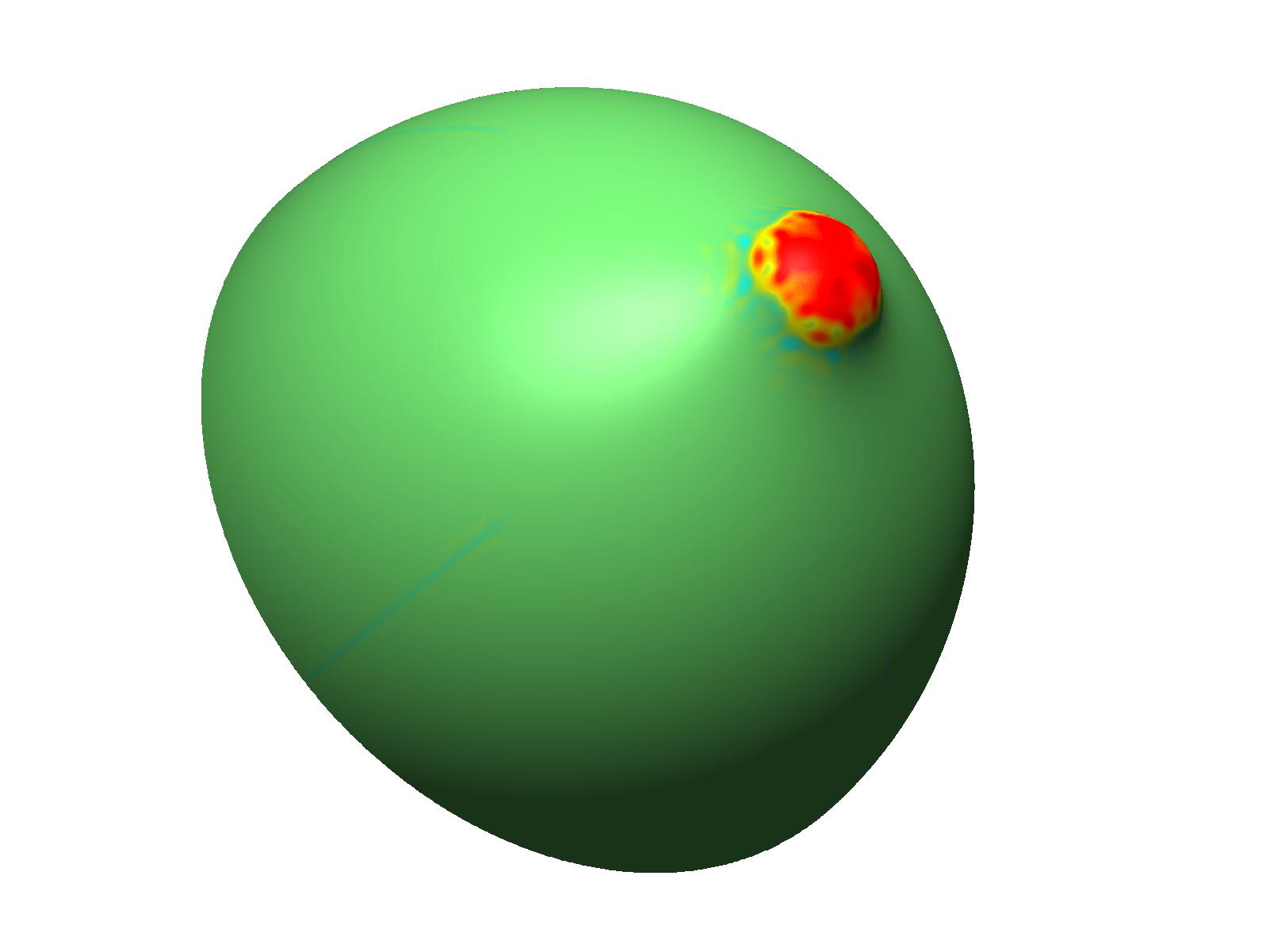}}
\put(-5.35,2.6){\includegraphics[height=31mm]{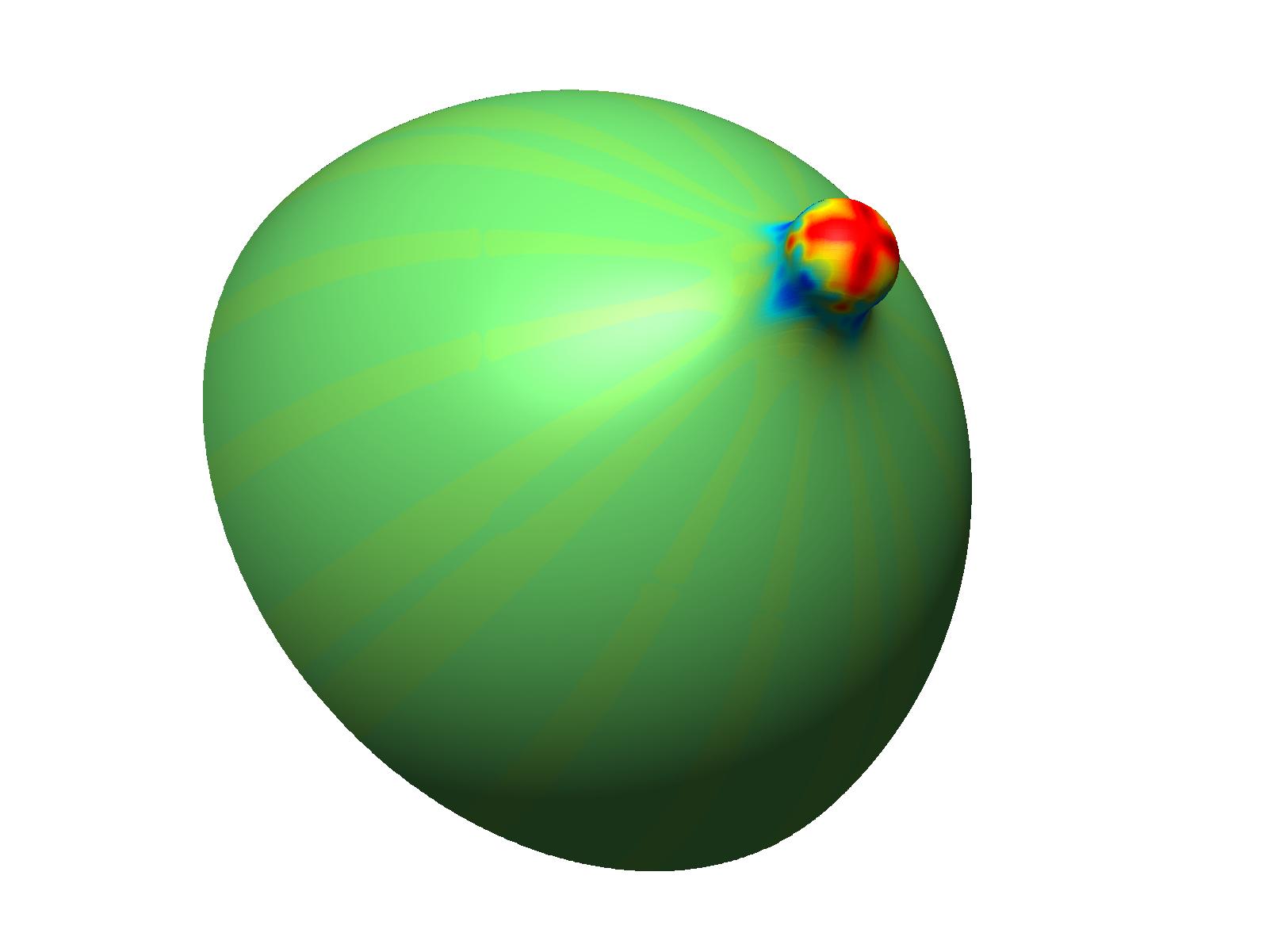}}
\put(-2.1,2.6){\includegraphics[height=31mm]{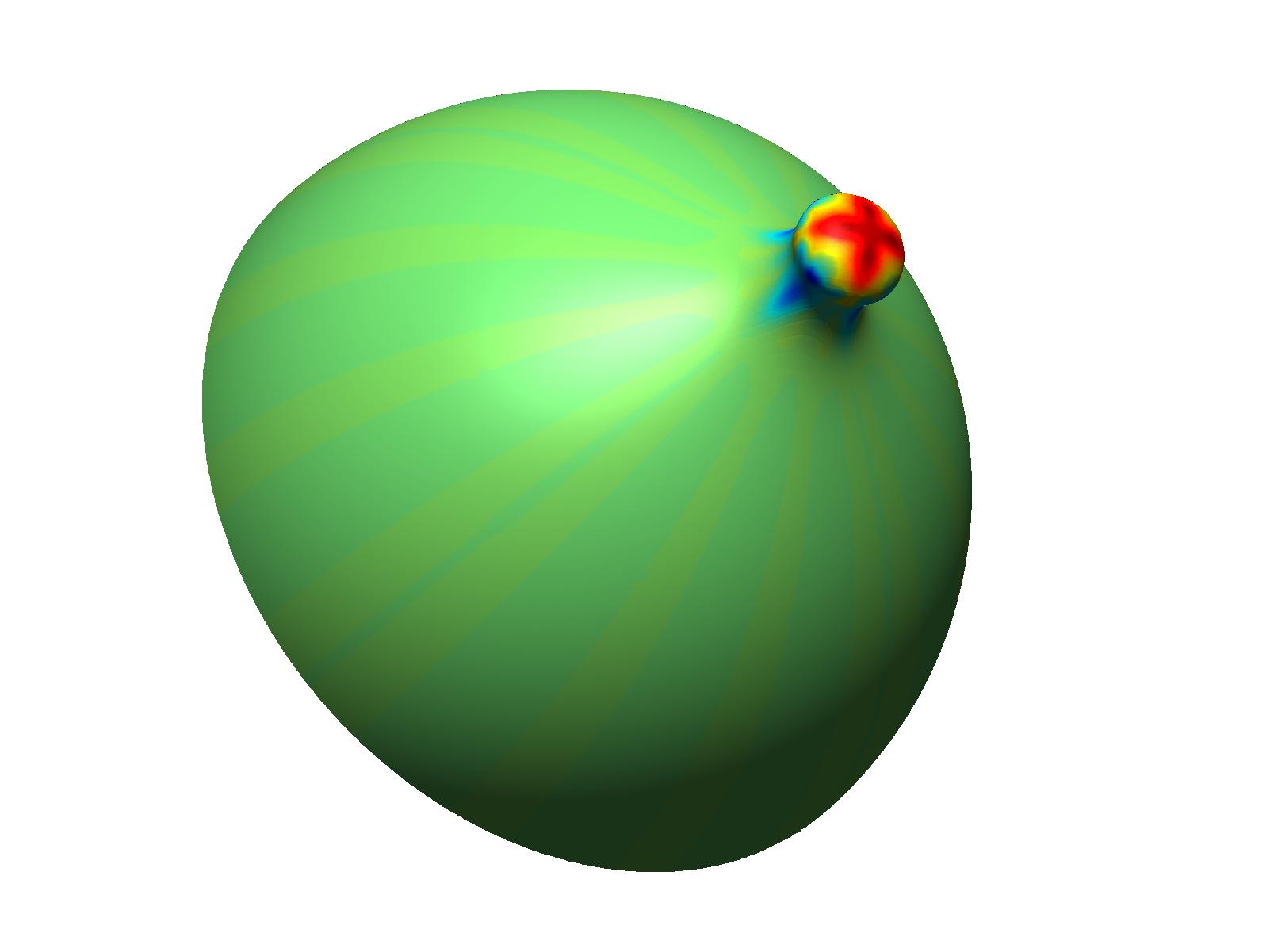}}
\put(1.15,2.6){\includegraphics[height=31mm]{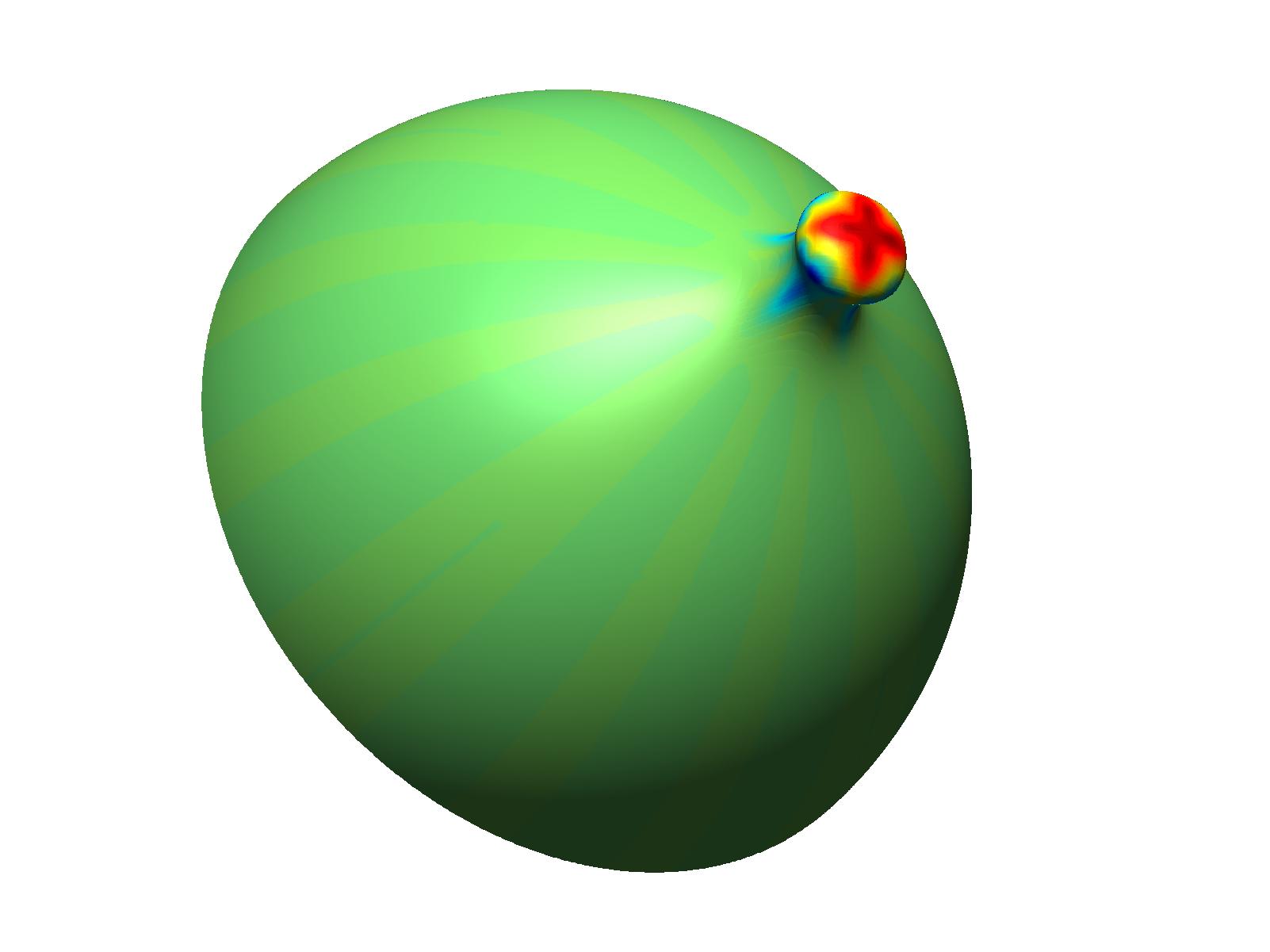}}
\put(4.4,2.6){\includegraphics[height=31mm]{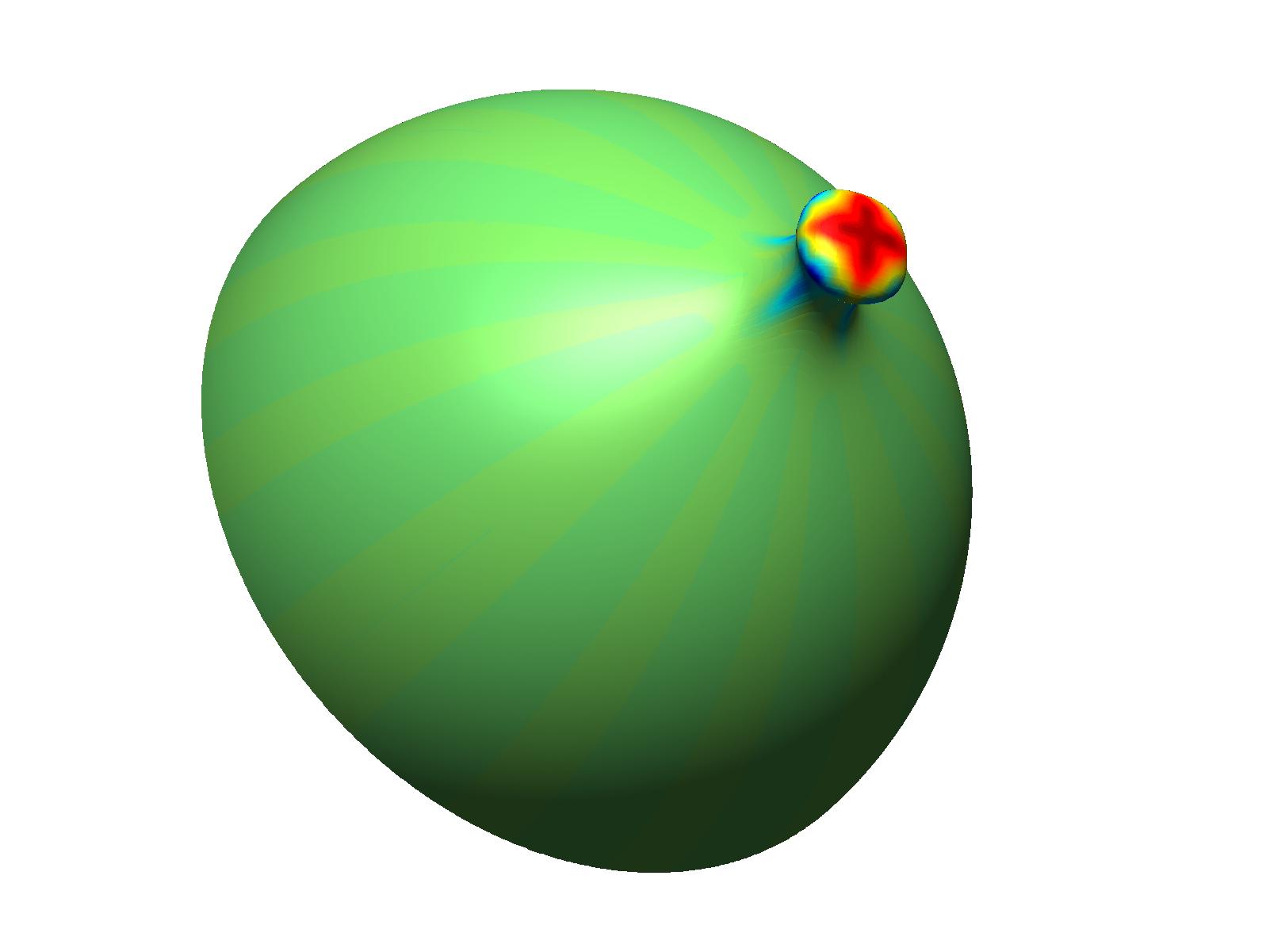}}
\end{picture}
\caption{Cell budding: Axisymmetric case at $\bar H_0=-5,-10,-15,-20,-25$ (left to right): 3D and side view of deformation and surface tension $\bar\gamma$.}
\label{f:bud_Ay}
\end{center}
\end{figure}

First, the axisymmetric case (case 1) is examined in Fig.~\ref{f:bud_Ay}. 
As seen $\gamma$ becomes maximum within the protein patch. This maximum is constant across the patch for low $H_0$. This changes for increasing $H_0$, where a distinct maximum appears in the center.\footnote{It can be seen that the distribution of $\gamma$ is not exactly axisymmetric. This is due to the inexact enforcement of axisymmetry noted in footnote 4.} At a certain level of $H_0$ rupture would occur here, depending on the strength of the lipid bilayer. 

\begin{figure}[h]
\begin{center} \unitlength1cm
\begin{picture}(0,5.6)
\put(3.95,-.4){\includegraphics[height=31mm]{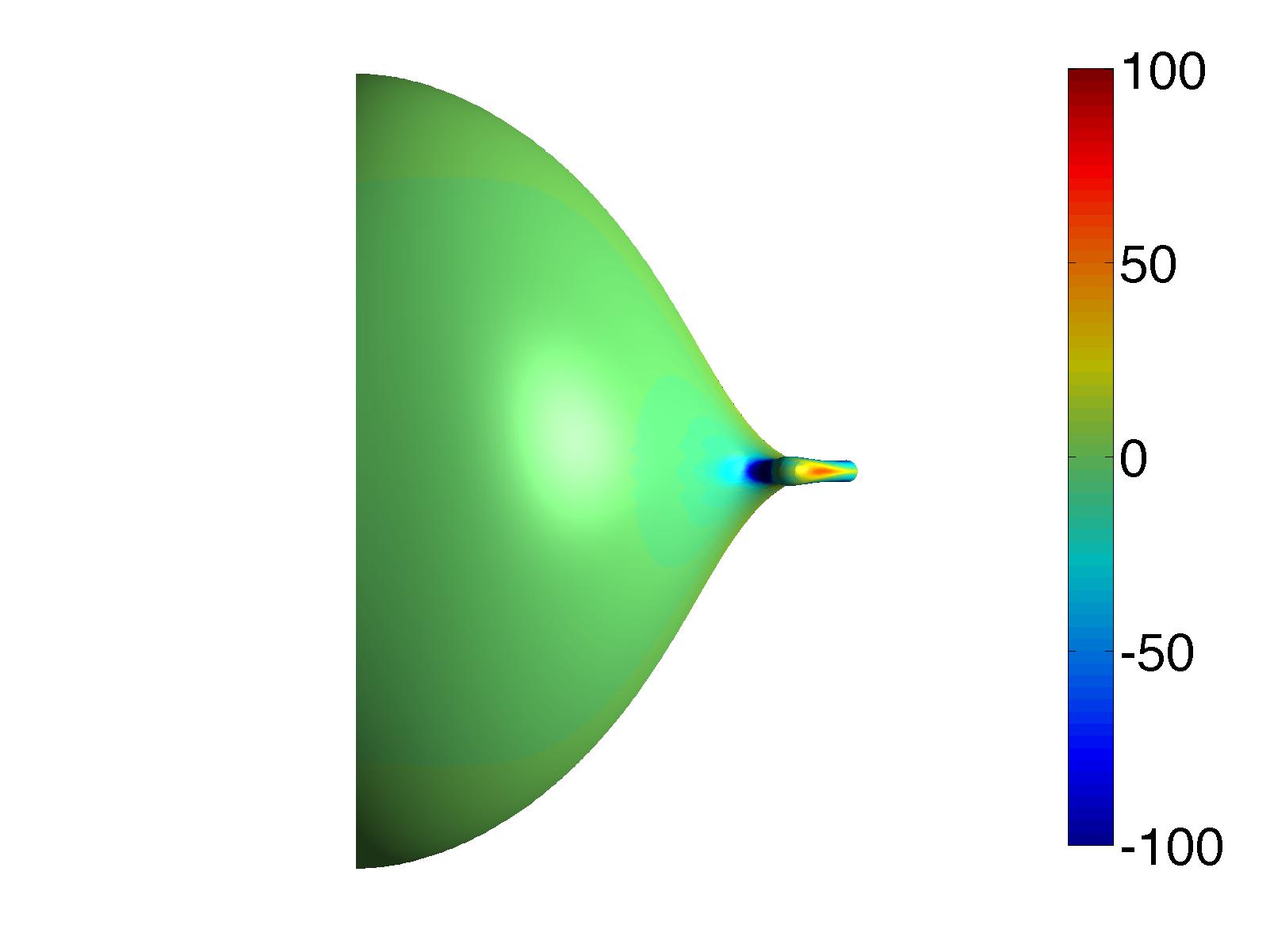}}
\put(0.75,-.4){\includegraphics[height=31mm]{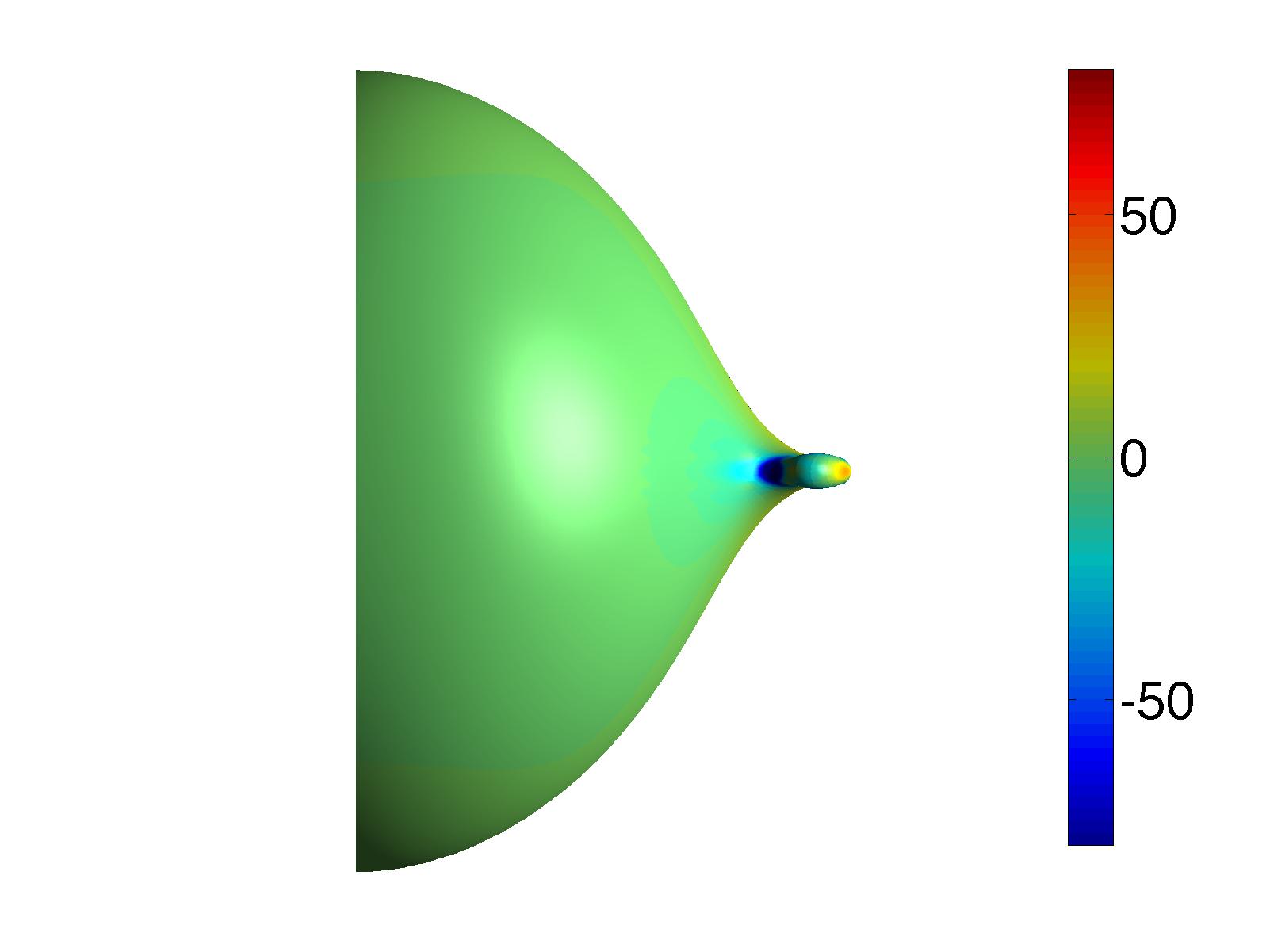}}
\put(-2.5,-.4){\includegraphics[height=31mm]{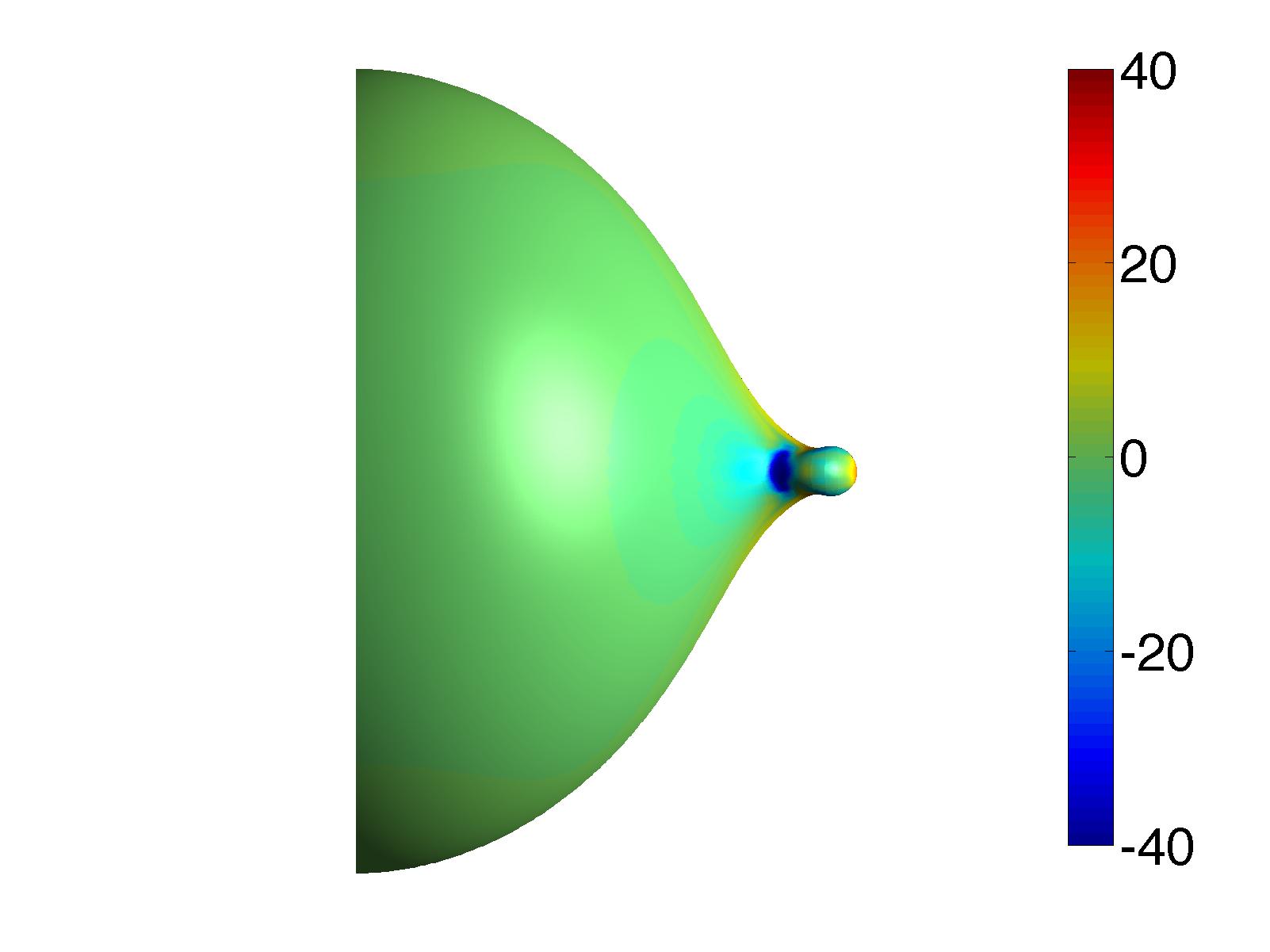}}
\put(-5.75,-.4){\includegraphics[height=31mm]{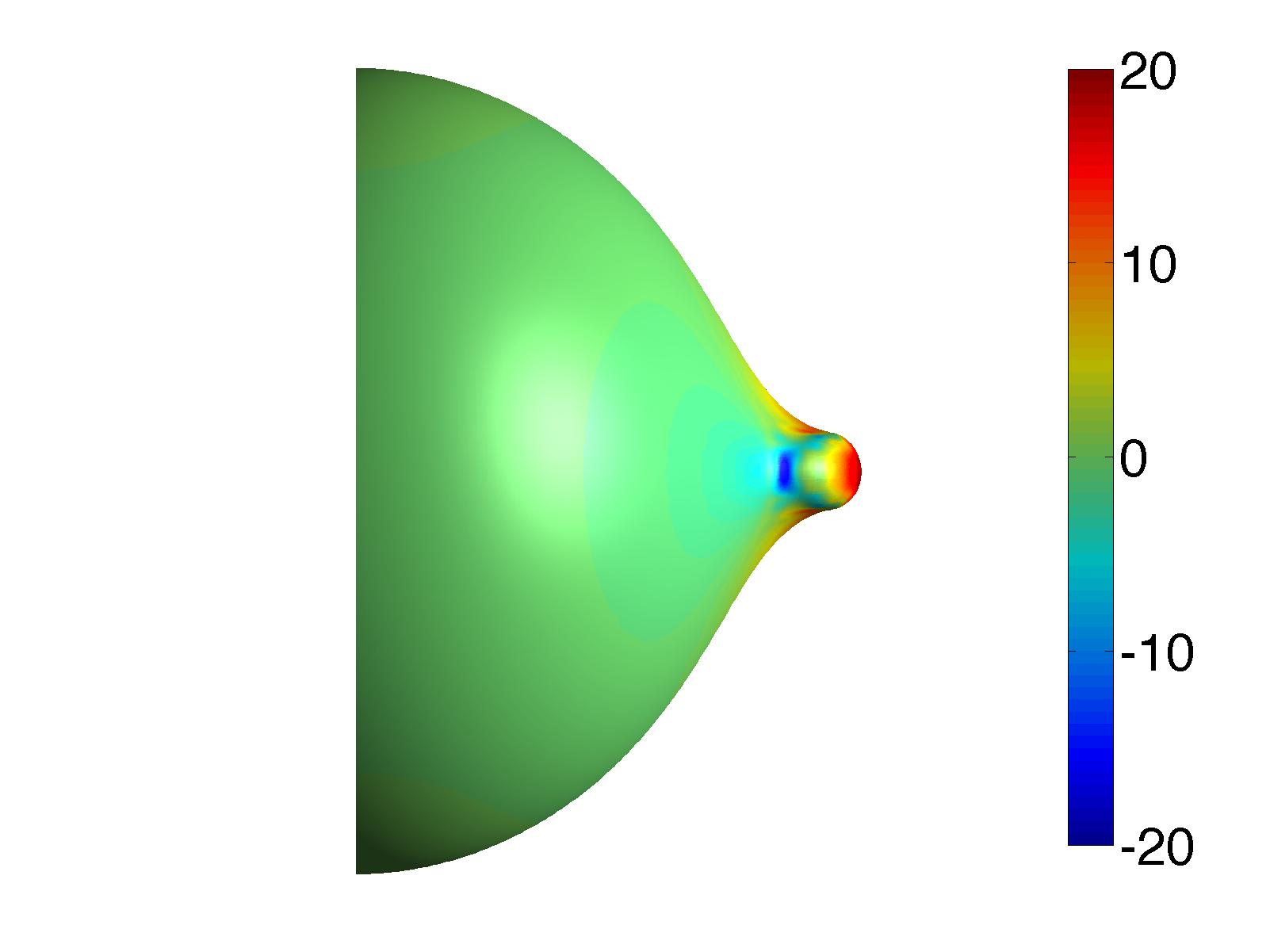}}
\put(-9.0,-.4){\includegraphics[height=31mm]{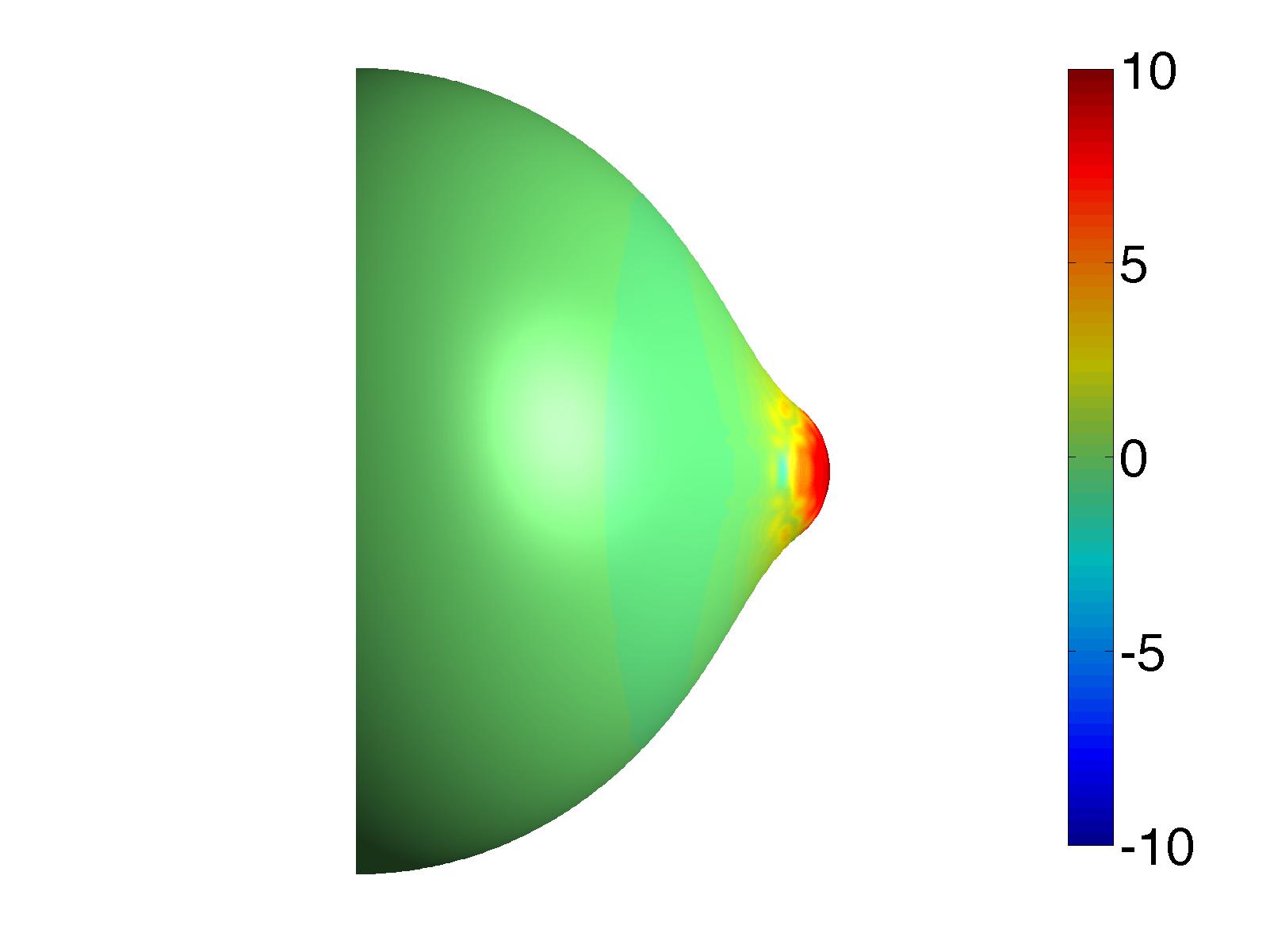}}
\put(-8.6,2.6){\includegraphics[height=31mm]{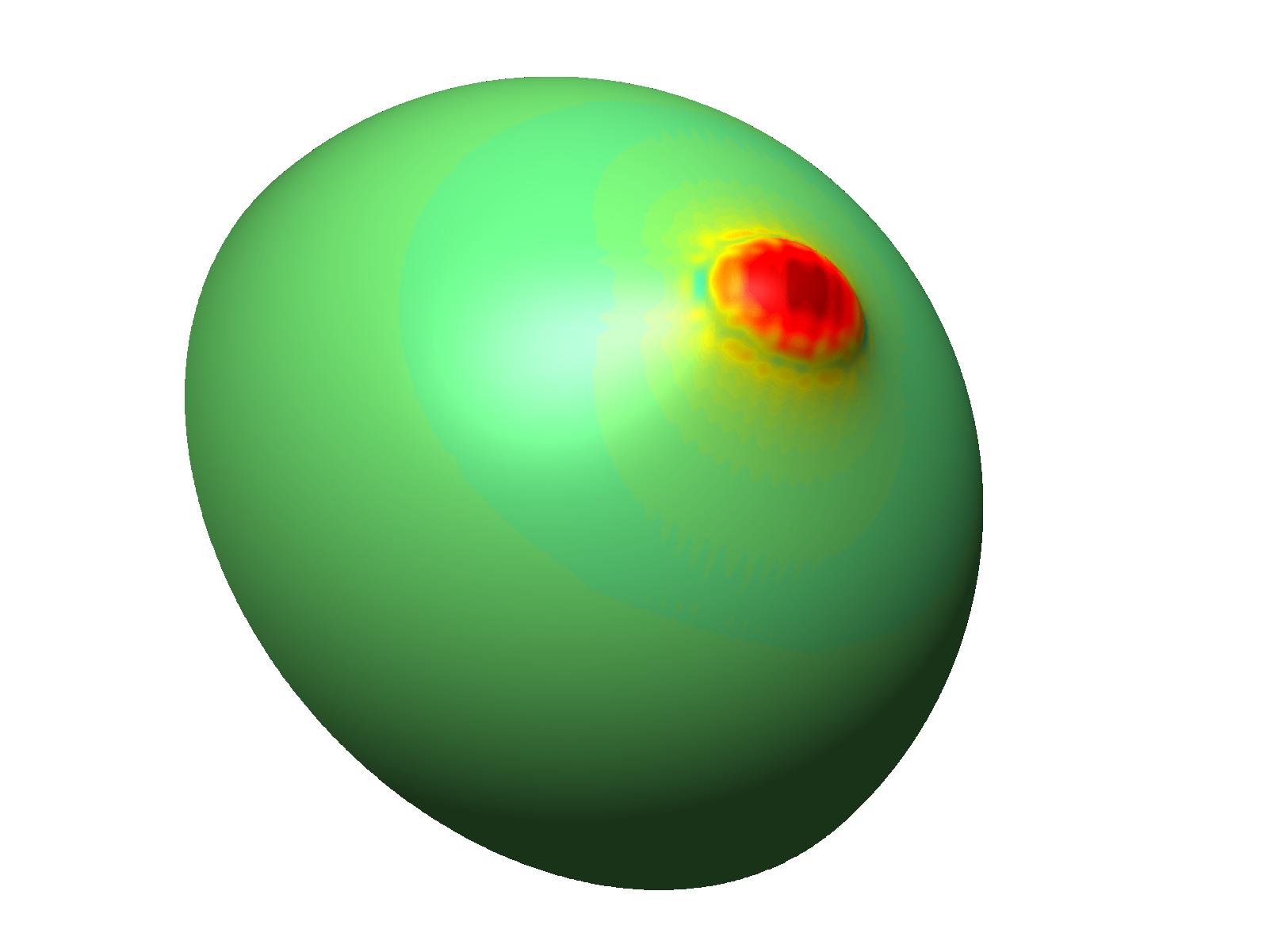}}
\put(-5.35,2.6){\includegraphics[height=31mm]{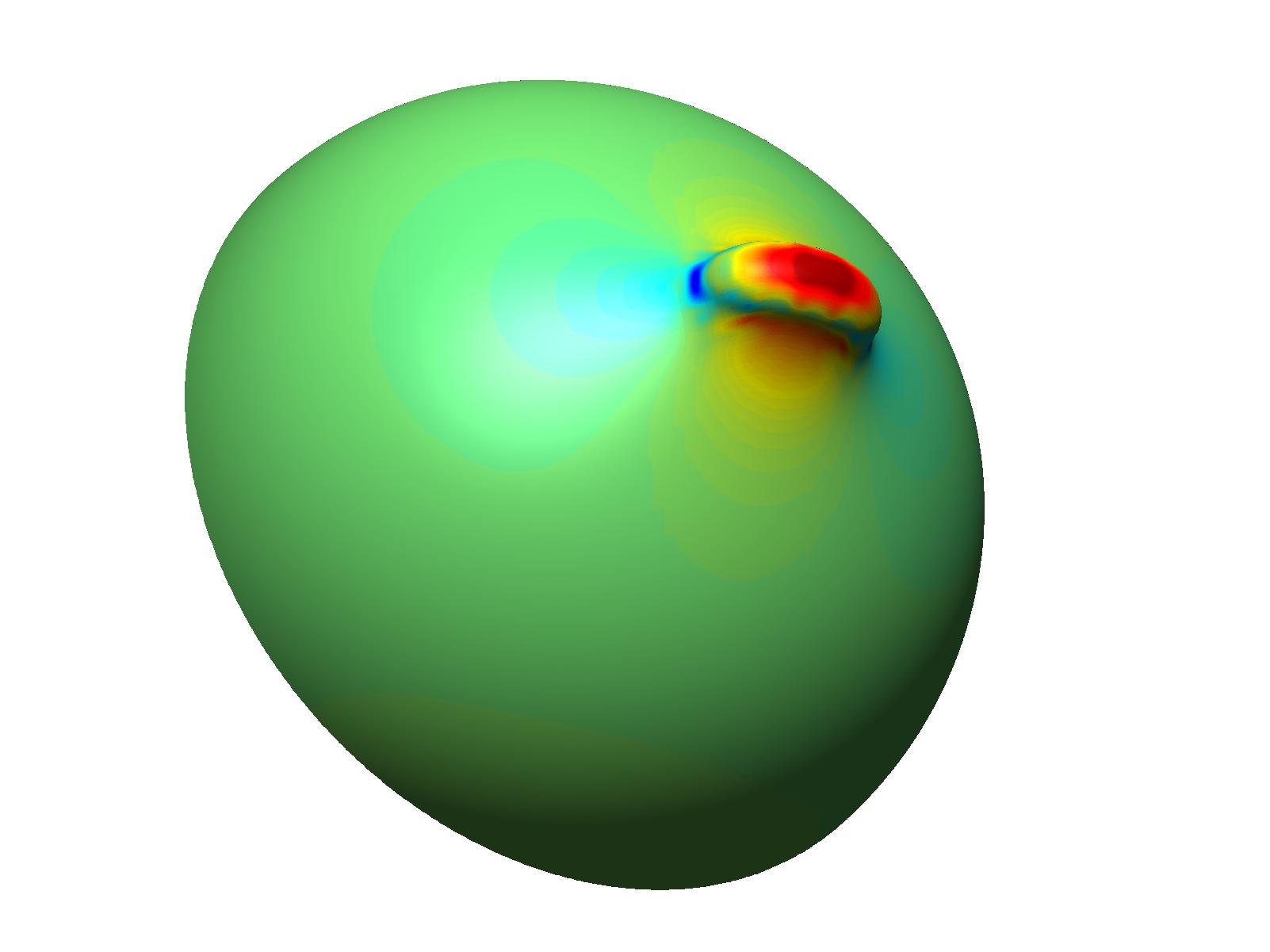}}
\put(-2.1,2.6){\includegraphics[height=31mm]{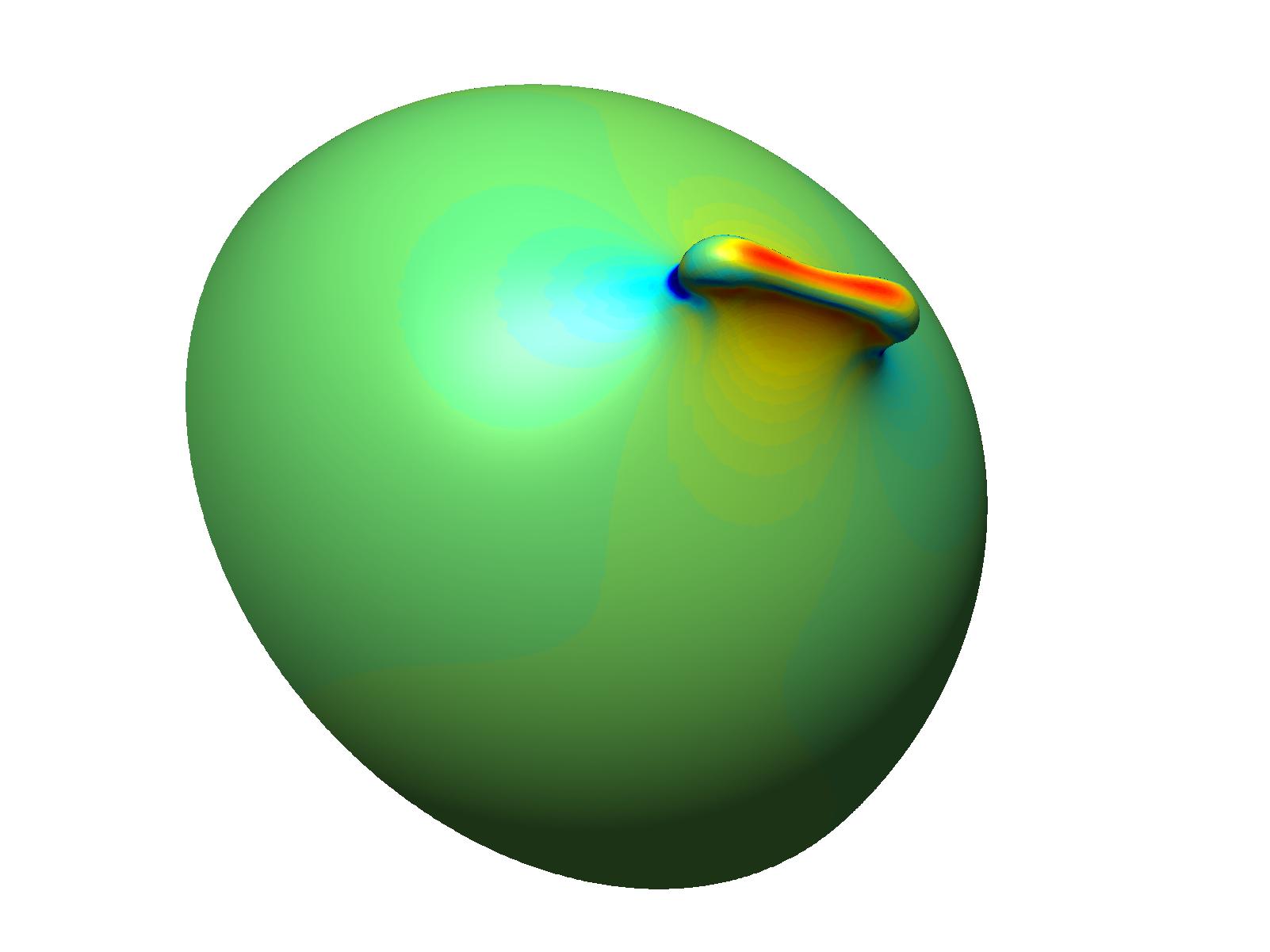}}
\put(1.15,2.6){\includegraphics[height=31mm]{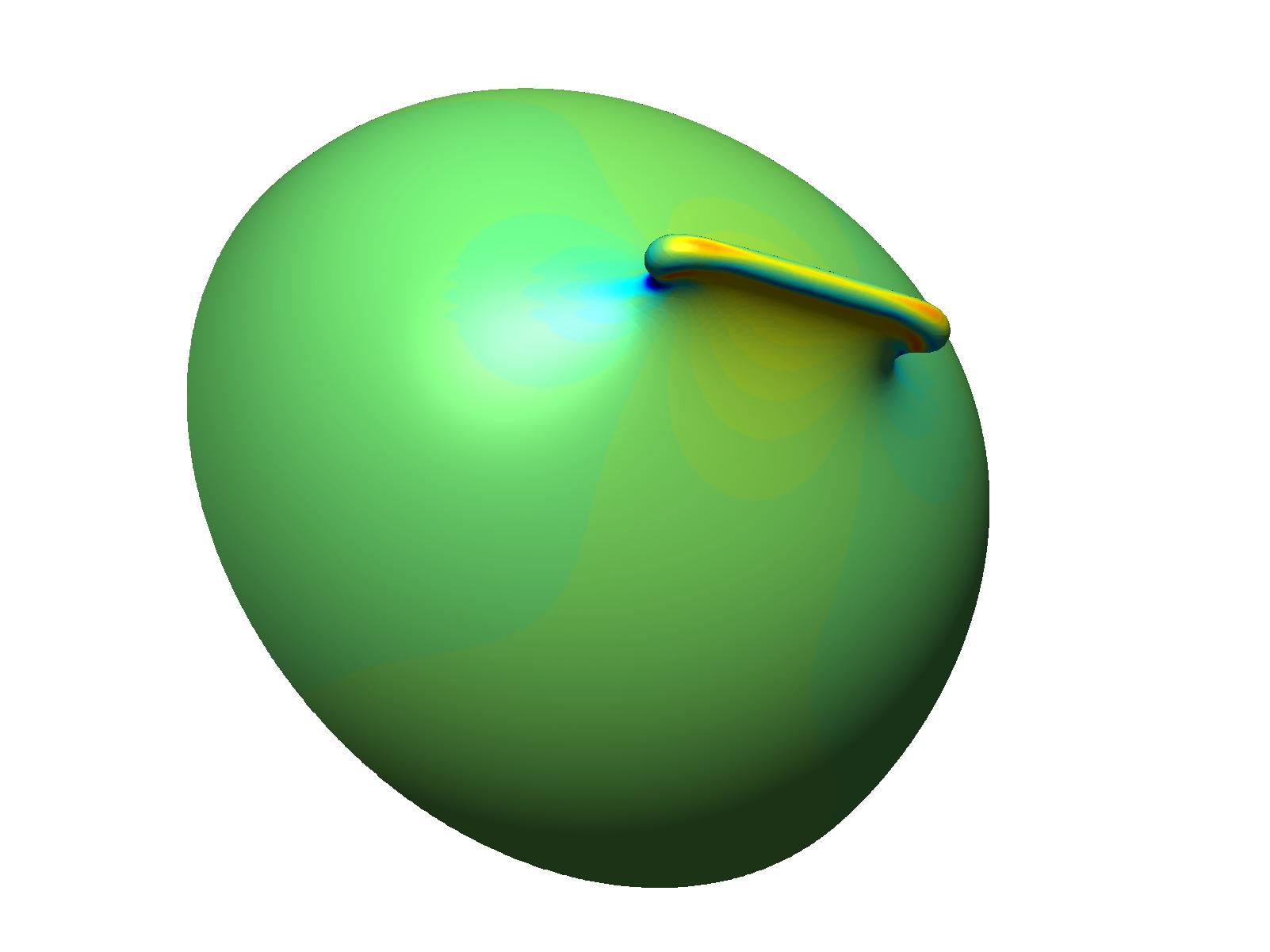}}
\put(4.4,2.6){\includegraphics[height=31mm]{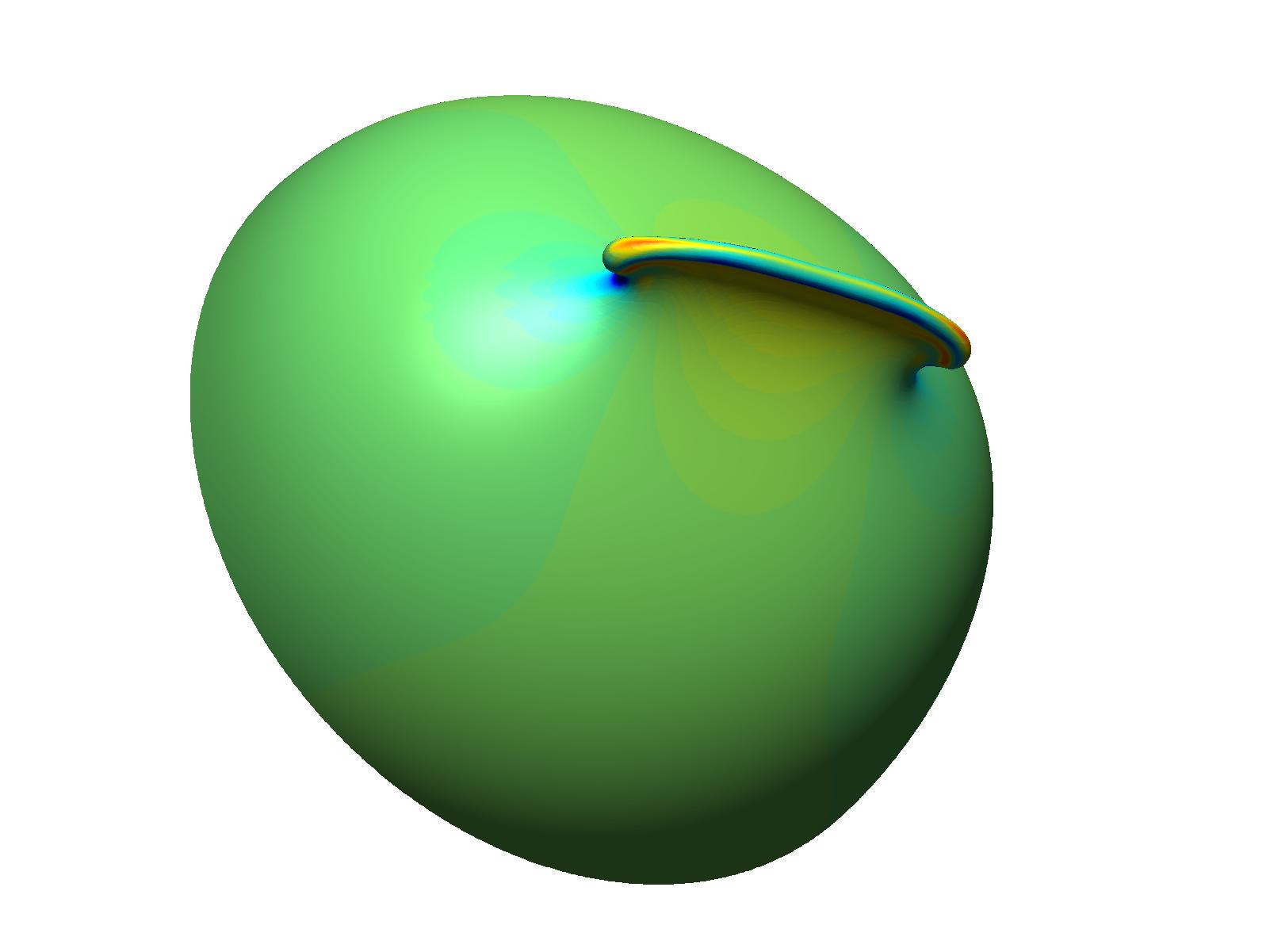}}
\end{picture}
\caption{Cell budding: Shear resistant, imperfect case at $\bar H_0=-5,-10,-15,-20,-25$ (left to right): 3D and side view of deformation and surface tension $\bar\gamma$.}
\label{f:bud_Imuy}
\end{center}
\end{figure}

\begin{figure}[h]
\begin{center} \unitlength1cm
\begin{picture}(0,5.6)
\put(3.95,-.4){\includegraphics[height=31mm]{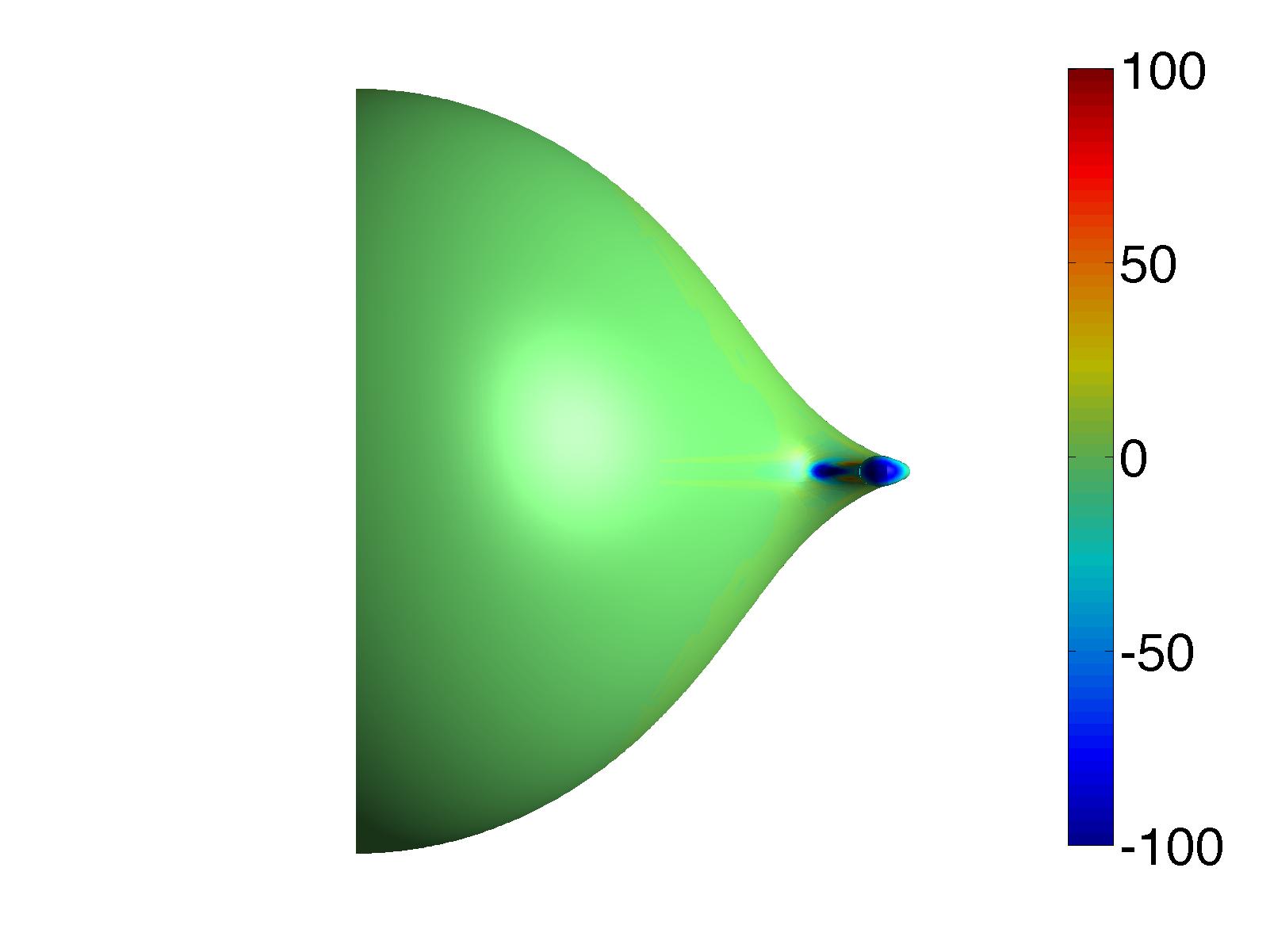}}
\put(0.75,-.4){\includegraphics[height=31mm]{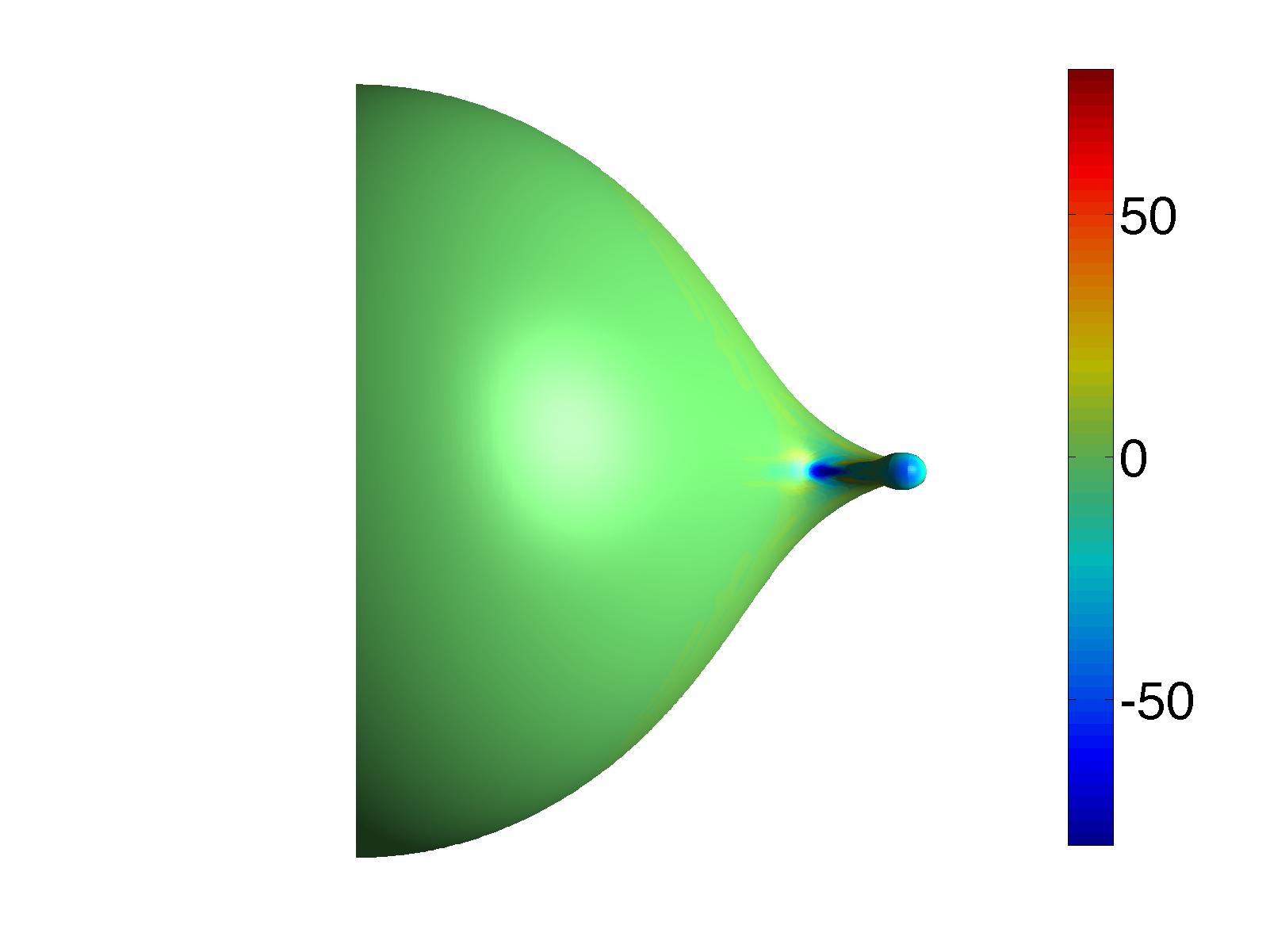}}
\put(-2.5,-.4){\includegraphics[height=31mm]{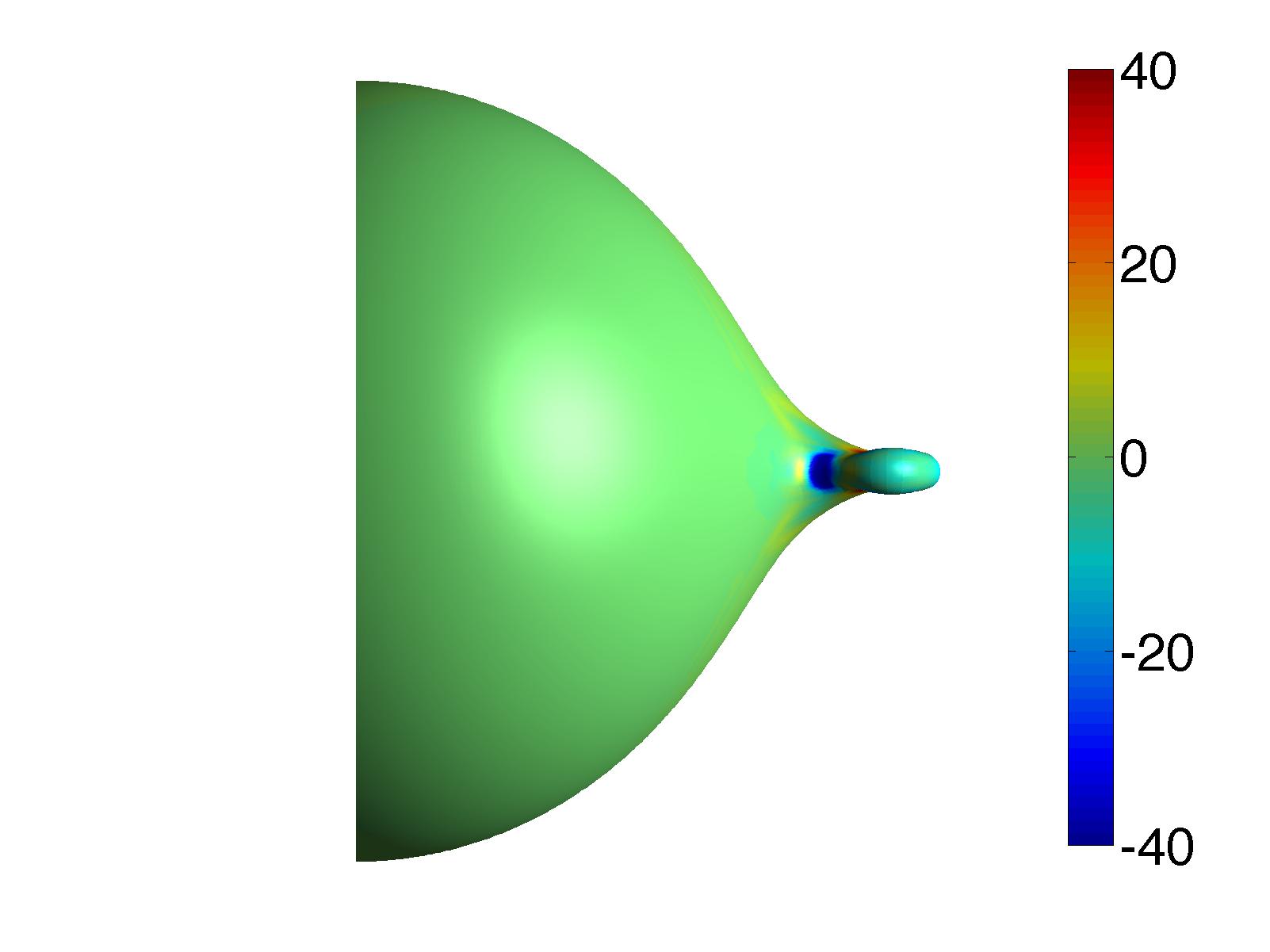}}
\put(-5.75,-.4){\includegraphics[height=31mm]{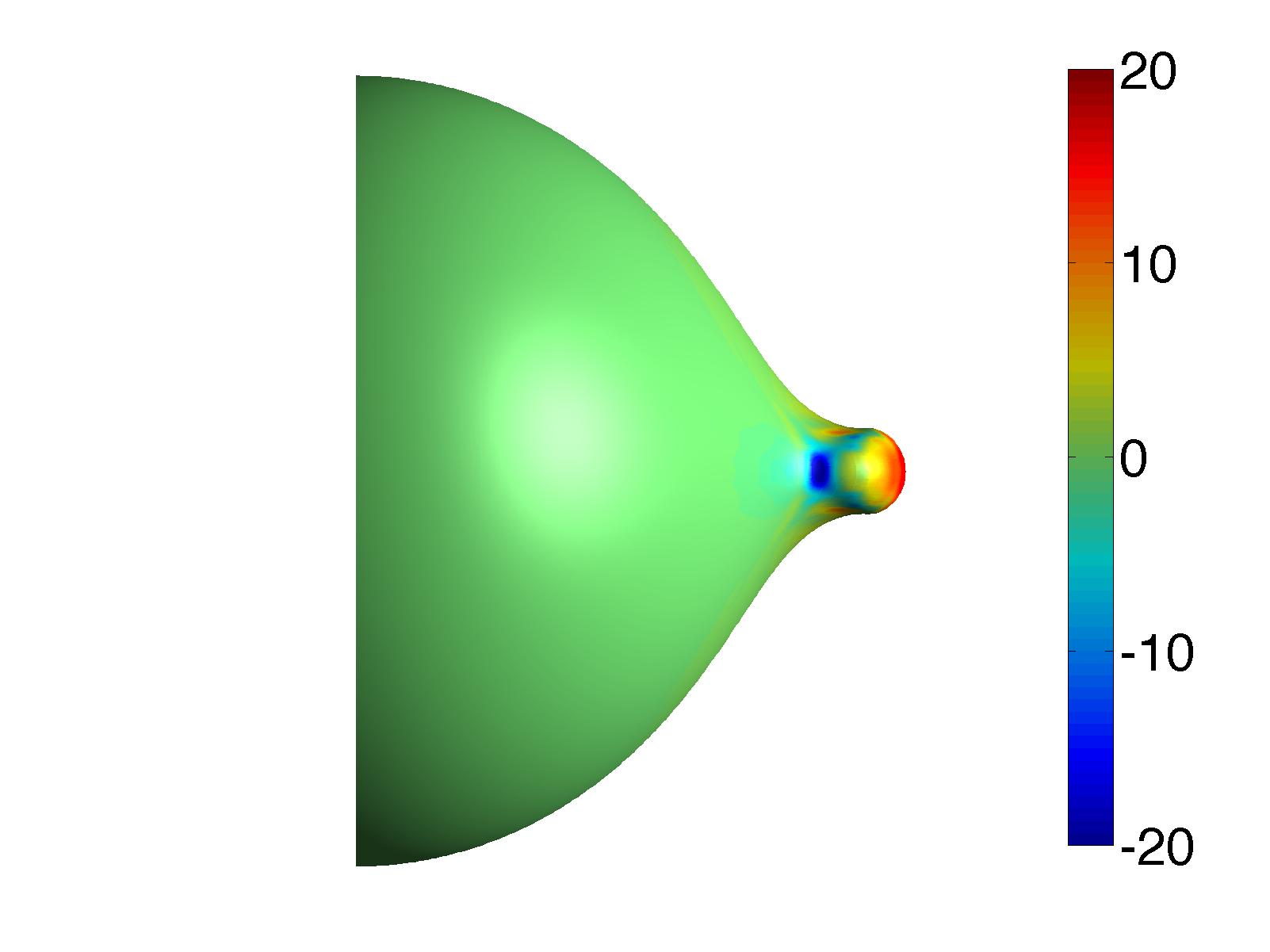}}
\put(-9.0,-.4){\includegraphics[height=31mm]{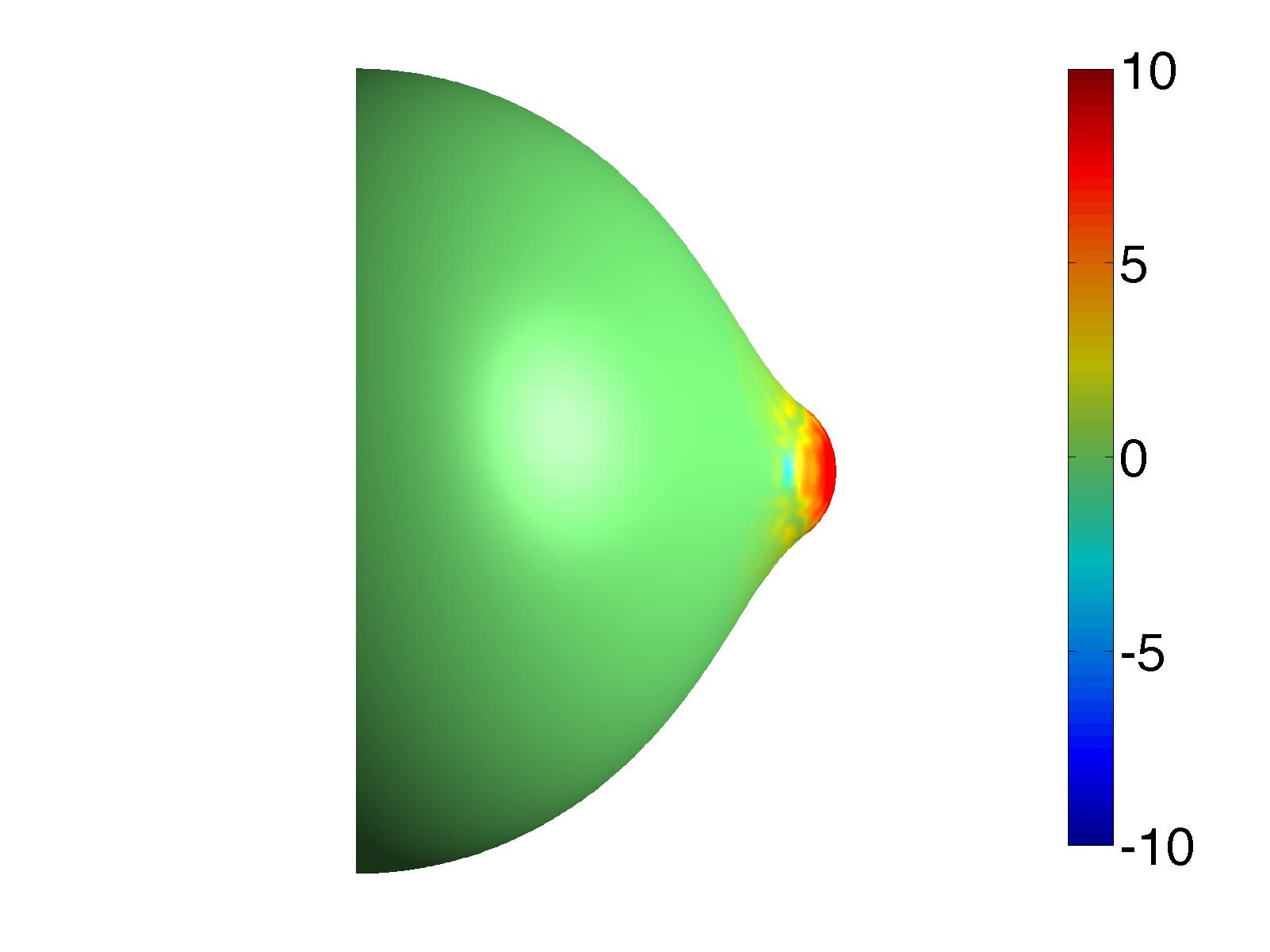}}
\put(-8.6,2.6){\includegraphics[height=31mm]{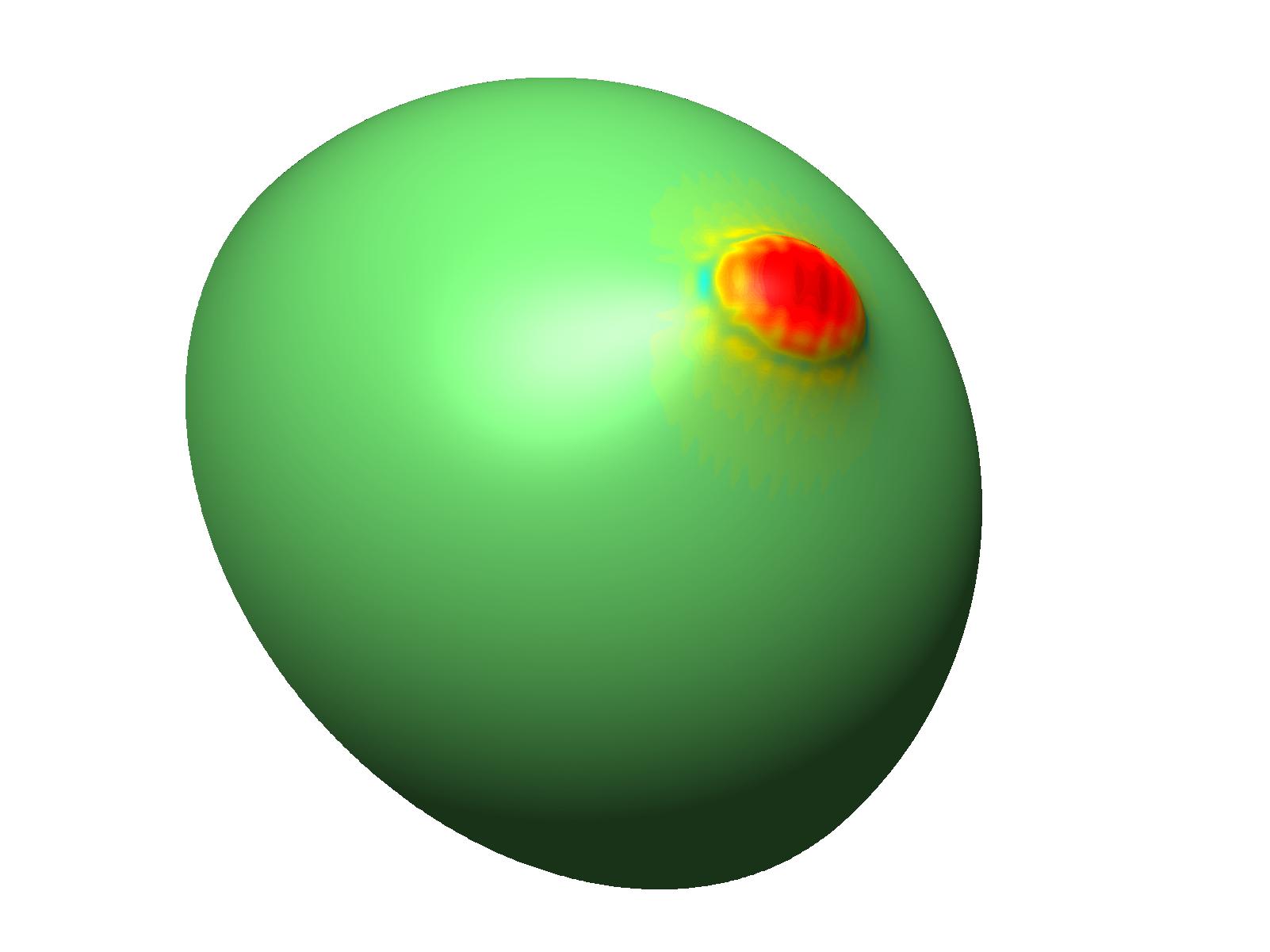}}
\put(-5.35,2.6){\includegraphics[height=31mm]{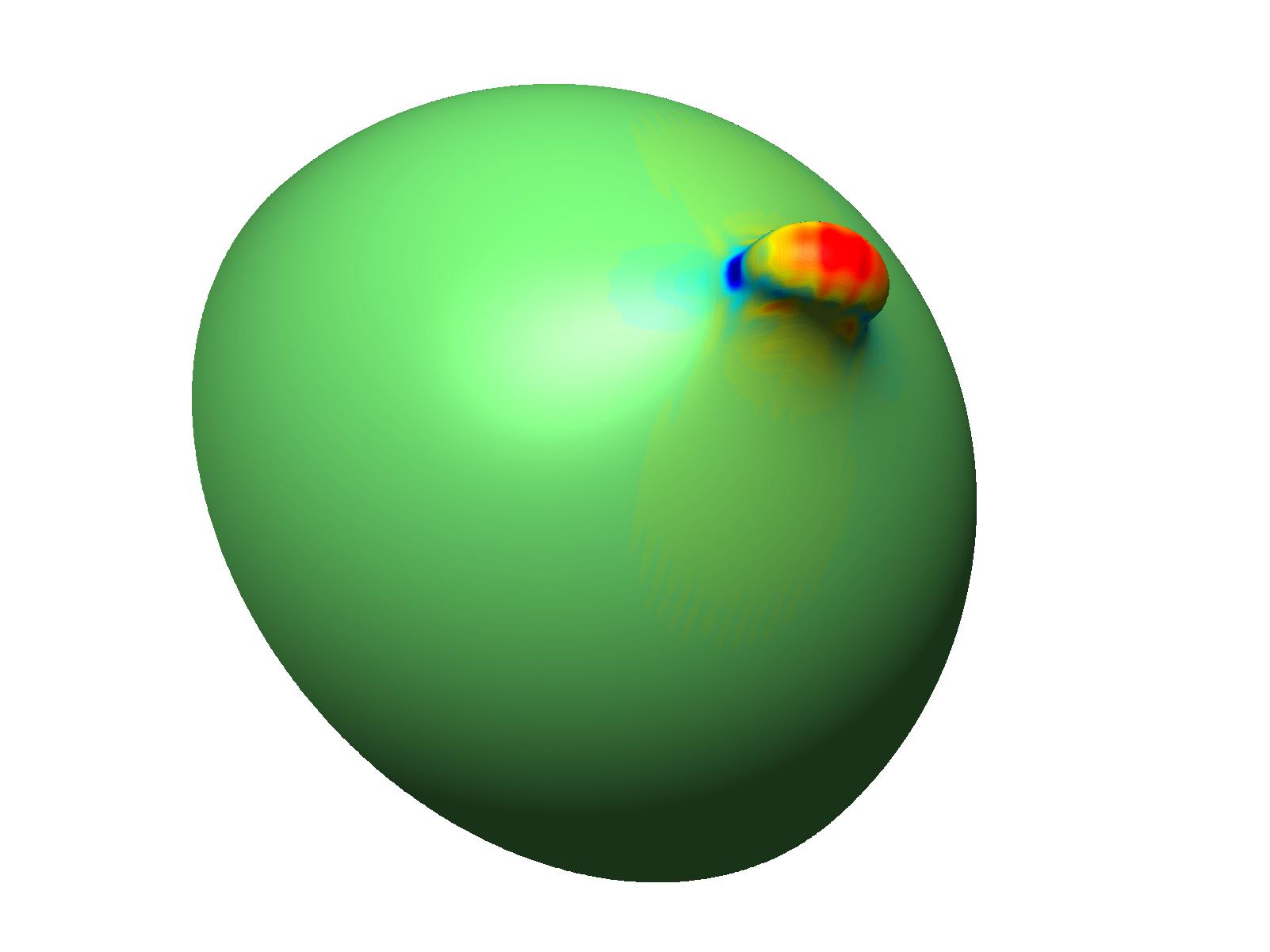}}
\put(-2.1,2.6){\includegraphics[height=31mm]{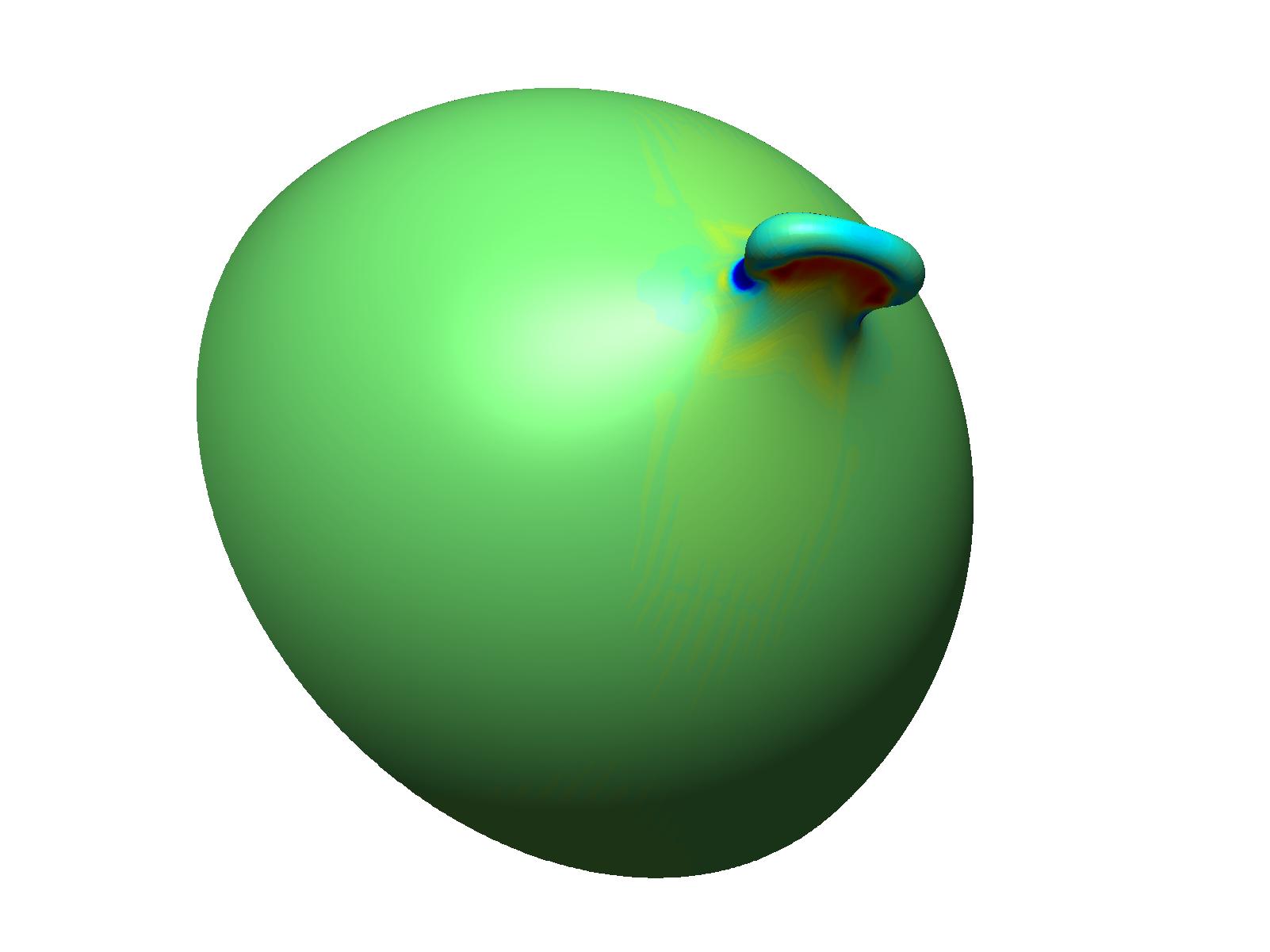}}
\put(1.15,2.6){\includegraphics[height=31mm]{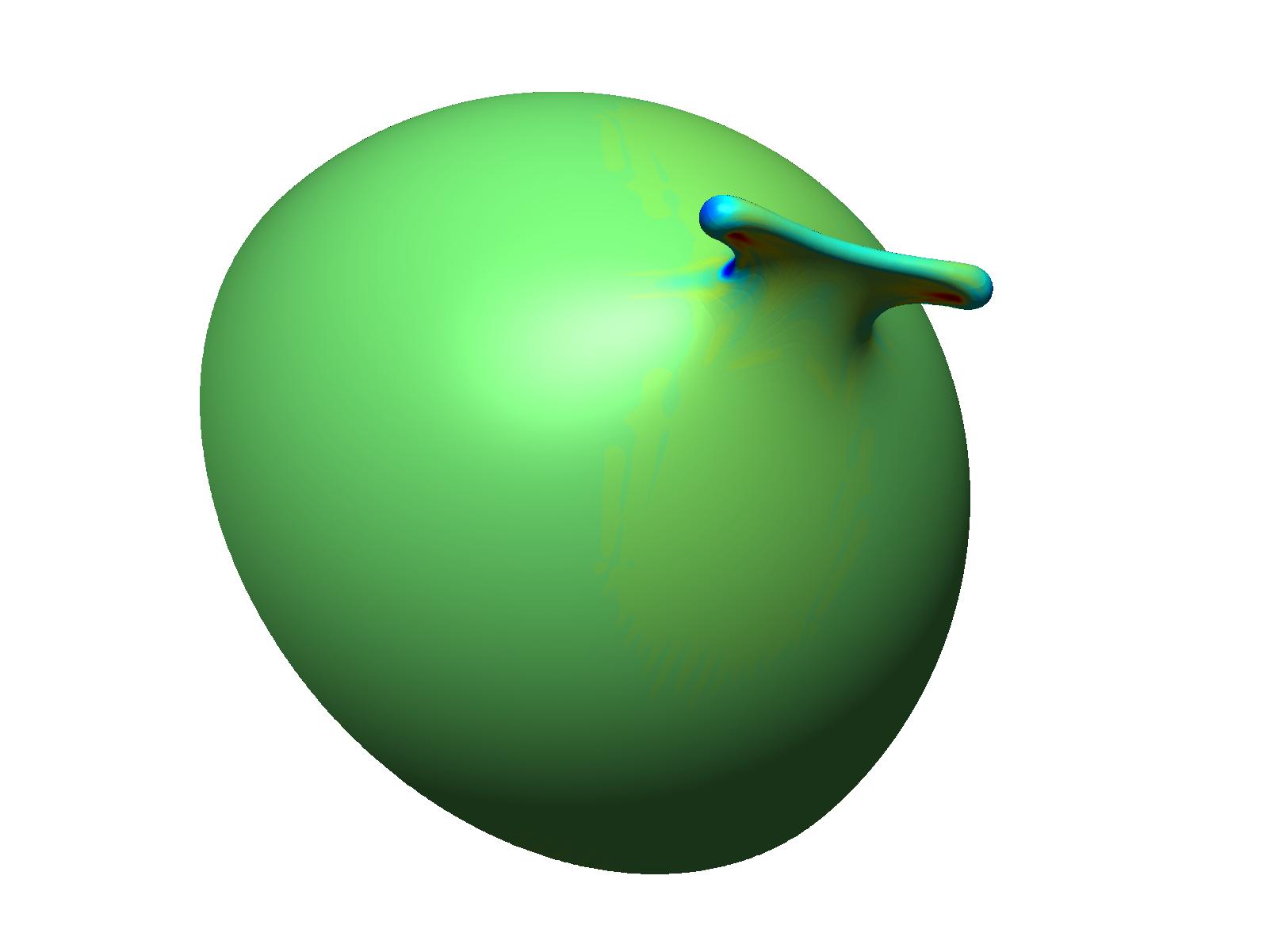}}
\put(4.4,2.6){\includegraphics[height=31mm]{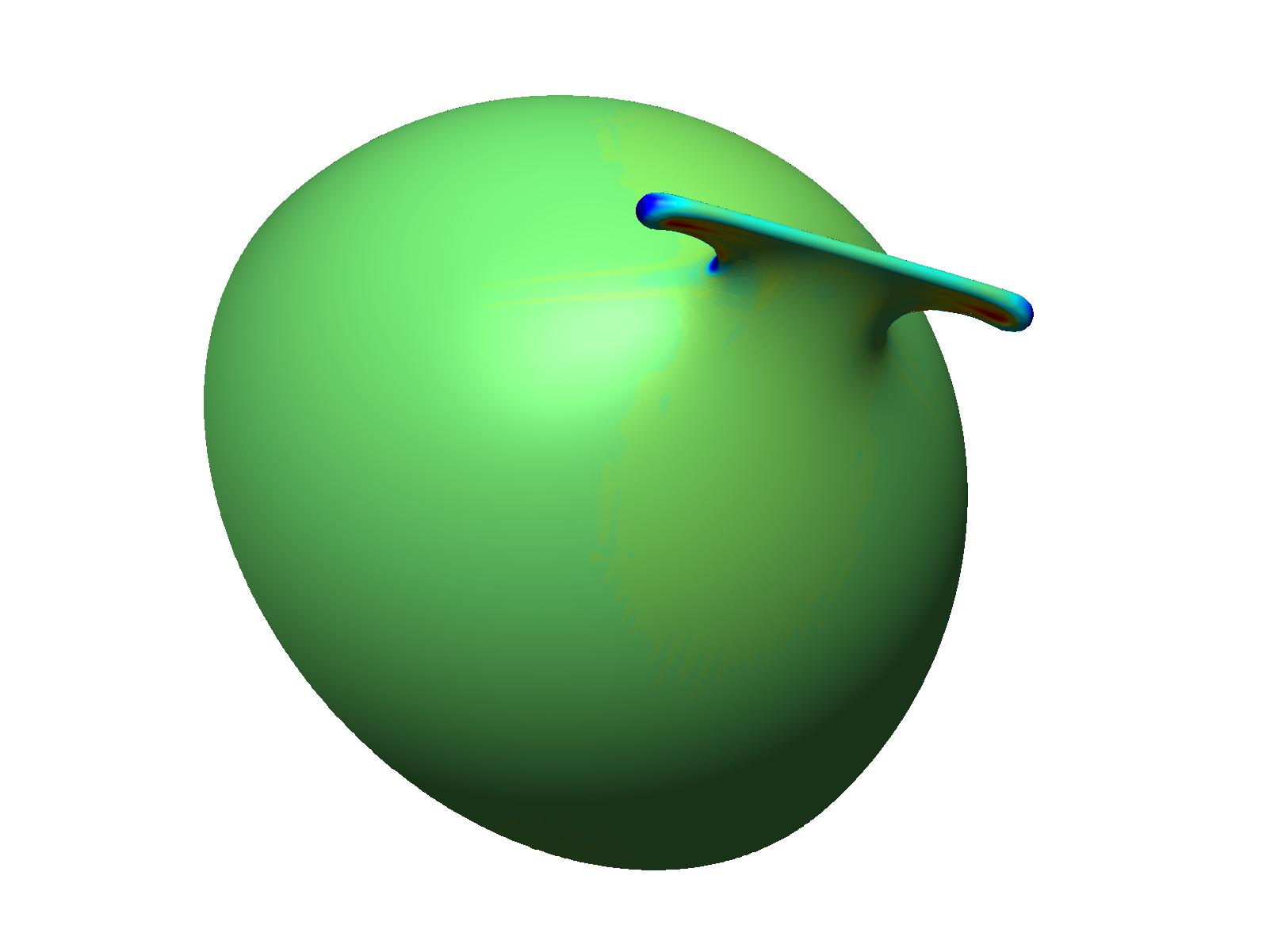}}
\end{picture}
\caption{Cell budding: Viscous, imperfect case at $\bar H_0=-5,-10,-15,-20,-25$ (left to right): 3D and side view of deformation and surface tension $\bar\gamma$.}
\label{f:bud_Inuy}
\end{center}
\end{figure}

Second, the shear stiff case (case 4) is examined in Fig.~\ref{f:bud_Imuy}. 
As the plate-like bud appears, the maximum of $\gamma$ moves away from the center and is concentrated at the end of the plate. 
The behavior is similar for the viscous case (case 5) shown in Fig.~\ref{f:bud_Inuy}. 
Both case~4 and 5 show that the extrema of $\gamma$ are concentrated in very small regions associated with large curvatures. However, these peak values are still much lower than the more distributed peak values of case~1.
The lower values are not surprising, as the system has much lower energy for non-axisymmetric shapes. For the non-axisymmetric cases with shear stiffness (case 4) and viscosity (case 5), the resulting surface tensions are of similar magnitude. Only the shapes are different.

\subsubsection{Effective shear stiffness}

\begin{figure}[h]
\begin{center} \unitlength1cm
\begin{picture}(0,5.6)
\put(-8.4,-.4){\includegraphics[height=31mm]{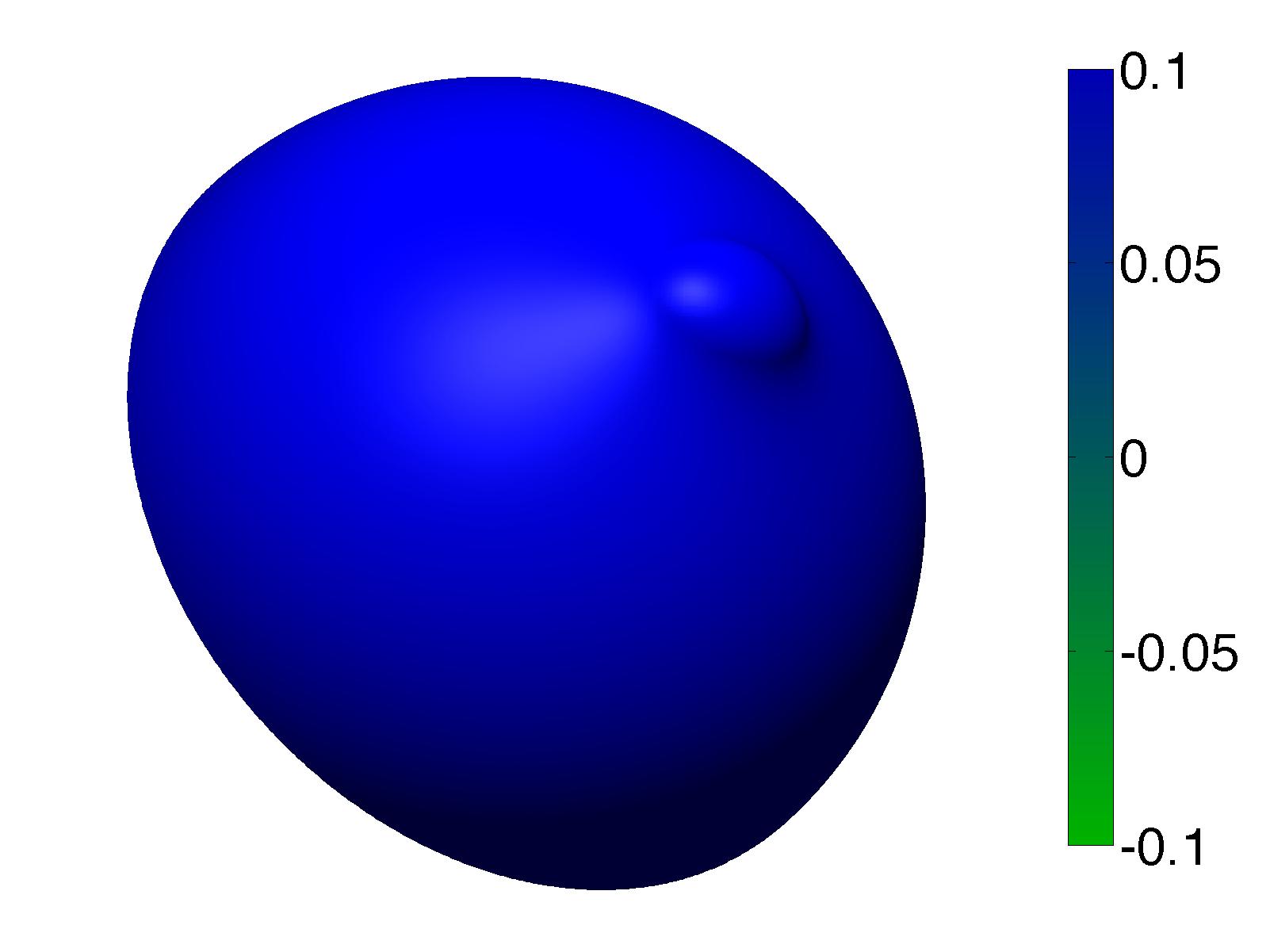}}
\put(-5.3,-.4){\includegraphics[height=31mm]{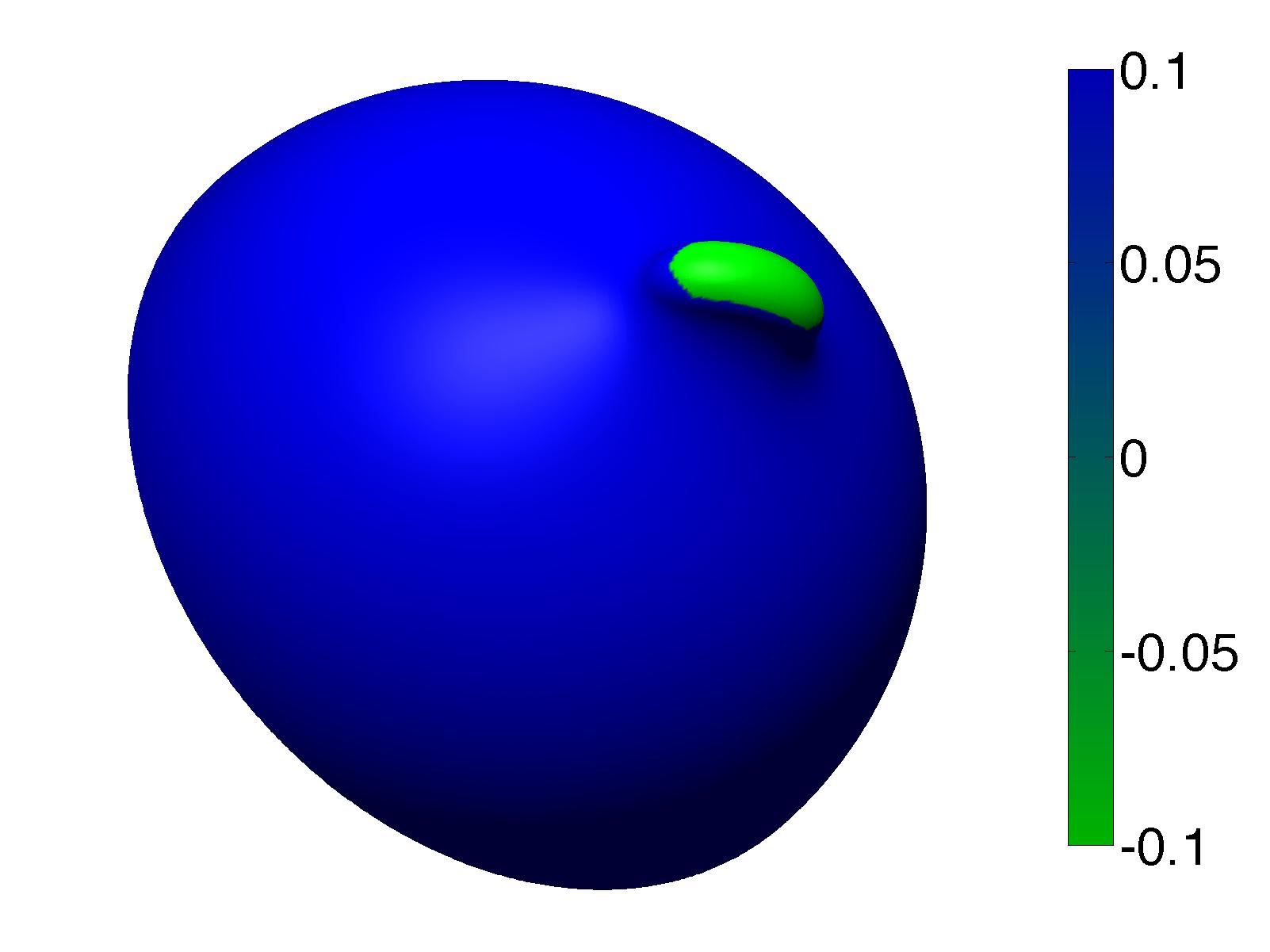}}
\put(-2.2,-.4){\includegraphics[height=31mm]{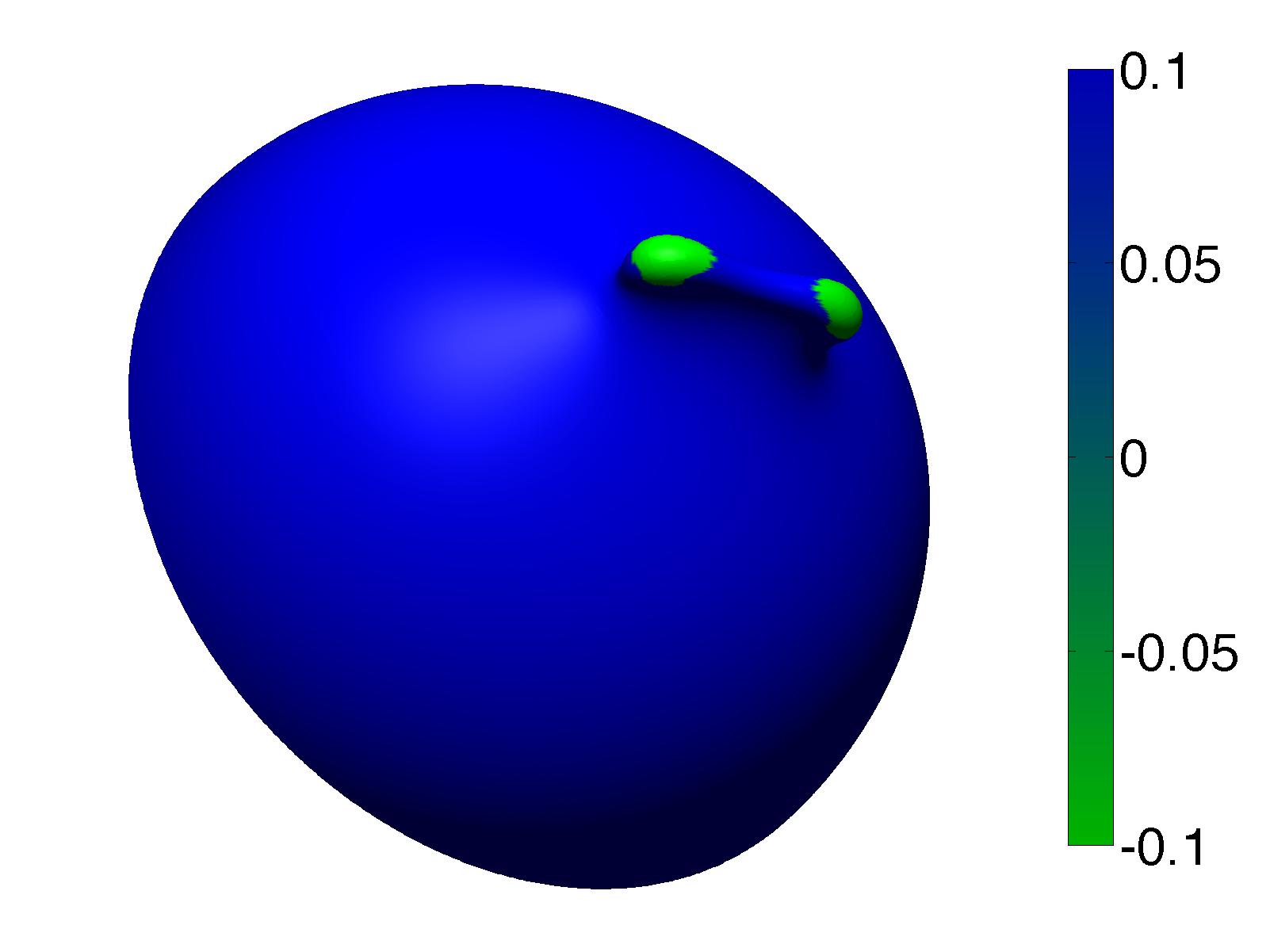}}
\put(0.9,-.4){\includegraphics[height=31mm]{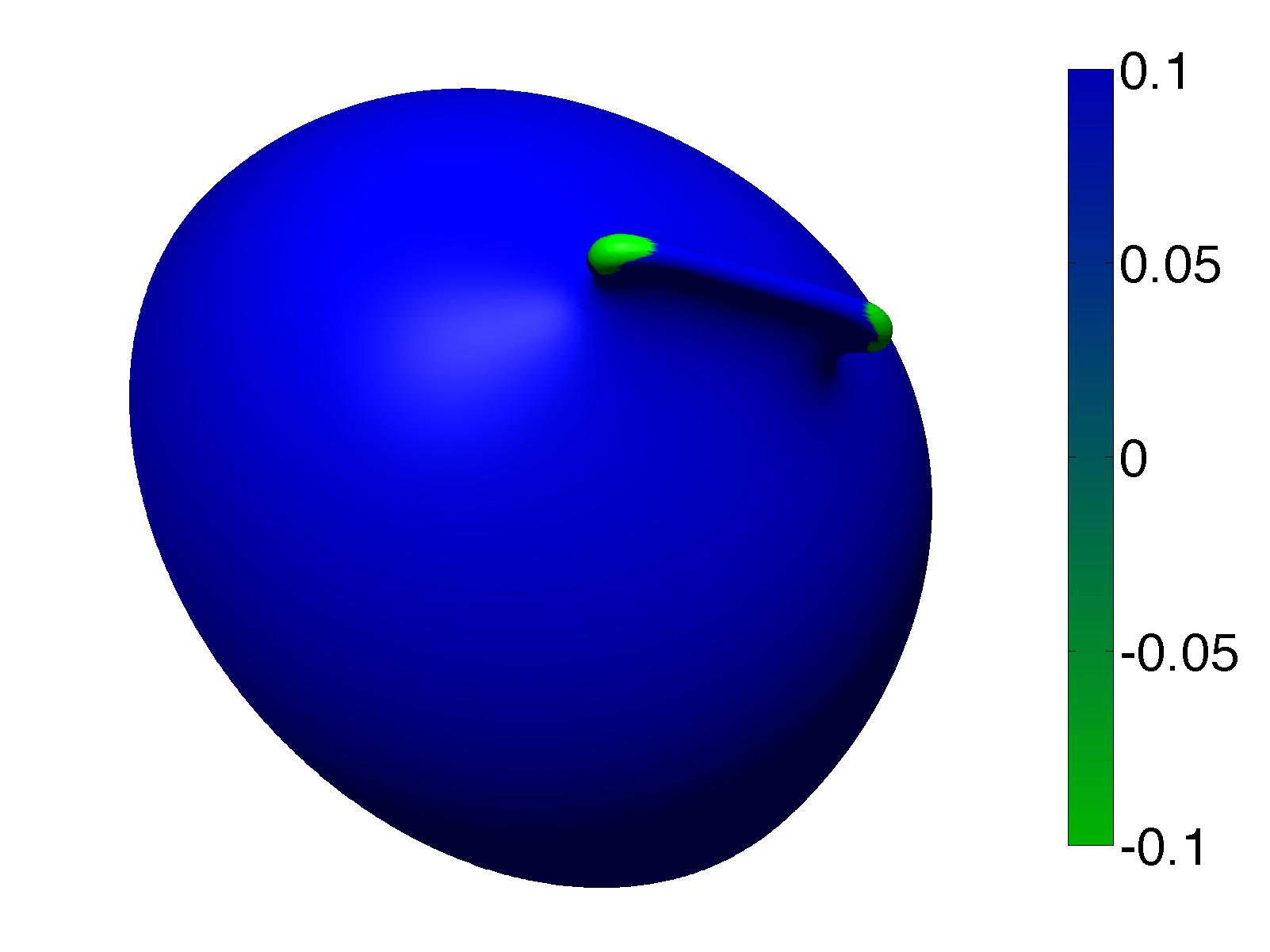}}
\put(4.0,-.4){\includegraphics[height=31mm]{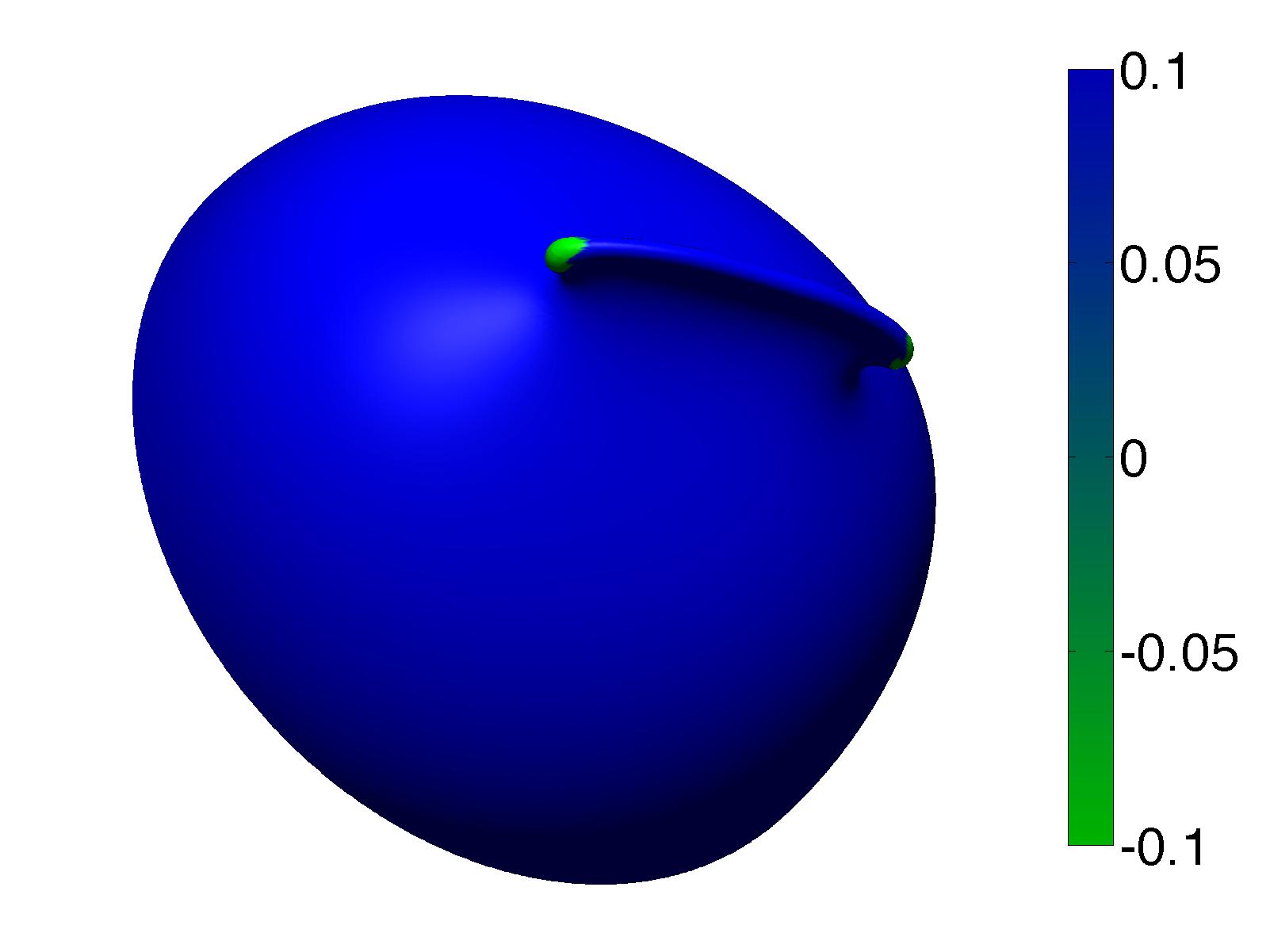}}
\put(-8.4,2.6){\includegraphics[height=31mm]{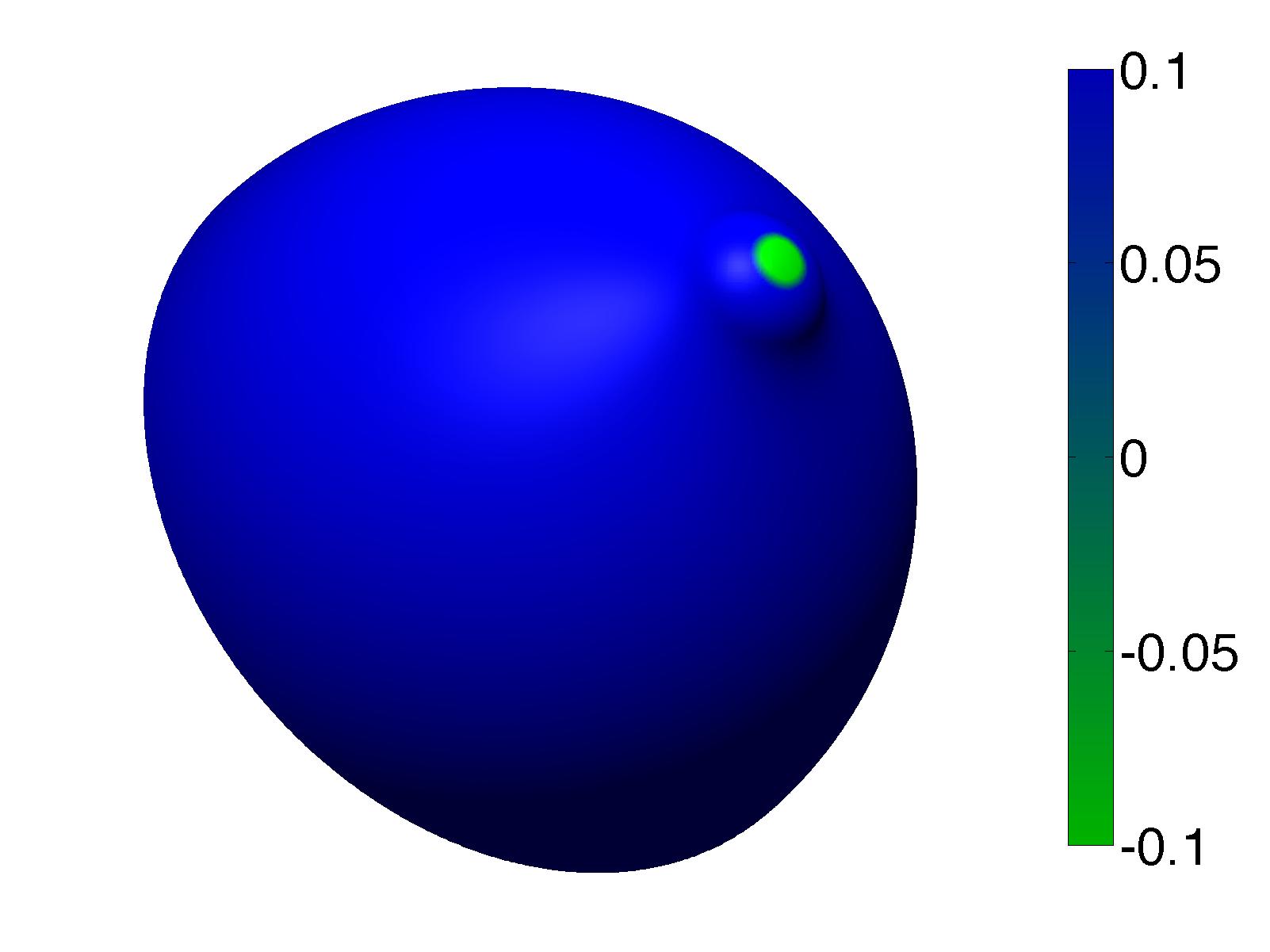}}
\put(-5.3,2.6){\includegraphics[height=31mm]{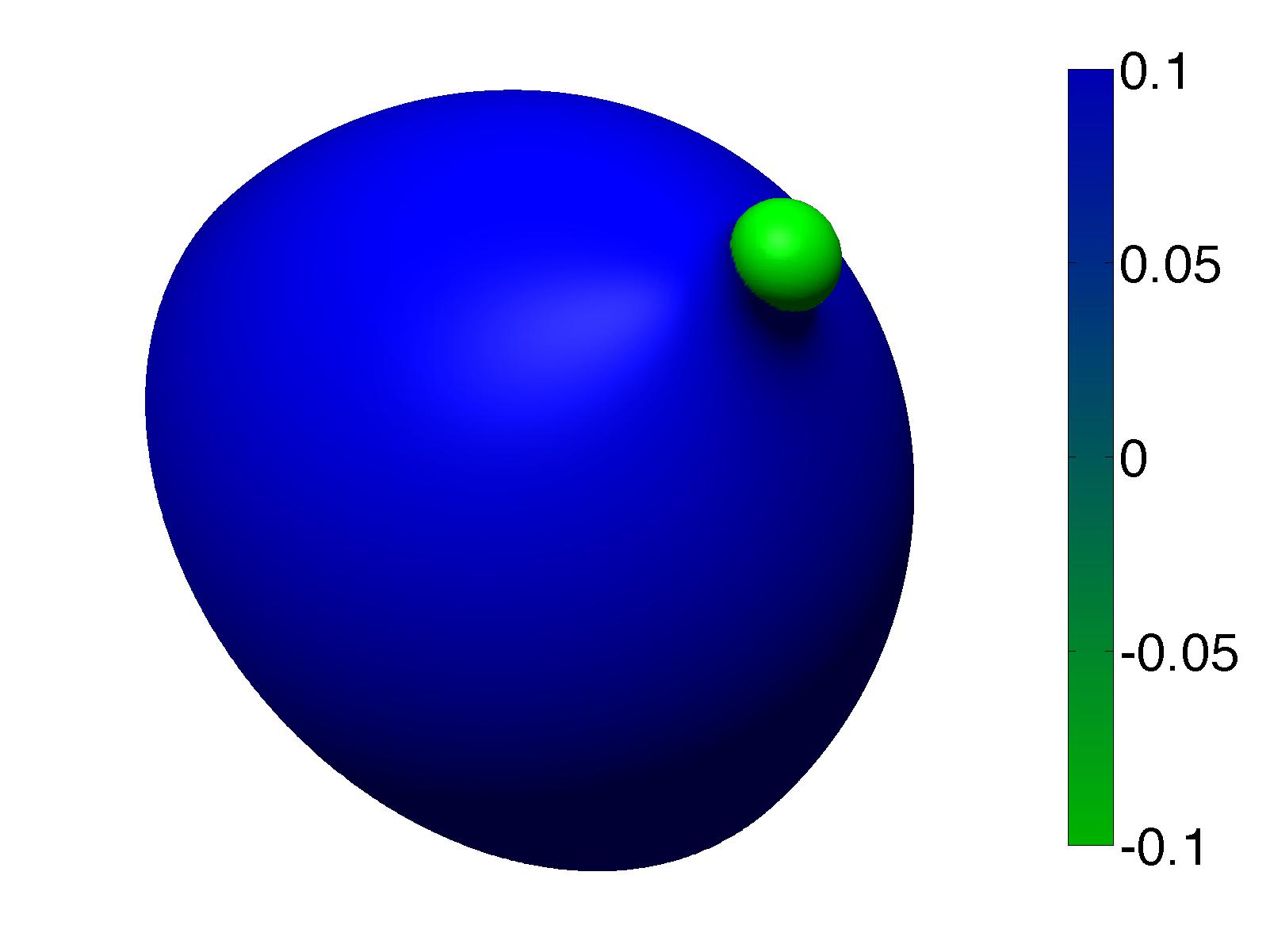}}
\put(-2.2,2.6){\includegraphics[height=31mm]{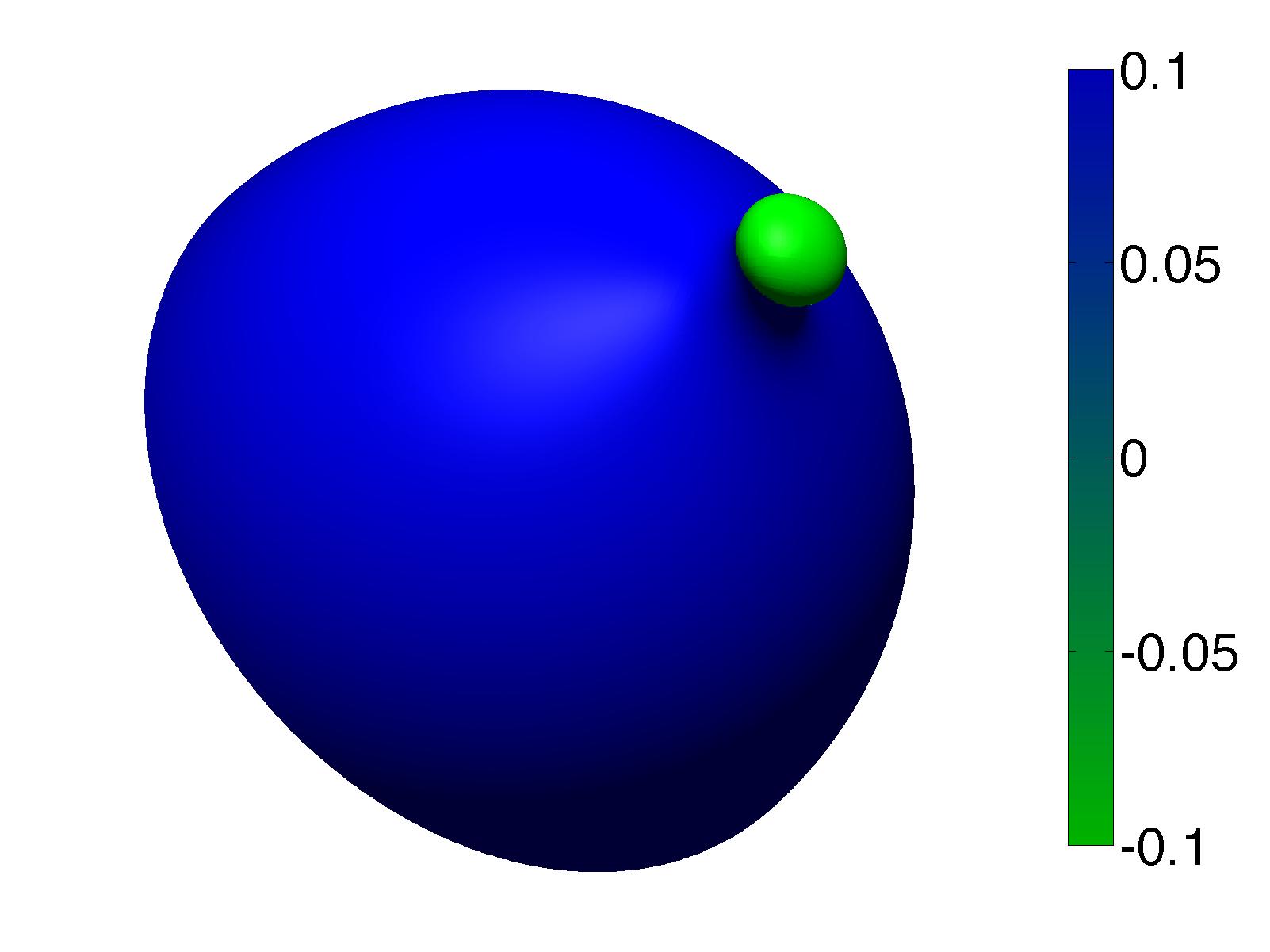}}
\put(0.9,2.6){\includegraphics[height=31mm]{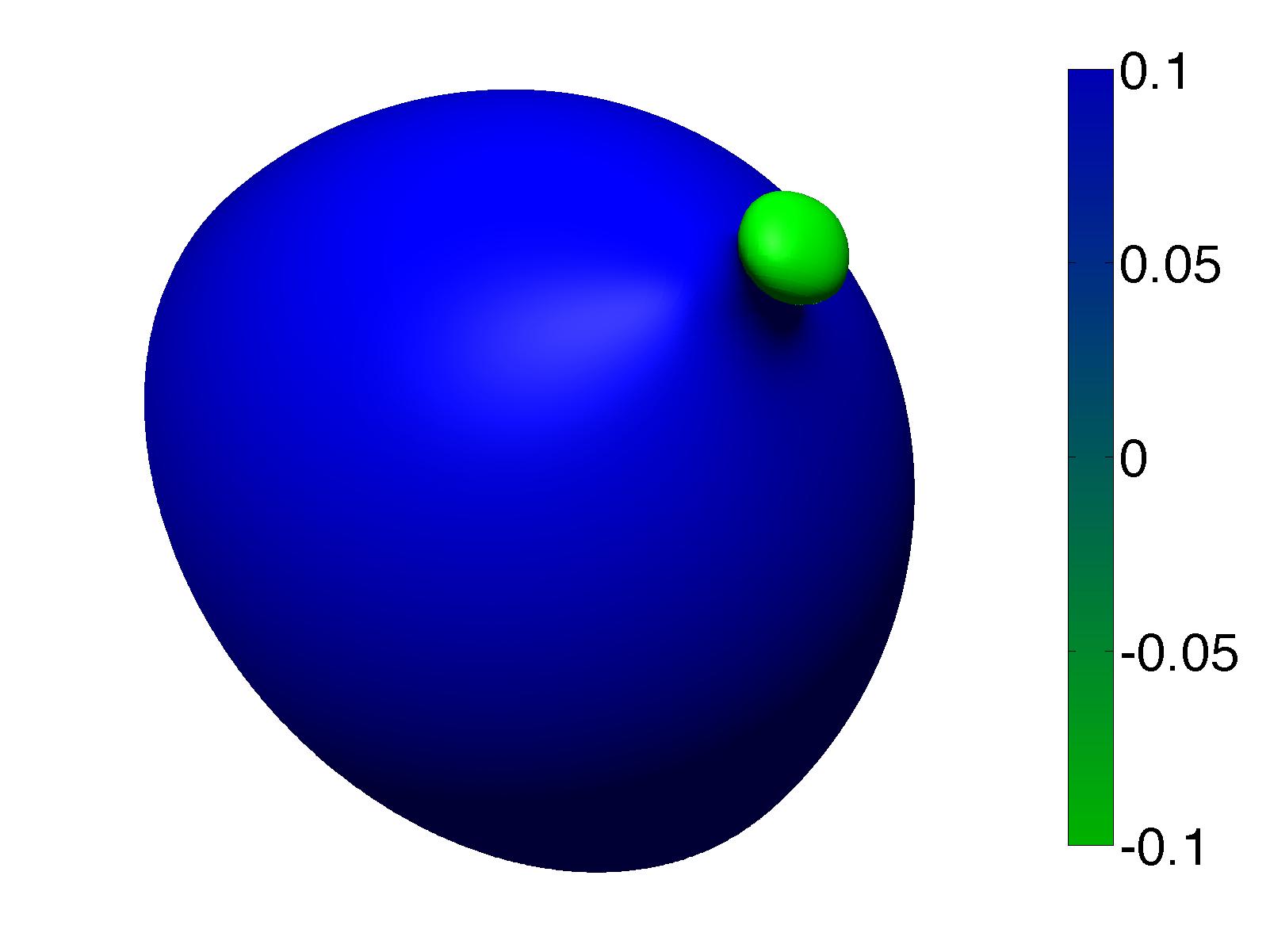}}
\put(4.0,2.6){\includegraphics[height=31mm]{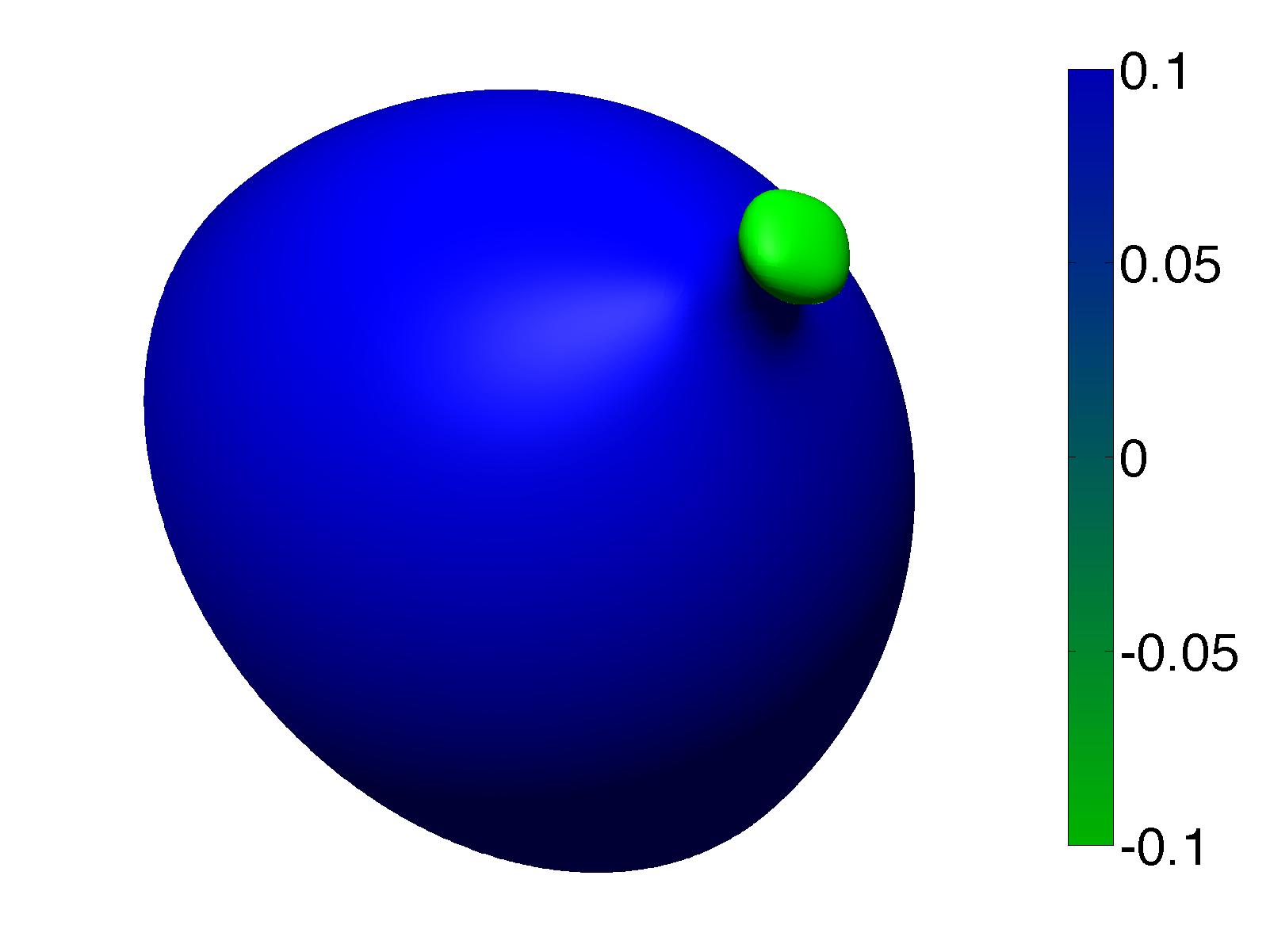}}
\end{picture}
\caption{Cell budding: sign of $\bar\mu_\mathrm{eff}$ for cases 1 (top) and 4 (bottom) at $\bar H_0=-5,-10,-15,-20,-25$ (left to right).}
\label{f:bud_mueff}
\end{center}
\end{figure}

Fig.~\ref{f:bud_mueff} shows the sign of the effective shear stiffness $\mu_\mathrm{eff}$ for cases 1 and 4.\footnote{For case 4, the physical shear stiffness $\bar\mu=10$ needs to be added to Eq.~\eqref{e:mueff}.}
It can be seen that for both cases, $\mu_\mathrm{eff}$ is negative in some regions.
The minimum values of $\bar\mu_\mathrm{eff}$ reaches down to $-150$ in case 1 and to $-50$ in case 4. 
Since the simulations run stably, even though $\mu_\mathrm{eff}<0$, there must be another stabilizing effect. 
It is probable that this related to the geometry: The figure shows that the regions of negative $\mu_\mathrm{eff}$ are all convex.
The condition $\mu_\mathrm{eff}<0$ is therefore not sufficient for unstable behavior.
Since stabilization may be provided naturally, the computations can sometimes be performed with no stabilization scheme. In practice it is however recommended to  
add one of the stabilization schemes proposed in Sec.~\ref{s:stab}. For example, when both $H$ and $\kappa$ are zero neither the Helfrich model nor the geometry can be expected to stabilize the structure. The stabilization parameters ($\mu$ and $n_t$) can be picked such that the stabilization scheme does not affect the physical behavior. This is the case for the stabilization chosen here.

\subsubsection{Influence of the area-compressibility}

As a final study, we investigate the influence of the area-compressibility (parameter $K$) and compare the behavior of model \eqref{e:W_c}, that depends on $K$, with \eqref{e:W_i}, where $K$ is infinite.
For this purpose, model \eqref{e:W_i} is discretized according to Sec.~\ref{s:areac}.
In theory, the behavior of model \eqref{e:W_c}, 
should approach that of model \eqref{e:W_i} as $K\rightarrow\infty$.

\begin{figure}[h]
\begin{center} \unitlength1cm
\begin{picture}(0,5.6)
\put(-8.4,-.4){\includegraphics[height=31mm]{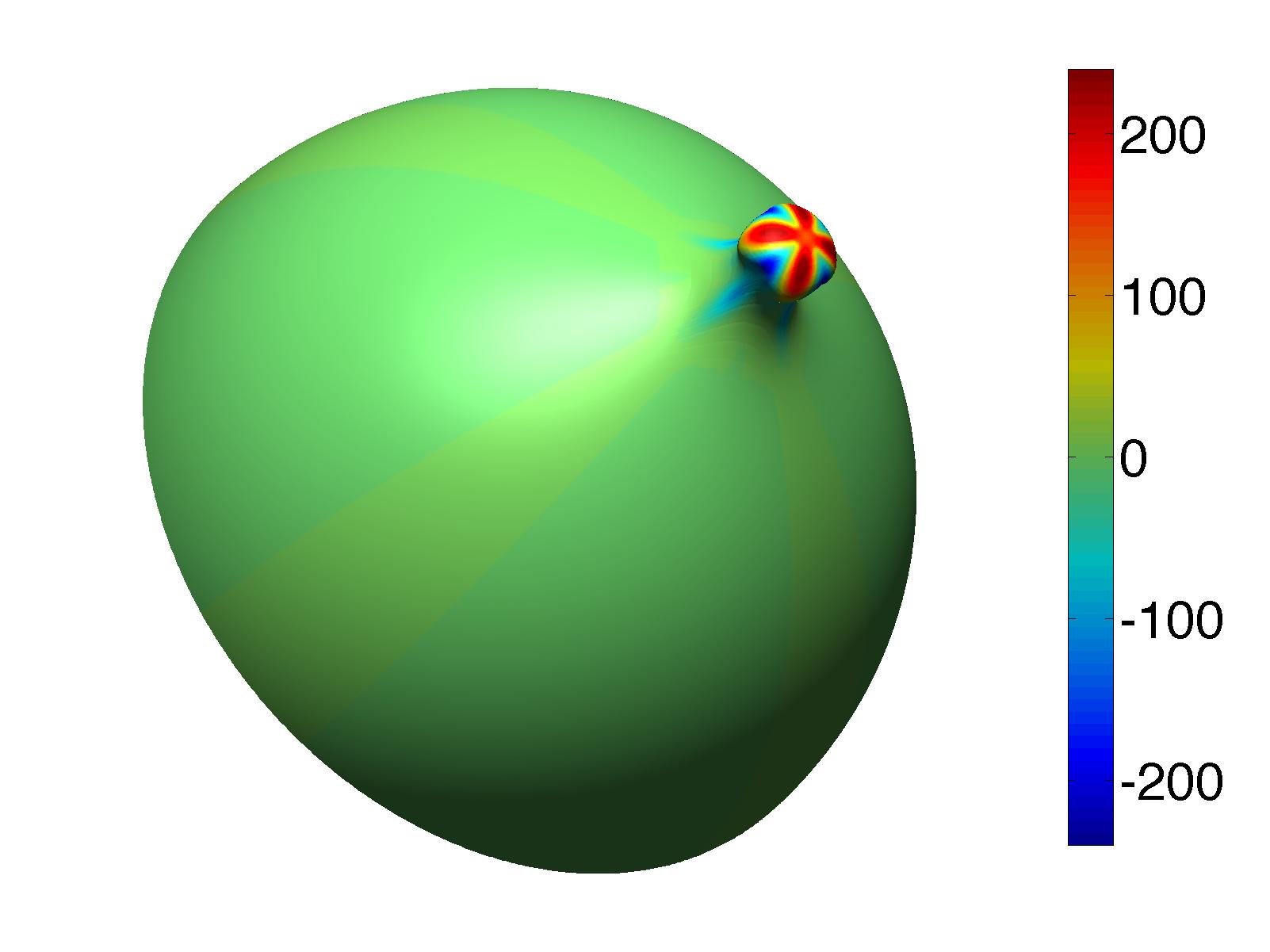}}
\put(-5.3,-.4){\includegraphics[height=31mm]{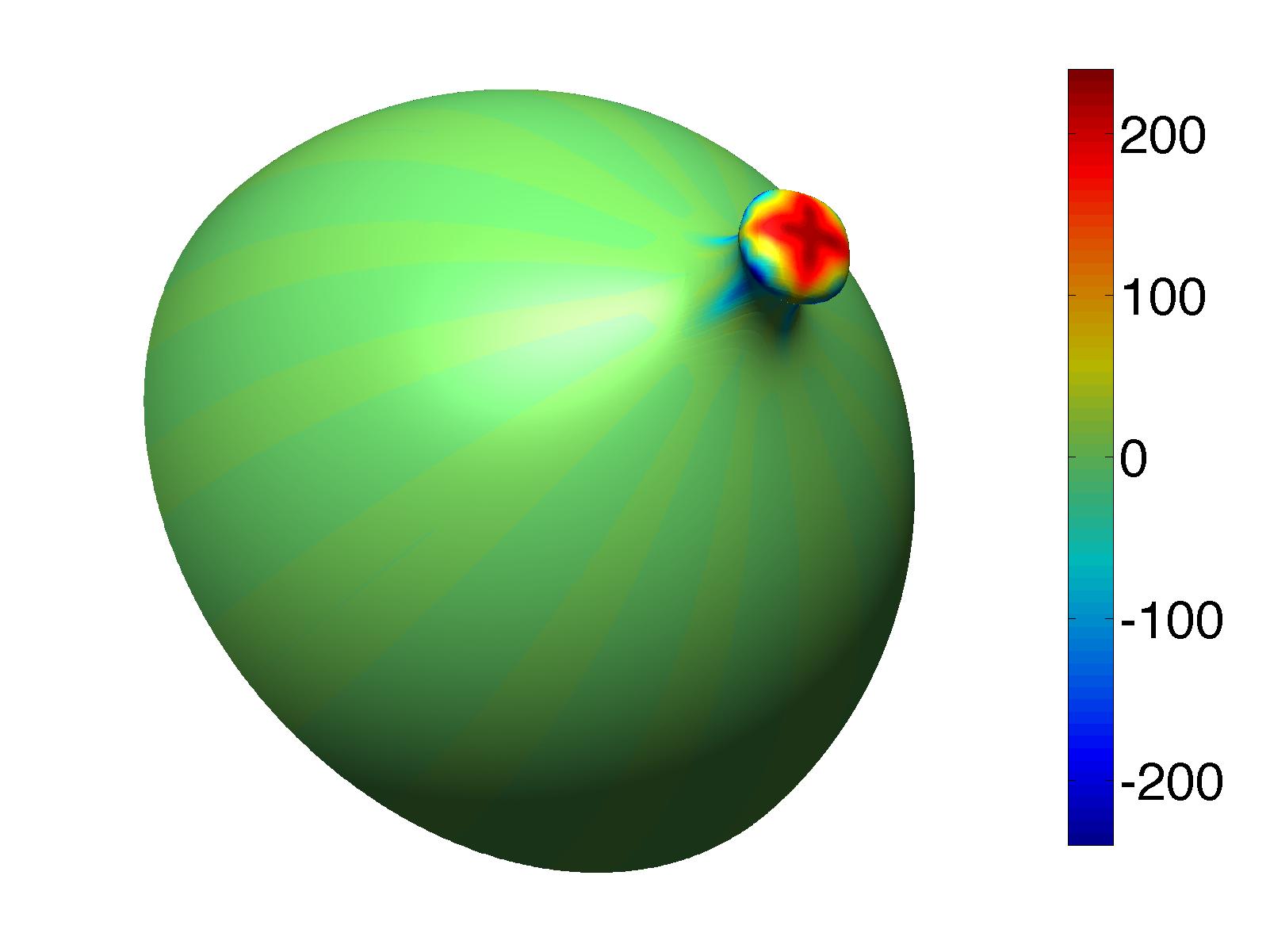}}
\put(-2.2,-.4){\includegraphics[height=31mm]{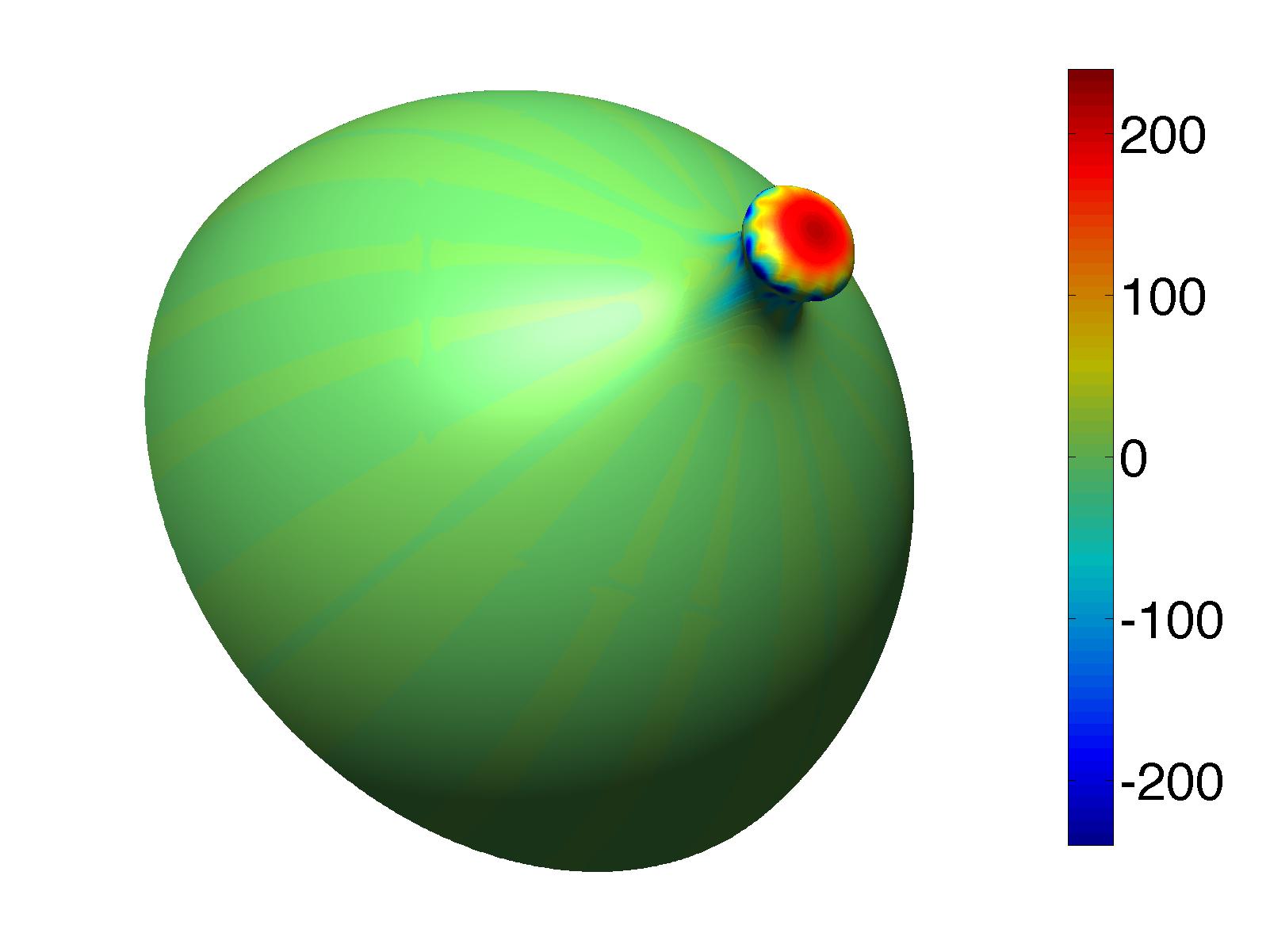}}
\put(0.9,-.4){\includegraphics[height=31mm]{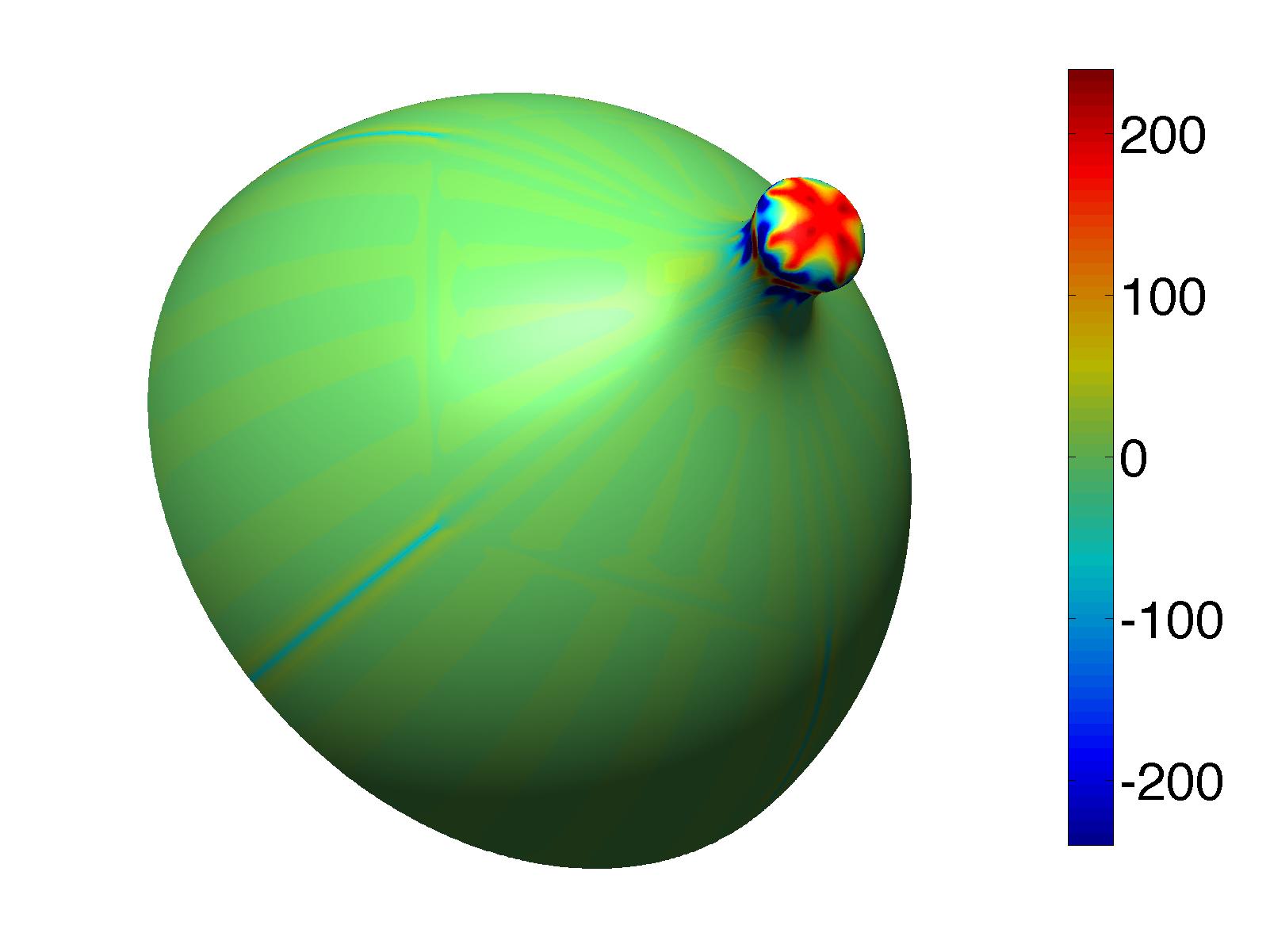}}
\put(4.0,-.4){\includegraphics[height=31mm]{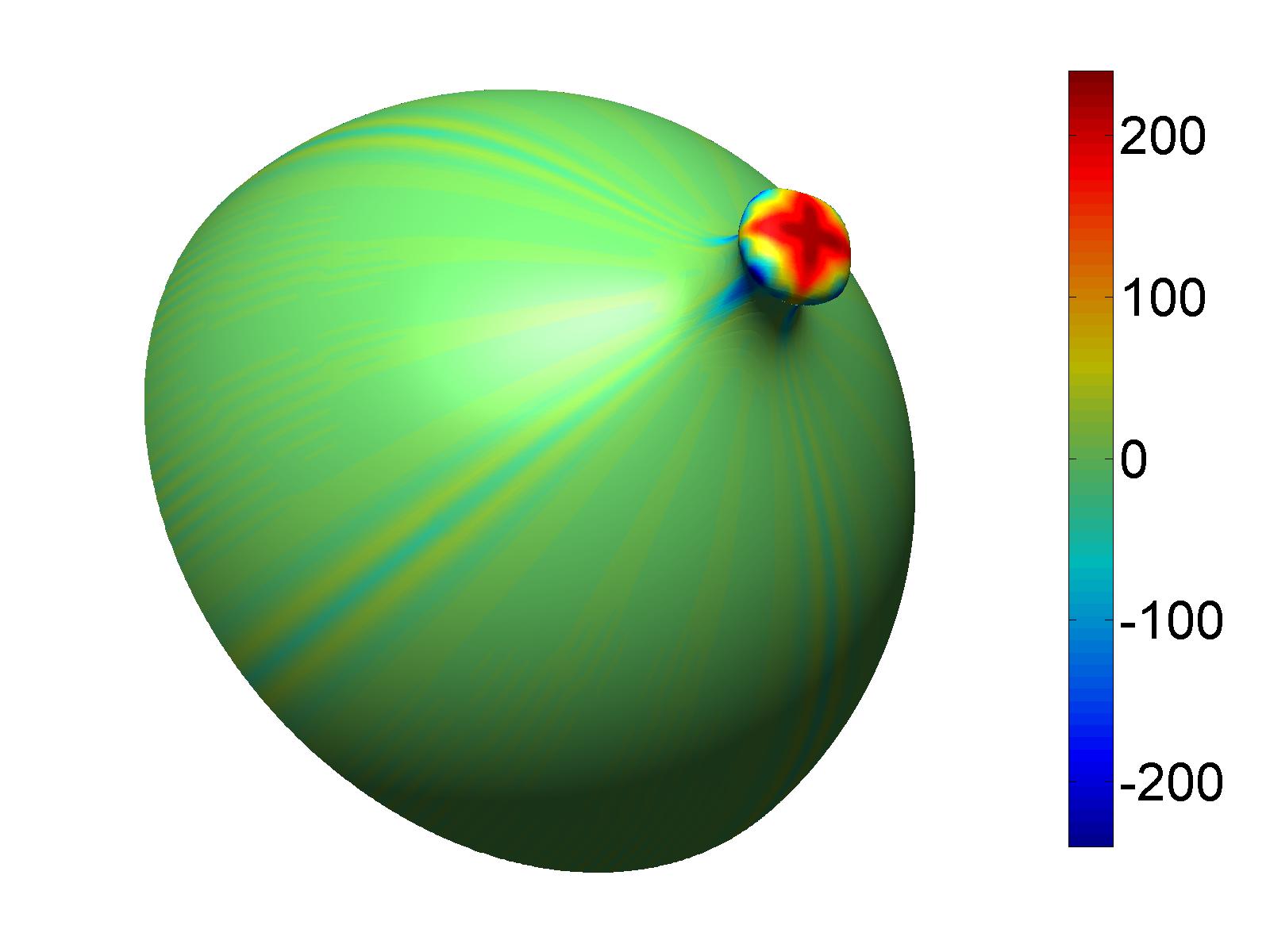}}
\put(-8.4,2.6){\includegraphics[height=31mm]{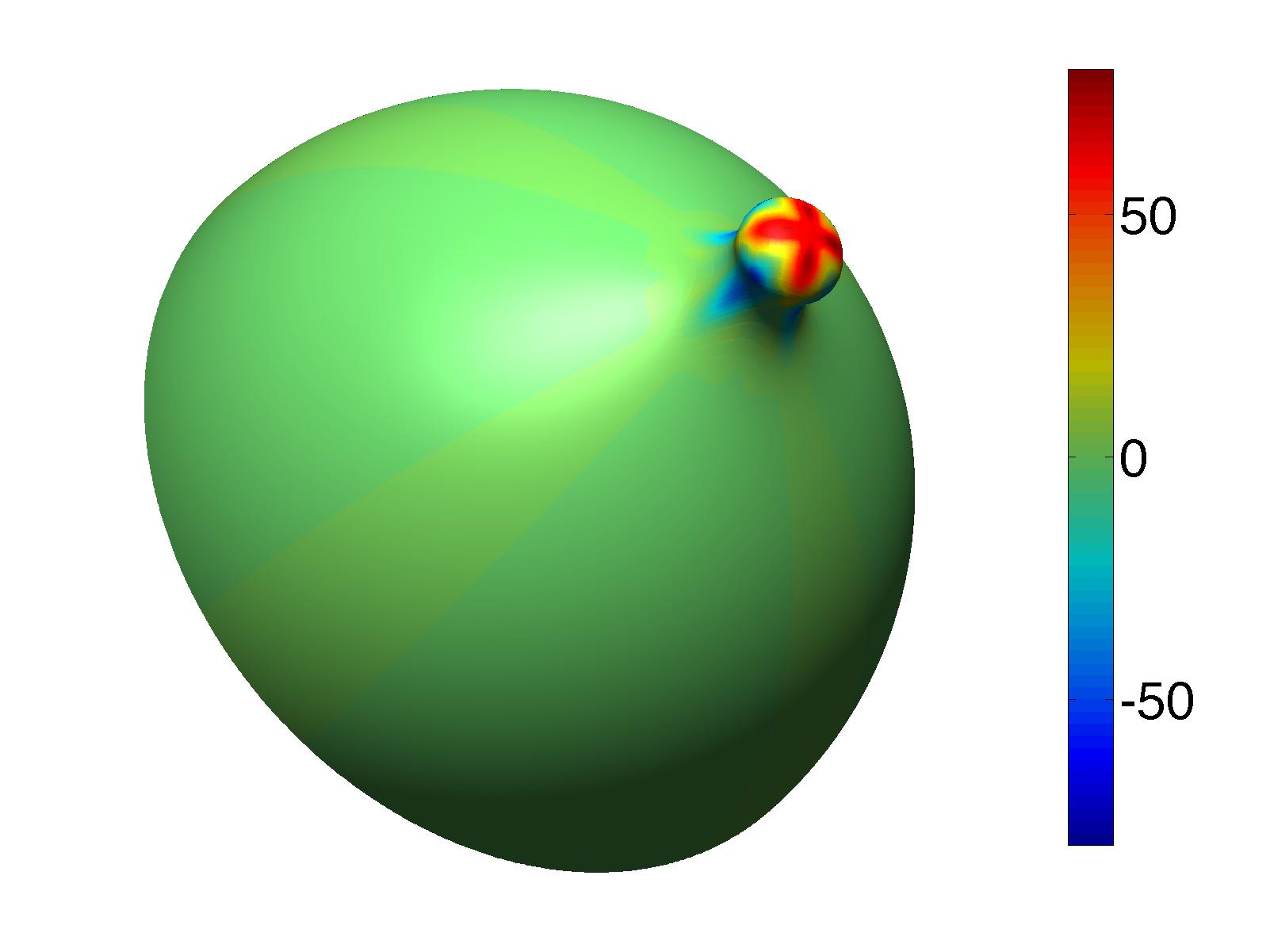}}
\put(-5.3,2.6){\includegraphics[height=31mm]{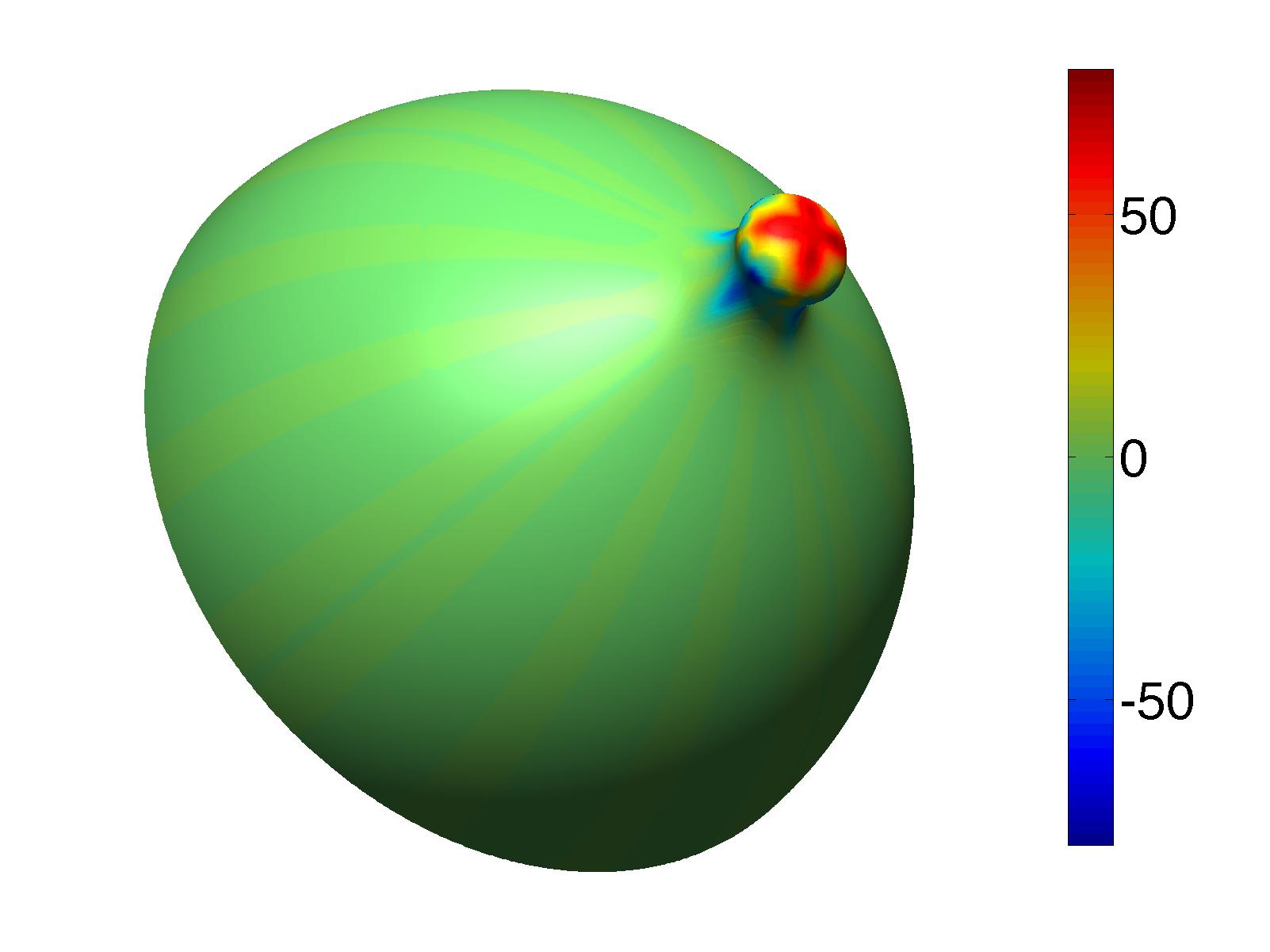}}
\put(-2.2,2.6){\includegraphics[height=31mm]{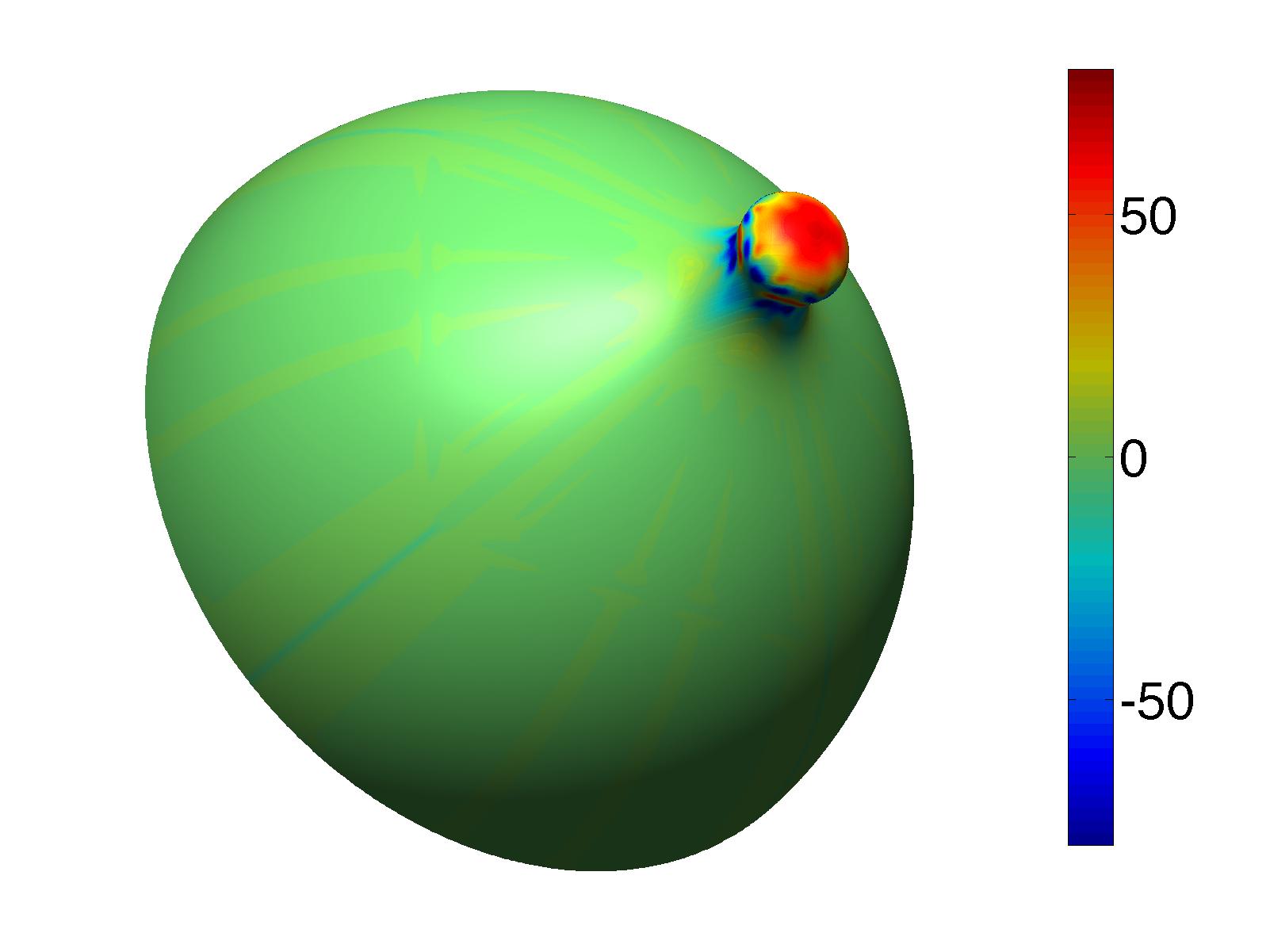}}
\put(0.9,2.6){\includegraphics[height=31mm]{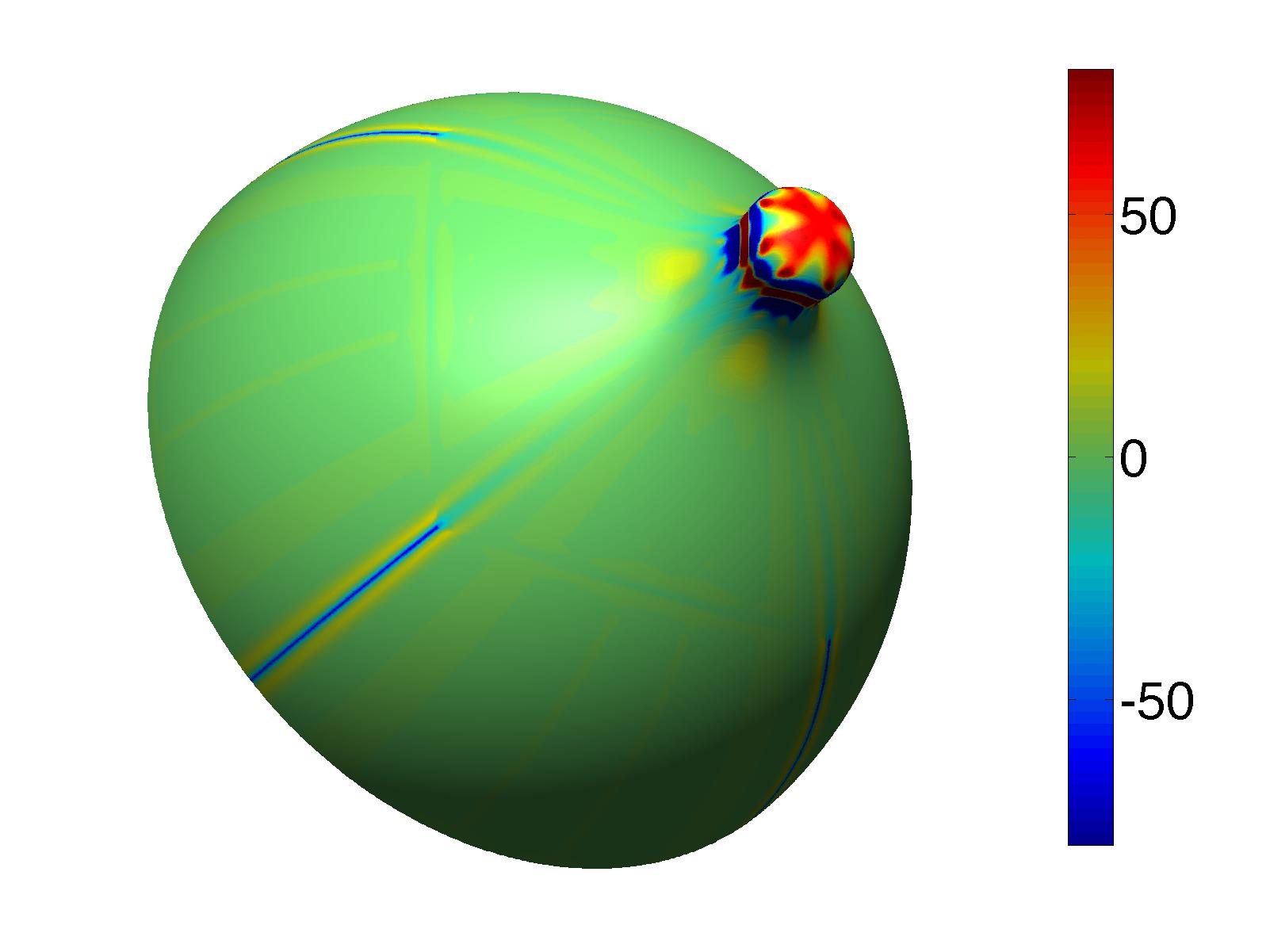}}
\put(4.0,2.6){\includegraphics[height=31mm]{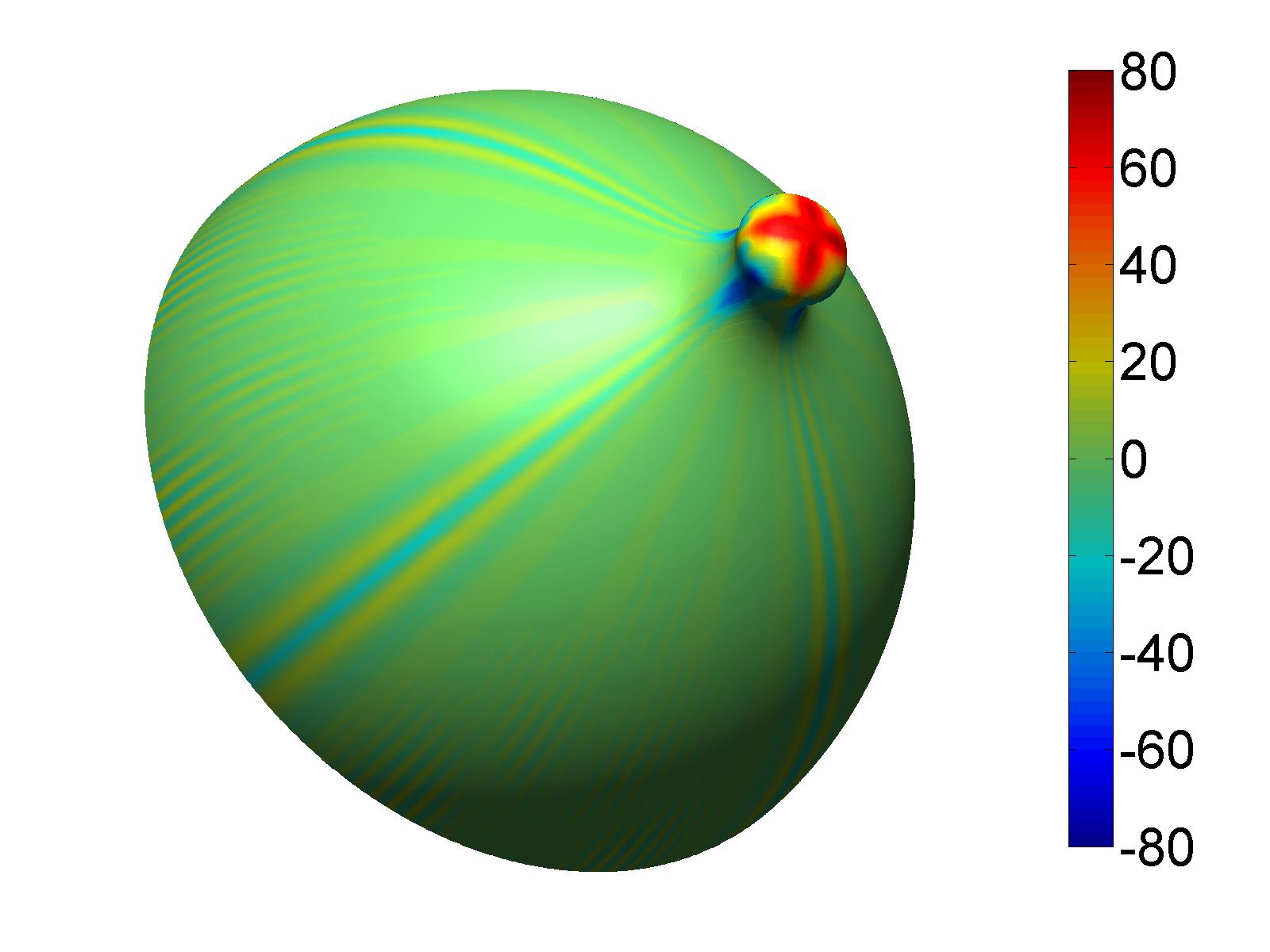}}
\end{picture}
\caption{Cell budding: influence of $K$ for the axisymmetric case (case 1) for $\bar H_0=-15$ (top) and $\bar H_0=-25$ (bottom) considering $\bar K=10^3,\,10^4,\,10^5,\,10^6,\,\infty$ (left to right). The color shows~$\bar\gamma$. 
}
\label{f:bud_KAy}
\end{center}
\end{figure}

For case 1, there is no significant difference between the two models as Fig.~\ref{f:bud_KAy} shows.
Only for the lowest $K$, differences appear.
Between all other cases, the bud shape only changes minimally. 
However, increasing $K$ in formulation \eqref{e:W_c} leads 
to oscillations in $\gamma$.

\begin{figure}[h]
\begin{center} \unitlength1cm
\begin{picture}(0,5.6)
\put(-8.4,-.4){\includegraphics[height=31mm]{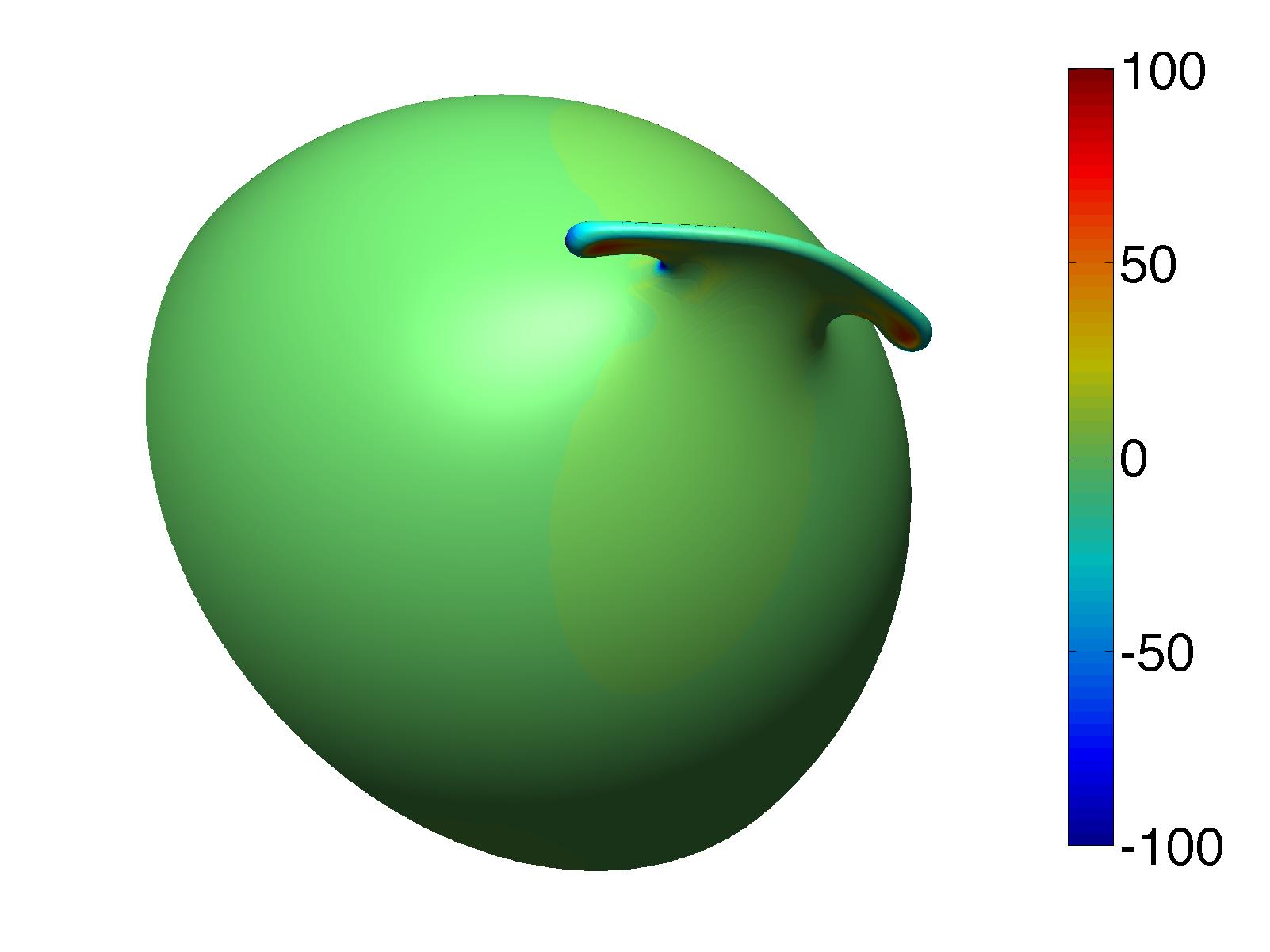}}
\put(0.9,-.4){\includegraphics[height=31mm]{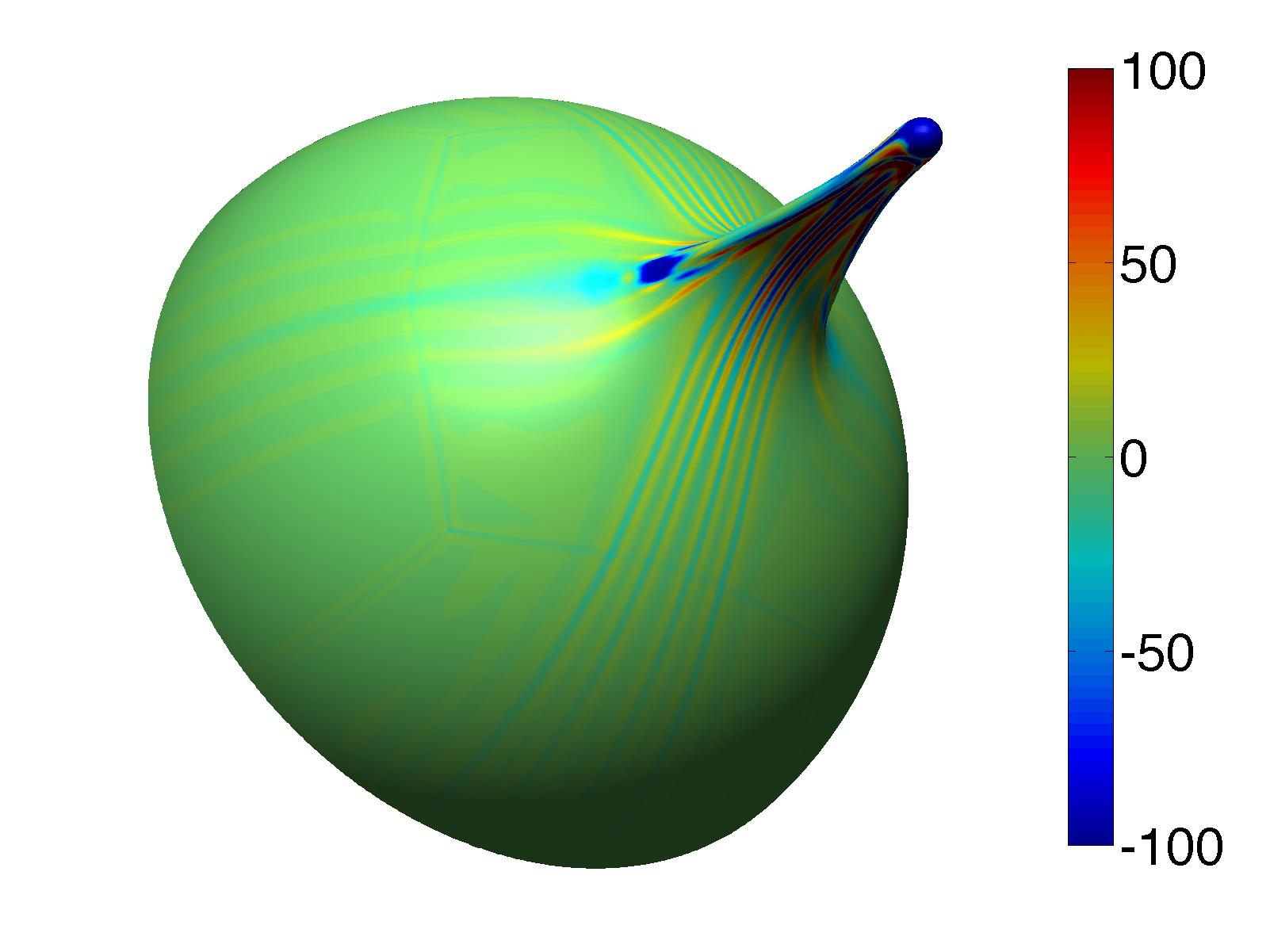}}
\put(-2.2,-.4){\includegraphics[height=31mm]{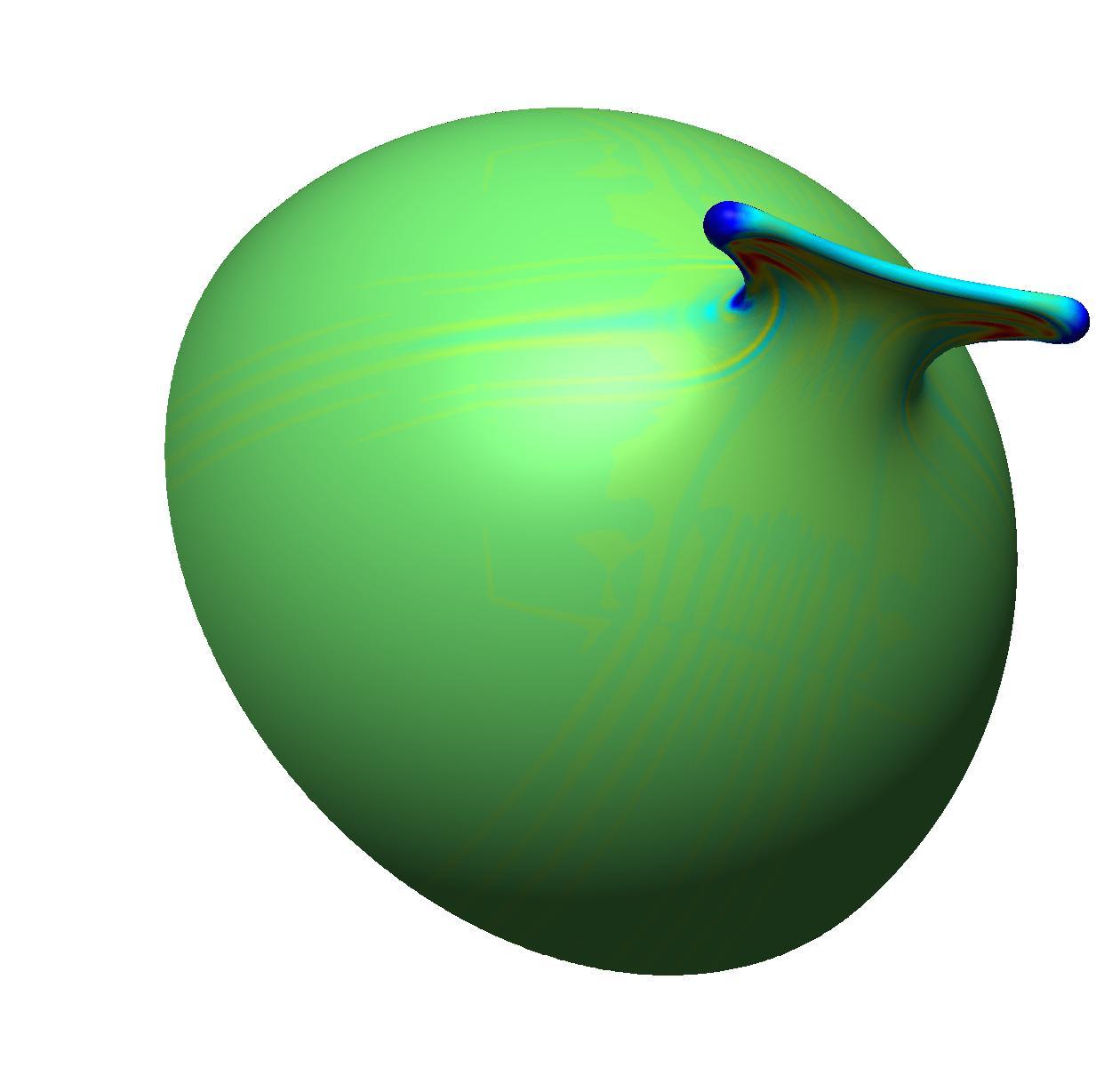}}
\put(-5.3,-.4){\includegraphics[height=31mm]{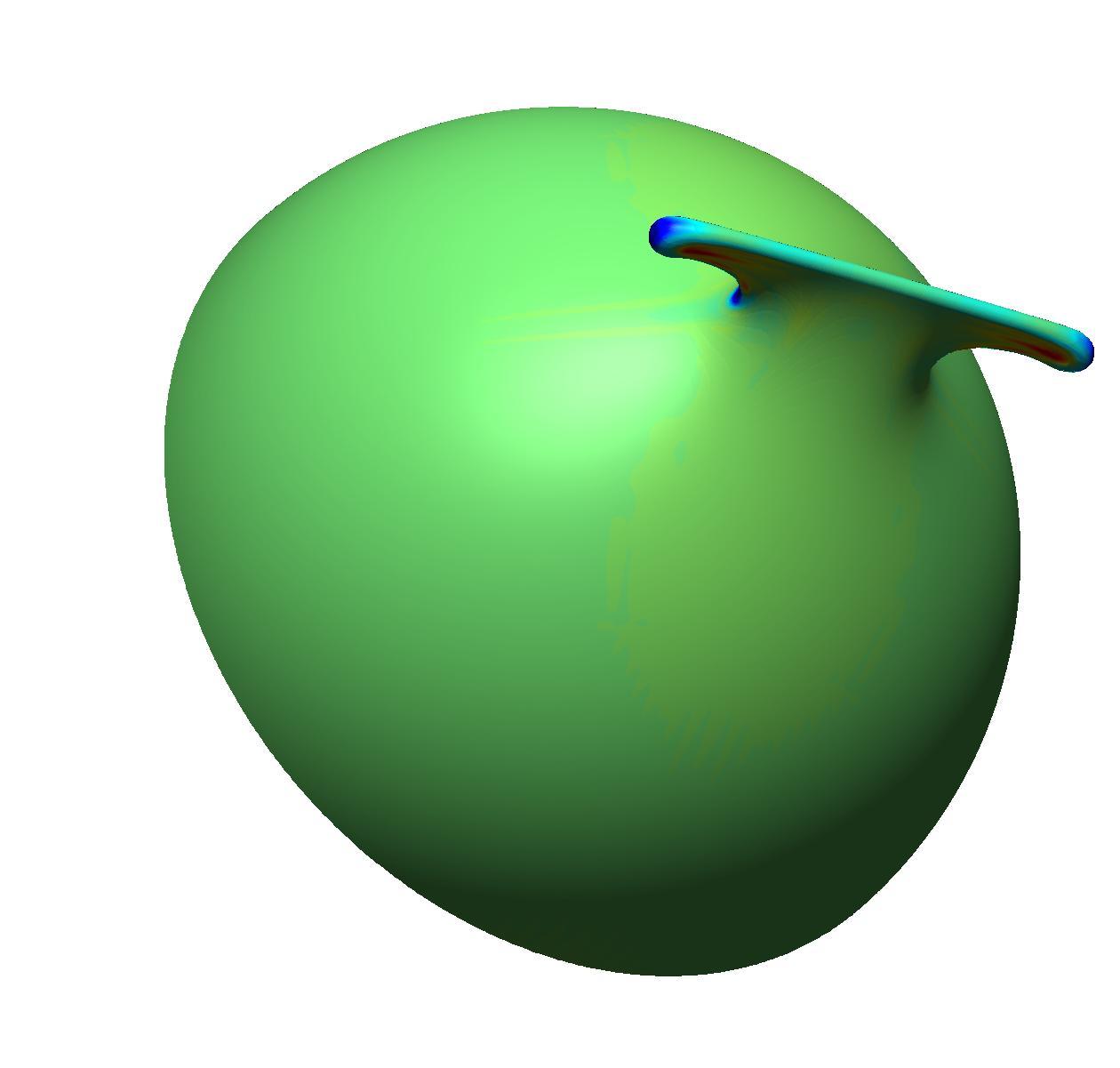}}
\put(4.0,-.4){\includegraphics[height=31mm]{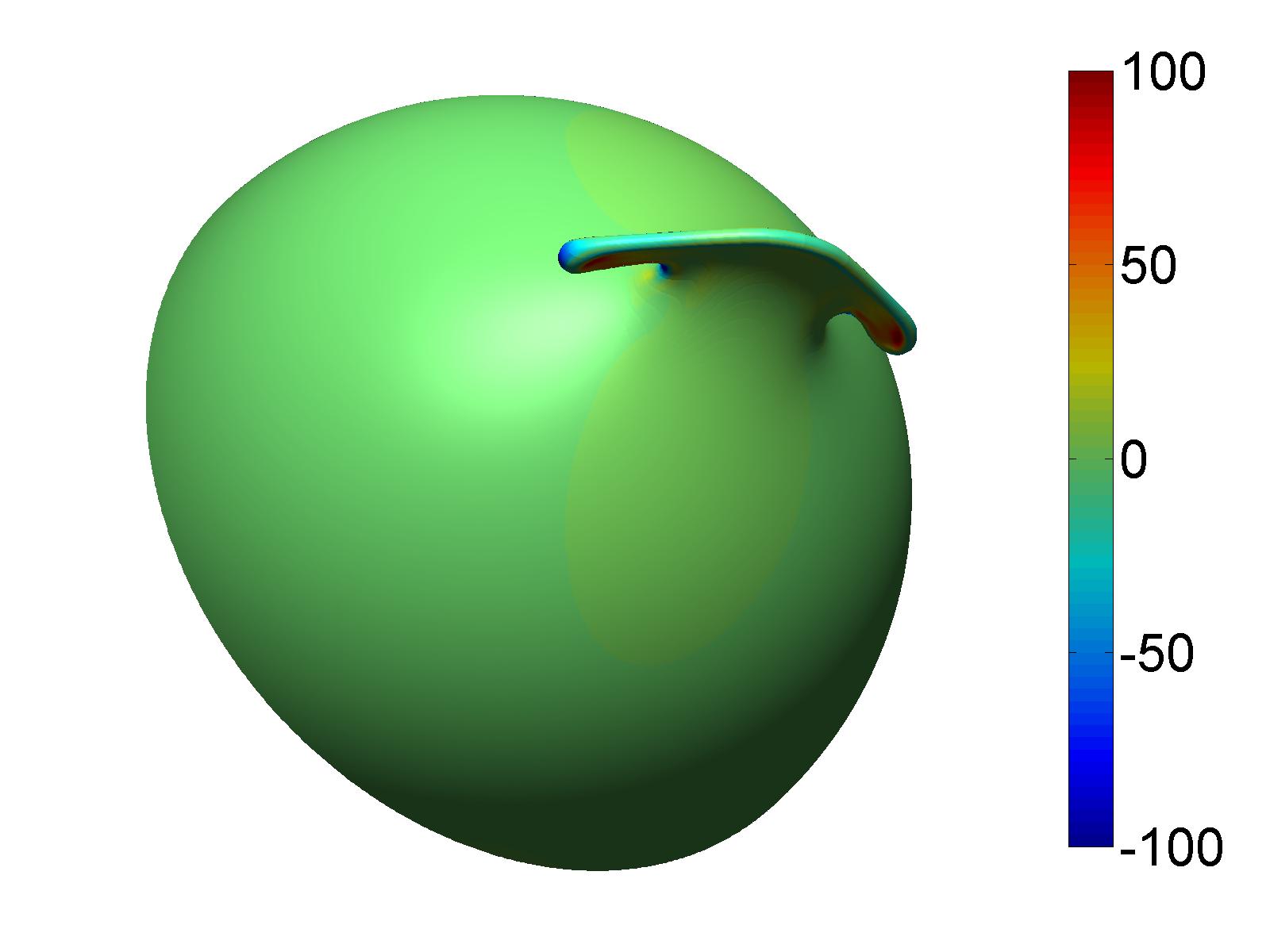}}
\put(-8.4,2.6){\includegraphics[height=31mm]{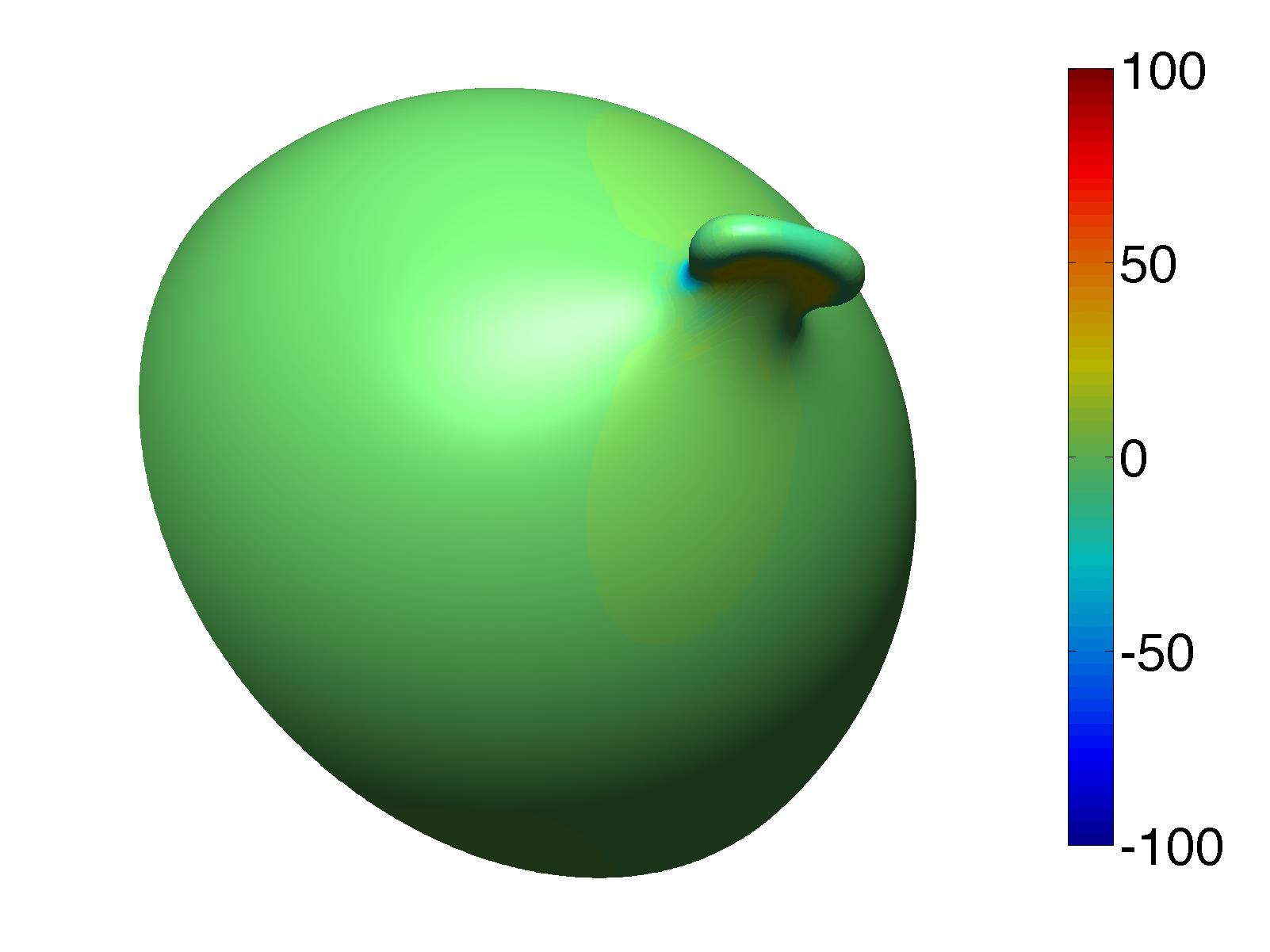}}
\put(-5.3,2.6){\includegraphics[height=31mm]{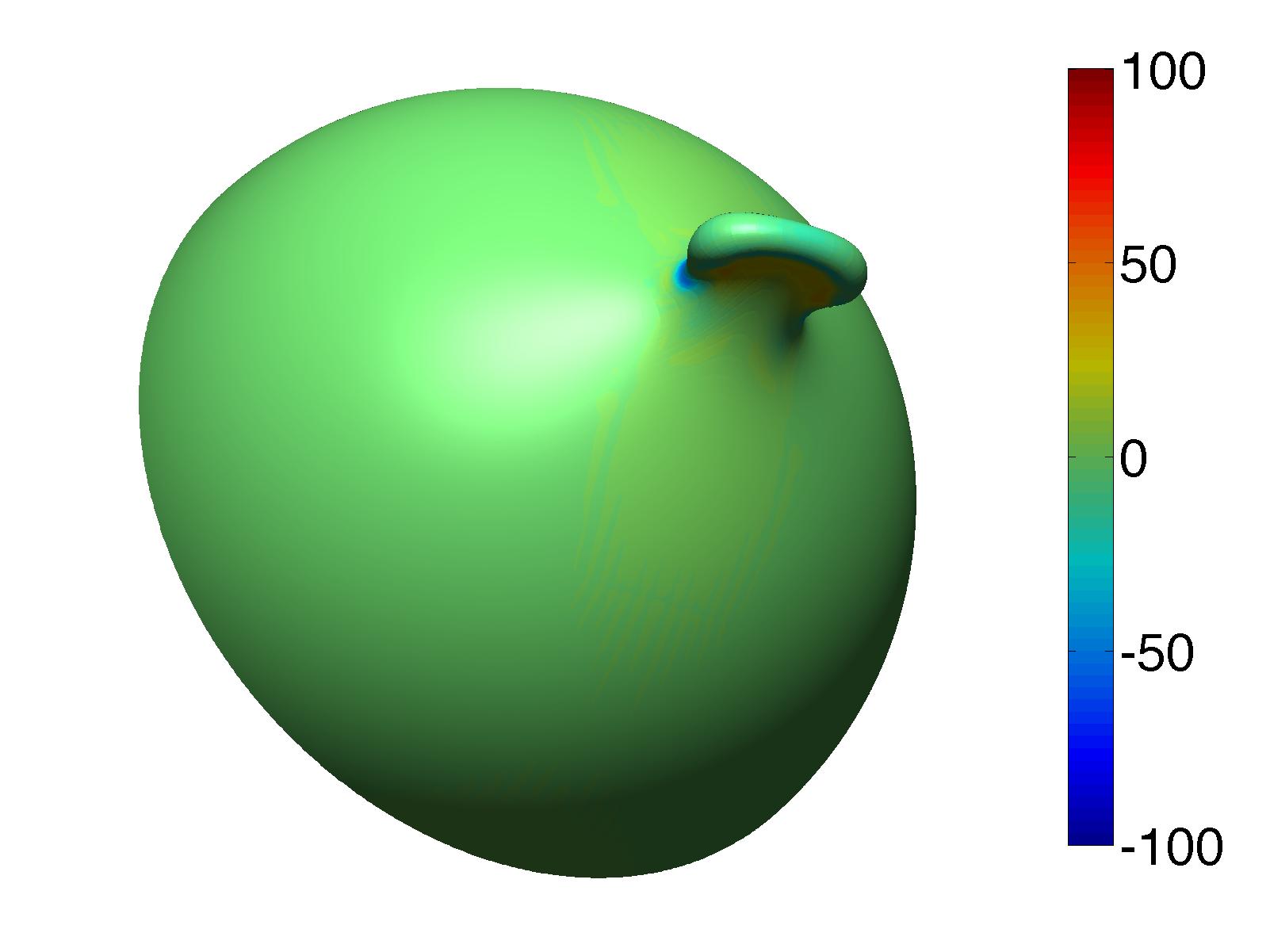}}
\put(-2.2,2.6){\includegraphics[height=31mm]{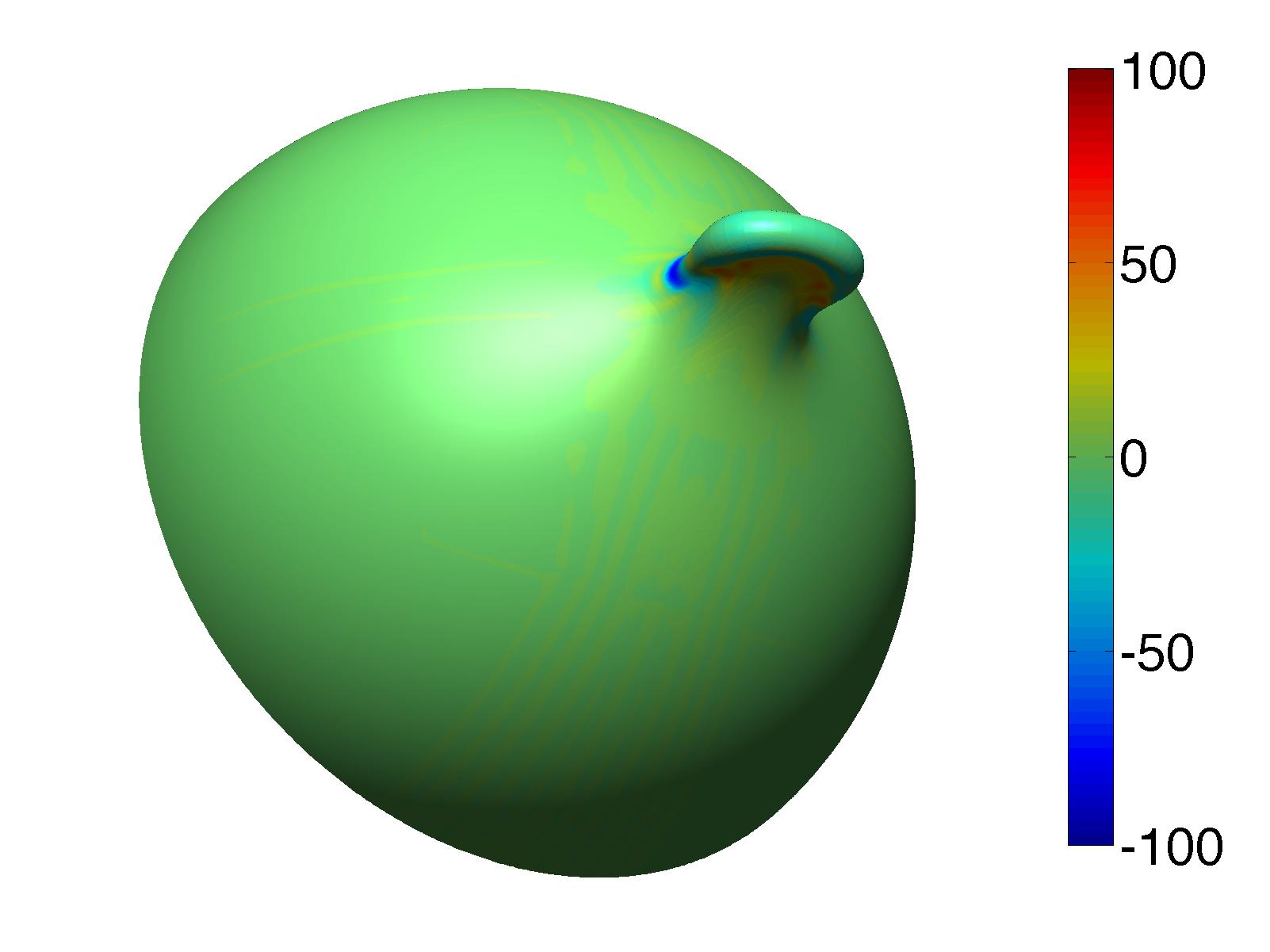}}
\put(0.9,2.6){\includegraphics[height=31mm]{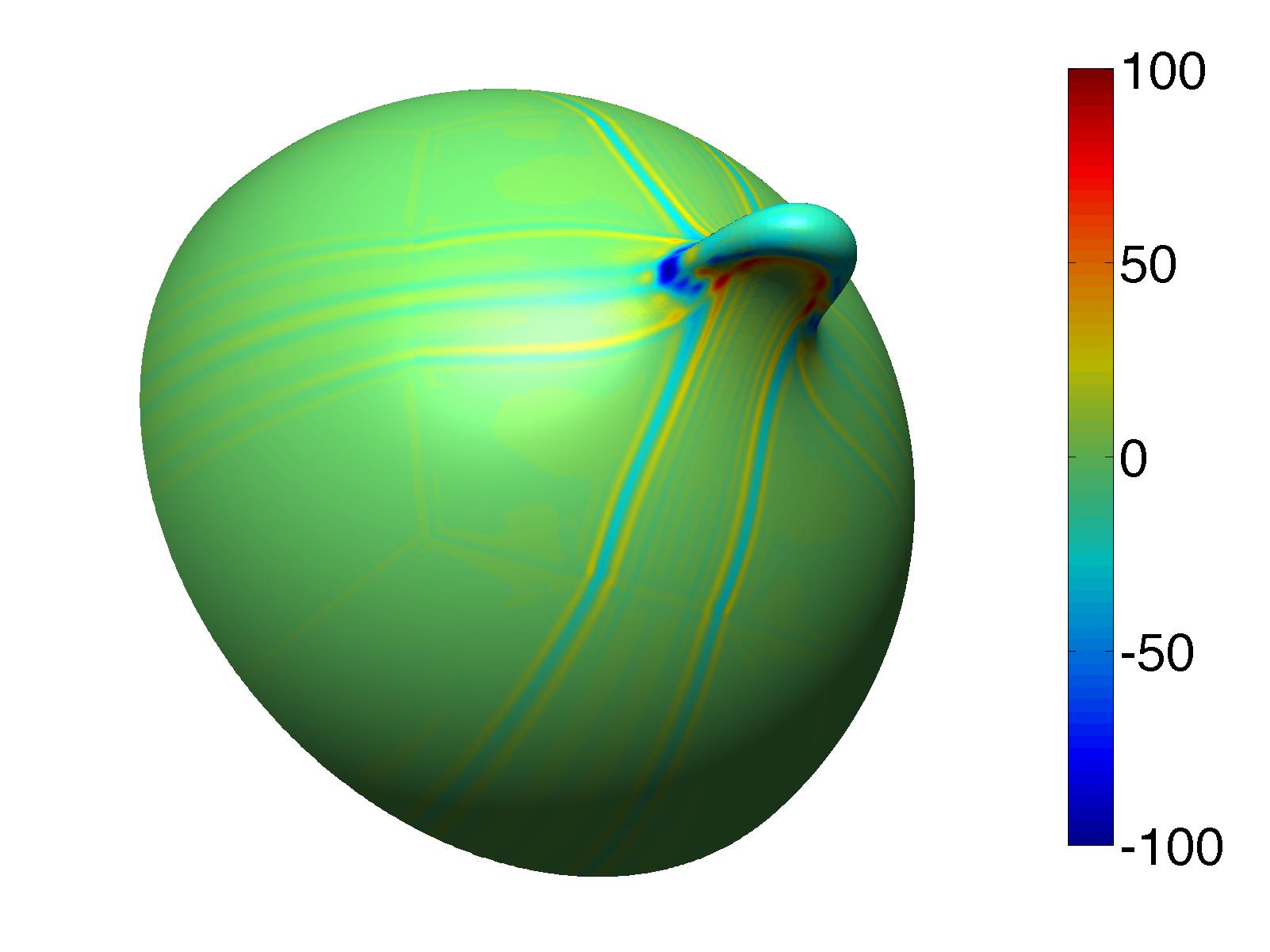}}
\put(4.0,2.6){\includegraphics[height=31mm]{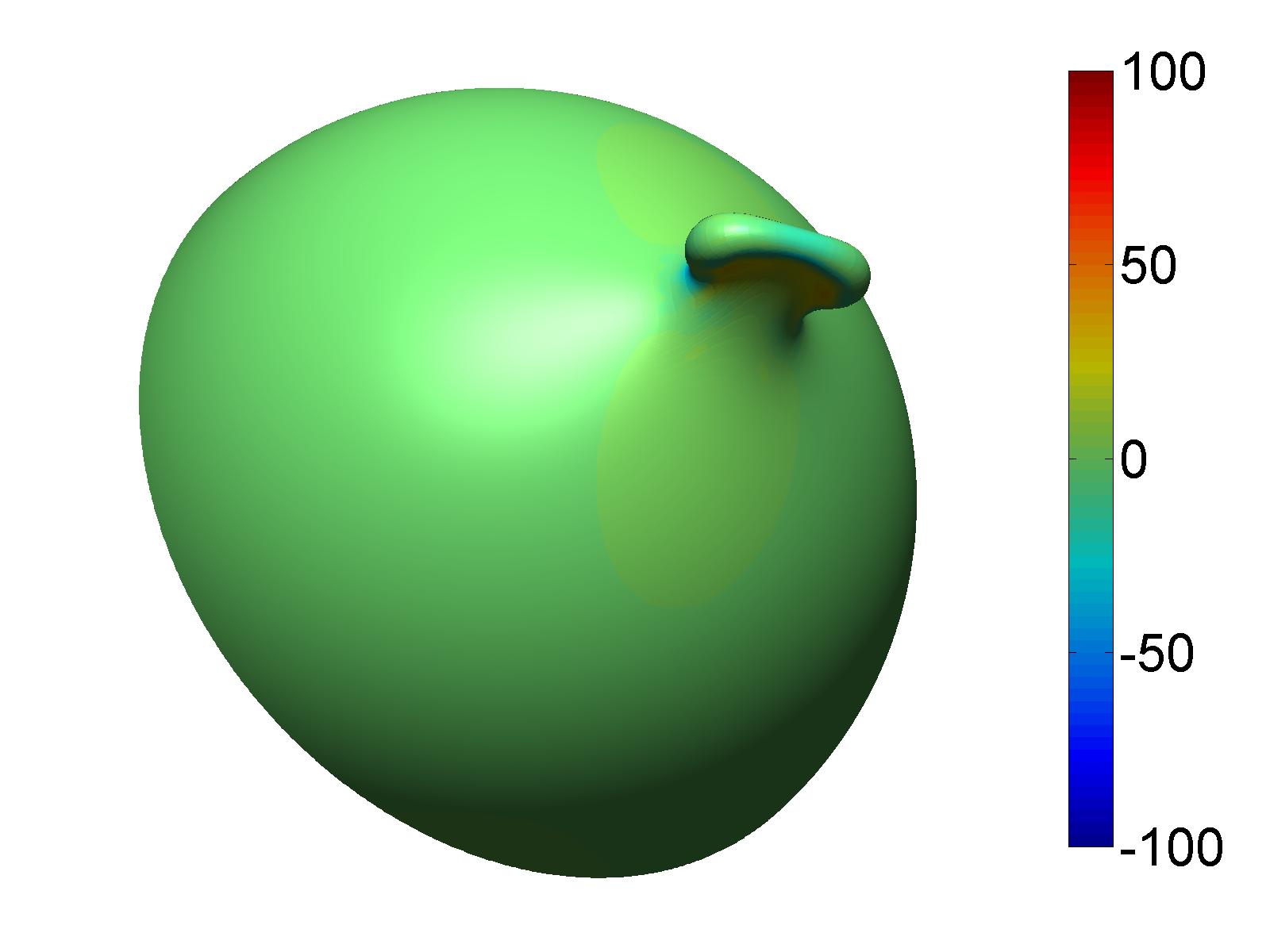}}
\end{picture}
\caption{Cell budding: influence of $K$ for the viscous case (case 5) for $\bar H_0=-15$ (top) and $\bar H_0=-25$ (bottom) considering $\bar K=10^3,\,10^4,\,10^5,\,10^6,\,\infty$ (left to right). The color shows $\bar\gamma$. 
}
\label{f:bud_Ky}
\end{center}
\end{figure}

\begin{figure}[h]
\begin{center} \unitlength1cm
\begin{picture}(0,2.7)
\put(-8.4,-.4){\includegraphics[height=31mm]{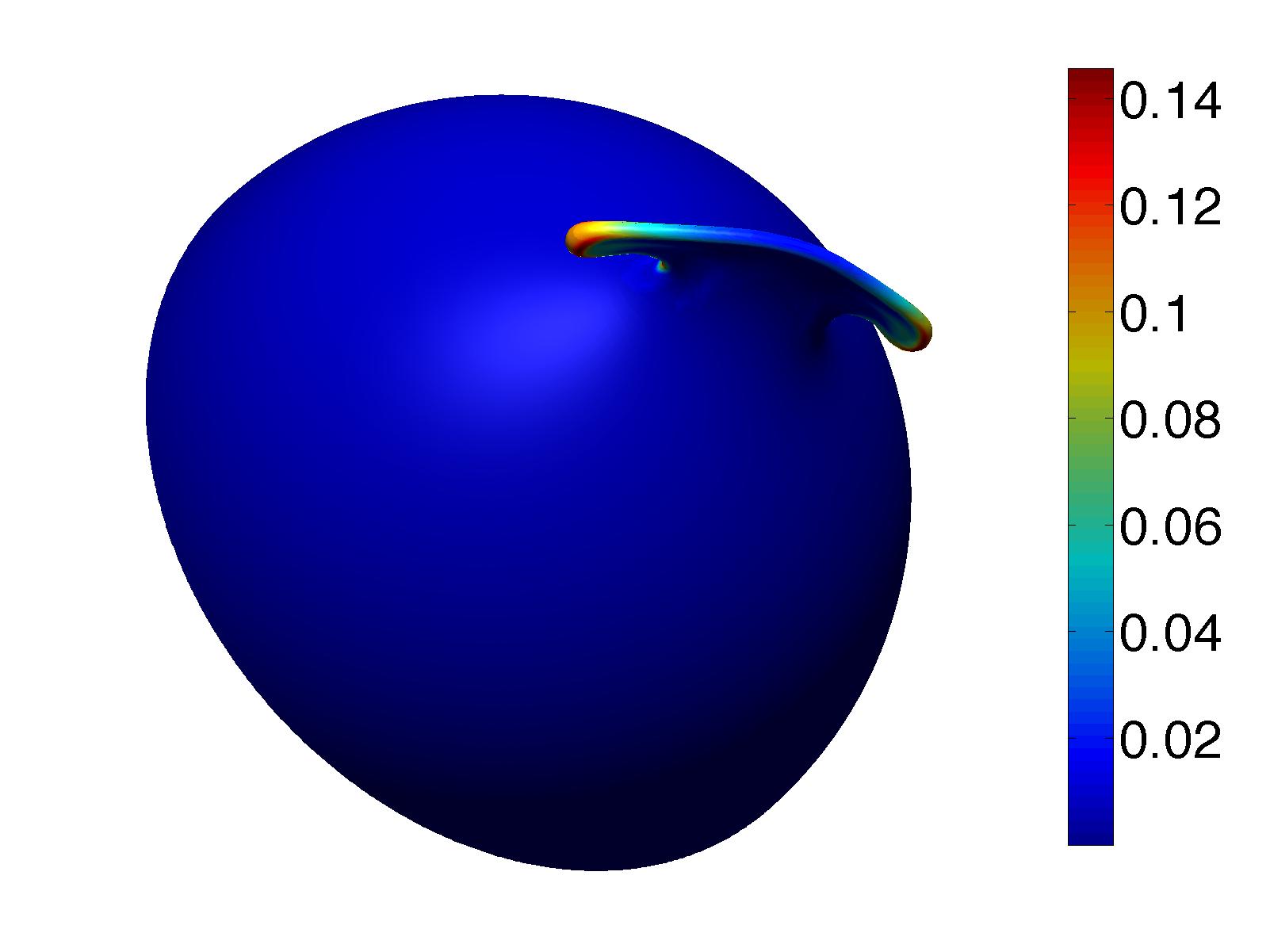}}
\put(-4.3,-.4){\includegraphics[height=31mm]{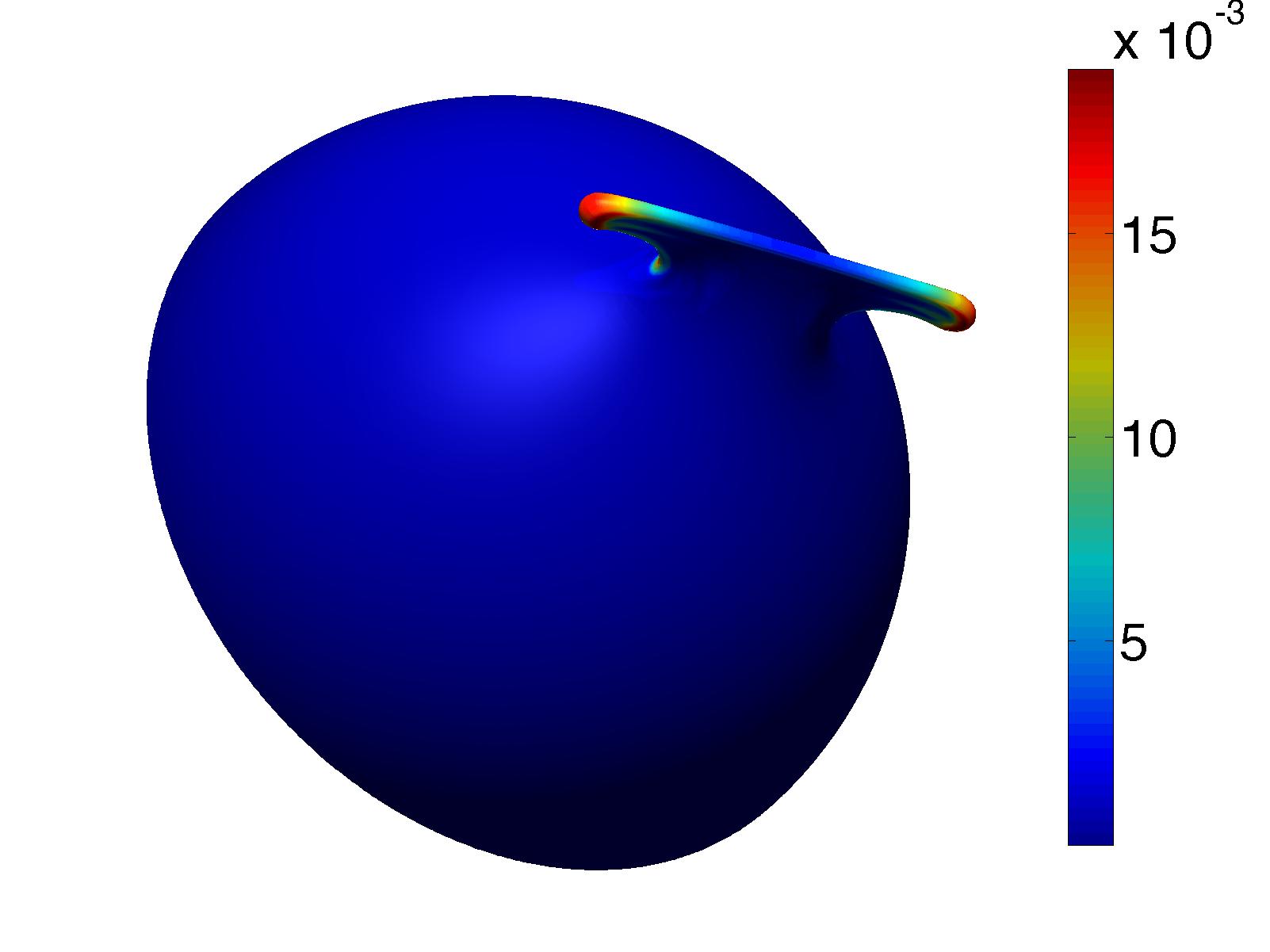}}
\put(-0.2,-.4){\includegraphics[height=31mm]{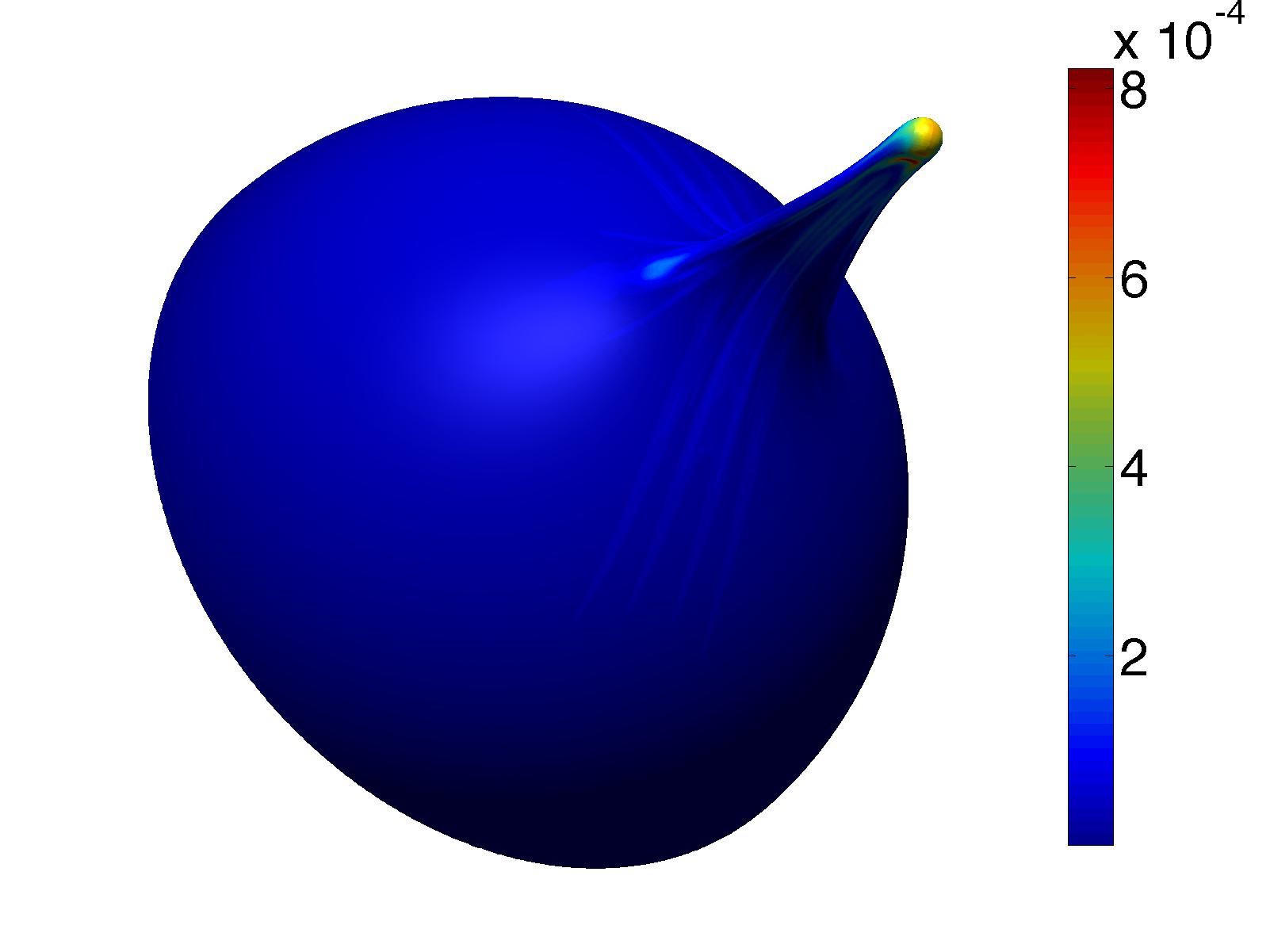}}
\put(3.9,-.4){\includegraphics[height=31mm]{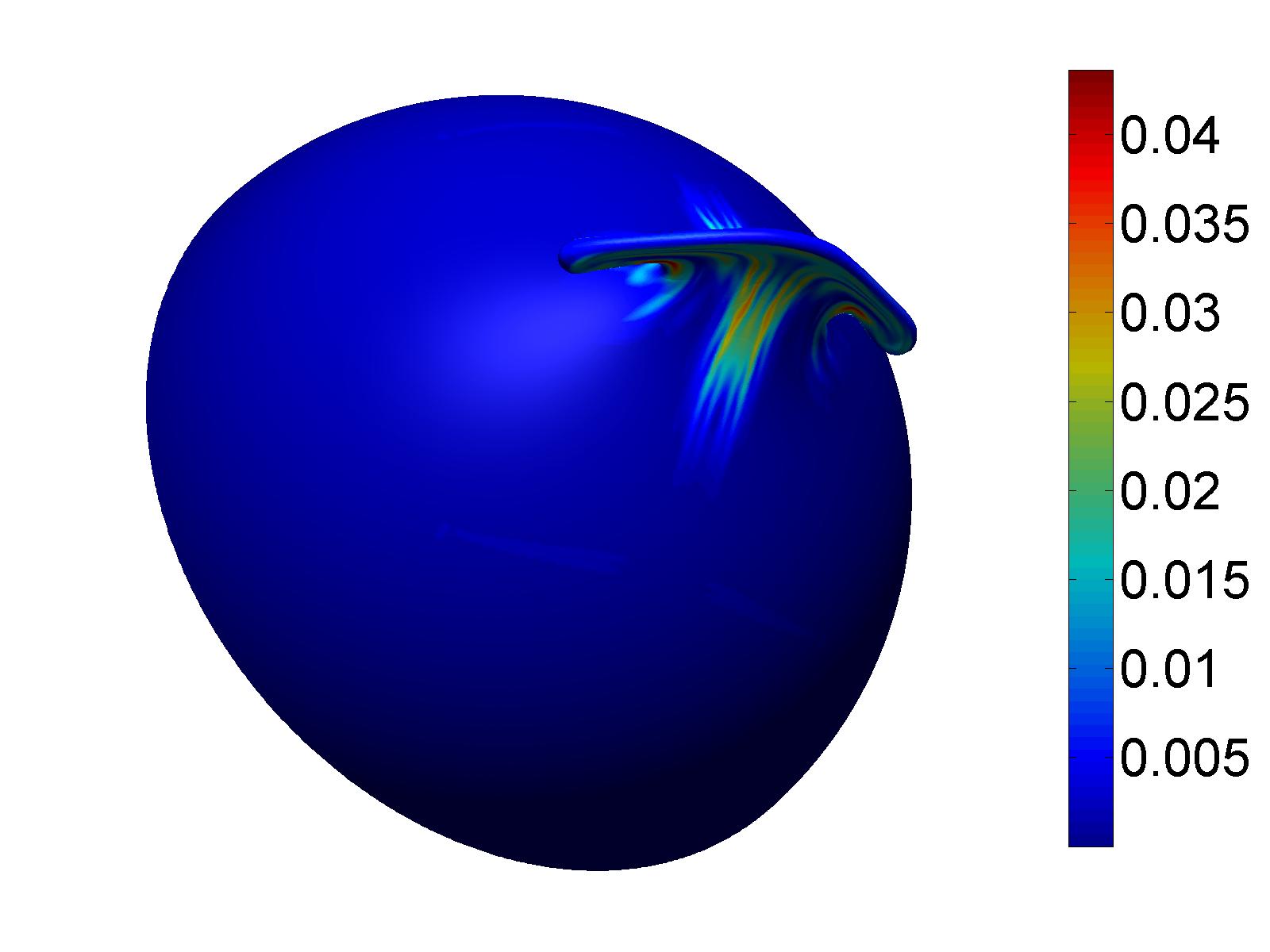}}
\end{picture}
\caption{Cell budding: influence of $K$ for case 5 for $\bar H_0=-25$ considering $\bar K=10^3,\,10^4,\,10^6,\,\infty$ (left to right). The color shows the relative error in $J$ given by $|J-1|$.}
\label{f:bud_Je}
\end{center}
\end{figure}

\begin{figure}[h]
\begin{center} \unitlength1cm
\begin{picture}(0,6)
\put(-8,0){\includegraphics[height=60mm]{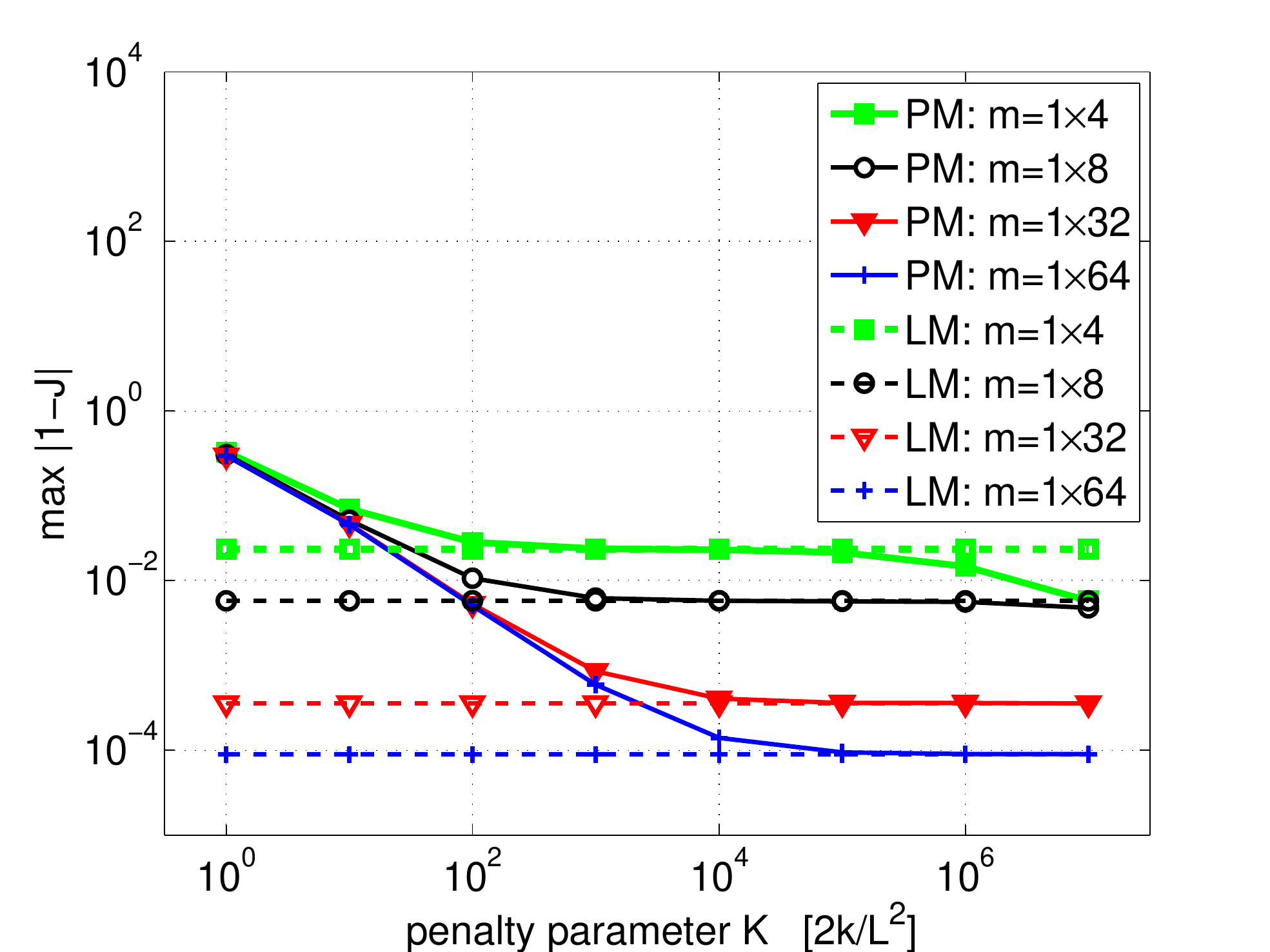}}
\put(0.4,0){\includegraphics[height=60mm]{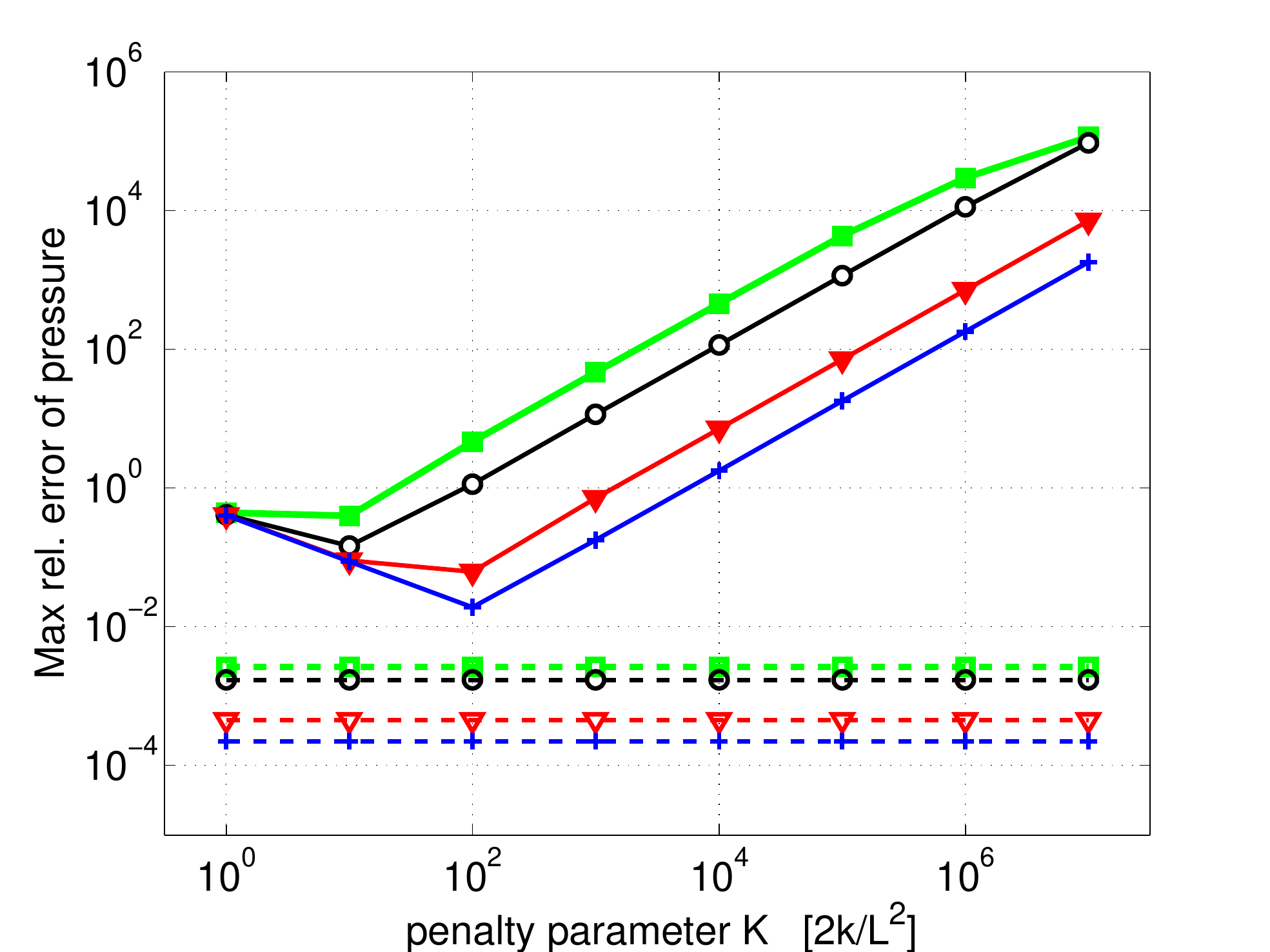}}
\put(-7.8,0){a.}
\put(0.6,0){b.}
\end{picture}
\caption{Pure bending: accuracy of the penalty regularization \eqref{e:W_c} compared to the Lagrange multiplier formulation \eqref{e:W_i}: a.~error in stretch $J$; b.~error in pressure $q$. As seen, $J$ can be more accurate, while $q$ does not converge with $K$. There is rather an optimum $K$ that minimizes the pressure error.}
\label{f:bud_err}
\end{center}
\end{figure}

For case 5, a strong dependency on $K$ appears, as 
Fig.~\ref{f:bud_Ky} shows.
Now, the deformation 
clearly does not converge with increasing $K$ in \eqref{e:W_c}.
Instead, the solution for $\bar K=10^3$ is closest to the solution from model \eqref{e:W_i}.
The figure also shows that oscillations appear in the solution from model \eqref{e:W_c} as $K$ increases. 
Essentially, the problem of model \eqref{e:W_c} is that even though $J$ converges to 1, as Fig.~\ref{f:bud_Je} shows, the pressure $q$ does not converge. 
This can be seen in Fig.~\ref{f:bud_err} for the pure bending problem of Sec.~\ref{s:strip}, which has an analytical solution for $q$.
As the figure shows, there is an optimal value of $K$, where the pressure error is minimal and beyond which it diverges.
This is different to examples in computational contact mechanics, where both kinematic and kinetic variables can converge with increasing penalty parameter\citep{spbf}.
So even though model \eqref{e:W_c} is simpler and more efficient, it has to be used with care, as the solution can become wrong.
If an implementation of \eqref{e:W_i} is not available, the mesh convergence of model \eqref{e:W_c} should be examined

\section{Conclusion}\label{s:concl}

This paper presents a general computational formulation for lipid bilayers based on general thin-shell kinematics, the Helfrich bending model, and 3D, LBB-conforming, $C^1$-continuous NURBS-based finite elements.
The rotational continuity of the formulation is ensured by using a rotational constraint across the patch boundaries in the FE mesh.
Two cases are considered in order to model the in-plane membrane response: area-compressibility and area-incompressibility, and for both suitable FE formulations are presented.
Since the formulation lacks shear stiffness, several shear stabilization schemes are proposed for quasi-static computations.
They are based on adding either numerical stiffness (class $\mA$), numerical viscosity (class $\ma$) or performing a projection of the solution onto the shell manifold (class $\mP$).
The numerical viscosity scheme can also be used to model physical viscosity as the example in Fig.~\ref{f:bud_Inu} illustrates, while the numerical stiffness scheme can also be used to  model physical stiffness as the example in Fig.~\ref{f:bud_Imu} illustrates.
It is further shown that the Helfrich bending model provides intrinsic shear stiffness as long as the surface curvature is non-zero.
Altogether, four different computational examples are considered in order to verify the formulation and to study its physical behavior.
The last example shows that the 3D budding behavior of lipid bilayers -- as described by the Helfrich model -- can become very complex, even though the model is purely mechanical and does not account for other effects, such as diffusion or temperature.
Computational challenges arise especially for area-incompressible, non-axisymmetric bilayer shapes.
For these a penalty regularization can give misleading results.

The paper shows that the proposed computational model is not only capable of capturing complex deformations, but is also a 
suitable tool to analyse and understand the stresses and energetics of lipid bilayers.
Still, more work is needed in order to further advance the computational modeling of lipid bilayers.
For example, the modeling of surface flow, protein diffusion, multicomponent lipids and thermal fluctuations would be useful extensions of the present work.
\\

\bigskip

{\Large{\bf Acknowledgements}}

The authors are grateful to the German Research Foundation (DFG) for supporting this research under grants  GSC 111 and SA1822/5-1.
They acknowledge support from the University of California at Berkeley and from Director, Office of Science, Office of Basic Energy Sciences, Chemical Sciences Division, of the U.S.~Department of Energy under contract No.~DE AC02-05CH11231.
Further, they thank the graduate student Yannick Omar for checking the theory.

\bigskip

\appendix

\section{Linearization of $\mf^e_\mathrm{into}$}\label{s:mfe}

Alternatively to Eq.~(\ref{e:fintio}.2), $\mf^e_\mathrm{into}$ can be also written as
\eqb{l}
\mf^e_\mathrm{into} = -\ds\int_{\Omega_0^e} \tau^{\alpha\beta}\,\mN^\mrT\,(\bn\otimes\bn)\,\mN_{,\alpha\beta}\,\dif A\,\mx_e~.
\label{e:fint}
\eqe
The linearization of $\mf^e_\mathrm{into}$ thus becomes
\eqb{l}
\mf^e_\mathrm{into}(\mx_e+\Delta\mx_e) \approx \mf^e_\mathrm{into}(\mx_e) + \Delta\mf^e_\mathrm{into}~,
\eqe
with
\eqb{lll} 
\Delta\mf^e_\mathrm{into} = 
\mi \ds\int_{\Omega_0^e} \tau^{\alpha\beta}\,\mN^\mrT\,(\bn\otimes\bn)\,\mN_{,\alpha\beta}\,\dif A\,\Delta\mx_e
- \ds\int_{\Omega_0^e} \Delta\tau^{\alpha\beta}\,b_{\alpha\beta}\,\mN^\mrT\,\bn\,\dif A \\[4mm]
\mi \ds\int_{\Omega_0^e} \tau^{\alpha\beta}\,\mN^\mrT\,(\bn\otimes\Delta\bn)\,\ba_{\alpha,\beta}\,\dif A
- \ds\int_{\Omega_0^e} \tau^{\alpha\beta}\,b_{\alpha\beta}\,\mN^\mrT\,\Delta\bn\,\dif A~.
\eqe
Inserting
\eqb{l}
\Delta\tau^{\alpha\beta} = c^{\alpha\beta\gamma\delta}\,\ba_\gamma\cdot\mN_{,\delta}\,\Delta\mx_e 
\eqe
\citep{membrane}
and
\eqb{l}
\Delta\bn = -\ba^\gamma\,(\bn\cdot\Delta\ba_\gamma)
\eqe
\citep{wriggers-contact}, we get
\eqb{l} 
\Delta\mf^e_\mathrm{into} = \mk_\mathrm{into}^e\,\Delta\mx_e~, 
\eqe
with the tangent matrix
\eqb{lll}
\mk^e_\mathrm{into} = 
\mi \ds\int_{\Omega_0^e} \tau^{\alpha\beta}\,\mN^\mrT\,(\bn\otimes\bn)\,\mN_{,\alpha\beta}\,\dif A
- \ds\int_{\Omega_0^e} c^{\alpha\beta\gamma\delta}\,b_{\alpha\beta}\,\mN^\mrT\,(\bn\otimes\ba_\gamma)\,\mN_{,\delta}\,\dif A \\[4mm]
\plus \ds\int_{\Omega_0^e} \tau^{\alpha\beta}\,\Gamma^\gamma_{\alpha\beta}\,\mN^\mrT\,(\bn\otimes\bn)\,\mN_{,\gamma}\,\dif A
+ \ds\int_{\Omega_0^e} \tau^{\alpha\beta}\,b_{\alpha\beta}\,\mN^\mrT\,(\ba^\gamma\otimes\bn)\,\mN_{,\gamma}\,\dif A~.
\eqe
Here, $\Gamma^\gamma_{\alpha\beta} := \ba^\gamma\cdot\ba_{\alpha,\beta}$ defines the Christoffel symbol of the second kind. 
The tensor components $c^{\alpha\beta\gamma\delta}$ are given in \citet{shelltheo} for various material models.
For an efficient implementation, the contractions appearing above should be worked out analytically. 

\section{Linearization of $\mg^e$}\label{s:mge}

The vector $\mg^e$ is independent of the Lagrange multiplier and thus only depends on $\mx_e$. The linearization of $\mg^e$ thus is
\eqb{l}
\mg^e(\mx_e+\Delta\mx_e) \approx \mg^e(\mx_e) + \Delta\mg^e~,
\eqe
with
\eqb{l} 
\Delta\mg^e = \ds\int_{\Omega^e_0}\mL^T\,\Delta g\,\dif A~, 
\eqe
and \citep{shelltheo}
\eqb{l}
\Delta g = \Delta J = \ds\frac{J}{2}a^{\alpha\beta}\,\Delta a_{\alpha\beta}~.
\eqe
Inserting the discretisation of $\Delta a_{\alpha\beta}$ \citep{solidshell}, and exploiting the symmetry of $a^{\alpha\beta}$, we get the approximation
\eqb{l} 
\Delta g \approx J\,\ba^\alpha\cdot\mN_{,\alpha}\,\Delta\mx_e~,
\eqe
such that
\eqb{l} 
\Delta\mg^e = \mk_g^e\,\Delta\mx_e~, 
\eqe
with the tangent matrix
\eqb{l}
\mk_g^e := \ds\int_{\Omega^e_0}\mL^T\,\ba^\alpha\cdot\mN_{,\alpha}\,J\,\dif A~.
\eqe

\bibliographystyle{apalike}
\bibliography{bibliography}

\end{document}